%% file: 00main.tex
\documentclass[twocolumn,secnumarabic,amssymb,notitlepage,floatfix,nobibnotes,aps,rmp]{revtex4-1}

\pdfoutput=1 

\usepackage[utf8]{inputenc} 
\usepackage{lineno}
\usepackage{graphicx}
\usepackage{amssymb}
\usepackage{amsmath}
\usepackage{color}
\usepackage[caption=false]{subfig}
\usepackage{booktabs}
\usepackage{tabularx}
\usepackage{rotating}
\usepackage{mathtools}
\usepackage{wasysym}
\usepackage{mathrsfs}
\usepackage{latexsym}
\usepackage{multirow}

\usepackage{url}
\usepackage[colorlinks, citecolor=blue,anchorcolor=blue,menucolor=blue, 
linkcolor=blue,filecolor=blue,runcolor=blue,urlcolor=blue,frenchlinks=true]{hyperref}

\newcommand{\nn}{\nonumber}
\newcommand{\as}{\alpha_s}
\newcommand{\bn}{{\bar n}}
\newcommand{\cO}{\mathcal O}
\newcommand{\qt}{\mathbf{q_T}}
\newcommand{\bt}{\mathbf{b_T}}
\newcommand{\kt}{\mathbf{k_T}}
\newcommand{\rt}{\mathbf{r_T}}
\newcommand{\lqcd}{\Lambda_\mathrm{QCD}}

\newcommand{\sla}[1]{\slash\!\!\!{#1}}

\def\bstar{{\color{blue}$\bigstar$}}
\def\bcirc{\raisebox{-1pt}{\scalebox{1.5}{\color{blue}$\circ$}}}
\def\rsquare{{\raisebox{-1pt}{\color{red}$\blacksquare$}}}

\DeclareUnicodeCharacter{2212}{-} %

\def\toprule{\hline}
\def\bottomrule{\hline}
\def\midrule{\hline}

\begin{document}
%% \linenumbers
%% \modulolinenumbers[5]
%% \pagewiselinenumbers
%% \setlength\linenumbersep{2pt}
%
\begin{widetext}
\null
\vspace{-0.5cm}
\thispagestyle{empty}  %
 \begin{flushright} %
 \small 
 Nikhef 2020-018; 
 MIT-CTP/5213;
MSUHEP-20-012; 
\\ 
 IFJPAN-IV-2020-3;
 SMU-HEP-20-03
 \end{flushright} %
\vspace{0.5cm}

\begin{center}
{\Large \bf Parton distributions and lattice QCD calculations: \\ toward 3D 
structure}
\vspace{0.5cm}
\\
\today
\\
\vspace{0.5cm}

{\normalsize %
Martha~Constantinou\rlap,$^{1,*}$         %
Aurore~Courtoy\rlap,$^{2}$               %
Markus A.~Ebert\rlap,$^{3}$               %
Michael~Engelhardt\rlap,$^{4,*}$            %
Tommaso Giani\rlap,$^{5}$               %
Tim~Hobbs\rlap,$^{6,7}$               %
Tie-Jiun Hou\rlap,$^{8}$               %
Aleksander~Kusina\rlap,$^{9}$               %
Krzysztof~Kutak\rlap,$^{9}$               %
Jian Liang\rlap,$^{10}$                  %
Huey-Wen~Lin\rlap,$^{11,12,*,\dagger}$       %
Keh-Fei Liu\rlap,$^{10}$                  %
Simonetta~Liuti\rlap,$^{13}$               %
C\'edric~Mezrag\rlap,$^{14}$               %
Pavel~Nadolsky\rlap,$^{6}$               %
Emanuele~R.~Nocera\rlap,$^{15,*}$        %
Fred~Olness$^{6,*}$                       %
Jian-Wei~Qiu\rlap,$^{7}$               %
Marco~Radici\rlap,$^{16}$               %
Anatoly~Radyushkin\rlap,$^{7,17}$               %
Abha Rajan\rlap,$^{18}$                  %
Ted~Rogers\rlap,$^{7,17}$               %
Juan~Rojo\rlap,$^{15,19}$               %
Gerrit~Schierholz\rlap,$^{20}$               %
C.-P.~Yuan\rlap,$^{11}$               %
Jian-Hui~Zhang\rlap,$^{21}$               %
Rui~Zhang\rlap,$^{11,12}$                    %
\quad (${}^*$editors)
}
\\
\vspace{0.5cm}
{\it \footnotesize
~$^{1}$Department of Physics, Temple University, 1925 N. 12th Street, Philadelphia, PA 19122-1801, USA\\
~$^{2}$Instituto de F\'isica, Universidad Nacional Aut\'onoma de M\'exico,Apartado Postal 20-364, 01000 Ciudad de M\'exico, Mexico\\
~$^{3}$Center for Theoretical Physics, Massachusetts Institute of Technology, Cambridge, MA 02139, USA\\
~$^{4}$Department of Physics, New Mexico State University, Las Cruces, NM 88003, USA\\
~$^{5}$  Higgs Centre for Theoretical Physics,  University of Edinburgh, Peter Guthrie Tait Road, Edinburgh EH9 3FD, United Kingdom\\
~$^{6}$Department of Physics, Southern Methodist University, Dallas, TX 75275, USA\\
~$^{7}$Thomas Jefferson National Accelerator Facility,  Newport News, VA 23606, USA\\
~$^{8}$Department of Physics, College of Sciences, Northeastern University, Shenyang 110819, China\\
~$^{9}$Institute of Nuclear Physics Polish Academy of Sciences, PL-31342 Krakow, Poland\\
~$^{10}$Department of Physics and Astronomy, University of Kentucky, Lexington, KY 40506, USA \\
~$^{11}$Department of Physics and Astronomy,  Michigan State University, East Lansing, MI 48824, USA\\
~$^{12}$Department of Computational Mathematics, Science and Engineering, Michigan State University, East Lansing, MI 48824, USA\\
~$^{13}$Physics Department, University of Virginia, 382 MCormick Road, Charlottesville, VA 22904, USA\\
~$^{14}$Irfu, CEA, Universit\'e Paris Saclay, F-91191, Gif-sur-Yvette, France \\
~$^{15}$Nikhef Theory Group, Science Park 105, 1098 XG Amsterdam, The Netherlands\\
~$^{16}$INFN - Sezione di Pavia, I27100 Pavia, Italy\\
~$^{17}$Department of Physics, Old Dominion University, Norfolk, VA 23529, USA\\
~$^{18}$Physics Department, Brookhaven National Laboratory, Upton, New York 11973, USA\\
~$^{19}$Department of Physics and Astronomy, VU University Amsterdam, Amsterdam, The Netherlands\\
~$^{20}$Deutsches Elektronen-Synchrotron DESY, 22603 Hamburg, Germany\\
~$^{21}$Center of Advanced Quantum Studies, Department of Physics, Beijing Normal University, Beijing 100875, China \\
}
\end{center}

\vspace{1.5cm}

The strong force which binds hadrons is described by the theory of
Quantum Chromodynamics (QCD).  Determining the character and
manifestations of QCD is one of the most important and
challenging outstanding issues necessary for a comprehensive
understanding of the structure of hadrons.
Within the context of the QCD parton picture, the Parton Distribution
Functions (PDFs) have been remarkably successful in describing a wide
variety of processes.  However, these PDFs have generally been
confined to the description of collinear partons within the
hadron.
New experiments and facilities provide the opportunity to
additionally explore the transverse structure of hadrons
which is described by Generalized Parton Distributions (GPDs)
and Transverse Momentum Dependent Parton Distribution Functions (TMD PDFs). 
In our previous report~\cite{Lin:2017snn}, we compared and contrasted the two
main  approaches  used to determine the collinear PDFs: the first based on
perturbative QCD factorization theorems, and the second
based on lattice QCD calculations.
In the present  report, we provide an update
of recent progress on the collinear PDFs, and also
expand the scope to encompass the generalized PDFs (GPDs and TMD PDFs).
We review the current state of the various calculations,
and consider what new data might be available in the near future. 
We also  examine how a shared effort can foster dialog
between the PDF and Lattice QCD communities,
and yield improvements for these generalized PDFs.

\vspace{1cm}
\begin{center}
\textit{The authors welcome comments and suggestions regarding the report content.}
\\
${}^{\dagger}$Corresponding author:   Huey-Wen Lin \quad  \texttt{hwlin@pa.msu.edu}
\end{center}
\newpage
\end{widetext}

\tableofcontents

\newpage
\setlength{\columnseprule}{0.4pt}  %

\input{01sec0intro}

\cleardoublepage
\input{02sec0pdf}
\cleardoublepage
\input{03sec0gpd}

\cleardoublepage
\input{04sec0tmd}
\newpage
\input{05sec0outlook}
\goodbreak
\newpage
\section*{Acknowledgments}

We are grateful to Autumn Hewitt, Nicole Kokx and Brenda Wenzlick, for their help in the organization of the workshop. The workshop was partly supported by Michigan State University and  US National Science Foundation via the grant under PHY 1653405 ``CAREER: Constraining Parton Distribution Functions for New-Physics Searches''
We thank Kostas~Orginos for his early-stage involvement in organizing the workshop, and Alberto~Accardi, Jorge~Benel, Nikhil~Karthik, Zhouyou~Fan,  Jian~Liang, Paul~Reimer,  Dave~Soper, Michael~Wagman, Zhite~Yu	
for their participation and discussions during the PDFLattice2019 workshop. 
We thank  Joseph Karpie and Gunnar Bali for providing numbers for Figs.\ref{fig:GlobalLatPDF} and \ref{fig:LatGFF}, respectively, and Phiala~Shanahan for remaking Fig.~\ref{fig:CS} for this paper.  
We thank H.~Moutarde,  Ingo~Schienbein, \hbox{P.~Sznajder}, and Yong~Zhao as well as members of the 
Lattice Hadron Physics Collaboration (LHPC) and  the Lattice TMD Collaboration
for useful discussion  during the preparation of this report.

We are pleased to acknowledge support from:
the U.S.~Department of Energy,
  grants 
  DE-SC0020405 [M.~Constantinou],
  DE-SC0011090 [M.~Ebert],
  DE-FG02-96ER40965 [M.~Engelhardt];
  DE-SC0010129 [T.~Hobbs, F.~Olness, P.~Nadolsky],
  DE-SC0016286 [S.~Liuti],
  DE-AC05-06OR23177 [J.W.~Qiu], 
  DE-FG02-97ER41028 [A.~Radyushkin], 
  DE-SC0012704 [A.~Rajan],
  DE-SC0018106 [T.~C.~Rogers],
\& the DOE Topical Collaboration on TMDs [M.~Engelhardt, J.W.~Qiu];
Jefferson Science Associates, LLC under  U.S.~DOE Contract DE-AC05-06OR23177
[J.W.~Qiu, A.~Radyushkin, T.C.~Rogers];
the US National Science Foundation grants
PHY-1714407 [M.~Constantinou], 
PHY~1653405 [H.~Lin];
the European Research Council (ERC)
under the European Union’s Horizon 2020 research and innovation program
  grant agreement N.~647981, 3DSPIN [M.~Radici],
   grant agreement N.~824093 [C.~Mezrag],
  and 
  the  Marie Sklodowska-Curie Action grant agreement N.~752748, ParDHonS [E.~Nocera];
the Universidad Nacional Autónoma de México grant DGAPA-PAPIIT IA101720 [A.~Courtoy];
the Scottish Funding Council grant H14027 [T.~Giani];
the Narodowe Centrum Nauki grant DEC-2017/27/B/ST2/01985 [K.~Kutak];
the Kosciuszko Foundation [A.~Kusina];
the  Research  Corporation  for  Science  Advancement
  through the Cottrell Scholar Award [H.~Lin];
 the Alexander von Humboldt Foundation through a Feodor Lynen Research Fellowship [M.~Ebert];
and
the National Natural Science Foundation of China (Grant No.~11975051)
and Fundamental Research Funds for the Central Universities [J.~Zhang];
the JLab EIC Center through an EIC Center Fellowship [T.~Hobbs];
and
the LDRD program of Brookhaven National Laboratory [A.~Rajan].

\cleardoublepage%
\bibliographystyle{apsrmp4-2}
\bibliography{main}

\end{document}

%% file: 01sec0intro.tex
\goodbreak

\section{Introduction and motivation}
\label{01sec:intro}
The Standard Model (SM) of particle physics has been remarkably successful 
in describing the nature of the elementary constituents and their interactions. 
Of the four fundamental forces, the strong interaction as described 
by the theory of quantum chromodynamics (QCD)
is particularly intriguing, as it exhibits a wealth of phenomena 
(confinement,  spontaneous chiral symmetry breaking, anomalies, instantons)
and binds the quarks and gluons that form the observable hadrons. 
However, it is precisely these properties of QCD that make it challenging 
to characterize the structure of the strongly interacting hadrons in the 
context of conventional perturbation theory. 

On the experimental front, there are a number of planned and proposed new facilities on the horizon
which can provide a wealth of new data to help us better characterize the strong interaction;
this will improve our knowledge of 
the fundamental forces and constituents of nature and lead to new discoveries. 
To fully utilize this new experimental information, 
it is imperative that our theoretical tools advance to keep pace; 
this is the focus of our report.

The QCD-based parton picture has provided a reliable computational framework 
for describing the physics of hadrons. However, an essential 
ingredient for these calculations are the parton distribution functions (PDFs),
which encode the properties of quarks and gluons inside the hadrons.
Ultimately, the accuracy of any calculation based on the QCD parton model
depends on the accuracy of the PDFs; 
thus, the PDFs are crucial for current LHC measurements
as well as future HL-LHC, LHeC, and EIC investigations. 

Unfortunately  the PDFs are, at present, often the element that limits the precision.
For example, the focus of the upcoming LHC runs
will be to improve the event statistics with increased luminosity;
thus, the path to future discoveries will be delineated with
high-precision comparisons as we search for discrepancies between the
data and the Standard-Model predictions.
Hence, our ability to
fully characterize the Higgs boson and constrain any new-physics signatures
ultimately comes down to how accurately we can determine the underlying
PDFs.

We also note here that the PDFs are imperative for some astrophysical studies, including
high-energy cosmic rays which are used to compute atmospheric neutrino backgrounds. 

At present, the computation of PDFs from first principles
is complex, computationally intensive, 
and it requires diligence and ingenuity to minimize uncertainties.

There are two main communities with complementary 
approaches to determining the PDFs:
1)~determination by global analysis of experimental measurements,
and
2)~direct computation using lattice QCD.
The global QCD analyses use large experimental data sets in the
context of the QCD parton model (along with factorization theorems)
to extract process-independent PDFs. 
The lattice-QCD approach directly computes the QCD path integral on a
discretized finite-volume Euclidean space-time grid
to extract moments of PDFs, quasi-PDFs, and other quantities. 
A general overview of these approaches was presented in 
the 2017 PDFLattice report~\cite{Lin:2017snn};
this work grew out of a 2017 workshop\footnote{%
Parton Distributions and Lattice Calculations in the LHC era (\textbf{PDFLattice2017})
22--24 March 2017, Oxford University, UK. 
\hbox{\url{http://www.physics.ox.ac.uk/confs/PDFlattice2017}}
}   %
which
brought the two communities together to address common questions and
followed up on the very fruitful discussions and interactions that took
place during the workshop.
That first document served as a common reference
to standardize notation and conventions to facilitate communication
between these communities.

This current report grew out of a follow-up 2019 workshop;\footnote{%
Parton Distributions and Lattice Calculations (\textbf{PDFLattice2019})
25--27 September 2019, Michigan State University (MSU) Kellogg Biological Station, USA.
\hbox{\url{https://indico.cern.ch/event/804857/}}
}   %
it updates and extends the scope of the previous studies in a number of
dimensions.

In particular, the prospect of future LHeC and EIC facilities enables
us to contemplate expanding beyond the collinear PDF framework and considering also generalized parton distributions (GPDs) and
transverse-momentum--dependent (TMD) PDFs.
Together with the collinear PDFs, we can obtain a complete three-dimensional 
description of the hadron structure.

The conventional PDFs $f(x)$ describe the probability of finding a parton with momentum fraction $x$ inside a hadron moving in the infinite momentum frame; in this case, the parton is \textit{assumed} to be moving collinear to the hadron with zero transverse momentum. 
If we desire additional information about the parton, we could ask about the transverse momentum $\kt $ 
and the impact parameter $\bt $  of the parton relative to the parent hadron. 
Both  $\kt $ and $\bt $ are 2-dimensional vectors measured relative to the axis defined by the hadron momentum. 
Combining this information together yields the 5-dimensional Wigner distribution $W(x,\kt ,\bt )$ which, in a sense,
provides complete information on the parton. 
If we integrate the Wigner distribution over the transverse momentum, we obtain a Generalized Parton Distribution (GPD)
$f_{\textsc{GPD}}(x,\bt )$, 
while if we integrate  over the impact parameter we obtain a Transverse Momentum Dependent (TMD) PDF
$f_{\textsc{TMD}}(x,\kt )$.
Thus, by expanding the scope of our investigations beyond collinear PDFs to include GPDs and TMD PDFs
we can extract more detailed information on the partons which comprise the hadron.

Not only do the  GPDs describe the spatial distributions of quarks and gluons in the plane transverse 
to the hadron momentum, but they also provide a mechanism to study the rich spin structure of the hadrons. %
Together with the collinear information encoded in the $x$ momentum fraction variable, 
the GPDs  provide a three-dimensional (3D) tomographic image
of the structure of hadrons in terms of QCD’s quarks and gluons.

Likewise, the TMD PDFs   encode not only the transverse momentum distribution of the hadronic components, but 
can also provide the polarization degrees of freedom. The information about transverse degrees of freedom  is essential in the context of QCD factorization theorems for multiscale, noninclusive collider observables. %
The transverse momentum degrees of freedom of partons can be addressed within so called transverse-momentum--dependent %
factorization, where essentially the transverse momentum is much smaller than the hard scale of the final state. A typical process where one applies this factorization is Drell-Yan production.
Another approach (valid at low $x$) is addressed within the color-glass condensate (CGC) model, %
which is an effective model formulated within QCD. 
The broad objective of our work is to advance tools that can 
perform detailed comparisons
between data and QCD theory, as well as combine information from the two sides, to gain an enhanced understanding of the various PDFs.
This work is ongoing and forms the basis for next-generation PDF input to facilitate precision measurements and new discoveries at the
future hadron facilities, as we
validate our understanding of the SM and search for
deviations which might signal evidence of new physics.

The outline of this paper is as follows.
In Section~2 we review the essential elements of the global QCD analysis and lattice-QCD methods 
for collinear PDFs, and highlight some of the recent developments.
In Section~3 we introduce the GPDs
and discuss areas where lattice QCD and machine learning (ML) can contribute.
In Section~4 we introduce TMD PDFs, including the new lattice quasi-TMD PDFs, and discuss recent advances. 
In Section~5 we discuss future interactions between the global-analysis
and lattice-QCD communities and offer some conclusions.

This document presents work that appears in the literature until June 1, 2020 (published, or on the arXiv). The discussion is extended to conference proceedings for recent work that has not been published elsewhere. To keep this document at a reasonable length, we present selected aspects of each publication discussed in the main text and we encourage the interested reader to consult the referred work.

%% file: 02sec0pdf.tex
\goodbreak
\section{Parton Distribution Functions}
\label{02sec:pdf}

In the framework of leading-twist collinear factorization~\cite{Collins:1989gx}, PDFs are the momentum densities of partons carrying a fraction $x$ of the momentum of their parent nucleon. Depending on the polarization of the parton with respect to that of the nucleon, three collinear PDF species can be defined: the unpolarized PDF, $f$, for an unpolarized parton in an unpolarized nucleon; the helicity PDF, $\Delta f$, for the net amount of partons polarized along or opposite a longitudinally polarized nucleon; and the transversity PDF, $\delta f=h_1^f$, for the net amount of quarks polarized along or opposite a transversely polarized nucleon. In canonical field theory, PDFs are equivalently defined in terms of matrix elements of bi-local operators, {\it e.g.} for quarks
\begin{align}
  f(x) 
  & =
  \frac{1}{4\pi}
  \int dy^- e^{-ixP^+y^-}
  \langle P,S |
  \bar{\psi}_f\gamma^+\mathcal{W}\psi_f
  |P,S\rangle \,,
  \nonumber
  \\
  \Delta f(x) 
  & = \frac{1}{4\pi}
  \int dy^-e^{-ixP^+y^-}
  \langle P, S|
  \bar{\psi}_f\gamma^+\gamma^5\mathcal{W}\psi_f|
  P, S\rangle \,,
  \nonumber
  \\
  h_1^f(x)
  & =
  \frac{1}{4\pi}
  \int dy^- e^{-ixP^+y^-}
  \langle P,S |
  \bar{\psi}_f
  i\sigma^{1+}\gamma^5 \mathcal{W}\psi_f
  |P,S\rangle \,,
  \label{eq:PDF_def}
\end{align}
where $\bar\psi_f=\bar\psi_f(0,0,\mathbf{0}_\perp)$ and $\psi_f=\psi_f(0,y^-,\mathbf{0}_\perp)$ are the (quark) fields, $y^\pm=\frac{1}{2}(y^0 \pm y^3)$ are the spacetime coordinates along the lightcone direction, $\gamma$ are the Dirac matrices, $\sigma$ are the Pauli matrices, $\mathcal{W}$ is the Wilson line ensuring the Gauge invariance of the operator, and $P$ and $S$ are the momentum and spin of the proton. All four-vectors are expressed using light-cone coordinates. 

Expressions similar to those in Eq.~\eqref{eq:PDF_def} also hold for upolarized and helicity gluon PDFs. They
all include renormalization of all fields and coupling in the $\overline{\rm MS}$ scheme (see~\cite{Kovarik:2019xvh} for details), a fact that makes PDF depend on the (renormalization) scale $\mu$. This dependence is purely perturbative, and can be computed by means of DGLAP evolution equations~\cite{Dokshitzer:1977sg,Altarelli:1977zs,Gribov:1972ri}. Nuclear modifications, occurring when a nucleon is bound in a nucleus, do not alter these definitions: nuclear and nucleon PDFs are defined by the same leading-twist operators entering Eq.~\eqref{eq:PDF_def} (though acting on different states), therefore it is customary to assume that nuclear modifications can be absorbed into PDFs.

In this Section, we review recent progress in the determination of unpolarized, helicity and transversity PDFs from the two methods delineated in the Introduction: global QCD analyses of experimental data, on the one hand; and lattice QCD, on the other hand. We focus on updates occurred since the publication of the 2017 PDFLattice report~\cite{Lin:2017snn}. We then investigate possible intersections between the two methods, specifically three aspects that we deem of particular relevance: moments of PDFs, the use of lattice data in global fits, and the unpolarized and helicity strange PDF. 

\input{02sec1global}

\input{02sec3latticePDFs}

\input{02sec4intersection}

%% file: 02sec1global.tex
\goodbreak
\subsection{Global PDFs Updates}
\label{02sec1}

A global QCD analysis of PDFs consists in modeling PDFs by means of a suitable parametrization, which is then optimized by comparing PDF-dependent predictions for one or more physical scattering processes involving initial-state nucleons to (multiple sets of) their actual measurements. In this sense, a PDF determination can be considered a nonlinear-regression problem, in which one has to learn the relative weights of a set of functions from data. Key to this procedure are QCD factorization theorems~\cite{Collins:1989gx}, according to which the cross section for a class of sufficiently inclusive hadronic processes can be determined by folding the PDFs with perturbatively computable partonic cross sections.

\begin{figure*}[!t]
\centering
\ \ \ \ \ \ \ \ \ \ 
Unpolarized 
\ \ \ \ \ \ \ \ \ \ \ \ \ \ \ \ \ \ \ \ \ \ \ \ \ \ \ \ \ \ \ \ \ \ \ \ \ \ 
Helicity  
\ \ \ \ \ \ \ \ \ \ \ \ \ \ \ \ \ \ \ \ \ \ \ \ \ \ \ \ \ \ \ \ \ \ \ \ \ \
Transversity\\
\includegraphics[width=\textwidth]{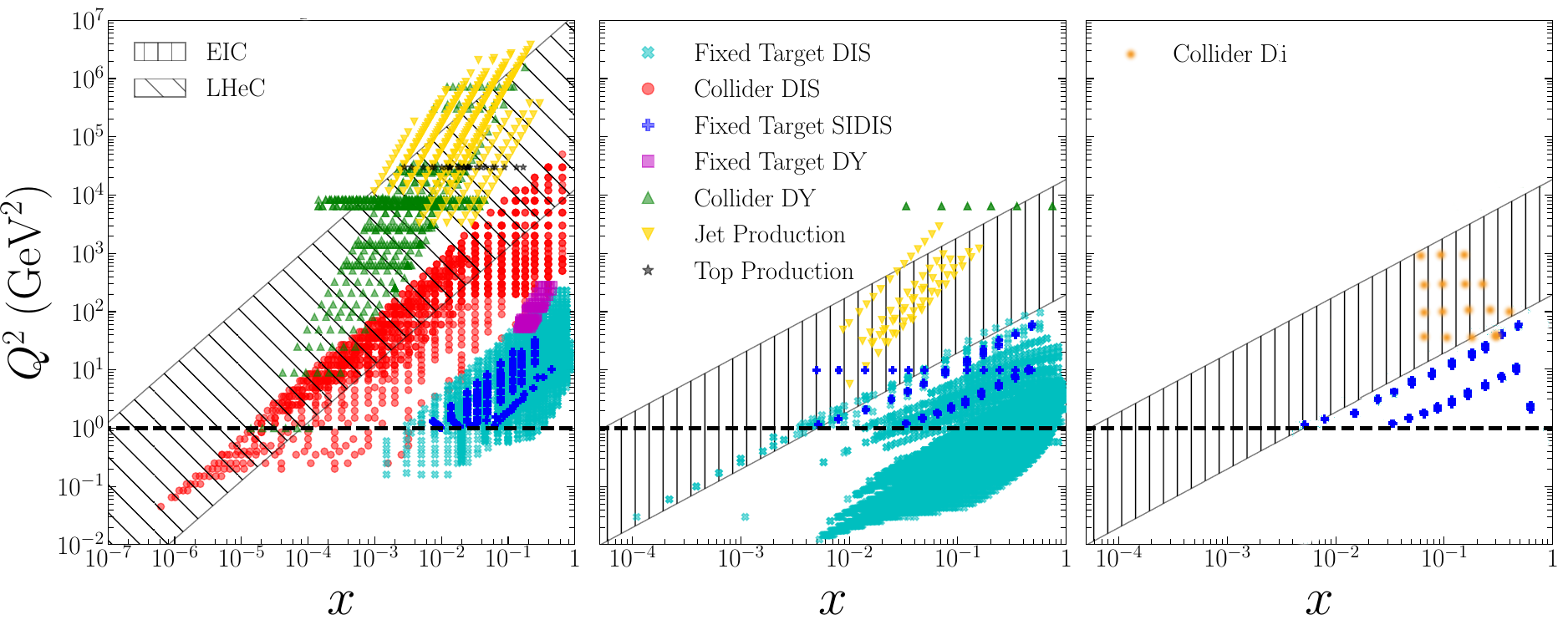}\\
\caption{The kinematic coverage in the $(x,Q^2)$ plane of the hadronic cross-section data for the processes commonly included in global QCD analyses of collinear unpolarized, helicity, and transversity PDFs. The extended kinematic ranges attained by the LHeC and the EIC are also displayed. See Fig.~1 of Ref.~\cite{Ethier:2020way} for unpolarized nuclear PDFs.}
\label{fig:kin_cov}
\end{figure*}

The accuracy and precision of a PDF determination largely depend on the data set and on the theoretical and statistical sophistication of the analysis. The input data sets are remarkably different across the three PDF species defined in Eq.~\eqref{eq:PDF_def}; while there are thousands of measurements available to determine the unpolarized PDFs, there are typically only hundreds of them for the helicity, and tens for the transversity. The kinematic coverage and the variety of hadronic processes decreases accordingly: unpolarized and helicity PDFs benefit, although to a different extent, from measurements in inclusive and semi-inclusive deep-inelastic scattering (DIS and SIDIS), Drell-Yan (DY), and a variety of proton-proton (pp) scattering processes (including gauge-boson, jet, and heavy-particle production). The transversity PDF, instead, is limited by its nature to processes coupled to a chiral-odd final state; in a purely collinear framework, that is di-hadron production in SIDIS or in pp collisions (Di). This state of affairs is summarized in Fig.~\ref{fig:kin_cov}, where we display the kinematic coverage, in the $(x, Q^2)$-plane ($Q=\mu$ is the characteristic scale of the data), of the hadronic cross section data for the processes commonly included in modern QCD analyses of unpolarized, helicity and transversity PDFs. The dashed line is at $Q=1$~GeV. Data below this value are usually not included in QCD analyses of PDFs because QCD becomes largely non-perturbative, due to the growth of the strong coupling. The extended kinematic ranges attained by a future Large Hadron-electron Collider (LHeC)~\cite{AbelleiraFernandez:2012cc} or a polarized Electron-Ion Collider (EIC)~\cite{Accardi:2012qut} are also displayed. A corresponding plot for unpolarized nuclear PDFs can be found in Fig.~1 of Ref.~\cite{Ethier:2020way}.

The remainder of this section will present a summary of recent developments in the determination of unpolarized, helicity and transversity PDFs, and will outline their respective features. We refer the interested reader to Ref.~\cite{Ethier:2020way} for a recent, more extended review of the subject.

\subsubsection{Unpolarized PDFs}

The continued excellent performance of the CERN Large Hadron Collider (LHC)~\cite{Evans:2008zzb} has brought the determination of unpolarized PDFs in a precision era. The recent completion of Run~II has accumulated data from an integrated luminosity of approximately 150~fb$^{-1}$, a fact that shrank the statistical uncertainty of the measurements to unprecedented small values, typically 1\% or less. This state of affairs challenges PDFs as a tool to perform precision tests of the SM and searches of new physics; PDFs must become comparably precise. Next-to-next-to-leading order (NNLO) accuracy has become standard, and large efforts were also devoted to update PDF sets with an increasing amount of LHC measurements, to develop statistical tools to investigate the impact and/or inconsistencies of the data, and to represent residual theoretical uncertainties ({\it e.g.}~beyond NNLO) in the PDFs. On the other hand, PDFs remain key to understand the nonperturbative structure of the proton, for instance to investigate the role of higher-twist terms in the operator product expansion, or the shape of the ratio of $d$ to $u$ flavor at large $x$. In the following, we discuss these aspects by summarizing the progress of the leading PDF sets.

\begin{figure}[!tb]
\centering
\includegraphics[width=0.48\textwidth]{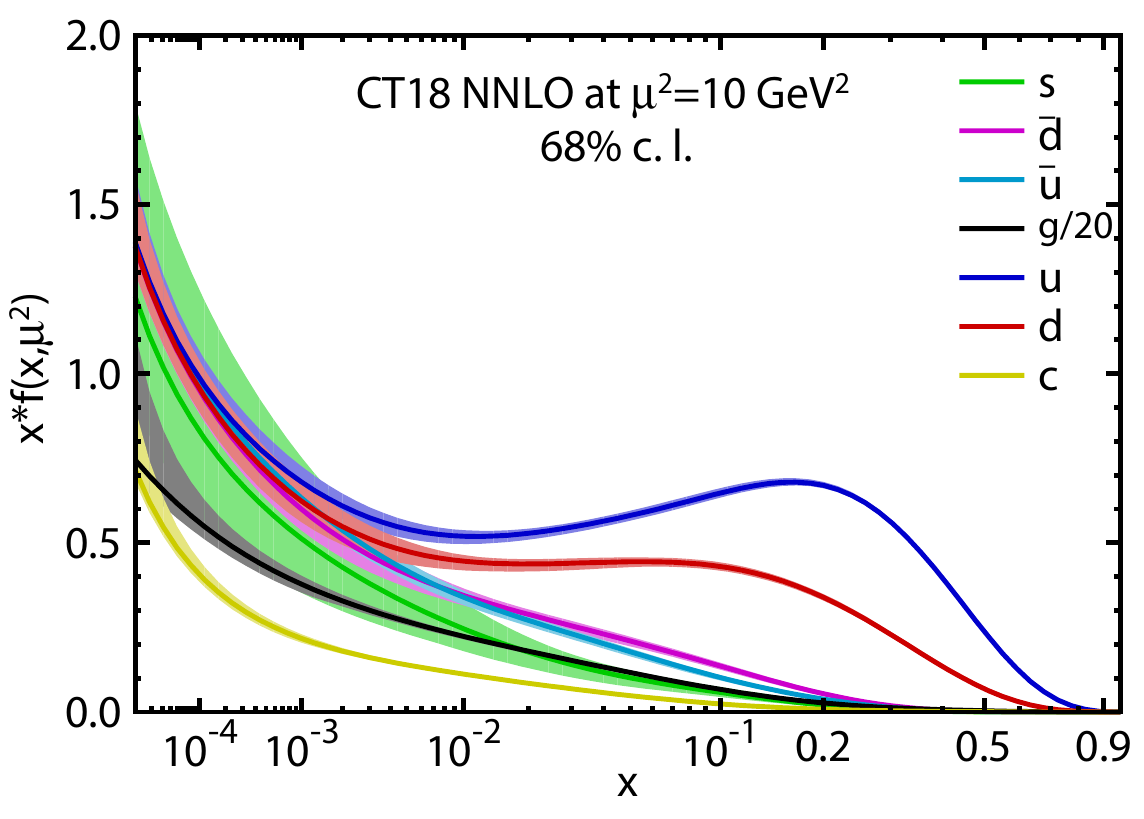}
\caption{The {\sc CT18} PDFs at $\mu^2=10$~GeV$^2$ for the $xu$, $x\bar{u}$, $xd$, $x\bar{d}$, $xs=x\bar{s}$, and $xg$ PDFs. Error bands correspond to the 68\% confidence level. Figure from~\cite{Kovarik:2019xvh}.}
\label{fig:ct18}
\end{figure}

The CTEQ-TEA collaboration has recently released their new {\sc CT18} PDF set~\cite{Hou:2019efy}, accurate up to NNLO. More than 700 data points from 12 new LHC data sets were included in the new fits. Among them, the most important data sets that drove changes in the quark and antiquark PDFs are the LHCb $W$- and $Z$-boson data at 7 and 8~TeV. Very mild changes in the fitted PDFs were observed with the addition of the ATLAS 8-TeV and CMS 7- and 8-TeV $W$- and $Z$-boson data. Changes in the gluon PDF were instead driven by the inclusion of CMS and ATLAS jet data, and of ATLAS 8-TeV $Z$-boson transverse-momentum distributions, which both made the central value and the uncertainty of the gluon decrease in the region $0.1<x<0.4$. The subsequent inclusion of top-pair--production data from ATLAS and CMS, in particular of single- and double-differential distributions, was demonstrated to have little impact in the {\sc CT18} fit. Finally, the central value of the sea-quark ratio $\left(s+\bar{s}\right)/\left(\bar{u} +\bar{d}\right)$ increased in the small-$x$ region, due predominantly to the inclusion of LHCb data. A methodology to study the sensitivity and the impact of specific data sets to the PDFs without the need of performing a fit was also developed~\cite{Wang:2018heo,Schmidt:2018hvu,Hou:2019gfw}. A variant of the baseline {\sc CT18} fit, called {\sc CT18Z}, was determined with the additional inclusion of the ATLAS 7-TeV $W^\pm/Z$ rapidity distributions and by taking the factorization and renormalization scales in the DIS process to be $x$-dependent. The fit represents the maximal deviation from the nominal {\sc CT18} fit, notably in the gluon and strange-quark PDFs. The resulting PDFs for the {\sc CT18} set, which are representative of the qualitative status of current unpolarized PDFs, are shown in Fig.~\ref{fig:ct18}. 

The latest general-purpose PDF determination from the MMHT collaboration is {\sc MMHT14}~\cite{Harland-Lang:2014zoa}, which was later extended to include HERA~I--II legacy measurements~\cite{Harland-Lang:2016yfn}, jet-production measurements~\cite{Harland-Lang:2017ytb}, and differential measurements in top-pair production~\cite{Bailey:2019yze} from the LHC. These intermediate updates demonstrated that experimental correlations across systematic uncertainties have been improperly estimated for some of the ATLAS jet and differential top data sets. The features of a new preliminary general-purpose PDF set were presented in Ref.~\cite{Thorne:2019mpt}, which included new LHC data sets, notably the particularly precise 7-TeV ATLAS $W$- and $Z$-boson measurements, which increase the ratio of strange to non-strange light sea quarks at low $x$, whilst still allowing for a positive light-sea-quark asymmetry, albeit with a maximum at slightly lower $x$. The {\sc MMHT} fit has also been updated with an improved and extended parametrization based on Chebyshev polynomials.  

The NNPDF collaboration released their latest general-purpose PDF set in Ref.~\cite{Ball:2017nwa}. This was later extended to include direct photon~\cite{Campbell:2018wfu}, single-top~\cite{Nocera:2019wyk}, and dijet-production measurements~\cite{AbdulKhalek:2020jut} from the LHC. A reassessment of the impact of top-pair differential distributions measured by ATLAS at 8~TeV was also presented in Ref.~\cite{Amoroso:2020lgh}, which demonstrated the different impact of absolute and normalized distributions in the fit, and the importance of fitting charm in their description. The NNPDF collaboration has also developed a statistical procedure to represent theory uncertainties in PDFs~\cite{Ball:2018lag}, and applied it to missing higher-order corrections (MHOU) in the strong-coupling expansion of theoretical predictions~\cite{AbdulKhalek:2019bux,AbdulKhalek:2019ihb}, and to nuclear uncertainties in observables obtained from scattering off nuclear targets~\cite{Ball:2018twp}.  The procedure consists in supplementing the experimental covariance matrix with a theoretical covariance matrix estimated by way of an educated guess. In the case of MHOU, correlated uncertainties were estimated at next-to-leading order (NLO) by varying the factorization and renormalization scales according to various prescriptions; in the case of nuclear corrections, correlated uncertainties were estimated as the difference between theoretical predictions obtained either with a free-proton or nuclear PDF. The representation of such uncertainties in PDFs is likely to become mandatory in the future, because their size is comparable to that determined from the uncertainty of the data. The inclusion of such theoretical uncertainties was demonstrated to improve the description of the data, while increasing PDF uncertainties only mildly.

\begin{figure}[!tb]
\centering
\includegraphics[width=0.49\linewidth]{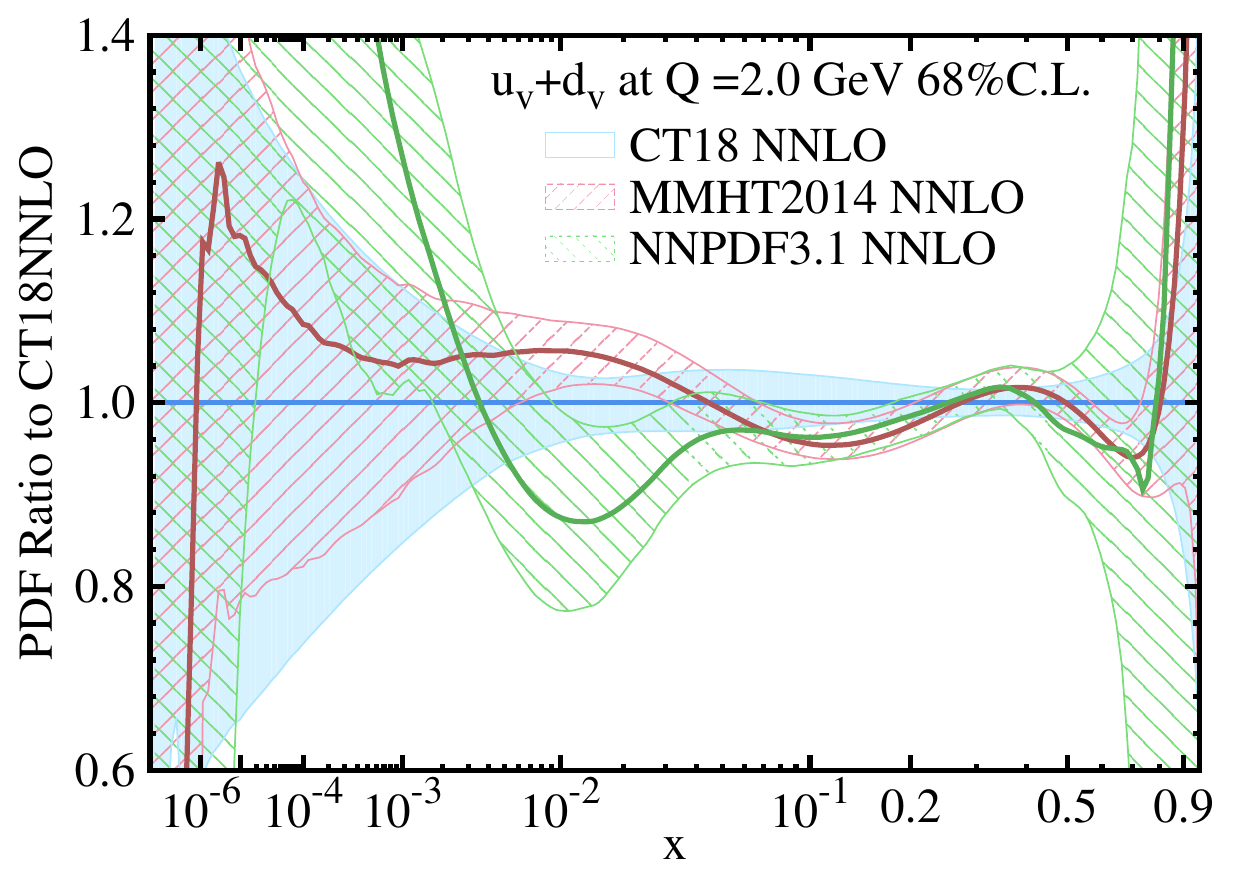}
\includegraphics[width=0.49\linewidth]{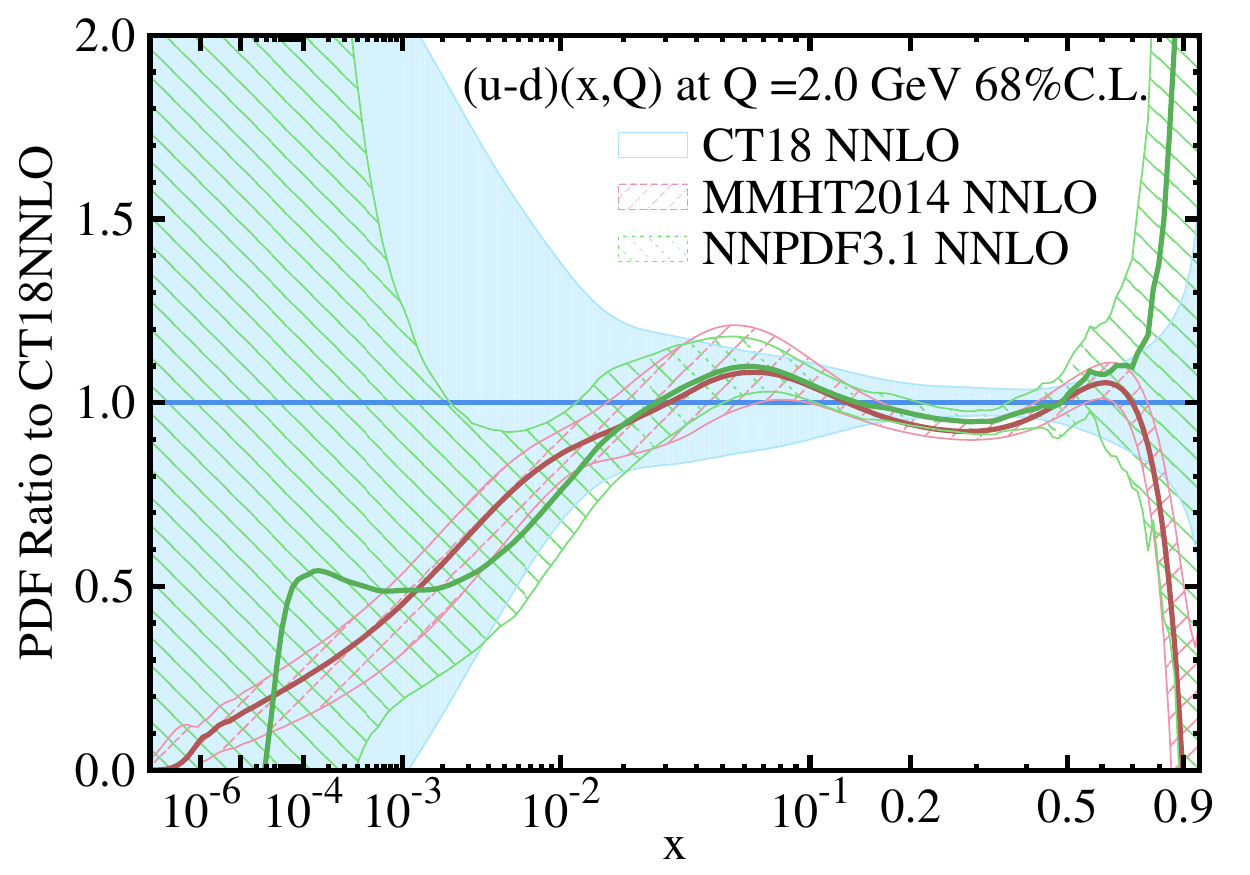}\\
\includegraphics[width=0.49\linewidth]{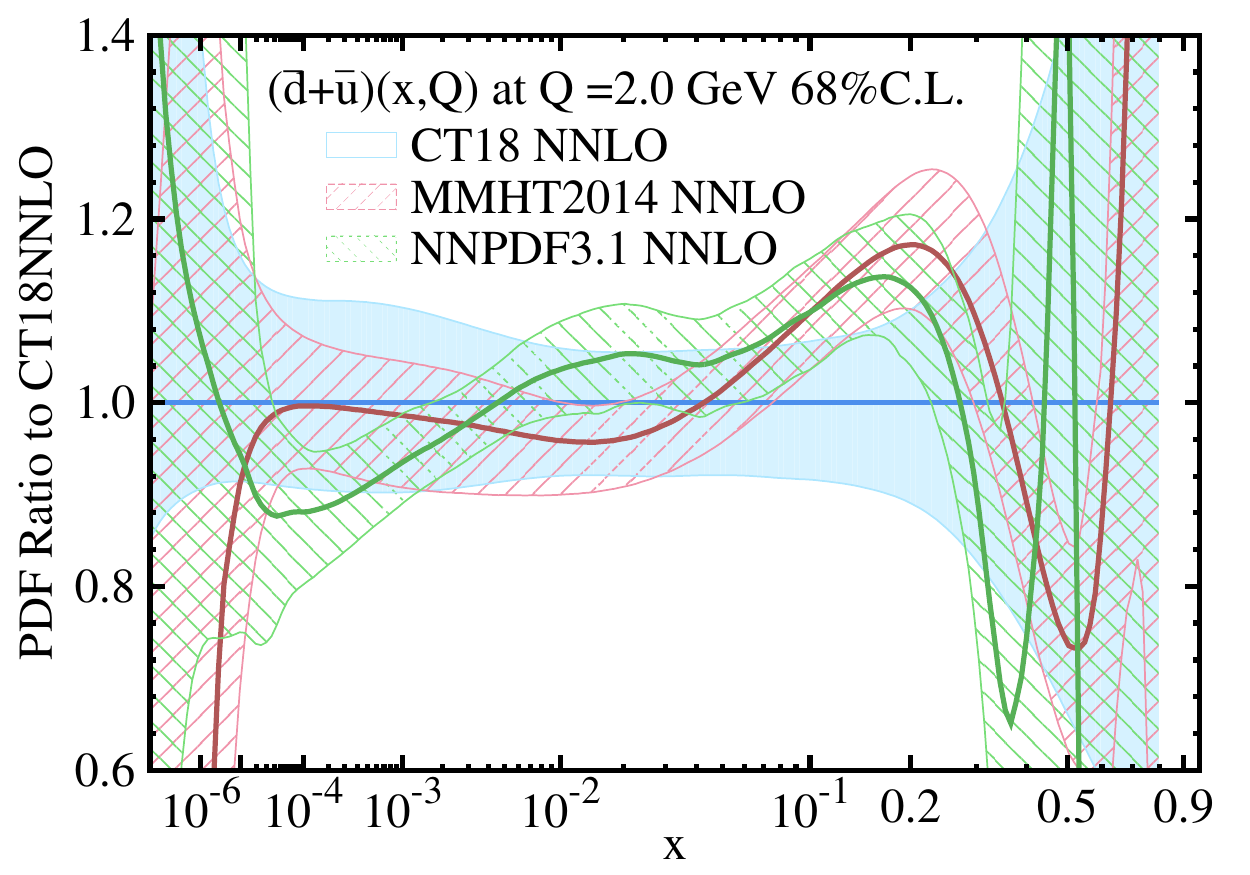}
\includegraphics[width=0.49\linewidth]{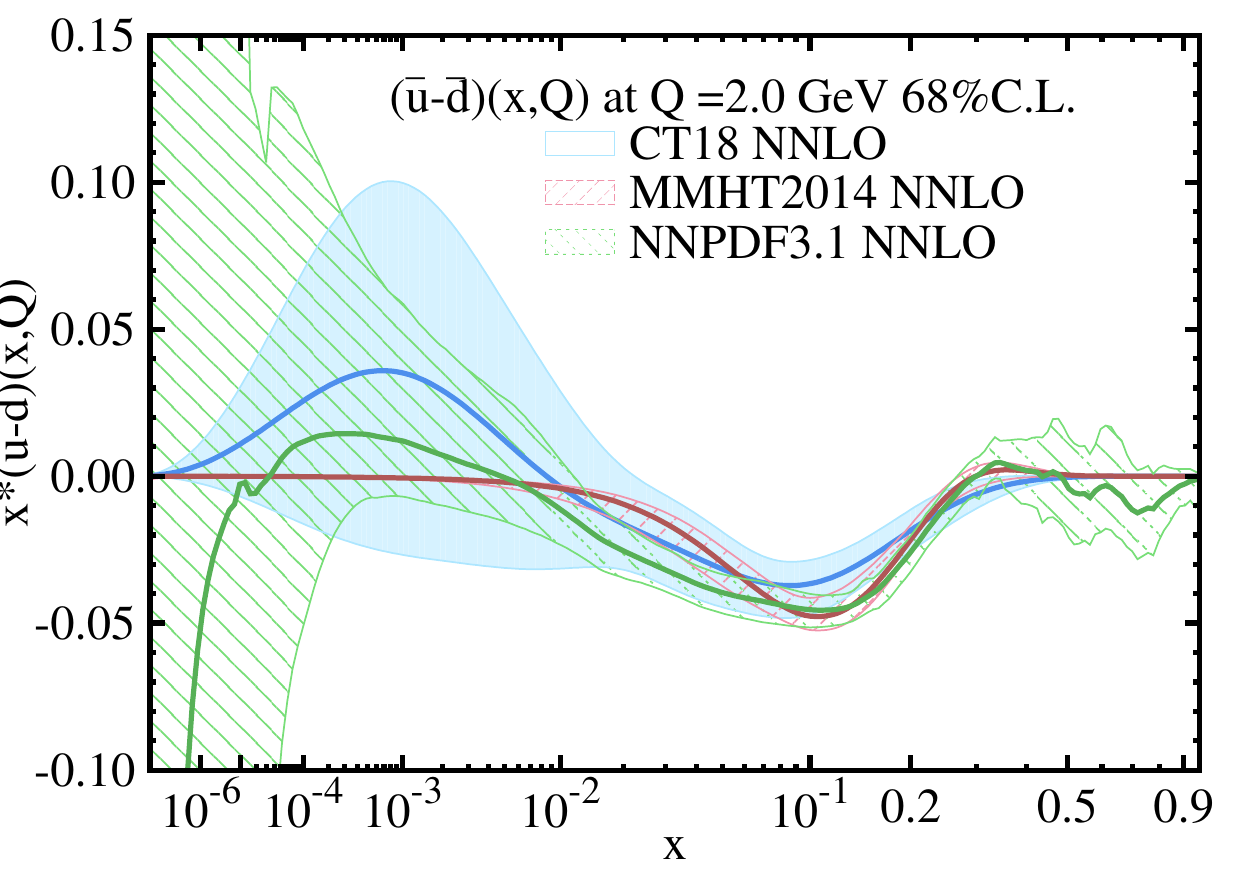}\\
\includegraphics[width=0.49\linewidth]{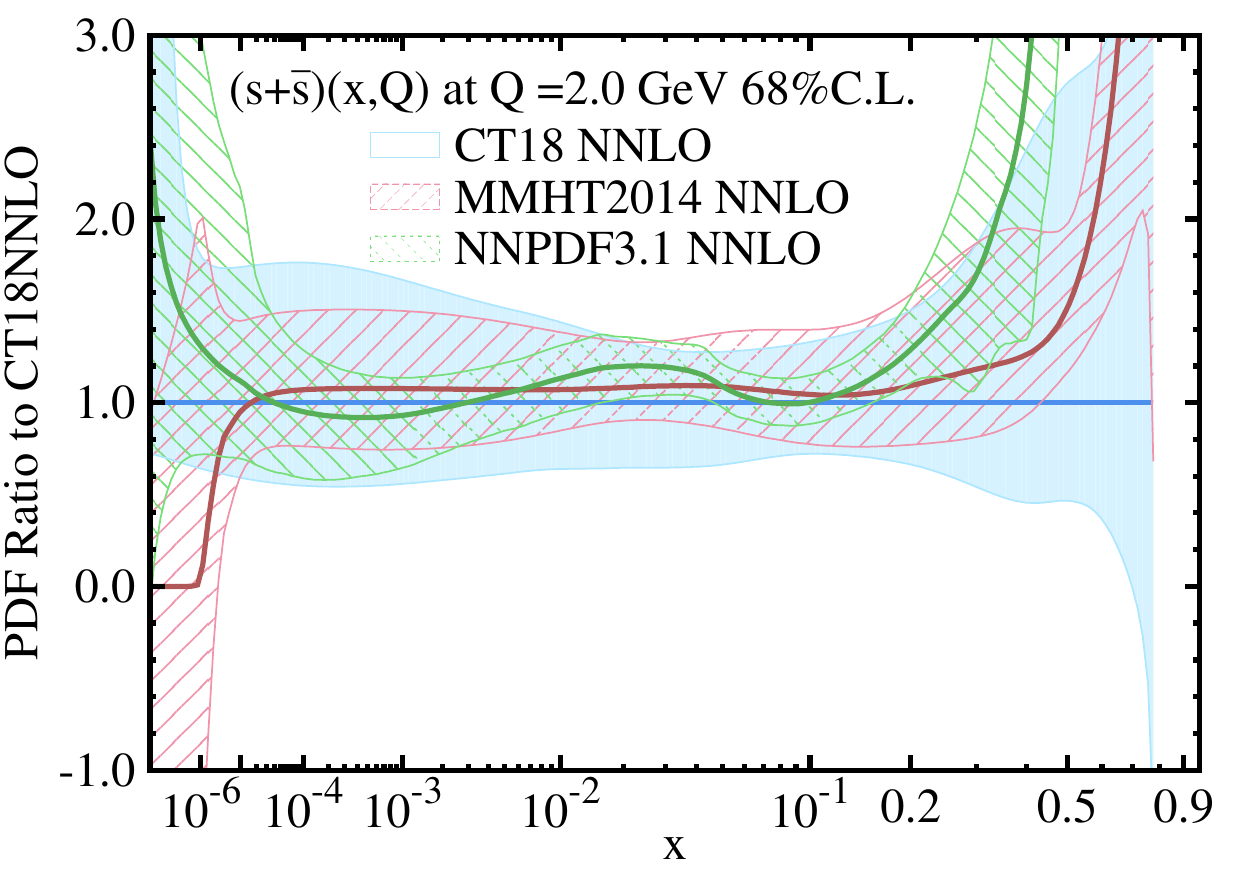}
\includegraphics[width=0.49\linewidth]{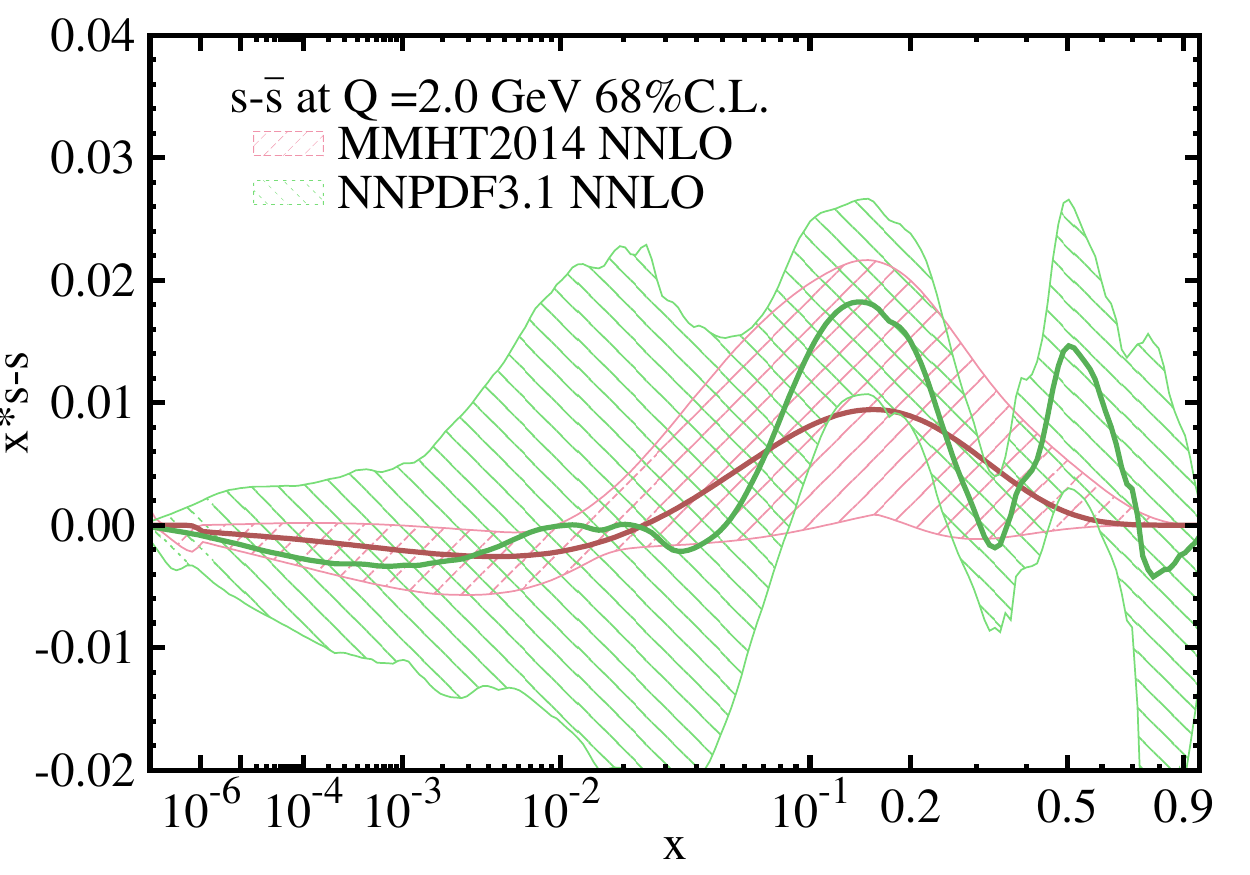}\\
\includegraphics[width=0.49\linewidth]{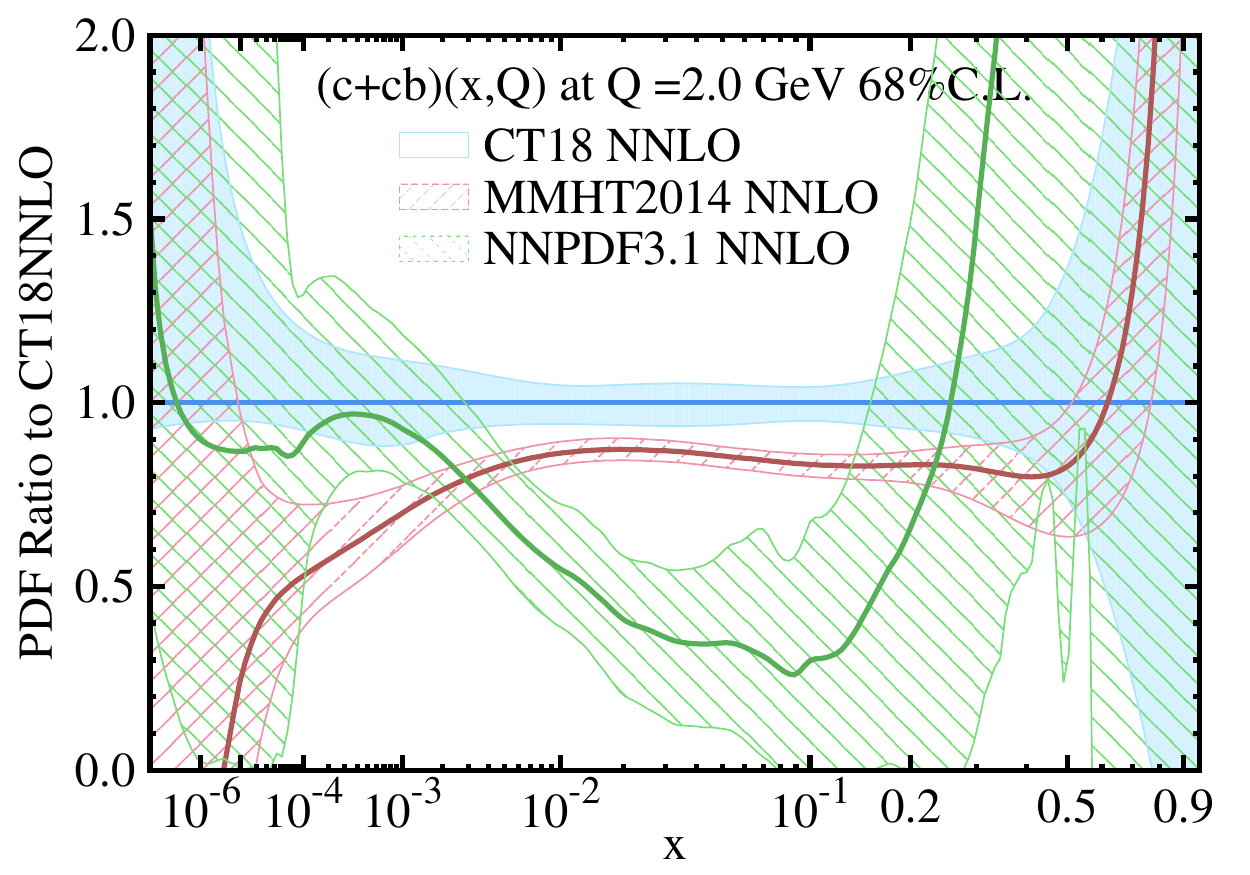}
\includegraphics[width=0.49\linewidth]{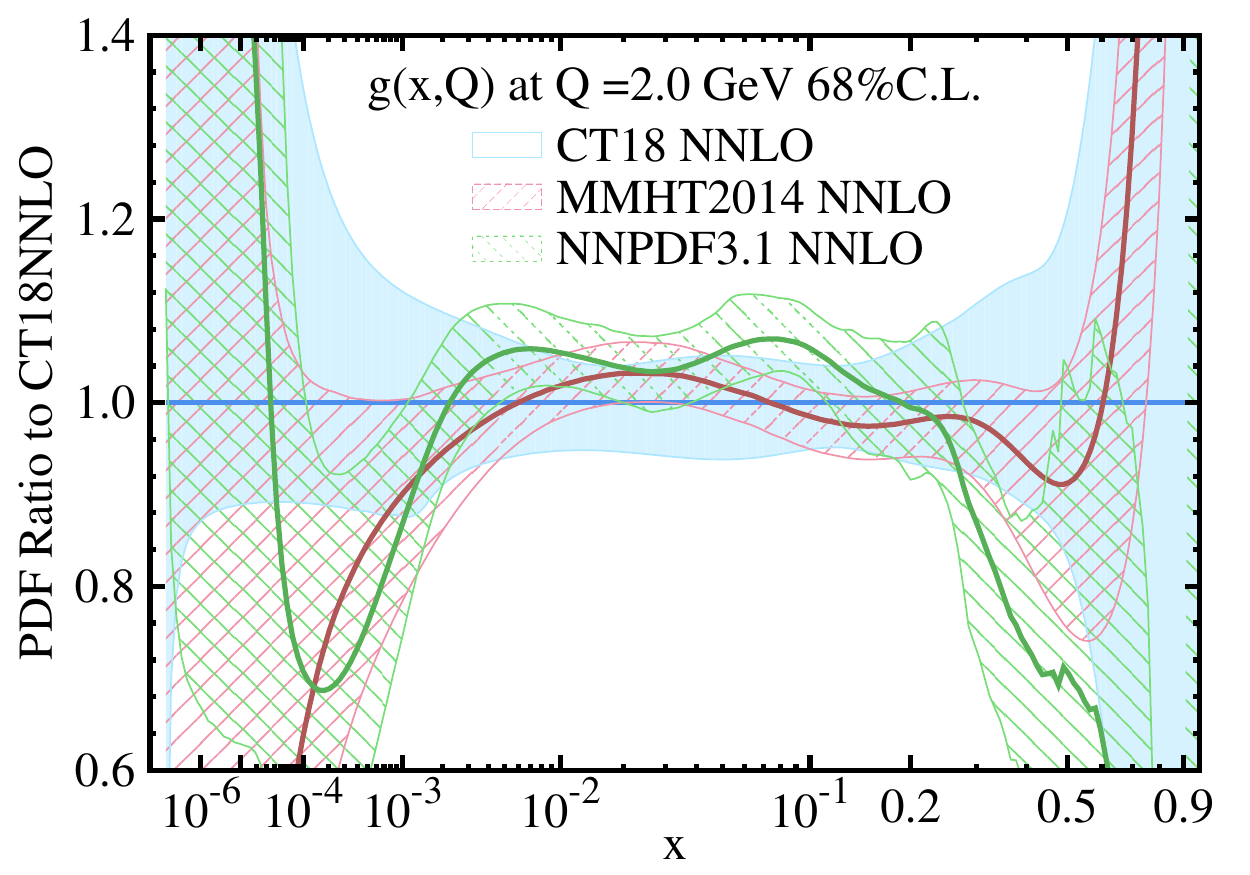}
\caption{The {\sc CT18}, {\sc MMHT14} and {\sc NNPDF3.1} PDFs at $Q=2$~GeV for various flavor combinations.}
\label{fig:allpdf}
\end{figure}

In Fig.~\ref{fig:allpdf} we compare the {\sc CT18}, {\sc MMHT14} and {\sc NNPDF3.1} PDF sets at a scale $Q=\mu=2$~GeV. Specifically, we display the following PDF combinations from top to bottom and left to right: $u_v+d_v=u-\bar u+d-\bar d$, $u-d$, $\bar u+\bar d$, $\bar d -\bar u$, $s+\bar s$, $s-\bar s$, $c+\bar c$ and $g$. Note the special scale on the $x$ axis. While the three global analyses produce similar total valence distributions $u_v+d_v$ for $0.05\lesssim x\lesssim 0.5$, their predictions on other flavor combinations could differ by $10\%$ or more, as in $\bar u - \bar d$, $\bar u + \bar d$, $s + \bar s$, $c+\bar c$ and $g$. In particular, the $c+\bar c$ PDF combination is largely different between {\sc NNPDF3.1} and the other sets, given that charm is parametrized on the same footing as other PDFs in the {\sc NNPDF3.1} set, while it is generated perturbatively in the others. Finally, note that the difference $s-\bar s$ is not displayed for {\sc CT18} because they assume $s=\bar s$; {\sc MMHT14} and {\sc NNPDF3.1} determine $s$ and $\bar s$ PDFs independently.

Beside the three general-purpose PDF sets described above, other unpolarized PDF determinations have been produced or updated recently, namely {\sc ABMP}, {\sc CJ}, {\sc JAM} and {\sc HERAPDF}. These PDF sets are based on a reduced set of measurements and/or on peculiar theoretical assumptions. As such, they are more limited in scope. 

The {\sc ABMP16}~\cite{Alekhin:2017kpj} PDF set is the only unpolarized PDF set determined in a schemes with a fixed number of flavors: for 3, 4 and 5 active flavors separately. It was recently supplemented with an extended set of single-top and top-pair measurements from the Tevatron and the LHC and an increasing number of DY data, notably recent ATLAS gauge-boson--production distributions at 5 and 7~TeV and double-differential distributions for $Z$-boson production from ATLAS and CMS. More stringent kinematic cuts have been applied, which reduce the impact of higher-twist terms included in the analysis.

\begin{figure}[!tb]
\centering
\includegraphics[width=\linewidth]{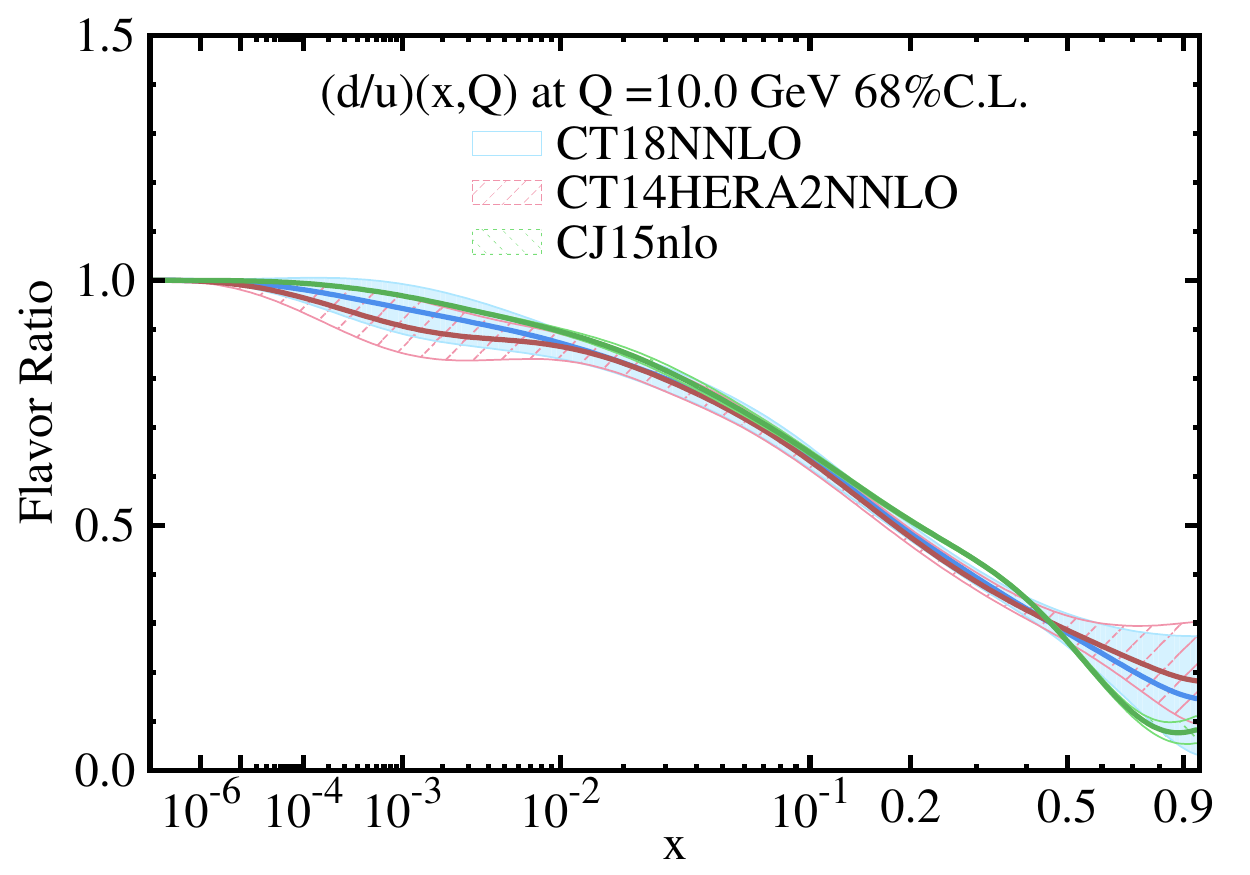}
\caption{The $u/d$ ratio for the {\sc CT18}, {\sc CT14HERA2} and {\sc CJ15} PDFs at $Q=10$~GeV.
\label{fig:cj}}
\end{figure}

The {\sc CJ15}~\cite{Accardi:2016qay} analysis determined the ratio $d/u$ at NLO by including fixed-target DIS data at low $Q$ and high $x$, which are otherwise excluded by kinematic cuts in other PDF analyses. In order to reduce theoretical bias, higher-twist contributions and nuclear effects important in this kinematic region were included in the analysis. In Fig.~\ref{fig:cj} the $d/u$ ratio obtained from the {\sc CJ15} analysis at NLO is compared to the result obtained from the {\sc CT14HERA2}~\cite{Hou:2016nqm} and {\sc CT18} analyses at NNLO. The plot shows that the {\sc CT18} and {\sc CJ15} central predictions converge to different values at large $x$, respectively $0.1$ and $0.2$. Nevertheless, the CT18 and CJ15 results are compatible within uncertainties, although the {\sc CJ15} error band is much smaller than the {\sc CT18} one, possibly because the tolerance criterion is not adopted in the {\sc CJ15} analysis.

The JAM collaboration performed the first {\it universal fit}, {\sc JAM19}~\cite{Sato:2019yez}, a simultaneous determination of PDFs and fragmentation functions (FFs) at NLO from a global analysis of DIS, DY, SIDIS and electron-positron annihilation data. This fit is unique in its kind, as it allows investigation of the unpolarized sea-quark PDFs from semi-inclusive processes, whereby bias from the unknown FFs is minimised as much as possible. The fit revealed in particular a strong suppression of the strange PDF from kaon production, in contrast to what is found in all other PDF determinations, which deserves further investigation.

Finally, the HERAPDF2.0 set was updated~\cite{Cooper-Sarkar:2019pkf} with the HERA legacy inclusive DIS data set supplemented with (multi-)jet production data in DIS from ZEUS and H1. Theoretical predictions were computed at NNLO with NNLOJET~\cite{Gehrmann:2018szu} for the first time. The new NNLO fit favors a
value of the strong coupling $\alpha_s(M_Z)=0.1150$ which is lower than that found at NLO, $\alpha_s(M_Z)=0.1183$. Consequently, two new HERA2PDF2.0Jets NNLO fits, with different $\alpha_s$ values were made available.  

As a general remark, we note that the small-$x$ region of unpolarized PDFs is phenomenologically relevant for the production of final states, for example dijets in the forward-rapidity region of HERA, RHIC and LHC. In $pp$ collision, a final state of rapidity around 4 can probe momentum regions as low as $x\sim 10^{-4}$. Furthermore, in heavy-ion collisions it is believed that the small-$x$ gluons form an over-occupied system needed for later thermalization of quark-gluon plasma. A precise understanding of the gluon PDF at low $x$ is therefore of uttermost importance, and represents a domain where all of the analyses discussed above can be improved, for instance along th elines of Ref.~\cite{Ball:2017otu}. The current approach to studying the small-$x$ gluon density is formulated upon the Balitsky-Kovchegov equation~\cite{Balitsky:1995ub,Kovchegov:1999yj} and its finite-$N_c$ 
generalization~\cite{JalilianMarian:1997jx,JalilianMarian:1997gr,JalilianMarian:1997dw}. The former can be solved using standard methods for integro-differential equations, while the latter, being a functional equation, can be solved by means of lattice-QCD methods, after reformulation as an Langevin equation~\cite{Rummukainen:2003ns}. The two equations are both nonlinear and lead to a saturation scale which signals the strong process of recombination of gluons and gluon saturation. Both equations describe the perturbative evolution of Wilson lines; therefore, they are valid strictly in the perturbative regime of QCD. However, one needs to regulate the nonperturbative large sizes of the Wilson lines, which are eventually probed during the evolution. This is one of the limitations of the approach. It would be of great advantage to further develop the lattice-QCD approach to attack the small-$x$ behaviour of the gluon PDF.

\subsubsection{Nuclear PDFs}

Should the unpolarized hard-scattering processes occur off nuclei instead of free nucleons, it is customary to assume that they can still be described in terms of factorization theorems~\cite{Barshay:1975zz}. Nuclear effects are then reabsorbed into PDFs, which differ from the unpolarized PDFs discussed in the previous section, and which are called nuclear PDFs. Recent progress in the phenomenological determination of nuclear PDFs has been mainly driven by the availability of hard probes in proton-lead collisions at the LHC. Representative examples of such processes include dijet~\cite{Eskola:2019dui} and $D$-meson production~\cite{Kusina:2017gkz} to constrain the gluon nuclear PDFs at small and large $x$, respectively, and $W,Z$ production to constrain nuclear modifications to quark flavor separation~\cite{Kusina:2016fxy}. These observables complement the existing information from neutral-current and charged-current DIS structure functions as well as from proton-lead collision data from RHIC (see Ref.~\cite{Ethier:2020way} for an overview). Several other processes could be added to global nuclear PDF fits in the future, such as photon production~\cite{Campbell:2018wfu}, low-mass DY, and possibly even top-quark pair production.

Several groups have presented phenomenological determinations of nuclear PDFs. These differ in the input data set, in the accuracy of the theoretical computations, in the flavor assumptions, and in the methodological details of the analysis, such as the parametrization of the nucleus dependence, and the representation of uncertainties. The nuclear-PDF analysis including the most extensive data set to date is {\sc EPPS16}~\cite{Eskola:2016oht}, which is the only one to take into account jet- and $W,Z$-production data from proton-lead collisions at the LHC. Other recent nuclear PDF sets, based on a more limited data set than {\sc EPPS16}, are {\sc nCTEQ15}~\cite{Kovarik:2015cma}, {\sc TuJu19}~\cite{Walt:2019slu}, and {\sc nNNPDF1.0}~\cite{AbdulKhalek:2019mzd}.

We refer the interested reader to Refs.~\cite{Ethier:2020way,Kovarik:2019xvh} for a detailed comparison of all of these nuclear PDF sets. Here, we note that the determination of nuclear PDFs is less advanced than that of their free-proton counterparts at least in four respects. First, the limited amount of experimental measurements, which results in large uncertainties on nuclear PDFs, especially for the gluon and the individual quark flavors. Second, the use, to some extent, of model-dependent assumptions, in particular to mimic shadowing and anti-shadowing effects as a function of $x$ and of the atomic number $A$. Third, the consistency between nuclear PDFs and their $A=1$ limit (that is, the free-proton PDFs) both in terms of central values and uncertainties, as well as the treatment of correlations between nuclear and proton PDFs. Fourth, possible data inconsistencies, for example between neutral- and charged-current structure functions and between some of the LHC observables.

Future facilities, specifically the EIC, will address the first of these limitations by extending the kinematic reach of the measurements down to $x\sim 10^{-9}$, as anticipated in several impact studies~\cite{Aschenauer:2017oxs,AbdulKhalek:2019mzd,HannuPaukkunenfortheLHeCstudygroup:2017ric}. Reliable calculations from lattice QCD could add valuable information to address some of the remaining limitations. The calculation of nuclear PDFs using lattice QCD is in general more demanding than the computation of free-proton PDFs, in a way that grows with the number of nucleons in the nucleus. For this reason, only few pioneering results on very light nuclei are currently available. In particular, the NPLQCD collaboration computed the first moment of the unpolarized gluon distribution in nuclei up to $A=3$~\cite{Winter:2017bfs} (for deuteron and $^3$He). These results are obtained at a higher-than-physical quark mass; as such, they should be viewed as a starting point for future developments.

\subsubsection{Helicity PDFs}

Helicity PDFs describe the net amount of momentum densities of partons aligned along or opposite a longitudinally polarized nucleon. Their relevance is related to the fact that the contribution of quarks, antiquarks and gluons to the nucleon spin is quantified by the first moments of the corresponding helicity PDFs, according to the canonical decomposition of the proton's total angular momentum~\cite{Leader:2013jra}.

The most recent analyses of polarized PDFs are {\sc DSSV14}~\cite{deFlorian:2014yva}, {\sc NNPDFpol1.1}~\cite{Nocera:2014gqa} and {\sc JAM17}~\cite{Ethier:2017zbq}. Since publication, the {\sc DSSV14} analysis was updated with a Monte-Carlo variant~\cite{deFlorian:2019zkl}, which also studied the impact of recent dijet measurements from STAR~\cite{Adamczyk:2016okk,Adam:2018pns}. The {\sc NNPDFpol1.1} PDF set was also updated with STAR measurements, including the same dijet measurements and additional $W$-boson production data~\cite{Nocera:2017wep}. The {\sc JAM17} analysis is unique in its kind, because it realises a universal fit, that is, a simultaneous determination of the helicity PDFs and of the FFs of pions and kaons by means of an analysis of the SIDIS data. Because of this fact, theoretical assumptions on the SU($N_f$) symmetry, which relate the first moments of the isotriplet and octet helicity PDF combinations to the weak decays of hadrons, can be relaxed. Therefore, theoretical bias in the determination of helicity sea-quark distributions is alleviated, in particular on strangeness (see Sect.~\ref{02sec5}).

\begin{figure}[!t]
\centering
\includegraphics[height=0.85\linewidth,width=0.75\linewidth]{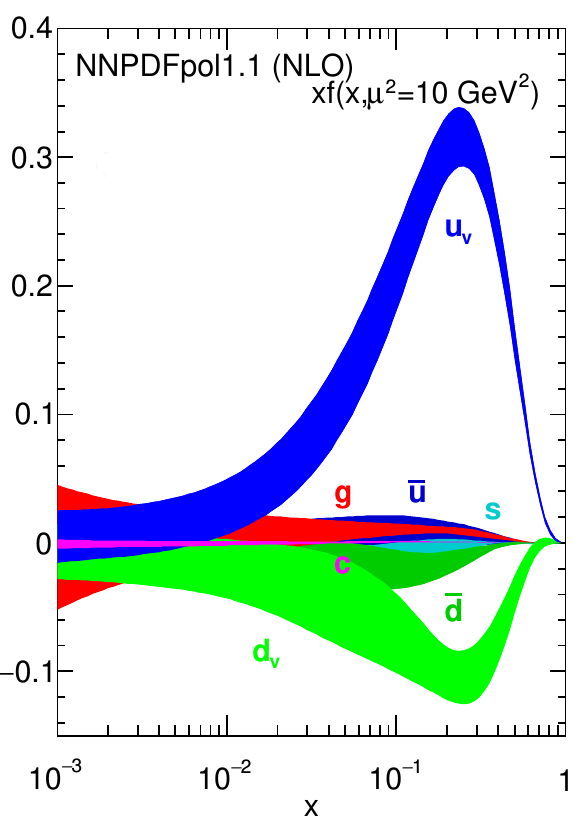}
\caption{The helicity PDFs from the {\sc NNPDFpol1.1} parton set at $\mu^2=10$~GeV$^2$. Figure from~\cite{Tanabashi:2018oca}.}
\label{fig:helicity_PDFs}
\end{figure}

All of these analyses are accurate to NLO. The PDFs from the {\sc NNPDFpol1.1} sets are displayed, at a scale $\mu^2=10$~GeV$^2$ in Fig.~\ref{fig:helicity_PDFs} for illustration purposes. The up quark valence distribution is the most constrained helicity PDF, primarily thanks to measurements of the proton structure function $g_1$ over a relatively broad range of $x$ and $Q^2$; see Fig.~\ref{fig:kin_cov}. The corresponding down-quark distribution, which is opposite in sign, is smaller in magnitude and shows somewhat larger uncertainties. Nevertheless, a fair agreement between various PDF sets has been achieved for the valence polarizations, with differences originating from theoretical, experimental, and methodological choices generally being smaller than the PDF uncertainty (representing the uncertainty of the data). The polarization of the sea quarks is significantly smaller than the polarization of the valence quarks. It is also more dependent on the measurements included in each PDF set, and on the flavor assumptions with which the data are analyzed. Specifically, while all PDF sets show a fairly consistent anti-down polarization, the anti-up distribution is positive in the {\sc NNPDFpol1.1} PDF set, due to $W$-boson production data, while it changes sign in the {\sc DSSV14} PDF set, due to SIDIS data. This difference is somewhat relieved in the {\sc JAM17} result, where the simultaneous determination of helicity PDFs and FFs generally leads to PDFs with larger uncertainties. The situation is more involved for the strange helicity PDF, which remains largely unconstrained from DIS data alone (see Sect.~\ref{02sec5}). Finally, the gluon helicity distribution has been established to be consistently sizable and positive in both the {\sc DSSV14} and {\sc NNPDFpol1.1} analyses thanks to the availability of precise jet, dijet, and pion-production spin-asymmetry measurements at RHIC. Such evidence, however, is limited to the region $0.02\lesssim x\lesssim 0.4$, outside of which the gluon helicity PDF is affected by large extrapolation uncertainties that prevent any definitive conclusion about its role in the proton spin decomposition. The extended kinematic region attained by a future polarized EIC (see Fig.~\ref{fig:kin_cov}) will be pivotal to clarify this picture, as demonstrated by several impact studies~\cite{Aschenauer:2012ve,Aschenauer:2015ata,Aschenauer:2013iia,Ball:2013tyh}.

\subsubsection{Transversity PDFs}
\label{sec:02sec1h1}

Transversity describes the correlation between the transverse polarizations of the nucleon and of its constituent partons. Because of its chiral-odd nature, it can be measured only in processes where it pairs to another chiral-odd object~\cite{Artru:1989zv,Jaffe:1991kp}. Historically, transversity was extracted for the first time in the framework of TMD factorization, by exploiting the Collins effect~\cite{Anselmino:2007fs} in SIDIS measurements, whereby transversity couples to the chiral-odd Collins FF $H_1^\perp$~\cite{Collins:1992kk}. This procedure may be prone to theoretical bias, because TMD evolution depends on nonperturbative parameters that are not very well constrained by experiment. Furthermore, it cannot be extended to $pp$ collision measurements, for which TMD factorization is explicitly broken~\cite{Rogers:2010dm}. This limitation can be overcome by considering the Collins effect for a hadron detected inside a jet~\cite{Adamczyk:2017ynk}. A hybrid framework can be established~\cite{Kang:2017btw}, where the TMD FF $H_1^\perp$ is paired to the collinear PDF $h_1$.

An alternative method was proposed to determine transversity in a fully collinear framework: it is based on the semi-inclusive production of two hadrons inside the same current jet with small invariant mass~\cite{Jaffe:1997hf,Bianconi:1999cd}. The elementary mechanism consists in the correlation between the transverse polarization of the quark directly fragmenting into the two hadrons and their transverse relative momentum~\cite{Collins:1994ax}. In this case, the dihadron SIDIS cross section (once integrated over partonic transverse momenta) contains a specific azimuthal modulation in the orientation of the plane containing the momenta of the two hadrons. The coefficient of this modulation is the simple product $h_1 H_1^\sphericalangle$, where $H_1^\sphericalangle$ is a chiral-odd dihadron FF (DiFF) quantifying the above correlation~\cite{Bianconi:1999cd,Radici:2001na,Bacchetta:2002ux}. The function $H_1^\sphericalangle$ can be independently determined from correlations between the azimuthal orientations of two hadron pairs in back-to-back jets produced in electron-positron annihilation~\cite{Boer:2003ya,Bacchetta:2008wb,Courtoy:2012ry,Matevosyan:2018icf}. Because of collinear factorization, it is possible to isolate the same combination $h_1 H_1^\sphericalangle$ also in $pp$ collisions~\cite{Bacchetta:2004it}, a fact that gives rise to an asymmetric azimuthal distribution of the final hadron pair when one of the two initial protons is transversely polarized~\cite{Radici:2016lam}.  

Experimental measurements of the asymmetric azimuthal distribution of $\pi^+ \pi^-$ pairs in SIDIS have been collected by the {\sc HERMES} collaboration for a proton target~\cite{Airapetian:2008sk} and by the {\sc COMPASS} collaboration for both proton and deuteron targets~\cite{Adolph:2012nw,Adolph:2014fjw}, summing up to 22 independent data points. The azimuthal asymmetry in the distribution of back-to-back $\pi^+ \pi^-$ pairs in $e^+ e^-$ annihilation was firstly measured by the {\sc Belle} collaboration~\cite{Vossen:2011fk}, a fact that opened the way to the first parametrization of $H_1^\sphericalangle$~\cite{Courtoy:2012ry}. In turn, this result was used in combination with the SIDIS data to extract the valence components of $h_1$~\cite{Bacchetta:2011ip,Bacchetta:2012ty,Radici:2015mwa} at leading-order (LO) accuracy. The {\sc STAR} collaboration recently released results for the predicted asymmetry of $\pi^+ \pi^-$ pairs produced from pp collisions with a transversely polarized proton at the center-of-mass energy of $\sqrt{s}=200$~GeV~\cite{Adamczyk:2015hri}. The first extraction of transversity from a global analysis of all of these data sets was, therefore, carried out in Ref.~\cite{Radici:2018iag} at LO, where the bootstrap method~\cite{Bacchetta:2012ty,Radici:2015mwa} was used to propagate the data uncertainty into the PDF uncertainty. We will call this analysis {\sc PV18} in the remainder of the paper. The {\sc STAR} collaboration subsequently released measurements at $\sqrt{s}=500$~GeV~\cite{Adamczyk:2017ynk}, that need to be included in future updates of the fit. The overall experimental information amounts to approximately 50 independent data points covering the range $0.008<x< 0.35$, about two orders of magnitude less than for the analysis of unpolarized PDFs, see Fig.~\ref{fig:kin_cov}.

From the theoretical point of view, the decomposition on the helicity basis of the quark-hadron helicity amplitude in terms of the three leading-twist PDFs (unpolarized, helicity and transversity) leads to the so-called Soffer inequality~\cite{Soffer:1994ww}
\begin{equation}
|h_1^q(x,Q^2)|\leq \frac{1}{2}\left[q(x,Q^2)+\Delta q(x,Q^2)\right] = F_\text{SB}^q (x,Q^2) \; ,
\label{eq:SB}
\end{equation}
for each quark $q$ and antiquark $\bar{q}$ separately. For valence quarks $q_v = q - \bar{q}$, the Soffer inequality reads
\begin{equation}
|h_1^{q_v}|\leq F_\text{SB}^{q} + F_\text{SB}^{{\bar q}} \equiv F_\text{SB}^{q+\bar{q}} \; ,
\label{eq:SBval}
\end{equation}
and is conserved under QCD evolution~\cite{Goldstein:1995ek,Vogelsang:1997ak}. Because the experimental data cover a limited range in $x$, Eq.~\eqref{eq:SBval} is a bound that limits the allowed parameter space in any reliable determination of transversity.

The global analysis of Ref.~\cite{Radici:2018iag} assumed the following functional form at the initial scale $Q_0^2 = 1$~GeV$^2$:
\begin{equation}
\label{eq:h1Q0}
x h_1^{q_v} (x, Q_0^2) = F^q (x) \  F_\text{SB}^{q+\bar{q}} (x, Q_0^2) \; ,
\end{equation}
where $F^q(x)$ contained five fitting parameters for each valence flavor $q_v$. Imposing the constraint $|F^q (x)| \leq 1$ for all $x$ automatically enforces the Soffer inequality at any scale. The low-$x$ behavior of $h_1^{q_v} (x)$ was further constrained by requiring that its zeroth Mellin moment, the tensor charge, is finite. The {\sc MSTW08}~\cite{Martin:2009iq} and {\sc DSSV08}~\cite{deFlorian:2009vb} parton sets were used as input for the unpolarized (at LO) and the helicity (at NLO) PDFs entering Eq.~\eqref{eq:SBval}. Using other parametrizations, for instance {\sc CT18}~\cite{Hou:2019efy} for the unpolarized PDF, or {\sc JAM15}~\cite{Sato:2016tuz} or {\sc JAM17}~\cite{Ethier:2017zbq}) for the helicity PDF, was shown to leave the fit of the transversity unaltered~\cite{Radici:2018iag,Benel:2019mcq}. 

The dependence of the transversity PDF on the Soffer bound implicit in Eq.~\eqref{eq:h1Q0} may result in a saturation of said bound at large $x$. If so, one may wonder whether the data is in tension with the bound itself, or whether the account of the Soffer bound is not flexible enough. This issue was investigated in the fit of Ref.~\cite{Benel:2019mcq}, which we call {\sc MEX19}, based only on SIDIS dihadron production data and on the bootstrap method. Four functional forms were considered, all consisting of sums of selected Bernstein polynomials of four different degrees that spanned almost the entire $x$ range and that were multiplied by $x^{1.25}$. An estimate of the statistical compatibility between the valence transversity distributions and the bound was performed through the reweighting of the data in the minimization procedure. Given the uncertainty on the deuteron data, the $h_1^{d_v}$ had to be constrained to an envelope designed based on the fall-off of PDFs at large $x$ using the Lagrange multiplier method.

\begin{figure}[!t]
\centering
\includegraphics[width=0.4\textwidth]{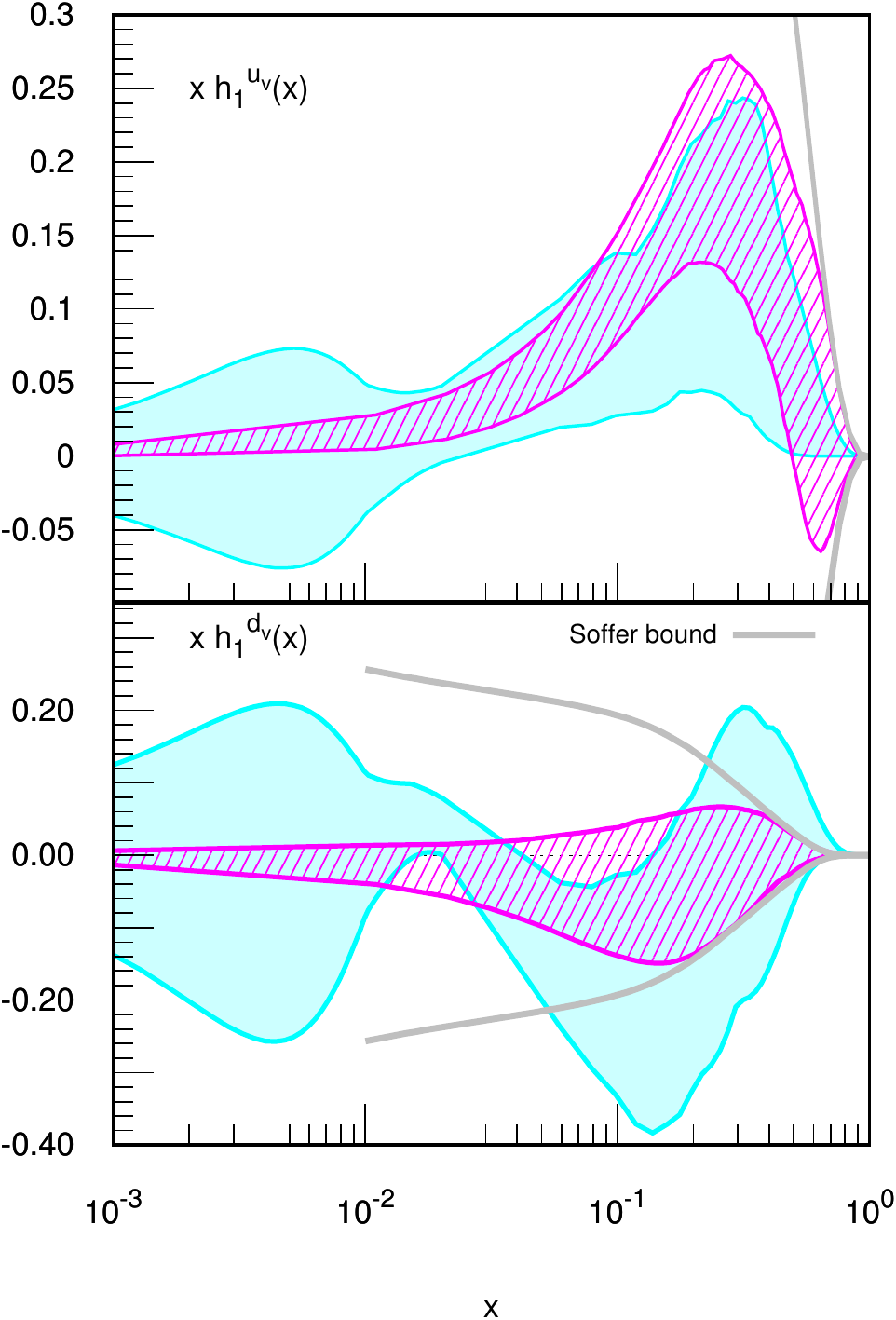} 
\caption{The transversity $x\, h_1(x)$ at $90\%$ CL. Upper (lower) plot for valence up (down) component. Gray lines represent the Soffer bound. Darker (pink) band for the {\sc PV18} global fit of~\cite{Radici:2018iag} at $Q^2 = 2.4$ GeV$^2$. Lighter (cyan) band for the {\sc MEX19} constrained analysis of~\cite{Benel:2019mcq} at the average scale of the data.}
\label{fig:xh1q}
\end{figure}

In Fig.~\ref{fig:xh1q} we show $x\, h_1(x)$ at $Q^2 = 2.4$~GeV$^2$. The upper (lower) panel refers to its valence up (down) component. The shaded (pink) band is the result of the {\sc PV18} global fit of Ref.~\cite{Radici:2018iag}. The error band corresponds to the 90\% confidence level (CL), and includes also a systematic theoretical error induced by the currently unconstrained gluon contribution to DiFF~\cite{Radici:2018iag}. It turns out that the transversity for the valence down quark is very sensitive to this uncertainty. The plain (cyan) band is the result of the {\sc MEX19} constrained fit of Ref.~\cite{Benel:2019mcq}; error bands correspond to the 90\% CL. The Soffer bound is represented in a limited range by a gray line using the unpolarized and helicity PDFs from the LO {\sc MSTW08}~\cite{Martin:2009iq} and the NLO {\sc DSSV08}~\cite{deFlorian:2009vb} sets, respectively. At small $x$, the error on the bound (still negligible in comparison to the current overall uncertainties on the transversity PDF) increases due to the uncertainties in both parametrizations. This is, however, inconsequential for valence distributions, which are suppressed by evolution. From Fig.~\ref{fig:xh1q} we observe that the two determinations are consistent within uncertainties, although the larger uncertainty on the valence down quark at large $x$ can be appreciated in the fit of Ref.~\cite{Benel:2019mcq} as a consequence of the Soffer bound.

Nevertheless, PDF shapes and uncertainties are comparable to those determined in TMD analyses of the Collins effect~\cite{Kang:2015msa,Anselmino:2013vqa}, including in a dedicated studies on the role of the Soffer bound on the tensor charge~\cite{DAlesio:2020vtw}. The Soffer bound was completely released in~\cite{Cammarota:2020qcw}. In all cases, the uncertainty bands remain large; future developments will depend on a detailed study of flavor dependence of DiFFs by using {\sc BELLE} data for dihadron multiplicities~\cite{Seidl:2017qhp}, and on including in the global fit also the {\sc STAR} data at higher center-of-mass energy~\cite{Adamczyk:2017ynk}. From the theoretical point of view, work is in progress to update the analyses to NLO accuracy.

%% file: 02sec3latticePDFs.tex
\goodbreak
\subsection{Lattice PDFs Updates}
\label{02sec3}

Lattice QCD regularizes QCD on a finite Euclidean lattice by means of numerical computations of QCD correlation functions in the path-integral formalism, using methods adapted from statistical mechanics. Of particular interest for this report, are novel methods to access information on PDFs using lattice QCD. This is a very challenging task, as distribution functions are light-cone quantities and cannot be calculated on a Euclidean lattice. A few methods have been proposed, some as early as the 1990s, and others very recently. This includes, studies based on the hadronic tensor~\cite{Liu:1993cv,Liu:1998um,Liu:1999ak}, auxiliary quark field approaches~\cite{Detmold:2005gg,Braun:2007wv}, Large-Momentum Effective Theory (LaMET)~\cite{Ji:2013dva,Ji:2014gla,Ji:2020ect} (quasi-PDFs), pseudo-PDFs~\cite{Radyushkin:2016hsy}, a method based on OPE~\cite{Chambers:2017dov}, and the Good Lattice Cross Sections approach~\cite{Ma:2014jla,Ma:2014jga,Ma:2017pxb}. We refer the reader to Ref.~\cite{Lin:2017snn,Cichy:2018mum,Ji:2020ect} for a thorough description of the methods. In this section we review recent progress on some of these methods that have been studied extensively within lattice QCD. The current stage of the field and the limited data for some methods, does not permit one to quantitatively compare the lattice results extracted from different methods. Having more data, in particular simulations at the physical point, will improve the current understanding of the sources of systematic uncertainties. This way, we can assess what are the cavities and technical limitations of each method, such as, the region of $x$ for which the lattice results are reliably extracted.

\subsubsection{PDFs at the physical point with quasi-distributions}

Pioneering works have shown great promise in obtaining quantitative results for the unpolarized, helicity and transversity quark and antiquark distributions~\cite{Lin:2014gaa,Lin:2014yra,Lin:2014zya,Alexandrou:2014pna,Alexandrou:2015rja,Chen:2016utp} using the quasi-PDFs approach~\cite{Ji:2013dva}, which utilized Large Momentum Effective Theory (LaMET)~\cite{Ji:2014gla} to match lattice calculable matrix elements, to their light-cone counterparts. The groups working on extracting $x$-dependent PDFs from lattice QCD have significantly improved their calculations. 
A recent review on the theory and lattice calculations can be found in Refs.~\cite{Cichy:2018mum,Ji:2020ect}.
There are several theoretical developments related to the renormalizability of quasi-PDFs~\cite{Ji:2015jwa,Ishikawa:2016znu,Chen:2016fxx,Ji:2017oey,Ji:2017rah}, and the appropriate matching formalism~\cite{Xiong:2013bka,Stewart:2017tvs,Izubuchi:2018srq}.
One of the major developments for the lattice calculation concerns the understanding of the renormalization pattern on the lattice~\cite{Constantinou:2017sej,Chen:2017mie}, as well as the development of a complete nonperturbative renormalization procedure~\cite{Alexandrou:2017huk,Chen:2017mzz} that removes all divergences of nonlocal operators.

\begin{figure*}[!t]
\includegraphics[width=.44\textwidth]{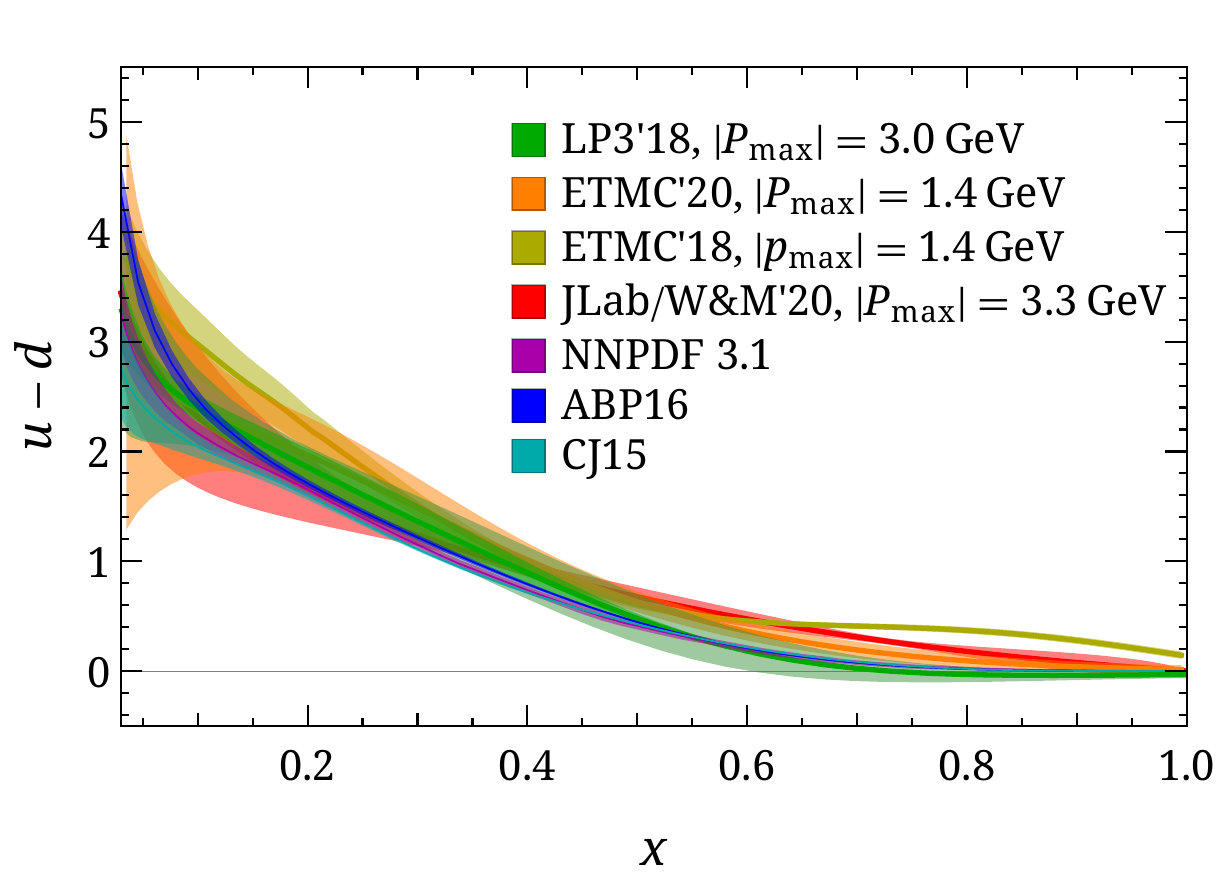}
\includegraphics[width=.46\textwidth]{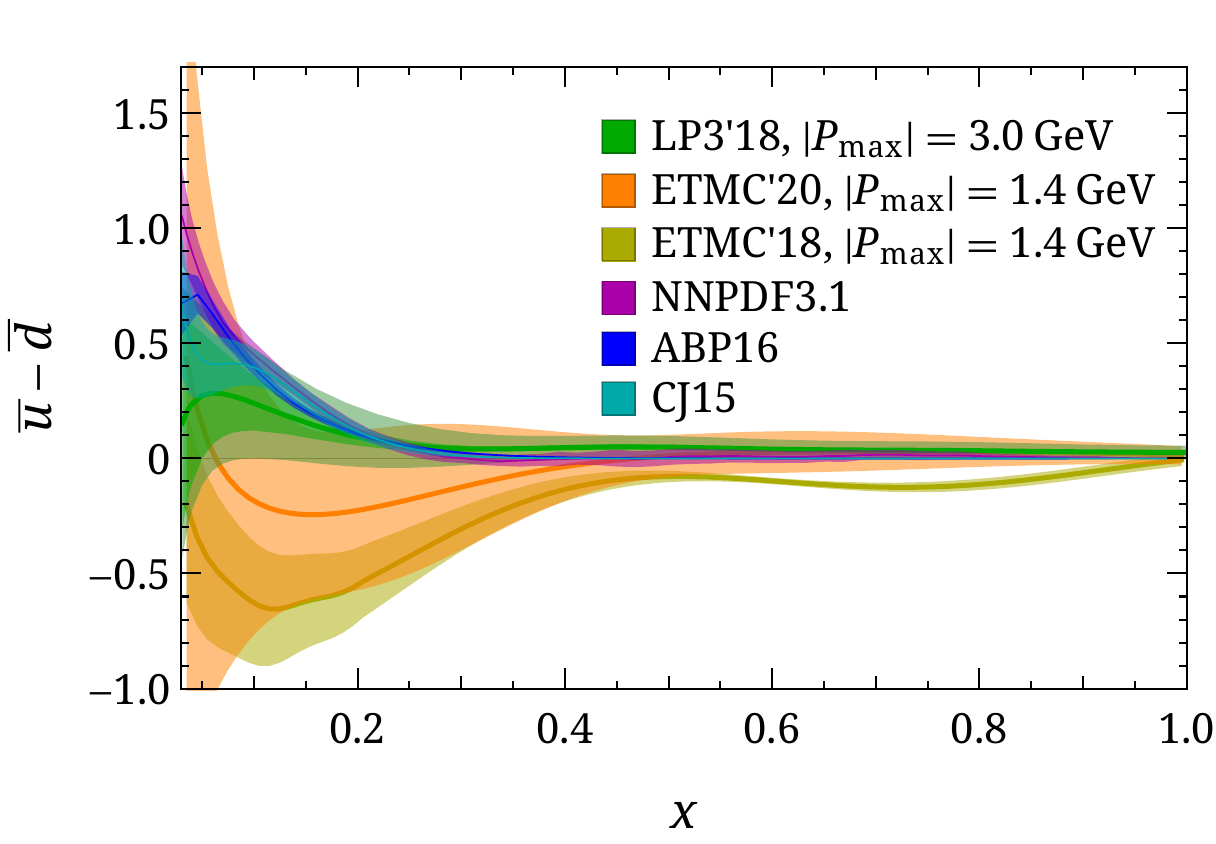}
\includegraphics[width=.44\textwidth]{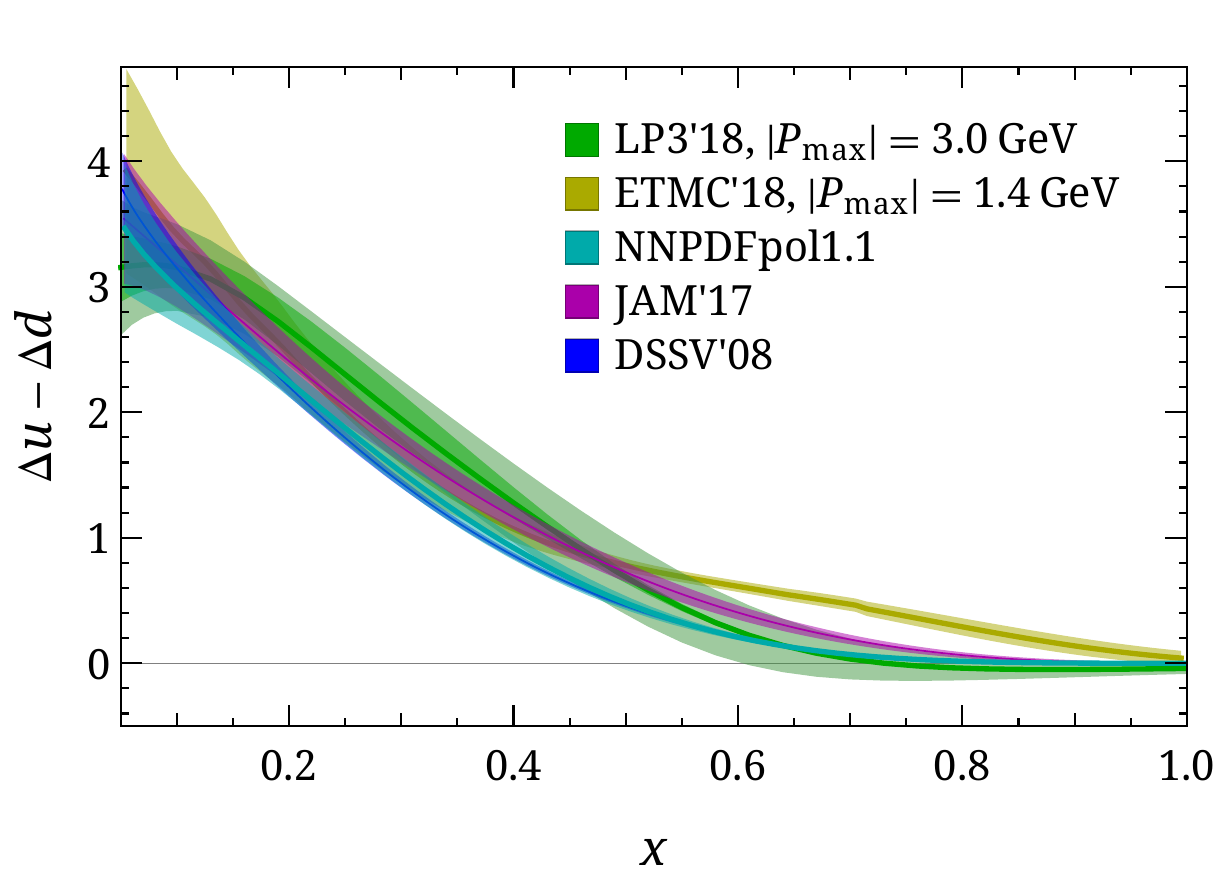}
\includegraphics[width=.46\textwidth]{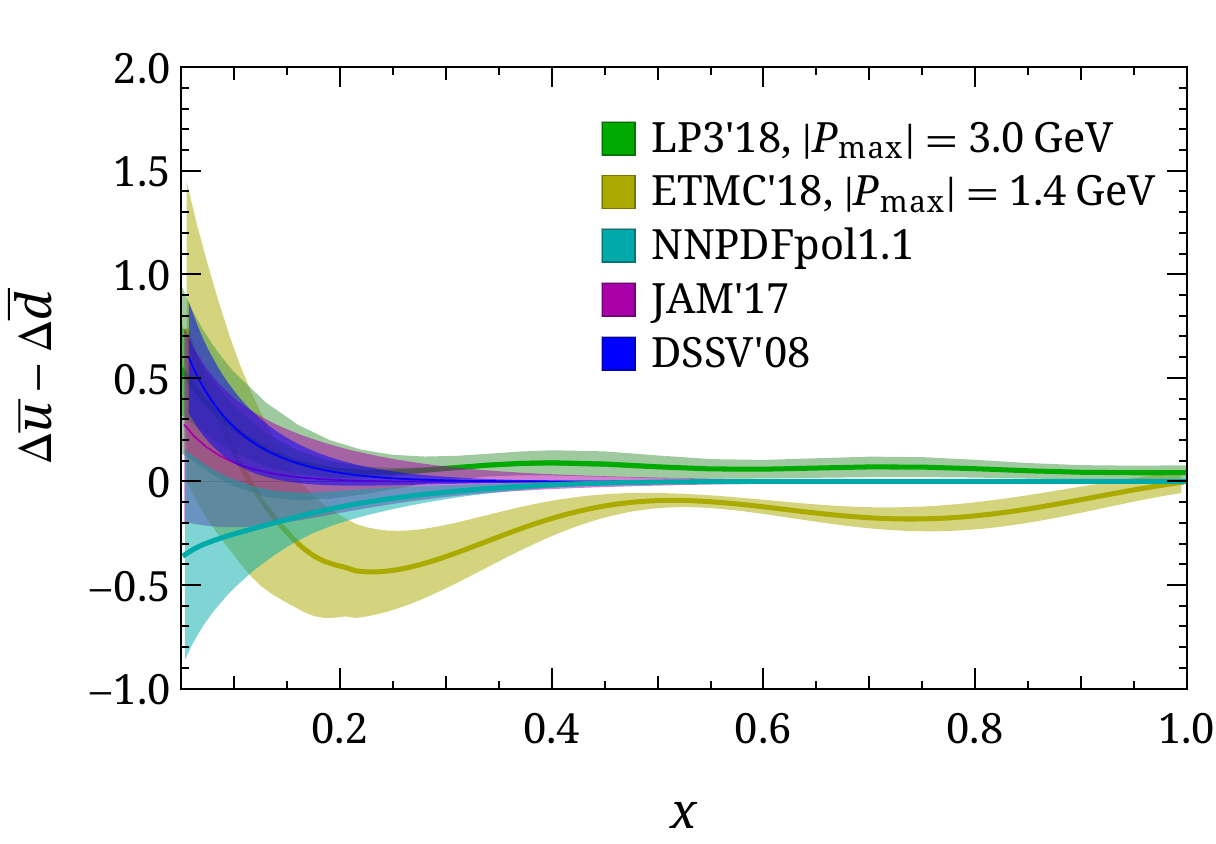}
\includegraphics[width=.44\textwidth]{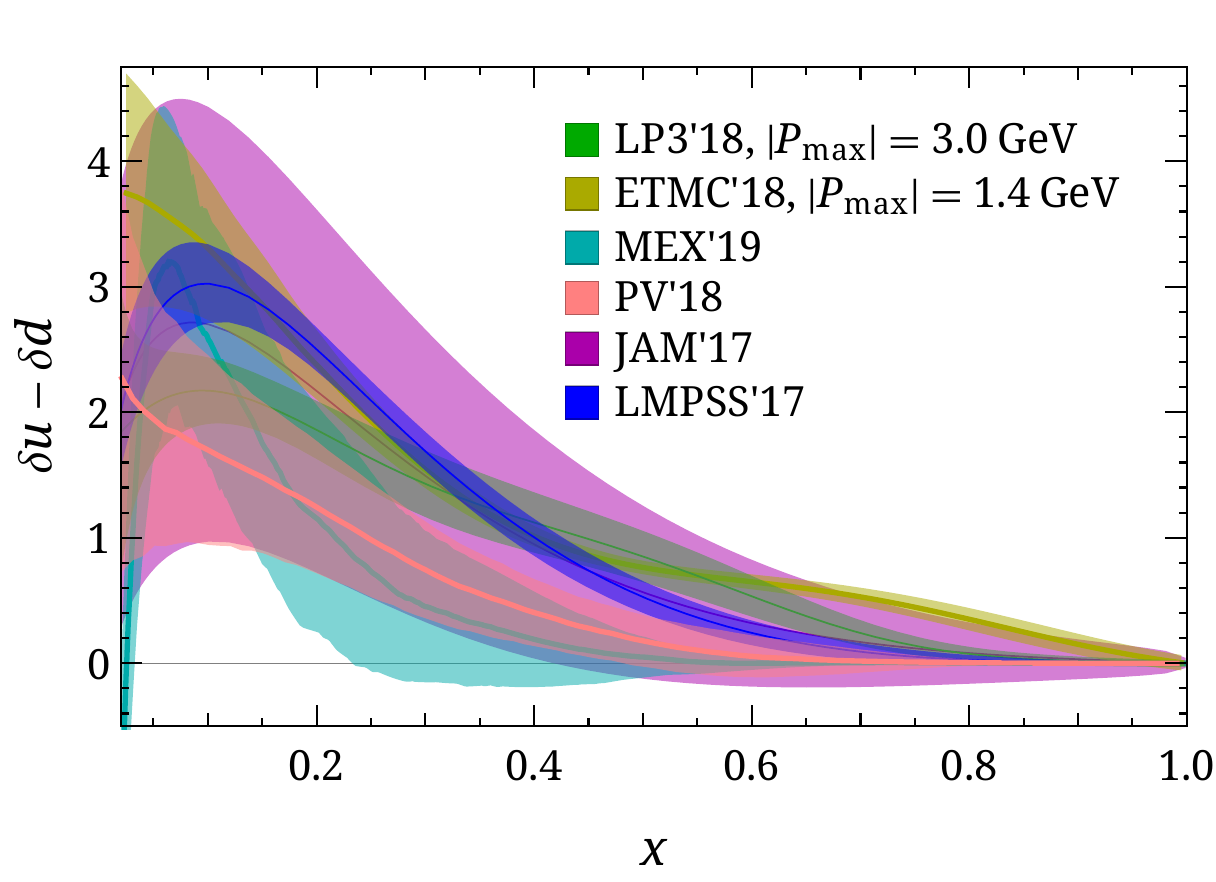}
\includegraphics[width=.46\textwidth]{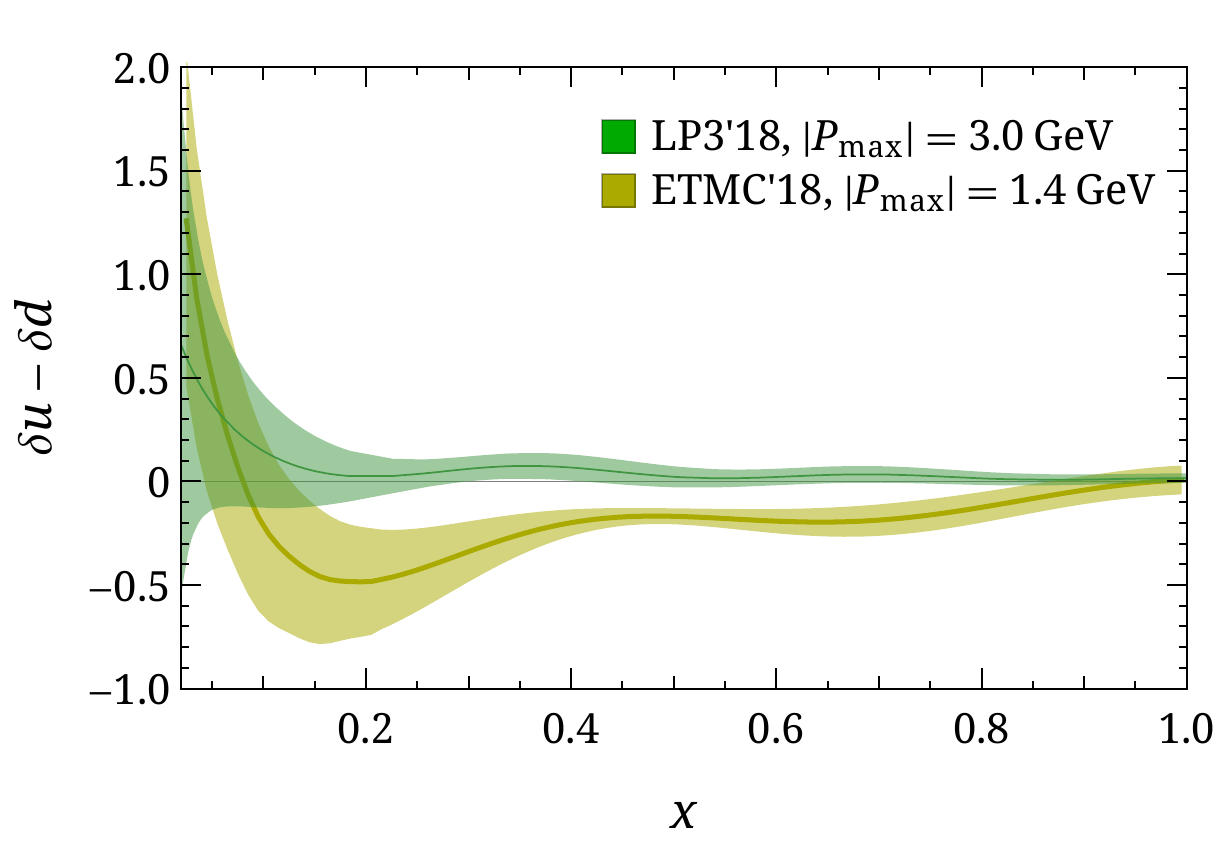}\\
\caption{
Summary of the lattice calculation of isovector unpolarized (top), helicity (middle) and transversity (bottom) 
with $LP^3$ and ETMC isovector quark (left column) and antiquark (right column) taken from~\cite{Chen:2018xof,Lin:2018qky,Liu:2018hxv,Alexandrou:2018pbm,Alexandrou:2018eet,Alexandrou:2019lfo,Bhat:2020ktg}, JLab/W\&M valence unpolarized distribution results from~\cite{Joo:2020spy}, and global fits from~\cite{Alekhin:2017kpj,Ball:2017nwa,Accardi:2016qay} (unpolarized), \cite{deFlorian:2009vb,Nocera:2014gqa,Ethier:2017zbq} (helicity), and~\cite{Lin:2017stx,Radici:2018iag,Benel:2019mcq} (transversity). Note that none of the current lattice calculations have taken the continuum limit ($a\rightarrow 0$) and have remaining lattice artifacts (such as finite-volume effects); disagreement in the obtained distributions is not unexpected.}
\label{fig:GlobalLatPDF}
\end{figure*}

The first unpolarized PDFs at the physical pion mass~\cite{Alexandrou:2017dzj,Lin:2017ani} using the quasi-PDFs approach were determined using small momentum. This, in addition with the challenges in the reconstruction of the $x$ dependence, may lead to the wrong sign of sea-flavor asymmetry to be seen following Fourier transformation. The upper plot of Fig.~\ref{fig:GlobalLatPDF} shows the newer PDF results on ensembles at the physical pion mass with momenta above 2~GeV, and then renormalized at 3~GeV~\cite{Lin:2018qky} (denoted as matched PDF). The associated error band includes systematic errors coming from variations in the renormalization scale, from the choice of $zP_z$ in the Fourier transform, from an estimation of lattice-spacing and finite-volume effects, from other nucleon matrix corrections, and from approximations made in the matching formula. As expected from the Fourier-transform study, the sea-flavor asymmetry is recovered with increased momentum. In the positive isovector quark region, the lattice results agree with CT14~\cite{Dulat:2015mca}, which is consistent with NNPDF3.1~\cite{Ball:2017nwa} and CJ15~\cite{Accardi:2016qay}, up to the small-$x$ region, where even larger $zP_z$ data are required for lattice calculations to have control over these regions.
The middle plot of Fig.~\ref{fig:GlobalLatPDF} shows LP$^3$'s isovector quark helicity PDF~\cite{Lin:2018qky} matched in the $\overline{\text{MS}}$-scheme at the scale $\mu=3$~GeV, extracted from LaMET at the largest proton momentum (3~GeV). The inner band denotes the statistical error, while the outer band also includes systematic uncertainties similar to those determined in the case of unpolarized PDFs. The lattice result is compared to two phenomenological fits, NNPDFpol1.1~\cite{Nocera:2014gqa} and JAM17~\cite{Ethier:2017zbq}. 
The bottom plot in Fig.~\ref{fig:GlobalLatPDF} shows the LP$^3$'s proton isovector transversity PDF~\cite{Liu:2018hxv} at the renormalization scale $\mu=\sqrt{2}$~GeV ($\overline{\text{MS}}$ scheme), extracted from lattice QCD and LaMET at $P_z=3$~GeV. The error band includes statistical errors (which fold in the excited-state uncertainty) and the same systematics mentioned in the unpolarized PDF case. The lattice result is compared to global fits by JAM17 and LMPSS17~\cite{Lin:2017stx}.

\begin{table*}[!t]
\begin{center}
\hspace*{-0.75cm}
\begin{tabular}{|c|c|c|c|c|c|c|c|c|}
\hline
Ref. & Sea quarks & Valence quarks  &  $N_{\Delta t}$ & method & $P_\text{max}$ (GeV) & a (fm) & $M_\pi$ (MeV) & $M_\pi L$  \\ \hline
ETMC'20 & 2f  twisted mass & twisted mass &  4 & pseudo-PDF & 1.38 & 0.09& 130 & 3.0\\
JLab/W\&M & 2+1 clover & clover & n/a & pseudo-PDF & 3.29 & 0.09& 172--358 & 5.08--5.47 \\
ETMC'18 & 2f  twisted mass & twisted mass &  4 &  quasi-PDF & 1.38 & 0.09& 130 & 3.0\\
LP3'18 & 2+1+1f HISQ & clover & 4 & quasi-PDF & 3 & 0.09 & 135& 4.0\\
LP3'17  & 2+1+1f HISQ & clover & 2 & quasi-PDF & 1.3   & 0.09 & 135& 4.0\\
\hline 
\end{tabular}
\end{center}
\caption{\label{tab:LQCD-PDF-parameters} The lattice parameters used in calculations of $x$-dependent PDFs near or extrapolated to physical pion mass. $N_{\Delta t}$ indicates the number of source-sink separation used in the lattice calculation to control the excited-state systematics which can be significant for nucleon structure. The work of the JLab/W\&M~\cite{Joo:2020spy} group uses the Feynman-Hellmann theorem, where the matrix elements are extracted from two-point--like lattice correlators, and $N_{\Delta t}$ does not apply in their approach. The references for each work are as follows:
ETMC'20~\cite{Bhat:2020ktg},
JLab/W\&M~\cite{Joo:2020spy},
ETMC'18~\cite{Alexandrou:2018pbm,Alexandrou:2018eet,Alexandrou:2019lfo},
LP3'18~\cite{Chen:2018xof,Lin:2018qky,Liu:2018hxv},
LP3'17~\cite{Lin:2017ani}.
}
\end{table*}

The ETMC Collaboration has also improved their calculations of PDFs, and refined the matching procedure by proposing a modified $\overline{\text{MS}}$ scheme, which satisfies particle number conservation. Note that this scheme corresponds to the quasi-PDFs, while the light-cone PDFs (see Eq.~\eqref{eq:PDF_def}) are always renormalized in the $\overline{\text{MS}}$ scheme. The modified $\overline{\text{MS}}$ scheme was applied to the unpolarized and helicity PDFs in Ref.~\cite{Alexandrou:2018pbm} and for the transversity PDF. The latter was presented for the first time at the physical point in Ref.~\cite{Alexandrou:2018eet}. The ensemble used has $N_f=2$ twisted-mass fermions with clover improvement~\cite{Abdel-Rehim:2015pwa}. The ensemble has a lattice spacing 0.093~fm, lattice spatial extent 4.5~fm, and pion mass 130~MeV. Three values were used for the momentum boost: 0.83, 1.11, and 1.38~GeV. In this work, increasing the momentum to values higher than 1.5~GeV led to an unreliable extraction of the ground state. The final PDFs are displayed in Fig.~\ref{fig:GlobalLatPDF} at scale 1.38~GeV and are compared with a selection of results from global fits. The unpolarized PDF has slope similar to the phenomenological one in the positive-$x$ region, but lies above it. The lattice results for the helicity PDF are compatible with phenomenological PDFs for $x\lesssim 0.4$--0.5. The lattice result for the transversity PDF is in agreement with the global fits, as well as with those constrained using the lattice determination of the tensor charge as input, for $x\lesssim 0.4$--0.5. Interestingly, the statistical accuracy of the aforementioned lattice results is better than either phenomenological determination.

Furthermore, ETMC studied several sources of systematic uncertainties directly at the physical point~\cite{Alexandrou:2019lfo}. This is essential, because the pion mass dependence is sizable for the nonlocal matrix elements (see, for example, Fig.~6 of Ref.~\cite{Alexandrou:2018pbm}). All sources of systematic uncertainties that can be studied were included in Ref.~\cite{Alexandrou:2019lfo}, that is: excited-state contamination, renormalization, reconstruction of the $x$-dependence, effects of truncating the Fourier transform, and different matching prescriptions. This in-depth study revealed that excited-state contamination is not negligible for fast-moving nucleons and should be properly reduced. The work also demonstrated that the discrete and limited number of data for nonzero $z$ are not sufficient for a reliable reconstruction of the $x$-dependence of PDFs. Antiquarks were particularly affected, because the sign of the sea asymmetry is highly prone to systematic uncertainties. 

\begin{figure}[!t]
\includegraphics[width=.44\textwidth]{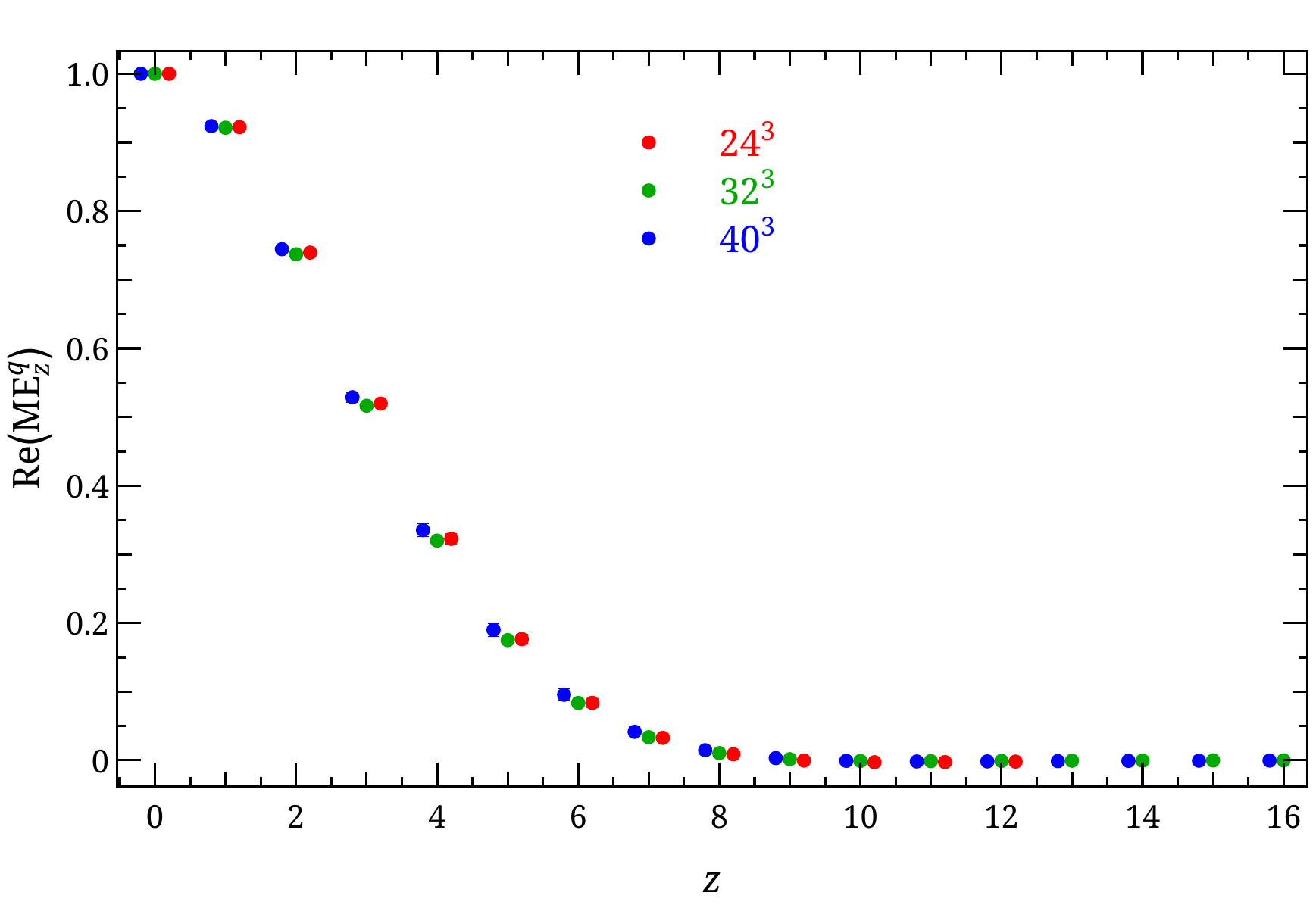}
\caption{The normalized isovector nucleon matrix elements for unpolarized PDFs $P_z \approx 1.3$~GeV as functions of $z$ at the three volumes ($M_\pi L =3.3$, 4.4 and 5.5 indicated by red, green and blue, respectively)~\cite{Lin:2019ocg}. 
\label{fig:FVPDF}}
\end{figure}

The first study of systematic uncertainties arising from finite-volume effects for the quasi-distributions was reported in Ref.~\cite{Lin:2019ocg} using a 220-MeV pion mass with $M_\pi^\text{val} L \approx 3.3$, 4.4 and 5.5, respectively. After a careful extraction of the bare matrix elements for the unpolarized and polarized distributions (see the right-hand side of Fig.~\ref{fig:FVPDF}), no volume dependence was observed in these ensembles within the statistical error. It was therefore concluded that finite-volume dependence does not play a significant role for the boosted nucleon matrix elements ($P_z\approx 1.3$~GeV larger) used for quasi-distributions within the range $M_\pi^\text{val} L \in \{3.3, 5.5\}$.

To calculate the small-$x$ ($x < 0.01$) behavior of the PDFs using existing lattice $x$-dependent methods, one needs to pursue boost momenta $P_z$ an order of magnitude higher than those currently being used. The minimum $x$ obtainable from lattice QCD is connected to 
the product of the length of the Wilson line and momentum boost, $zP_z$. 
It is also expected that the size of the momentum boost is connected to the reliability of the lattice results in the various $x$ regions. For example, higher values of $P_z$ result in fast decay of the matrix elements, and therefore, the truncation of the maximum value of $z$ in the Fourier transform matters less. There are a number of proposals to avoid to Fourier transformation by working in position space; this would work in an ideal world where there is arbitrarily precise data throughout the large-$zP_z$ region. However, in reality, the lattice data taken in the small-$zP_z$ region are not precise enough to even discern whether the parton distribution is flat across all $x$; one loses sensitivity to very distinct distributions in $x$-space, which appear very similar in $zP_z$ space. Furthermore, one still needs large boost momenta $zP_z$ to reliably obtain the distribution in the small-$x$ region, and very fine lattice spacing, $a<0.02$~fm to reduce the lattice discretization systematic error that goes like $(P_z a)^n$. There have been some exploratory studies of how to get around the problem: Ref.~\cite{Zhang:2019qiq} applied machine-learning algorithms to make predictions for different types of lattice LaMET data, such as kaon PDFs, meson distribution amplitudes and gluon PDFs. The data correlation and predictions of higher momenta and their dependence on the amount of training and bias-correction data, and machine-learning model parameters were studied in great detail. More studies done in this direction are mandatory to revolutionize the investigation of the small-$x$ nucleon structure.

\subsubsection{PDFs at the physical point with pseudo-distributions}

Besides quasi-PDFs, additional novel methods have been proposed for the extraction of lightcone distributions~\cite{Aglietti:1998ur,Detmold:2005gg,Braun:2007wv,Ma:2014jla,Ma:2017pxb,Liu:1993cv,Chambers:2017dov,Bali:2018spj,Detmold:2018kwu,Sufian:2019bol,Liang:2019frk}. One of these methods is the so-called pseudo-distributions approach, originally proposed by Radyushkin~\cite{Radyushkin:2016hsy,Radyushkin:2017cyf,Radyushkin:2017lvu,Radyushkin:2017sfi,Radyushkin:2018cvn,Radyushkin:2018nbf,Radyushkin:2019owq,Radyushkin:2019mye}. The raw data are the same as in the quasi-PDFs approach, but the analysis differs in several ways. In the pseudo-PDFs approach, the matrix elements are expressed in terms of $z^2$ and of the Ioffe time~\cite{Ioffe:1969kf} $\nu\equiv -p\cdot z$ (where $p$ is the nucleon momentum boost), and are called Ioffe-time distributions. One of the main differences between quasi- and pseudo-PDFs is the reconstruction of the $x$-dependence of the distributions: quasi-PDFs are defined as the Fourier transform of the matrix elements in $z$, while pseudo-PDFs as the transform in Ioffe time. As a consequence, pseudo-distributions have only canonical support ($ x\in[-1:1]$), while quasi-PDFs are allowed to be nonzero outside the canonical $x$-range. Another major difference of practical value is that the Ioffe time can increase by either increasing $z$ or $p$. Thus, lattice data at small values of the momentum boost are also desirable. 

The pseudo-distribution approach was investigated on the lattice soon after it was proposed. The valence unpolarized PDF of the nucleon was studied first in the quenched setup~\cite{Orginos:2017kos}, and later using dynamical simulations at $m_\pi=400$~MeV~\cite{Joo:2019jct}. The pion PDFs were also studied in Ref.~\cite{Joo:2019bzr}. Other work related to the pseudo-PDFs approach on the lattice can be found in Ref.~\cite{Karpie:2018zaz,Karpie:2019eiq}. A recent calculation including an ensemble near physical pion mass (170~MeV) is presented in Ref.~\cite{Joo:2020spy}. The maximum value of $P_3$ used for the ensembles with pion mass 358, 278, and 179~MeV is 2.46, 3.29, and 2.12~GeV, respectively.
In Fig.~\ref{fig:GlobalLatPDF} we show the valence PDF extrapolated to the physical pion mass using the aforementioned ensembles, as explained in Ref.~\cite{Joo:2020spy}. It is important to emphasize that, unlike the case of quasi-PDFs, the final PDFs do not rely on a sole value of $P_3$. The relevant parameter for the reconstruction of the PDFs is the Ioffe time, which can be increased by increasing either $z$ or $P_3$. Typically, the data at large $P_3$ are more noisy, and, therefore, do not drive the values of the final PDFs. The lattice results are compared to various determinations from global fits. The agreement of the former with the latter is excellent in the region $x<0.25$, while it deteriorates at large $x$. This discrepancy may be due to an underestimation of systematic uncertainties in the lattice result, which deserves further investigation.

Another notable lattice calculation using the pseudo-distribution approach, with simulations directly at the physical point, was presented in Ref.~\cite{Bhat:2020ktg}. A novel aspect is the extraction not only of the valence distribution $q_v(x)$, but also of the combination with antiquarks $q_v(x)+2\bar{q}(x)$. This led to the extraction of the total and sea-quark distributions, $q(x)=q_v(x)+\bar{q}(x)$ and $q_s(x)=\bar{q}(x)$. The lattice results are presented in Fig.~\ref{fig:GlobalLatPDF}. The purple band represents the statistical uncertainties of the lattice result, and the cyan band the combination of statistical and systematic uncertainties. Some of these uncertainties were computed explicitly (for example, the reconstruction of the $x$-dependence), while others (for example, for volume and discretization effects) an estimate was given, see Ref.~\cite{Bhat:2020ktg} for details. The lattice result is compared to a global fit of unpolarized PDFs, NNPDF3.1, with which it is in very good agreement. This fact demonstrates that a realistic determination of unpolarized PDFs from the lattice is possible.

\subsubsection{Twist-3 PDFs}

A new direction is currently being pursued with lattice calculations of the $x$-dependence of twist-3 PDFs, which are important to characterize the structure of hadrons beyond the leading twist (they contain information on quark-gluon-quark correlations~\cite{Balitsky:1987bk,Kanazawa:2015ajw}). The first calculation was completed recently~\cite{Bhattacharya:2020cen} for the twist-3 $g_T(x)$ distribution, including the computation of the one-loop matching kernel~\cite{Bhattacharya:2020xlt}. The quasi-distribution approach is used with three values of the momentum boost (0.83, 1.25, and 1.67~GeV). The pion mass of the ensemble is 260~MeV. This calculation not only predicts the $x$-dependence of $g_T(x)$ but serves as a test of the Wandzura-Wilczek (WW) approximation~\cite{Wandzura:1977qf}. In this approximation, the twist-3 $g_T(x)$ distribution may be obtained from its twist-2 counterpart $g_1(x)$
\begin{equation}
\label{eq:gT_WW}
g_T^\text{WW}(x)=\int_x^1 \frac{dy}{y} g_1(y)\,.
\end{equation}

\begin{figure}[!t]
\includegraphics[scale=0.5]{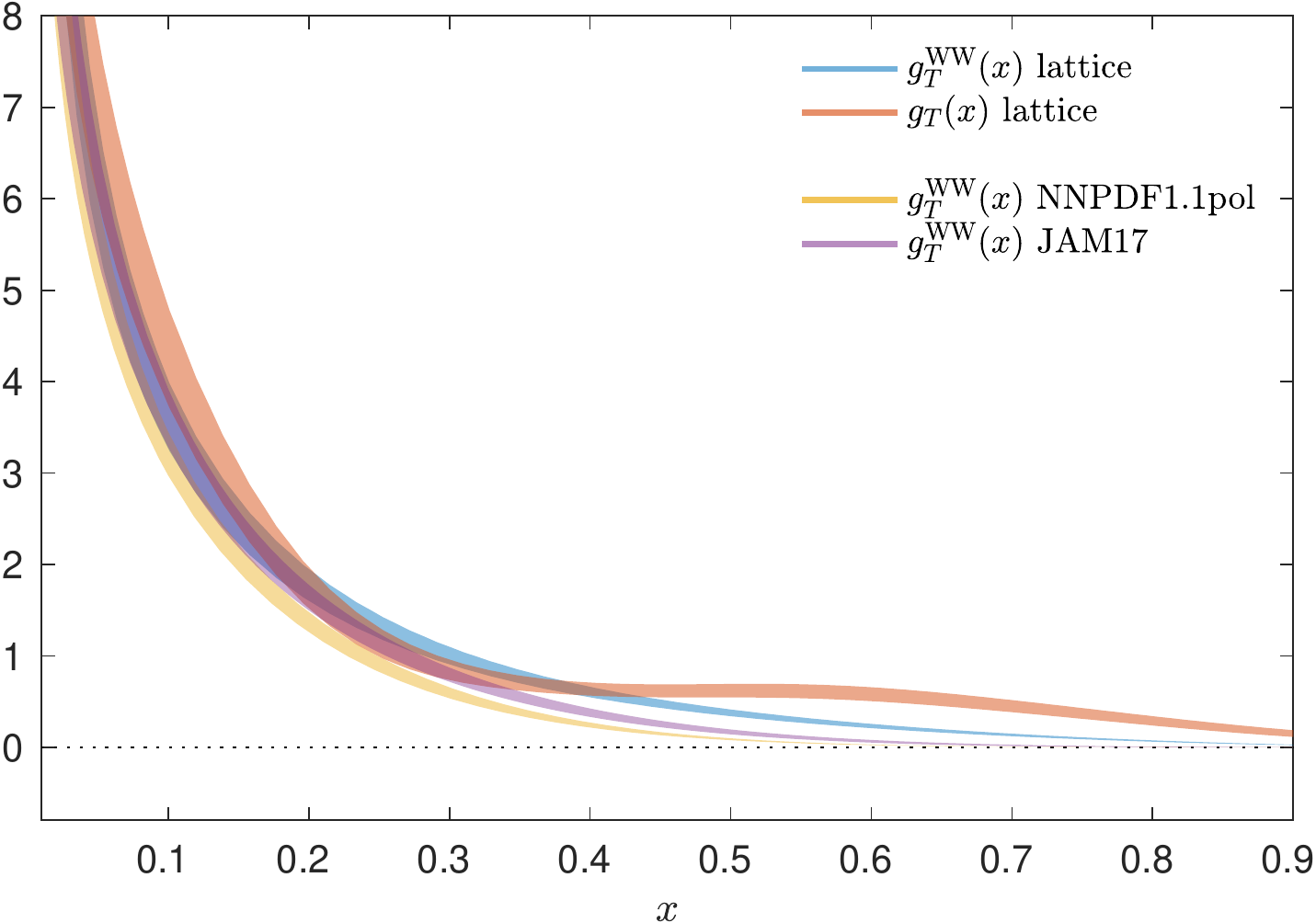}
	\caption{Comparison of $g_T(x)$ from lattice (red band) with its WW approximations: lattice-extracted $g_T^{\rm WW}$ (blue band) and calculated from global fits (NNPDF1.1pol~\cite{Nocera:2014gqa} orange band, JAM17~\cite{Ethier:2017zbq} purple band). The proton momentum is $P_3=1.67$ GeV.}
 	\label{fig:gT_WW}
\end{figure}

In Ref.~\cite{Bhattacharya:2020cen}, $g_T^\text{WW}$ is evaluated using the lattice data as a function of $x$, as shown in Fig.~\ref{fig:gT_WW} for $P_3=1.67$~GeV. It is also compared to the actual data of $g_T(x)$, and is found to be consistent for a large range of $x$. The large-$x$ region reveals some tension, which may be due to systematic uncertainties. It is also possible that this is an indication of violations of the WW approximation. It is also interesting to compare the lattice result with the WW approximation obtained using $g_1$ from global fits~\cite{Nocera:2014gqa,Ethier:2017zbq}. Good agreement is observed up to $x\approx 0.3$, while the discrepancy for larger $x$ values indicates possible systematic effects.

\subsubsection{Gluon PDFs}

The unpolarized gluon PDF is defined by the Fourier transform of the lightcone correlation in the hadron, 
\begin{eqnarray}\label{eq:LCgluon}
 g(x,\mu^2)&=&\int\frac{\text{d}y^-}{\pi x}e^{-ix y^-P^+} \enskip \times \nonumber\\
        &\times & \ \   \left\langle P|F^+_{\mu}(y^-)\mathcal{W}(y^{-},0)F^{\mu +}(0)|P\right\rangle \ ,
\end{eqnarray} 
where $y^{\pm}=\frac{1}{2}(y^0\pm y^3)$ is the spacetime coordinate along the lightcone direction, $|P\rangle$ is the hadron state with momentum $P_{\mu}=(P_0,0,0,P_z)$ and normalization $\langle P|P\rangle=1$, $\mu$ is the renormalization scale, $F_{\mu\nu}=T^aG^a_{\mu\nu}=T^a(\partial_{\mu}A^a_{\nu}-\partial_{\nu}A^a_{\mu}-gf^{abc}A^b_{\mu}A^c_{\nu})$ is the gluon field tensor, and $\mathcal{W}(y^{-},0)=\mathcal{P}\exp(-ig\int^{y^-}_0\text{d}\eta^-A^+(\eta^-))$ is the lightcone Wilson link from $y^{+}$ to 0 with $A^+$ as the gluon potential in the adjoint representation. Unfortunately, these time-separated and nonlocal operators cannot be directly calculated using lattice QCD.

\begin{figure}[!t]
\includegraphics[width=.45\textwidth]{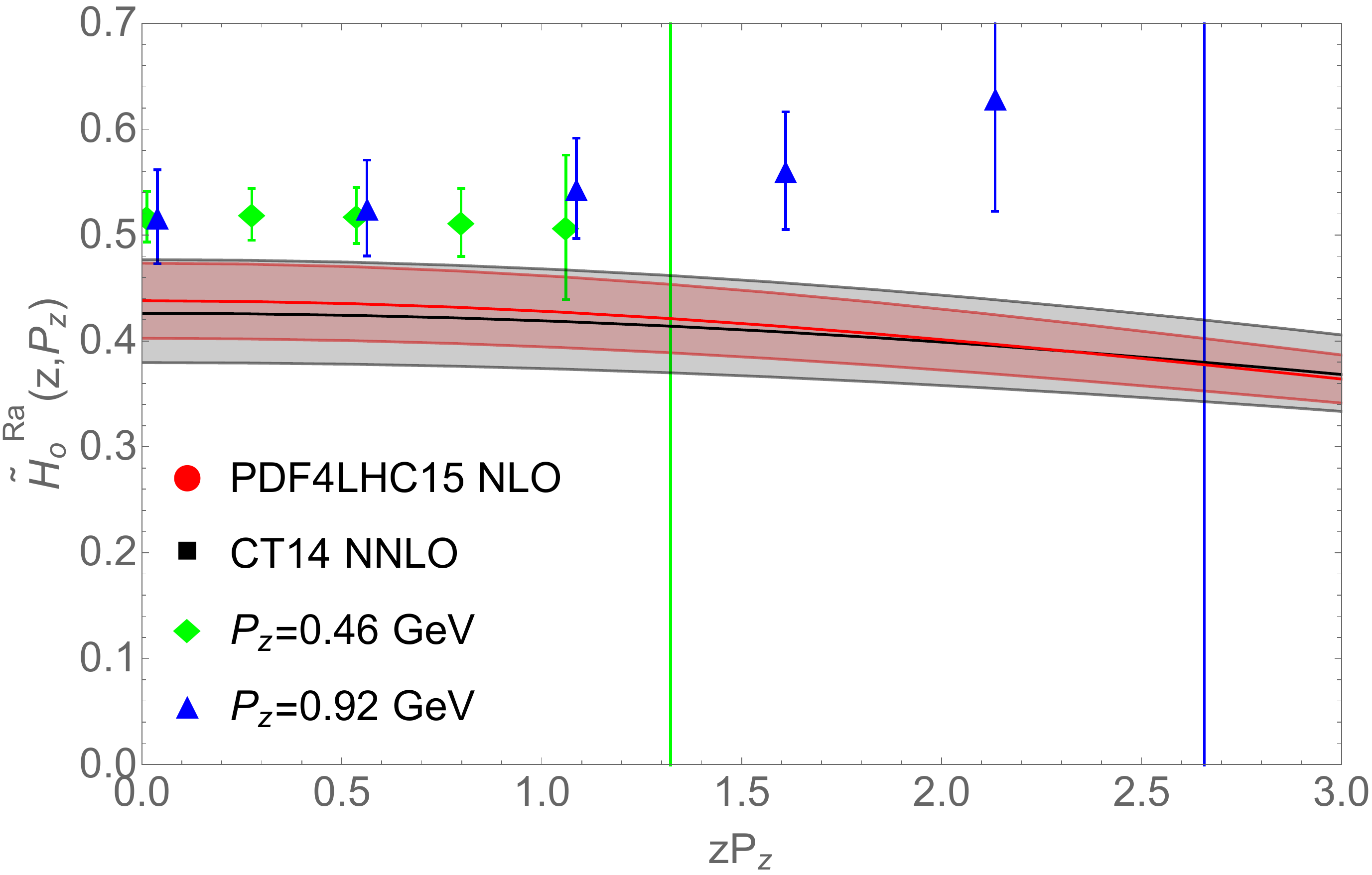}
\includegraphics[width=.45\textwidth]{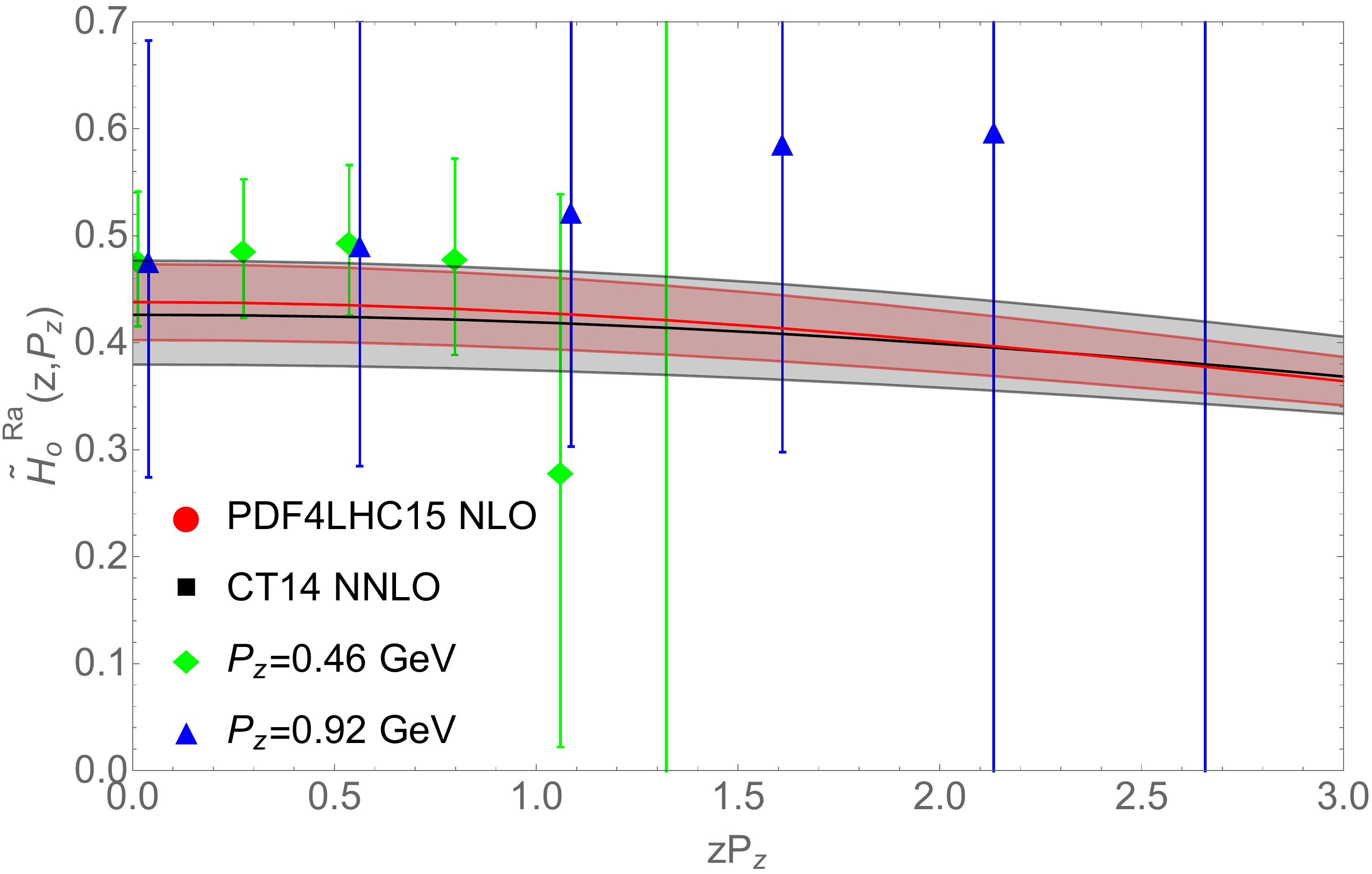}
\caption{The final results of gluon PDF matrix elements at 678~MeV (top) and 340~MeV (bottom) pion mass as functions of $zP_z$, compared with the FT of the gluon PDF from the global fits PDF4LHC15~\cite{Butterworth:2015oua} and CT14~\cite{Dulat:2015mca}. 
\label{fig:gluonPDF}}
\end{figure}

In the first exploratory study which applied the quasi-PDF approach to the gluon PDFs~\cite{Fan:2018dxu}, ensembles with unphysically heavy quark masses were used. Since gluon quantities are much noisier than quark disconnected loops, calculations with very high statistics are necessary to reveal a signal. The calculations were done using overlap fermions on gauge ensembles with 2+1 flavors of domain-wall fermion and spacetime volume $24^3\times 64$, $a=0.1105(3)$~fm, and $M_\pi^\text{sea}=330$~MeV. The gluon operators were calculated for all volumes and high statistics: 207,872 measurements were taken of the two-point functions with valence quarks at the light sea and strange masses (corresponding to pion masses 340 and 678~MeV, respectively).
The coordinate-space gluon quasi-PDF matrix element ratios are plotted in Fig.~\ref{fig:gluonPDF}, and compared to the corresponding Fourier transform of the gluon PDF based on two global fits at NLO: the {\sc PDF4LHC15} combination~\cite{Butterworth:2015oua} and the {\sc CT14} set~\cite{Dulat:2015mca}. Up to perturbative matching and power corrections $O(1/P_z^2)$, the lattice results are compatible with global fits within the statistical uncertainty at large $z$. The results at the lighter pion mass (at the unitary point) of 340~MeV are also shown in Fig.~\ref{fig:gluonPDF}. These are consistent with those from the strange point but have larger uncertainties. The gluon quasi-PDFs in the pion were also studied for the first time in Ref.~\cite{Fan:2018dxu} and features similar to those observed for the proton were revealed. Finally, there have been recent developments in improving the operators for the gluon-PDF lattice calculations~\cite{Balitsky:2019krf,Wang:2019tgg,Zhang:2018diq}, which will allow us to take the continuum limit for the gluon PDFs in future lattice calculations.

\subsubsection{Hadronic Tensor}

The hadronic tensor
\begin{equation}
\label{HT1}
W_{\mu\nu}=\frac{1}{4\pi}\int d^{4}ze^{iq\cdot z}\left\langle P,S\left|\left[J_{\mu}^{\dagger}(z)J_{\nu}(0)\right]\right|P,S\right\rangle
\end{equation}
characterizes the nonperturbative nature of the nucleon in scattering processes involving
nucleons such as DIS. It can be further decomposed into structure functions~\cite{Blumlein:2012bf}, which are commonly used to determine the PDFs in global fits through factorization theorems. A lattice calculation of the hadronic tensor provides a first-principle method to study PDFs as well as physical processes like the low-energy neutrino-nucleon scattering ~\cite{Liu:1993cv,Liu:1999ak,Liu:2016djw,Liang:2017mye,Liang_2020}.
This approach has several advantages. The hadronic tensor is scale independent, such that no renormalization is needed except for the finite lattice normalization, if local vector and axial currents are used for the lattice calculation. Furthermore, the structure functions are frame independent, therefore, no highly boosted nucleons are needed.

In the hadronic tensor approach, additional parton degrees of freedom are revealed graphically, among which, the connected sea (CS) antipartons which account for the Gottfried sum-rule violation~\cite{Liu:1993cv,Liu:2012ch}. It is worthwhile pointing out that the quasi-PDF in the negative $x$ region in the connected insertion is the CS antiparton in the hadronic tensor approach due to the fact that it is the antiparton in the connected insertion which is not affected by the gluon and disconnected sea (DS) distributions in evolution. Besides, in the practical lattice calculation, some diagrams (Wick contractions) contain only higher twist contributions, a fact that offers the possibility of studying higher twist effects on the lattice.

Numerically, a substantial challenge of this approach is to convert the hadronic tensor from the Euclidean space back to the Minkowski space, which involves solving an inverse problem of a Laplacian transform~\cite{Liu:2016djw,Liang:2017mye,Liang_2020}. In order to tackle this problem, three methods (i.e., the Backus-Gilbert method~\cite{BG1,Backus123,Hansen:2017mnd}, the maximum entropy method~\cite{rietsch1977maximum,Asakawa:2000tr} and the Bayesian reconstruction method~\cite{Burnier:2013nla}) have been implemented and tested~\cite{Liang:2017mye,Liang_2020}. It is believed that the Bayesian reconstruction method and its further improved version~\cite{Fischer:2017kbq,Kim:2018yhk} is the best method so far to solve the problem. On the other hand, the inverse problem is a common problem in PDF calculations, and can be simplified or avoided by using model-inspired fitting functions.

\begin{figure}[!t]
    \includegraphics[width=\linewidth]{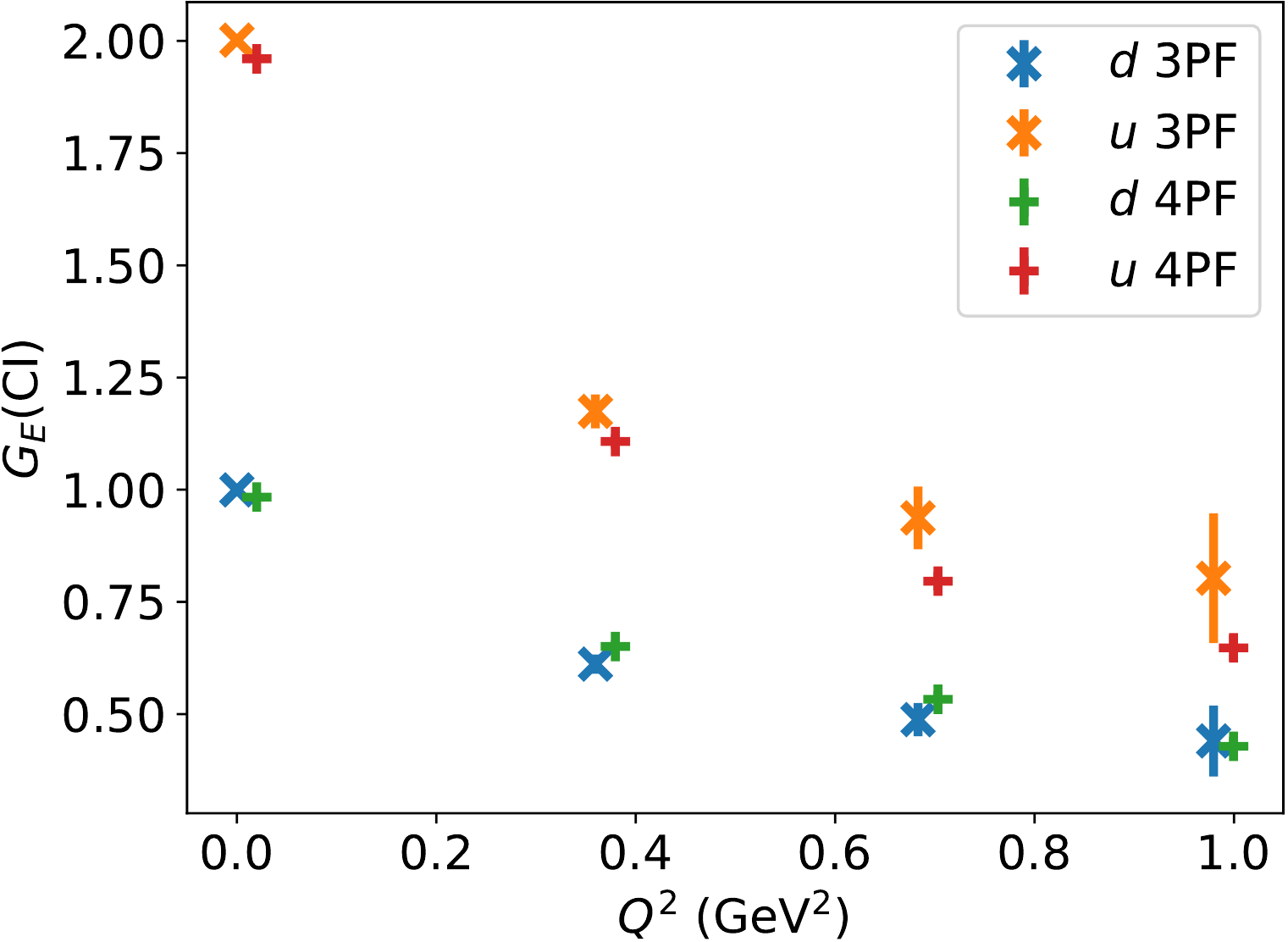}
    \caption{\small The electric form factors (connected insertions only) calculated by means of normal 3-point functions (3PF) and 4-point functions 
    (4PF) for both $u$ and $d$ quarks. \label{FF_comparison}}
\end{figure}

Another challenge of this approach is to access high momentum and energy transfers in such a way that the calculation can be extended into the DIS region. Numerical tests~\cite{Liang_2020} show that small lattice spacings are essential for this purpose. The need of fine lattices appears to be a common problem faced by the lattice PDF community. For the hadronic tensor calculation, one important point is that it works in all the energy ranges (from elastic scattering to inelastic scattering and on to DIS). Before lattices with very fine spacing are available, the hadronic tensor can be used to study the nucleon elastic form factors and low-energy scatterings such as the neutrino-nucleon scattering which is of significant physical importance. Fig.~{\ref{FF_comparison}} shows a numerical check of this approach in terms of form factors on the RBC-UKQCD 32Ifine lattice~\cite{Blum:2014tka}. The structure function of the elastic scattering from the hadronic tensor, a 4-point function, is the product of the elastic nucleon form factors for the currents involved. Fig.~{\ref{FF_comparison}} shows that the the electric form factors (connected insertions only) calculated by means of the  3-point functions for both $u$ and $d$ quarks are found to be consistent with errors with those deduced from the hadronic tensor. 

This lays a solid foundation for further calculations. Currently, lattices with lattice spacing $\sim$0.04~fm are suitable, for instance, to study the neutrino-nucleus scattering for DUNE experiments where the neutrino energy is between $\sim1$ to $\sim7$ GeV. At the same time, these lattices can also be used to explore the pion PDFs.In the future, working on lattices with lattice spacing of 0.03~fm or even smaller would be desirable for studying the nucleon substructure.

%% file: 02sec4intersection.tex
\goodbreak
\subsection{Intersection of Lattice and Global PDF Fits}
\label{02sec4}

In this Section, we review some aspects of potential interplay between lattice QCD computations and PDF fits. First, we update the benchmark of lattice QCD and global fit results for the lowest moments of unpolarized and helicity PDFs, in continuity with the previous ~\cite{Lin:2017snn}, and extend it to transversity. Second, we investigate the usage of lattice QCD data in fits of unpolarized PDFs. Third, we discuss how lattice QCD computations can shed light on the poorly known unpolarized and helicity strange PDFs.

\subsubsection{Moments}

One of the main outcomes of the 2017 PDFLattice white paper was a detailed comparison between lattice QCD and global fit results for the (lowest) moments of unpolarized and helicity PDFs. Specifically, we identified a set of benchmark moments, we appraised the various results available for them in the literature, and we provided corresponding benchmark values for lattice QCD and global fits. In the following, we present an update of this exercise.

\begin{table*}[!t]
\centering
\begin{tabularx}{\linewidth}{|c|X|X|l|}
\hline
& {\raisebox{-0.8pt}{\scalebox{1}{\color{blue}$\bigstar$}}}
& \bcirc
& \rsquare\\
\hline
DE & At least three lattice spacings$^a$ with at least two lattice spacings below 0.1 fm and a range of          lattice spacings that satisfies $[a_{\rm max}/a_{\rm min}]^2\geq 2$
   & At least two lattice spacings$^a$ with at least one point below 0.1 fm and a range of lattice spacings that satisfy $[a_{\rm max}/a_{\rm min}]^2\geq 1.4$
   & Otherwise\\
\hline
CE & One ensemble with a physical pion mass$^b$ {\it or} a chiral extrapolation with three or more pion masses,    with at least two pion masses below 250 MeV and at least one below 200 MeV
   & A chiral extrapolation with three or more pion masses$^b$, two of which are below 300 MeV.
   & Otherwise\\
\hline
FV & Ensembles with $M_{\pi,{\rm min}}L\geq 4$ {\it or} at least three volumes with spatial extent $L>2.5$       fm$^c$
   & Ensembles with $M_{\pi,{\rm min}}L\geq 3.4$ {\it or} at least two volumes with spatial extent      $L>2.5$ fm$^c$
   & Otherwise\\
\hline
RE & Non-perturbative renormalization$^d$
   & Perturbative renormalization (one loop or higher)
   & Otherwise \\
\hline
ES & At least three source-sink separations or a variational method to optimize the operator derived from at     least a 3$\times$3 correlator matrix, at every pion mass and lattice spacing
   & Two source-sink separations at every pion mass and lattice spacing, or three or more source-sink  separations at one pion mass below 300~MeV. For the variational method, an optimized operator        derived from a 2$\times$2 correlator matrix at every pion mass and lattice spacing, or a 3$\times$3   correlator matrix for one pion mass below 300~MeV
   & Otherwise \\
\hline
\end{tabularx}
\begin{tabularx}{\linewidth}{X}
$^a$ We assume that the lattice actions are $\mathcal{O}(a)$-improved, {\it i.e.} that the discretization errors vanish quadratically with the lattice spacing. For unimproved actions, an additional lattice spacing is required. These criteria must be satisfied in each case for at least one pion mass below 300~MeV. To receive 
a \bstar~or a \bcirc~either a continuum extrapolation must be performed, or the results must demonstrate no significant discretization effects over the appropriate range of lattice spacings.\\
$^b$ We define a physical pion mass ensemble to be one with $M_\pi=135\pm 10$~MeV for the above criteria.\\
$^c$ For calculations that use a mixed-action approach, {\it i.e.} with different lattice actions for the valence and sea quarks, we apply these criteria to the valence quarks; $M_{\pi,{\rm min}}$ is the lightest pion mass employed in the calculation.\\
$^d$ For $g_A$ we award a \bstar~also to calculations that use fermion actions for which $Z_A/Z_V=1$ or employ combinations of quantities for which the renormalization is unity by construction.\\
\end{tabularx}
\caption{\small Summary of the rating criteria for lattice QCD computations of PDF moments, see also~\cite{Lin:2017snn}.}
\label{tab:criteria}
\end{table*}

We consider the following benchmark moments of unpolarized, helicity and transversity PDFs, respectively
\begin{equation}
\langle x \rangle_{u^+ - d^+}\,,\quad
\langle x \rangle_{u^+}\,,\quad
\langle x \rangle_{d^+}\,,\quad
\langle x \rangle_{s^+}\,,\quad
\langle x \rangle_g\,;
\label{eq:unp_moms}
\end{equation}
\begin{align}
\label{eq:hel_moms}
g_A & \equiv \langle 1 \rangle_{\delta u^- - \Delta d^+}\,,\quad
\langle 1 \rangle_{\delta u^-}\,,\quad
\langle 1 \rangle_{\Delta d^+}\,,\quad
\langle 1 \rangle_{\Delta s^+}\,,\quad\\
& \qquad\qquad\qquad\qquad \langle x \rangle_{\Delta u^- - \Delta d^-}\,;
\nonumber
\end{align}
\begin{equation}
g_T \equiv \langle 1 \rangle_{\delta u^- - \delta d^-}\,,\quad
\langle 1 \rangle_{\delta u^-}\,,\quad
\langle 1 \rangle_{\delta d^-}\,.
\label{eq:tra_moms}
\end{equation}
The moments of unpolarized and helicity PDFs, Eqs.~\eqref{eq:unp_moms}-\eqref{eq:hel_moms}, correspond to the benchmark quantities already identified in the 2017 PDFLattice white paper, and are expressed using the conventional notation described in Appendix~A of~\cite{Lin:2017snn}. We now define the moments of transversity PDFs in a similar way
\begin{equation}
g_T\equiv\langle 1\rangle_{\delta u^- - \delta d^-}
= 
\int_0^1 dx \left[ h_1^{u^-} - h_1^{d^-} \right]\,,
\end{equation}
\begin{equation}
\langle 1\rangle_{\delta u^-}
= 
\int_0^1 dx h_1^{u^-}\,,
\qquad
\langle 1\rangle_{\delta d^-}
= 
\int_0^1 dx h_1^{d^-}\,,
\label{eq:defmomh1}
\end{equation}
where $h_1^{u^-}=h_1^u-h_1^{\bar u}$ and $h_1^{d^-}=h_1^d-h_1^{\bar d}$. We do not focus on quantities other than those listed above, because current lattice calculations of higher moments and/or moments of other PDF combinations are not sufficiently mature to allow for a meaningful comparison between lattice QCD and global fit results. 

In the 2017 PDFLattice white paper we appraised each source of systematic uncertainty in the lattice QCD results by means of a rating system, defined in Sect.~(3.1.2) of~\cite{Lin:2017snn}. Specifically, we considered the following sources of systematic uncertainties: discretization effects and extrapolation to the continuum limit (DE); chiral extrapolation to unphysical pion masses (CE); finite-volume effects (FV); choice of renormalization (RE); and excited state contamination (ES). The rating system, which is inspired to that adopted by the Flavor Lattice Averaging Group {\sc FLAG})~\cite{Aoki:2019cca}, awards a blue star (\bstar) for sources of uncertainty that are well controlled or very conservatively estimated, a blue circle (\bcirc) for sources of uncertainty that have been controlled or estimated to some extent, and a red square (\rsquare) for uncertainties that have not met our criteria or for which no estimate is given. The details of the rating system are further summarized in Tab.~\ref{tab:criteria} for the reader's convenience. Our criteria and the corresponding ratings are chosen to provide as fair an assessment of the various calculations as possible, and are not intended to discredit the merits of any of them. In this respect, our criteria are aspirational: where lattice QCD results do not meet these standards, we hope that the lattice community will work towards improved calculations and greater precision. Modifications to this rating system will occur as the lattice QCD results evolve.

\paragraph{Moments of unpolarized PDFs.}

Modern analyses of unpolarized PDFs benefit from a wide data set (see Fig.~\ref{fig:kin_cov}) and are performed at NNLO perturbative accuracy in the strong coupling as standard in most cases. For this reason, the phenomenological knowledge of lattice-calculable quantities --- including unpolarized PDF moments --- is well established. The pulls of a typical phenomenological analysis of unpolarized PDFs on the corresponding moments was comprehensively evaluated and mapped in Ref.~\cite{Hobbs:2019gob}.

\begin{table*}[!t]
\begin{tabularx}{\linewidth}{lXXlccccccc}
  \toprule
  Moment & Collaboraton & Reference & $N_f$ 
  & DE
  & CE
  & FV
  & RE
  & ES
  &
  & Value \\
  \midrule
  $\langle x \rangle_{u^+-d^+}$
  & ETMC\,20
  & \cite{Alexandrou:2020sml}
  & 2+1+1
  & \rsquare
  & \bstar
  & \bcirc
  & \bstar
  & \bstar
  & $^{**}$
  & 0.171(18) \\
  & PNDME\,20
  & \cite{Mondal:2020cmt}
  & 2+1+1
  & \bstar
  & \bstar
  & \bstar
  & \bstar
  & \bstar
  &
  & 0.173(14)(07) \\
  & ETMC\,19
  & \cite{Alexandrou:2019ali}
  & 2+1+1
  & \rsquare
  & \bstar
  & \bcirc
  & \bstar
  & \bstar
  & $^{**}$
  & 0.178(16)\\
  & Mainz\,19
  & \cite{Harris:2019bih}
  & 2+1
  & \bstar
  & \bcirc
  & \bstar
  & \bstar
  & \bstar
  &
  & 0.180(25)($^{+14}_{-6})$\\
  & $\chi$QCD\,18
  & \cite{Yang:2018nqn}
  & 2+1
  & \bcirc
  & \bstar
  & \bcirc
  & \bstar
  & \bstar
  & 
  & 0.151(28)(29)\\
  &  ETMC\,19
  & \cite{Alexandrou:2019ali}
  & 2
  & \rsquare
  & \bstar
  & \bcirc
  & \bstar
  & \bstar
  & $^{**}$
  &  0.189(23) \\
  & RQCD\,18
  & \cite{Bali:2018zgl}
  & 2
  & \bstar
  & \bstar
  & \bcirc
  & \bstar
  & \bstar
  &
  & 0.195(07)(15)\\
  \midrule
  $\langle x\rangle_{u^+}$
  & ETMC\,20
  & \cite{Alexandrou:2020sml}
  & 2+1+1
  & \rsquare
  & \bstar
  & \bcirc
  & \bstar
  & \bstar
  & $^{**}$
  & 0.359(30) \\
  & $\chi$QCD\,18
  & \cite{Yang:2018nqn}
  & 2+1
  & \bcirc
  & \bstar
  & \bcirc
  & \bstar
  & \bstar
  &
  & 0.307(30)(18)\\
  \midrule
  $\langle x\rangle_{d^+}$
  & ETMC\,20
  & \cite{Alexandrou:2020sml}
  & 2+1+1
  & \rsquare
  & \bstar
  & \bcirc
  & \bstar
  & \bstar
  & $^{**}$
  & 0.188(19) \\
  & $\chi$QCD\,18
  & \cite{Yang:2018nqn}
  & 2+1
  & \bcirc
  & \bstar
  & \bcirc
  & \bstar
  & \bstar
  &
  & 0.160(27)(40)\\
  \midrule
  $\langle x\rangle_{s^+}$
  & ETMC\,20
  & \cite{Alexandrou:2020sml}
  & 2+1+1
  & \rsquare
  & \bstar
  & \bcirc
  & \bstar
  & \bstar
  & $^{**}$
  & 0.052(12) \\
  & $\chi$QCD\,18
  & \cite{Yang:2018nqn}
  & 2+1
  & \bcirc
  & \bstar
  & \bcirc
  & \bstar
  & \bstar
  &
  & 0.051(26)(5)\\
  \midrule
  $\langle x\rangle_{g}$
  & ETMC\,20
  & \cite{Alexandrou:2020sml}
  & 2+1+1
  & \rsquare
  & \bstar
  & \bcirc
  & \bstar
  & \bstar
  & $^{**}$
  & 0.427(92) \\
  & $\chi$QCD\,18
  & \cite{Yang:2018nqn}
  & 2+1
  & \bcirc
  & \bstar
  & \bcirc
  & \bstar
  & \bstar
  &
  & 0.482(69)(48)\\
  & $\chi$QCD\,18a
  & \cite{Yang:2018bft}
  & 2+1
  & \rsquare
  & \bstar
  & \bstar
  & \bstar
  & \rsquare
  & 
  & 0.47(4)(11)\\
  \bottomrule
  \multicolumn{11}{l}{$^{**}$ No quenching effects are seen.}\\
\end{tabularx}
\caption{Lattice QCD values of the benchmark moments of unpolarized PDFs $\langle x \rangle_{u^+-d^+}$, $\langle x \rangle_{u^+}$, $\langle x \rangle_{d^+}$, $\langle x \rangle_{s^+}$ and $\langle x \rangle_{g}$, rated according to the criteria in Tab.~\ref{tab:criteria}. The numbers in parentheses refer to the statistical and systematic uncertainties, respectively, or to the combination of the two, if a single value is provided. All values are obtained at $\mu=2$ GeV.}
\label{tab:unp_moms_det}
\end{table*}
\begin{table}[!t]
\centering
 \begin{tabularx}{\linewidth}{Xccc}
    \toprule
    Moment & PDFLattice17 & CT18 & JAM19 \\
    \midrule
    $\langle x \rangle_{u^+-d^+}$
    & 0.161(18)
    & 0.156(7)
    & 0.157(2) \\
    $\langle x\rangle_{u^+}$
    & 0.352(12)
    & 0.350(5)
    & 0.363(1) \\
    $\langle x \rangle_{d^+}$
    & 0.192(6)
    & 0.193(5)
    & 0.206(2) \\
    $\langle x \rangle_{s^+}$
    & 0.037(3)
    & 0.033(9)
    & 0.018(2) \\
    $\langle x \rangle_g$
    & 0.411(8)
    & 0.413(8)
    & 0.403(2) \\
    \bottomrule
   \end{tabularx}
\caption{Values for global fit determinations of the benchmark moments of unpolarized PDFs (see text for details) for the PDFLattice17 average~\cite{Lin:2017snn}, and for the {\sc CT18}~\cite{Hou:2019efy} and {\sc JAM19}~\cite{Sato:2019yez} analyses. All values are shown at $\mu=2$ GeV.}
\label{tab:unp_moms_pdfs}
\end{table}

In comparison to the 2017 PDFLattice white paper, some new lattice QCD computations were performed, which we summarize in Tab.~\ref{tab:unp_moms_det} and rate according to the criteria defined in Tab.~\ref{tab:criteria}. New results, mostly for $N_f=2+1+1$ and $N_f=2+1$ dynamical flavors, were computed at a scale $\mu=2$~GeV, including for $\langle x\rangle_g$. On the global fit front, two new analyses of unpolarized PDFs were released, as discussed in Sect.~\ref{02sec1}: {\sc CT18}~\cite{Hou:2019efy} and {\sc JAM19}~\cite{Sato:2019yez}. We therefore compute the corresponding values for the benchmark moments $\langle x \rangle_{u^+-d^+}$, $\langle x \rangle_{u^+}$, $\langle x \rangle_{d^+}$, $\langle x \rangle_{s^+}$ and $\langle x \rangle_{g}$ with the highest perturbative accuracy allowed by each analysis, NNLO and NLO, respectively. We report these values at $\mu=2$~GeV in Tab.~\ref{tab:unp_moms_pdfs}, where we also display the PDFLattice17 benchmark value~\cite{Lin:2017snn} for comparison. We recall that this value was determined as the unweighted average of the NNPDF3.1~\cite{Ball:2017nwa}, CT14~\cite{Dulat:2015mca}, MMHT14~\cite{Harland-Lang:2014zoa}, ABMP16~\cite{Alekhin:2017kpj} (with $N_f=4$ flavors), CJ15~\cite{Accardi:2016qay} and HERAPDF2.0~\cite{Abramowicz:2015mha} analyses (see Tab.~3.5 of Ref.~\cite{Lin:2017snn} for their values). 

\begin{figure*}[!p]
\centering
\includegraphics[angle=270,clip=true, trim=0 9cm 0 3cm,width=\linewidth]{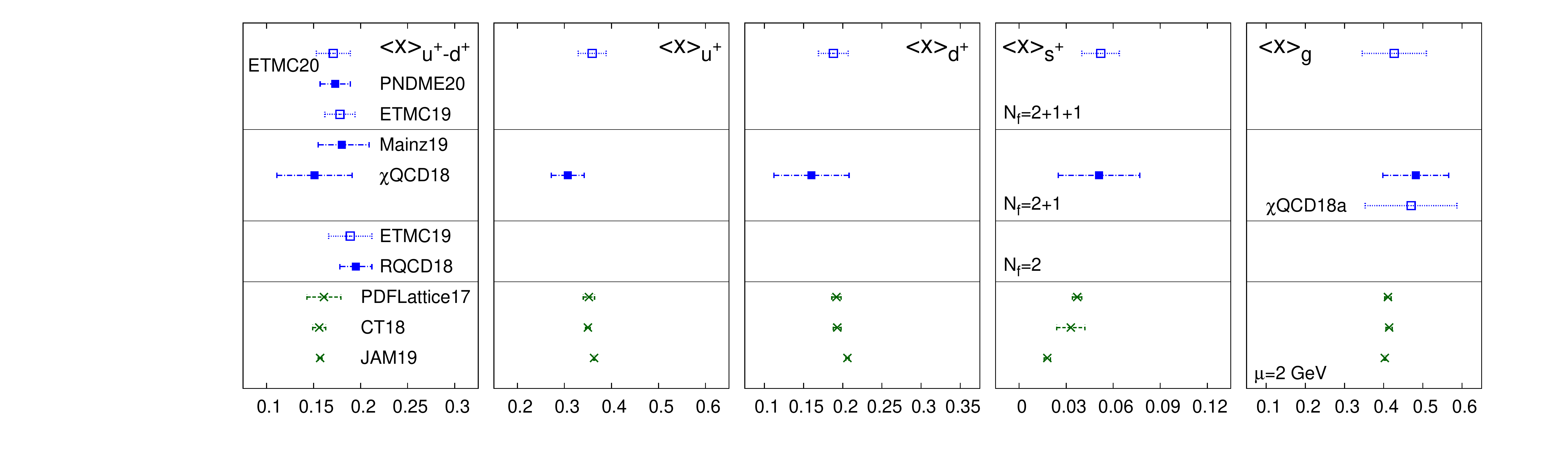}
\caption{\small The lattice QCD (in blue) and global fit (in green) values for the benchmark moments defined in Eq.~\eqref{eq:unp_moms}, as collected in Tabs.~\ref{tab:unp_moms_det} and \ref{tab:unp_moms_pdfs}. Lattice QCD results with stars or circles for all of the sources of systematic uncertainties in Tab.~\ref{tab:criteria} are denoted with filled squares, otherwise with empty squares. All values are at $\mu=2$ GeV.}
\label{fig:unp_moms}
\end{figure*}

The results for the benchmark moments $\langle x \rangle_{u^+-d^+}$, $\langle x \rangle_{u^+}$, $\langle x \rangle_{d^+}$, $\langle x \rangle_{s^+}$ and $\langle x \rangle_{g}$, obtained either from lattice QCD (Tab.~\ref{tab:unp_moms_det}) or from a global analysis of the experimental data (Tab.~\ref{tab:unp_moms_pdfs}), are graphically compared in Fig.~\ref{fig:unp_moms} at $\mu=2$~GeV. In each case, we also determine benchmark values by combining various lattice QCD results, on the one hand, and global fit results, on the other hand. For lattice QCD computations, whenever more than one result is available, we take the envelope of all of the available results, separately for each number of dynamical flavors. This choice is conservative, but seems optimal to us: it will not overestimate too much the uncertainty of the benchmark values in the cases in which lattice moments are in good mutual agreement (such as for $\langle x \rangle_{u^+ -d^+}$), but at the same time will avoid bias in the cases in which systematic uncertainties on the individual values had not been completely controlled. For global analyses of experimental data, given the overall consistency of the various results and the limited progress, we follow the same prescription adopted in the 2017 PDFLattice white paper, but we update it by replacing the {\sc CT14} values with their {\sc CT18} counterparts in the unweighted average. We do not include the {\sc JAM19} values in it, because the analysis is unique in its kind, given that it determines PDFs and FFs simultaneously. Our benchmark values are summarized in Tab.~\ref{tab:unp_moms}.

\begin{figure*}[!p]
\includegraphics[angle=270,clip=true, trim=0 9cm 0 3cm,width=\linewidth]{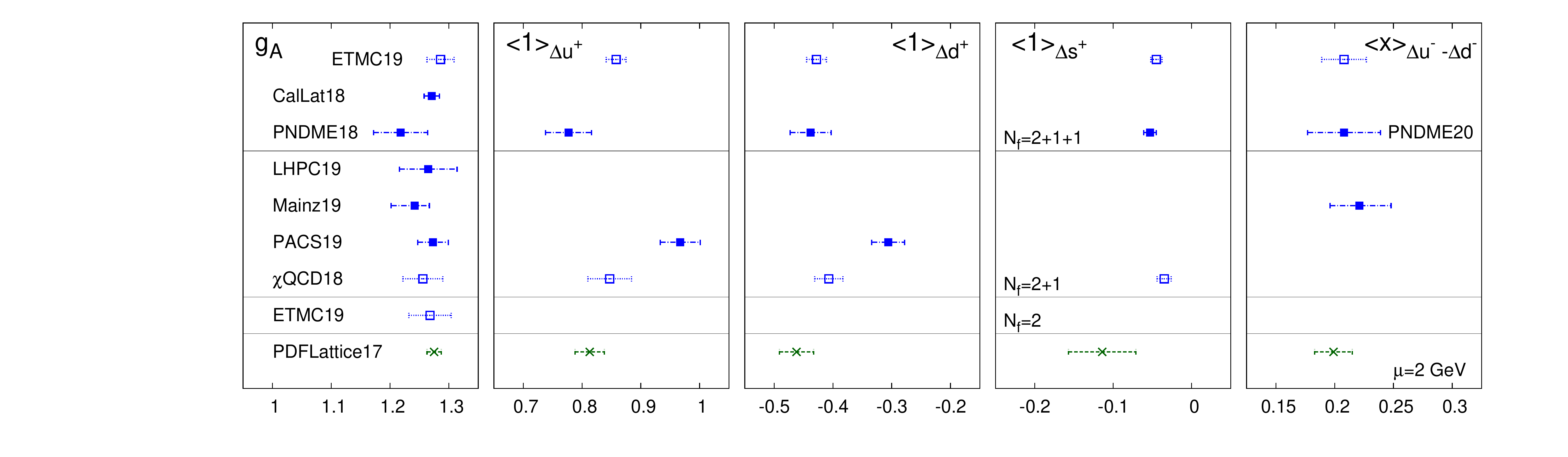}
\caption{\small Same as Fig.~\ref{fig:unp_moms}, but for the benchmark moments of helicity PDFs. See Tab.~\ref{tab:hel_moms_det} for references. }
\label{fig:hel_moms}
\end{figure*}
\begin{figure*}[!p]
\includegraphics[angle=270,clip=true, trim=0 9cm 0 3cm,width=\linewidth]{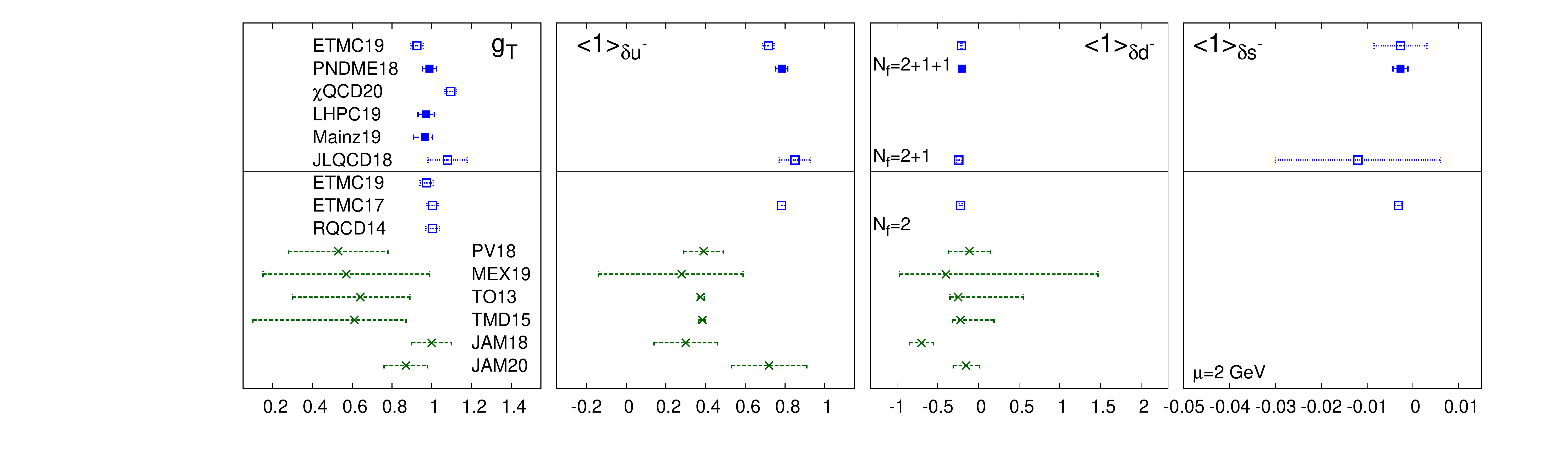}
\caption{\small Same as Fig.~\ref{fig:unp_moms}, but for the benchmark moments of transversity PDFs.}
\label{fig:momh1}
\end{figure*}
\begin{table}[!t]
\centering
 \begin{tabularx}{\linewidth}{Xcc}
    \toprule
    Moment & Lattice QCD & Global Fit\\
    \midrule
    \multirow{3}{*}{$\langle x \rangle_{u^+-d^+}$}
    & 0.153 --- 0.194$^{a}$ 
    & \multirow{3}{*}{0.161(18)}\\
    & 0.111 --- 0.209$^b$
    & \\
    & 0.166 --- 0.212$^c$
    & \\
    \midrule
    \multirow{2}{*}{$\langle x\rangle_{u^+}$}
    & 0.359(30)$^{a,\dag}$
    & \multirow{2}{*}{0.353(12)}\\
    & 0.307(35)$^{b,\dag}$
    & \\
    \midrule
    \multirow{2}{*}{$\langle x \rangle_{d^+}$}
    & 0.188(19)$^{a,\dag}$
    & \multirow{2}{*}{0.192(6)}\\
    & 0.160(48)$^{b,\dag}$
    & \\
    \midrule
    \multirow{2}{*}{$\langle x \rangle_{s^+}$}
    & 0.052(12)$^{a,\dag}$
    & \multirow{2}{*}{0.037(3)}\\
    & 0.051(26)$^{b,\dag}$
    & \\
    \midrule
    \multirow{2}{*}{$\langle x \rangle_g$}
    & 0.427(92)$^{a,\dag}$
    & \multirow{2}{*}{0.411(8)}\\
    & 0.353 --- 0.587$^{b}$
    & \\
    \bottomrule
    \multicolumn{3}{l}{$^a$ $N_f=2+1+1\quad ^b$ $N_f=2+1 \quad ^c$ $N_f=2$}\\
    \multicolumn{3}{l}{$^\dag$ Single lattice result}\\
   \end{tabularx}
\caption{\small  Benchmark values for lattice QCD calculations and global fit determinations of the benchmark moments of unpolarized PDFs (see text for details). All values are shown at $\mu=2$ GeV.}
\label{tab:unp_moms}
\end{table}

As is apparent from Fig.~\ref{fig:unp_moms} and Tabs.~\ref{tab:unp_moms_det}-\ref{tab:unp_moms}, the overall picture described in the 2017 PDFLattice white paper is somewhat confirmed: there is a fair agreement between lattice QCD and global fit results, with previous tension for $\langle x \rangle_{s^+}$ and $\langle x \rangle_g$ relieved by the new lattice results from ETMC20 and $\chi$QCD18. The uncertainty of lattice calculations is in general larger by one order of magnitude than for global fits, also because the constraint of momentum sum rule is usually imposed on the latter but not on the former. All these remarks are still valid if individual results for the benchmark moments are taken in lieu of averages. 

\paragraph{Moments of helicity PDFs.}
The nucleon axial charge $g_A$ has long been considered as one of the most important benchmark quantities for lattice QCD computations, given that it is experimentally well determined through neutron weak decays~\cite{Tanabashi:2018oca}. Also, $g_A$ is of central importance to QCD, as it is related to the Bjorken sum rule~\cite{Bjorken:1966jh,Bjorken:1969mm}.

In comparison to the 2017 PDFLattice white paper, several new lattice QCD computations of $g_A$ and of the flavor-diagonal axial charges appeared in the literature, which we summarize in Tab.~\ref{tab:hel_moms_det} and rate according to the criteria defined in Tab.~\ref{tab:criteria}. A pool of new results for $N_f=2+1$ and $N_f=2+1+1$ dynamical flavors were computed. Conversely, the determination of helicity PDFs in global fits did not witness significant changes since the publication of the 2017 PDFLattice white paper. We therefore refer the reader to the numbers reported in Tab.~3.6 of Ref.~\cite{Lin:2017snn}.

\begin{table*}[!t]
 \begin{tabularx}{\linewidth}{lXXlccccccc}
  \toprule
  Moment & Collaboration & Reference & $N_f$   
  & DE 
  & CE 
  & FV 
  & RE  
  & ES 
  &
  & Value \\
  \midrule
  $g_A$
  & ETMC\,19 & \cite{Alexandrou:2019brg}
  & 2+1+1
  & \rsquare
  & \bstar
  & \bcirc
  & \bstar
  & \bstar
  & $^{**}$
  & 1.286(23) \\
  & CalLat\,18 & \cite{Chang:2018uxx}
  & 2+1+1
  & \bcirc
  & \bstar
  & \bstar
  & \bstar
  & \bstar
  &
  & 1.271(13) \\
  & PNDME\,18 & \cite{Gupta:2018qil}
  & 2+1+1
  & \bstar
  & \bstar
  & \bstar
  & \bstar
  & \bstar
  &
  & 1.218(25)(30) \\
  & LHPC\,19    & \cite{Hasan:2019noy}
  & 2+1 
  & \bcirc
  & \bstar 
  & \bstar  
  & \bstar  
  & \bstar 
  & 
  & 1.265(49)\\
  & Mainz\,19 & \cite{Harris:2019bih}
  & 2+1
  & \bstar
  & \bcirc
  & \bstar
  & \bstar
  & \bstar
  &
  & 1.242(25)($^0_{-31}$) \\  
  & PACS\,19    & \cite{Shintani:2018ozy}
  & 2+1 
  & \bcirc
  & \bstar
  & \bstar 
  & \bstar  
  & \bstar 
  & 
  & 1.273(24)(5)(9)\\
  & $\chi$QCD\,18 & \cite{Liang:2018pis}
  & 2+1
  & \bcirc
  & \rsquare
  & \bstar
  & \bstar
  & \bstar
  &
  & 1.256(16)(30) \\
    & ETMC\,19 & \cite{Alexandrou:2019brg}
  & 2
  & \rsquare
  & \bstar
  & \bcirc
  & \bstar
  & \bstar
  & $^{**}$
  & 1.268(36) \\
  \midrule
  $\langle 1\rangle_{\Delta u^+}$
  & ETMC\,19   & \cite{Alexandrou:2019brg}
  & 2+1+1  
  & \rsquare
  & \bstar
  & \bcirc
  & \bstar  
  & \bstar  
  & $^{**}$
  & $0.858(17)$\\
  & PNDME\,18   & \cite{Gupta:2018qil}
  & 2+1+1  
  & \bstar
  & \bstar  
  & \bstar 
  & \bstar 
  & \bstar  
  & 
  & $0.777(25)(30)$\\
  & PACS\,19   & \cite{Shintani:2018ozy}
  & 2+1  
  & \bcirc
  & \bstar
  & \bstar 
  & \bstar 
  & \bstar 
  & 
  & $0.967(30)(16)$\\
  & $\chi$QCD\,18   & \cite{Liang:2018pis}
  & 2+1  
  & \bcirc
  & \rsquare 
  & \bstar 
  & \bstar 
  & \bstar 
  & 
  & $0.847(18)(32)$\\
\midrule  
$\langle 1\rangle_{\Delta d^+}$
  & ETMC\,19   & \cite{Alexandrou:2019brg}
  & 2+1+1  
  & \rsquare
  & \bstar 
  & \bcirc  
  & \bstar 
  & \bstar 
  & $^{**}$
  & $-0.428(17)$\\
  & PNDME\,18   & \cite{Gupta:2018qil}
  & 2+1+1  
  & \bstar
  & \bstar 
  & \bstar 
  & \bstar 
  & \bstar 
  & 
  & $-0.438(18)(30)$\\
  & PACS\,19   & \cite{Shintani:2018ozy}
  & 2+1  
  & \bcirc
  & \bstar
  & \bstar 
  & \bstar 
  & \bstar 
  & 
  & $-0.306(19)(21)$\\
  & $\chi$QCD\,18   & \cite{Liang:2018pis}
  & 2+1  
  & \bcirc
  & \rsquare 
  & \bstar
  & \bstar 
  & \bstar 
  & 
  & $-0.407(16)(18)$\\
\midrule
$\langle 1\rangle_{\Delta s^+}$
  & ETMC\,19   & \cite{Alexandrou:2019brg}
  & 2+1+1  
  & \rsquare
  & \bstar
  & \bcirc
  & \bstar
  & \bstar 
  & $^{**}$
  & $-0.0450(71)$\\ 
  & PNDME\,18   & \cite{Gupta:2018qil}
  & 2+1+1  
  & \bstar 
  & \bstar 
  & \bstar 
  & \bstar 
  & \bstar 
  & 
  & $-0.053(8)$\\
  & $\chi$QCD\,18  & \cite{Liang:2018pis}
  & 2+1 
  & \bcirc
  & \rsquare
  & \bstar
  & \bstar 
  & \bstar
  & 
  & $-0.035(6)(7)$\\
\midrule
$\langle x\rangle_{\Delta u^--\Delta d^-}$
  & PNDME\,20  & \cite{Mondal:2020cmt}
  & 2+1+1
  & \bstar
  & \bstar
  & \bstar
  & \bstar 
  & \bstar
  & 
  & $0.208(19)(24)$\\
  & ETMC\,19  & \cite{Alexandrou:2019brg}
  & 2+1+1
  & \rsquare
  & \bstar
  & \bcirc
  & \bstar 
  & \bstar
  & $^{**}$
  & $0.193(19)$\\
  & Mainz\,19  & \cite{Harris:2019bih}
  & 2+1 
  & \bstar 
  & \bcirc
  & \bstar
  & \bstar 
  & \bstar
  & 
  & $0.221(25)$($^{+10}_{-0}$)\\
 \bottomrule
 \multicolumn{11}{l}{$^{**}$ No quenching effects are seen.}\\
\end{tabularx}
\caption{Same as Tab.~\ref{tab:unp_moms_det}, but for the axial charge $g_A=\langle 1 \rangle_{\delta u^--\delta d^-}$ and the moments of helicity PDFs $\langle 1 \rangle_{\delta u^-}$, $\langle 1 \rangle_{\delta d^-}$, and $\langle 1 \rangle_{\delta s^-}$. All values are obtained at $\mu=2$ GeV. Note that $g_A$ is scale invariant.}
\label{tab:hel_moms_det}
\end{table*}

The results for the benchmark moments $g_A$, $\langle 1\rangle_{\Delta u^+}$, $\langle 1\rangle_{\Delta d^+}$,
$\langle 1\rangle_{\Delta s^+}$, and $\langle x\rangle_{\Delta u^- - \Delta_d^-}$, obtained either from lattice QCD (Tab.~\ref{tab:hel_moms_det}) or from the PDFLattice17 average of global fits~\cite{Lin:2017snn}, are graphically compared in Fig.~\ref{fig:hel_moms} at $\mu=2$~GeV. In each case, we also determine the benchmark values from the various results. For lattice QCD computations, we use the same prescription adopted in the case of unpolarized moments: whenever more than one result is available, we take the envelope of all of the available results, separately for each number of dynamical flavors. For global analyses of experimental data, given the lack of new results, we take the 2017 PDFLattice average. We recall that it was determined as the unweighted average of the {\sc NNPDFpol1.1}~\cite{Nocera:2014gqa}, {\sc DSSV08}~\cite{deFlorian:2009vb} and {\sc JAM15}~\cite{Sato:2016tuz} analyses. Our benchmark values are summarized in Tab.~\ref{tab:hel_moms} at $\mu=2$~GeV.

\begin{table}[!t]
  \begin{tabularx}{\linewidth}{Xcc}
    \toprule
    Moment & Lattice QCD & Global Fit \\
    \midrule
    \multirow{3}{*}{$g_A$} 
    & 1.179 --- 1.309$^a$
    & \multirow{3}{*}{\ 1.258(28)}\\
    & 1.202 --- 1.314$^{b}$ 
    & \\
    & 1.268(36)$^{c,\dag}$
    & \\
    \midrule
    \multirow{2}{*}{$\langle 1 \rangle_{\Delta u^+}$}     
    & 0.738 --- 0.875$^a$
    & \multirow{2}{*}{0.813(25)}\\
    & 0.810 --- 1.001$^{b}$
    & \\
    \midrule
    \multirow{2}{*}{$\langle 1 \rangle_{\Delta d^+}$}
    & -0.473 --- -0.403$^a$
    & \multirow{2}{*}{-0.462(29)}\\
    & -0.431 --- -0.278$^{b}$
    &\\
    \midrule
    \multirow{2}{*}{$\langle 1 \rangle_{\Delta s^+}$}     
    & -0.0538 --- -0.0379$^{a}$
    & \multirow{2}{*}{-0.114(43)}\\
    & -0.0035(9)$^{b,\dag}$
    & \\
    \midrule
    \multirow{2}{*}{$\langle x\rangle_{\Delta u^- - \Delta d^-}$}  
    & 0.174 --- 0.239$^{a}$ 
    & \multirow{2}{*}{0.199(16)}\\
    & 0.221($^{27}_{-25}$)$^{b,\dag}$
    & \\
    \bottomrule
    \multicolumn{3}{l}{$^{a}N_f=2+1+1 \quad ^{b}N_f=2+1 \quad ^{c}N_f=2$}\\
    \multicolumn{3}{l}{$^\dag$ Single lattice result}\\
   \end{tabularx}
\caption{\small Same as Tab.~\ref{tab:unp_moms}, but for helicity PDFs.}
\label{tab:hel_moms}
\end{table}

As is apparent from Fig.~\ref{fig:hel_moms} and Tabs.~\ref{tab:hel_moms_det}-\ref{tab:hel_moms} there is an overall fair agreement between lattice QCD computations and phenomenological fits, except for some discrepancies in the results for the individual flavor components of the axial charge. Contrary to the case of unpolarized moments, the uncertainties of the two are of comparable size. We note however that the uncertainties on the global fit result might be underestimated because of extrapolation into the small- and large-$x$ regions, where data are less abundant. We also remark that current uncertainties on $g_A$ do not discriminate among lattice QCD results obtained with a different numbers of dynamical flavours, thus confirming that SU$_f$(2) is a good symmetry of QCD. 

\paragraph{Moments of Transversity PDFs.}

The nucleon isovector tensor charge $g_T$ has recently received much attention, namely as a probe of beyond-the-Standard-Model (bSM) effects. On the one hand, it can contribute to neutron $\beta$ decay through a possible tensor coupling not included in the SM Lagrangian~\cite{Bhattacharya:2011qm}. On the other hand, it affects the contribution of quark electric dipole moments to the neutron electric dipole moment in a way which is relevant for searches of bSM sources of CP-violation~\cite{Dubbers:2011ns,Yamanaka:2017mef}. A precise determination of $g_T$ has therefore become crucial~\cite{Courtoy:2015haa}. 

Lattice calculations of the isovector tensor charge are currently the most precise lattice estimates of isovector charges because of small statistical fluctuations and a mild dependence on extrapolation parameters. Results from several collaborations are available, including for the flavor-diagonal charges $\langle 1 \rangle_{\delta u^-}$, $\langle 1 \rangle_{\delta d^-}$ and $\langle 1 \rangle_{\delta s^-}$, which are summarized (and rated according to the criteria in Tab.~\ref{tab:criteria}) in Tab.~\ref{tab:tra_moms_det} at $\mu=2$ GeV.

\begin{table*}[!t]
 \begin{tabularx}{\linewidth}{lXXlccccccc}
  \toprule
  Moment & Collaboration & Reference & $N_f$ 
  & DE
  & CE
  & FV
  & RE
  & ES
  &
  & Value \\
  \midrule
  $g_T$
  & ETMC\,19 & \cite{Alexandrou:2019brg}
  & 2+1+1 
  & \rsquare
  & \bstar
  & \bcirc
  & \bstar 
  & \bstar 
  & $^{**}$
  & 0.926(32) \\
  & PNDME\,18 & \cite{Gupta:2018qil}
  & 2+1+1
  & \bstar
  & \bstar
  & \bstar
  & \bstar
  & \bstar
  & $^*$
  & 0.989(32)(10)\\
  & $\chi$QCD\,20 &\cite{Horkel:2020hpi}
  & 2+1
  & \rsquare
  & \bstar
  & \bcirc
  & \bstar
  & \bstar
  & $\dag$
  & 1.096(30) \\
  & LHPC\,19 & \cite{Hasan:2019noy}
  & 2+1
  & \bcirc
  & \bstar
  & \bcirc
  & \bstar
  & \bstar
  & $^*$
  & 0.972(41)\\
  & Mainz\,19 & \cite{Harris:2019bih}
  & 2+1
  & \bstar
  & \bcirc
  & \bstar
  & \bstar
  & \bstar
  &
  & 0.965(38)($^{+13}_{-41}$)\\
  & JLQCD\,18 & \cite{Yamanaka:2018uud}
  & 2+1
  & \rsquare
  & \bcirc
  & \bcirc
  & \bstar
  & \bstar
  &
  & 1.08(3)(3)(9)\\
    & ETMC\,19 & \cite{Alexandrou:2019brg}
  & 2
  & \rsquare
  & \bstar
  & \bcirc
  & \bstar 
  & \bstar 
  & $^{**}$
  & 0.974(33) \\
  & ETMC\,17 & \cite{Alexandrou:2017qyt}
  & 2
  & \rsquare
  & \bstar
  & \rsquare
  & \bstar
  & \bstar
  &
  & 1.004(21)(02)(19)\\
  & RQCD\,14 & \cite{Bali:2014nma}
  & 2
  & \bcirc
  & \bstar
  & \bstar
  & \bstar
  & \rsquare
  &
  & 1.005(17)(29)\\
 \midrule
  $\langle 1\rangle_{\delta u^-}$
  & ETMC\,19 & \cite{Alexandrou:2019brg}
  & 2+1+1 
  & \rsquare
  & \bstar  
  & \bcirc
  & \bstar 
  & \bstar 
  & $^{**}$
  & 0.716(28) \\
  & PNDME\,18 & \cite{Gupta:2018qil}
  & 2+1+1
  & \bstar
  & \bstar
  & \bstar
  & \bstar
  & \bstar
  & $^*$
  & 0.784(28)(10)\\
  & JLQCD\,18 & \cite{Yamanaka:2018uud}
  & 2+1
  & \rsquare
  & \bcirc
  & \bcirc
  & \bstar
  & \bstar
  & 
  & 0.85(3)(2)(7)\\
  & ETMC\,17 & \cite{Alexandrou:2017qyt}
  & 2
  & \rsquare
  & \bstar
  & \rsquare
  & \bstar
  & \bstar
  &
  & 0.782(16)(2)(13)\\
 \midrule
  $\langle 1 \rangle_{\delta d^-}$
  & ETMC\,19 & \cite{Alexandrou:2019brg}
  & 2+1+1 
  & \rsquare
  & \bstar  
  & \bcirc
  & \bstar 
  & \bstar 
  & $^{**}$
  & -0.210(11) \\
  & PNDME\,18 & \cite{Gupta:2018qil}
  & 2+1+1
  & \bstar
  & \bstar
  & \bstar
  & \bstar
  & \bstar
  & $^*$
  & -0.204(11)(10)\\
  & JLQCD\,18 & \cite{Yamanaka:2018uud}
  & 2+1
  & \rsquare
  & \bcirc
  & \bcirc
  & \bstar
  & \bstar
  & 
  & -0.24(2)(0)(2)\\
  & ETMC\,17 & \cite{Alexandrou:2017qyt}
  & 2
  & \rsquare
  & \bstar 
  & \rsquare
  & \bstar
  & \bstar
  &
  & -0.219(10)(2)(13)\\
 \midrule
 $\langle 1 \rangle_{\delta s^-}$
 & ETMC\,19 & \cite{Alexandrou:2019brg}
 & 2+1+1 
 & \rsquare 
 & \bstar  
 & \bcirc 
 & \bstar 
 & \bstar 
 & $^{**}$
 & -0.0027(58) \\
 & PNDME\,18 & \cite{Gupta:2018qil}
 & 2+1+1
 & \bstar
 & \bstar
 & \bstar
 & \bstar
 & \bstar
 & $^*$
 & -0.0027(16)\\
 & JLQCD\,18 & \cite{Yamanaka:2018uud}
 & 2+1
 & \rsquare
 & \bcirc
 & \bcirc
 & \bstar
 & \bstar
 & 
 & -0.012(16)(8)\\
 & ETMC\,17 & \cite{Alexandrou:2017qyt}
 & 2
 & \rsquare
 & \bstar
 & \rsquare
 & \bstar
 & \bstar
 &
 & -0.00319(69)(2)(22)\\
 \bottomrule
 \multicolumn{11}{l}{$^{**}$ No quenching effects are seen.}\\
 \multicolumn{11}{l}{$^*$ Not fully $\mathcal{O}(a)$ improved by requiring an additional lattice spacing.}\\
 \multicolumn{11}{l}{$\dag$ The rating comes from the valence pion mass used in the calculation. In other calculations, valence an sea pions are the same.}\\
\end{tabularx}
\caption{\small Same as Tab.~\ref{tab:unp_moms_det}, but for the tensor charge $g_T=\langle 1 \rangle_{\delta u^--\delta d^-}$ and the transverse flavor diagonal charges $\langle 1 \rangle_{\delta u^-}$, $\langle 1 \rangle_{\delta d^-}$, and $\langle 1 \rangle_{\delta s^-}$. All values are obtained at $\mu=2$ GeV.}
\label{tab:tra_moms_det}
\end{table*}

The values of the tensor charge $g_T$ and of its flavor components $\langle 1\rangle_{\delta u^-}$ and $\langle 1 \rangle_{\delta d^-}$ obtained from the determination of transversity PDFs from experimental data are summarized in Tab.~\ref{tab:gTpheno}. As discussed in Sect.~\ref{sec:02sec1h1} two of these analyses are performed in a purely collinear framework, namely {\sc PV18}~\cite{Radici:2018iag} and {\sc MEX19}~\cite{Benel:2019mcq}, while the others are performed by means of the TMD formalism (see Sect.~\ref{04sec:tmd}). All of these analyses do not include information on the tensor charge from the lattice, except for {\sc JAM18}~\cite{Lin:2017stx}, where the fit is constrained by the average lattice value of $g_T$. All values in Tab.~\ref{tab:gTpheno} are computed at a scale $\mu=2$~GeV, except for the {\sc TO13} ($\mu=1$~GeV) and for the {\sc TMD15} ($\mu=\sqrt{10}$~GeV) analyses; error bars on the {\sc PV18}~\cite{Radici:2018iag} and {\sc JAM20}~\cite{Cammarota:2020qcw} results represent 90\% confidence levels.

\begin{table}[!t]
\begin{tabularx}{\linewidth}{Xccc}
\toprule
Collaboration & $g_T$ & $\langle 1 \rangle_{\delta u^-}$ & $\langle 1 \rangle_{\delta d^-}$  \\
\midrule
{\sc PV18}      & $0.53(25)$           & $0.39(10)$            & $-0.11(26)$             \\
{\sc MEX19}     & $0.57(42)$           & $0.28(^{+31}_{-42})$  & $-0.40(^{+87}_{-57})$   \\
{\sc TO13}      & $0.64(^{+25}_{-34})$ & $0.375(^{+18}_{-12})$ & $-0.25(^{+30}_{-10})$   \\
{\sc TMD15}     & $0.61(^{+26}_{-51})$ & $0.385(^{+16}_{-20})$ & $-0.22(^{+31}_{-10})$   \\
{\sc JAM18}     & $1.0(1)$             & $0.30(16)$            & $-0.70(15)$             \\
{\sc JAM20}     & $0.87(11)$           & $0.72(19)$            & $-0.15(16)$             \\
\bottomrule
\end{tabularx}
\caption{\small The values of the tensor charge $g_T$ and of its flavor components $\langle 1 \rangle_{\delta u^-}$ and $\langle 1 \rangle_{\delta d^-}$ obtained from the following determinations of transverssity PDFs: {\sc PV18}~\cite{Radici:2018iag}, from a global fit of data on inclusive di-hadron production in SIDIS and pp; {\sc MEX19}~\cite{Benel:2019mcq}, from the same mechanism but on SIDIS data only; {\sc TO13}~\cite{Anselmino:2013vqa}, from a parton-model analysis of Collins effect in SIDIS and $e^+ e^-$ data; {\tt TMD15}~\cite{Kang:2015msa}, from the same data set but analyzed in the TMD framework including evolution effects; {\sc JAM18}~\cite{Lin:2017stx}, as in  {\sc TO13}, but constrained to reproduce the lattice average value for $g_T$; and {\sc JAM20}~\cite{Cammarota:2020qcw} from a global fit of inclusive single-hadron production data. All values are computed at $\mu=2$ GeV, except for {\sc TO13} ($\mu=1$ GeV) and {\sc TMD15} ($\mu=\sqrt{10}$ GeV).}
\label{tab:gTpheno}
\end{table}

The results for the tensor charge $g_T$ and for its flavor components, obtained ether from lattice QCD or from a global analysis of the experimental data, are graphically compared in Fig.~\ref{fig:momh1} at $\mu=2$ GeV.
In each case, we also determine benchmark values by suitably combining various lattice QCD results, on the one hand, and global fit results, on the other hand. For lattice QCD computations, we use the same prescription adopted in the case of unpolarized moments: whenever more than one result is available, we take the envelope of all of the available results, separately for each number of dynamical flavors. For global analyses of experimental data, the spread of values is such that any combination, {\it e.g.} in the form of their unweighted average, seems unjustified; more conservatively, we therefore take the envelope of the various results. Note that no global fit results are available for the strange component of the tensor charge, $\langle 1 \rangle_{\delta s^-}$ due to the lack of data sensitive to it. All these values are summarized in Tab.~\ref{tab:tra_moms} at $\mu=2$ GeV.

\begin{table}[!t]
  \begin{tabularx}{\linewidth}{Xcc}
    & & \\
    \toprule
    Moment & Lattice QCD & Global Fit\\
    \midrule
    \multirow{3}{*}{$g_T$}
    & 0.894 --- 1.023$^a$
    & \multirow{3}{*}{0.10 --- 1.1}\\
    & 0.909 --- 1.175$^b$
    & \\
    & 0.941 --- 1.039$^c$
    & \\
    \midrule
    \multirow{3}{*}{$\langle 1 \rangle_{\delta u^-}$}
    & 0.688 --- 0.814$^a$
    & \multirow{3}{*}{-0.14 --- 0.91}\\
    & 0.85(8)$^{b,\dag}$
    & \\
    & 0.782(21)$^{c,\dag}$
    & \\
    \midrule
    \multirow{3}{*}{$\langle 1 \rangle_{\delta d^-}$}
    & -0.221 --- -0.189$^a$
    & \multirow{3}{*}{-0.97 --- 0.47}\\
    & -0.24(3)$^{b,\dag}$
    & \\
    & -0.219(17)$^{c,\dag}$
    & \\
    \midrule
    \multirow{3}{*}{$\langle 1 \rangle_{\delta s^-}$}
    & -0.0085 --- 0.0031$^a$
    & \multirow{3}{*}{---}\\
    & -0.012(18)$^{b,\dag}$
    & \\
    & -0.00319(72)$^{c,\dag}$
    & \\    
    \bottomrule
    \multicolumn{3}{l}{$^{a}N_f=2+1+1 \quad ^{b}N_f=2+1 \quad ^{c}N_f=2$}\\
    \multicolumn{3}{l}{$^\dag$ Single lattice result}\\
   \end{tabularx}
\caption{Same as Tab.~\ref{tab:unp_moms}, but for transversity PDFs.}
\label{tab:tra_moms}
\end{table}

As is apparent from Fig.~\ref{fig:momh1} and Tabs.~\ref{tab:gTpheno}-\ref{tab:tra_moms}, the results obtained from a global analysis of experimental data are affected by uncertainties which are significantly larger than those of their lattice counterparts as well as than those of their unpolarized and helicity companions. The reason for this state of affairs is the severe lack of data (see Fig.~\ref{fig:kin_cov}), and the consequent difficulty in estimating the extrapolation uncertainty that affects transversity outside the region covered by data. For instance, in the {\sc PV18} analysis, the Mellin moment was computed in the range $[x_{\mathrm{min}},1]$, where $x_{\mathrm{min}} = 10^{-6}$ for the adopted {\sc MSTW08LO} parametrization~\cite{Martin:2009iq} of the unpolarized PDF in Eq.~(\ref{eq:SB}). This constraint avoided extrapolation errors below $x_{\mathrm{min}}$ and ensured that the tensor charge is evaluated at 1\% accuracy~\cite{Radici:2018iag}. In the {\sc MEX19} analysis, extrapolation was controlled by choosing a parametrization that ensured the integrability of $h_1$ by construction. 

From Fig.~\ref{fig:momh1}, we also conclude that most determinations of the tensor charge from global fits are compatible with each other. The compatibility with lattice results is a little more involved. While it is obvious for $g_T^d$ (within large uncertainties), there is a clear tension (up to $2\sigma$) for  
$g_T^u$ and, consequently, for $g_T$. In particular, the {\sc JAM18} result is compatible with the lattice prediction for $g_T$ by construction, but it is largely incompatible for the flavor-diagonal tensor charges. The {\sc JAM20} result is the only one that is compatible with all of the lattice tensor charges; it clearly supersedes {\sc JAM18} and confirms how crucial is to pursue global fits from large independent data sets. However, the {\sc JAM20} analysis has some relevant limitations (for example, it does not include effects from TMD evolution, which might be important when connecting SIDIS data to $e^+ e^-$ and hadronic collision data at much higher scales); because of the still large error bars, this result is also compatible with other phenomenological outcomes, and in particular with the global fit {\sc PV18}. In conclusion, the issue of compatibility of tensor charges obtained from lattice calculations and from phenomenological fits still requires further studies. 

\subsubsection{Lattice data in global fits}

In order to extract PDFs from a set of experimental measurements, a factorization theorem connecting the theoretical predictions for these data to the PDFs is needed. For example, in the case of unpolarized DIS, the measured structure functions $F$ are expressed in terms of the unpolarized PDFs $f$ as
\begin{align}
\label{eq:DISfact}
F\left(x,Q^2\right) &= \sum_q\int_x^1 \frac{d\xi}{\xi}C_q\left(\frac{x}{\xi},\frac{Q^2}{\mu^2},\alpha_s\right)f_q\left(\xi,\mu^2\right) 
\nonumber \\ 
&\equiv \sum_q C_q\left(x,\frac{Q^2}{\mu^2},\alpha_s\right)\otimes f_q\left(x,\mu^2\right)\,,
\end{align}
where $C_q$ are the perturbatively computable Wilson coefficients and the symbol $\otimes$ denotes the convolution integral. One can parametrize the unknown PDFs at a given scale $Q_0$, and express the theoretical predictions as a function of the PDF parameters (called $\theta$ henceforth) after performing DGLAP evolution of the PDF to the scale of the data $Q$ via the Altarelli-Parisi kernel $\Gamma$:
\begin{align}
\label{eq:DGLAPevolution}
f_q\left(x,Q^2; \theta\right) = \Gamma\left(Q^2,Q_0^2,\alpha_s\right) \otimes f_q\left(x,Q_0^2; \theta\right)\,.
\end{align}
In a global fit, the procedure is applied simultaneously to every process for which experimental data and a factorization theorem are available. The PDF parameters are then determined by optimizing an appropriate figure of merit, usually the log-likelihood $\chi^2$.

\begin{figure}[!t]
\centering
\includegraphics[width=\linewidth]{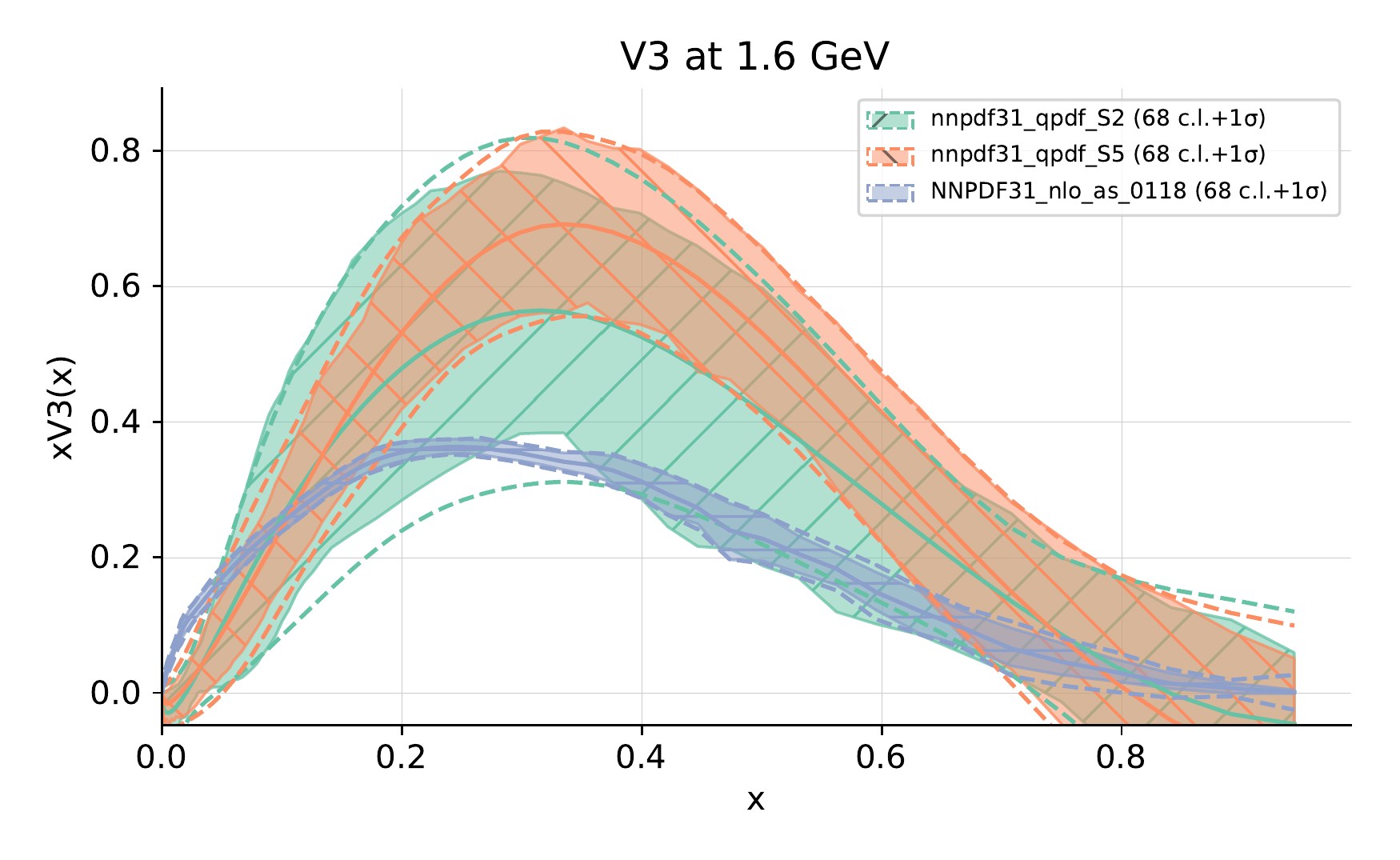}\\
\includegraphics[width=\linewidth]{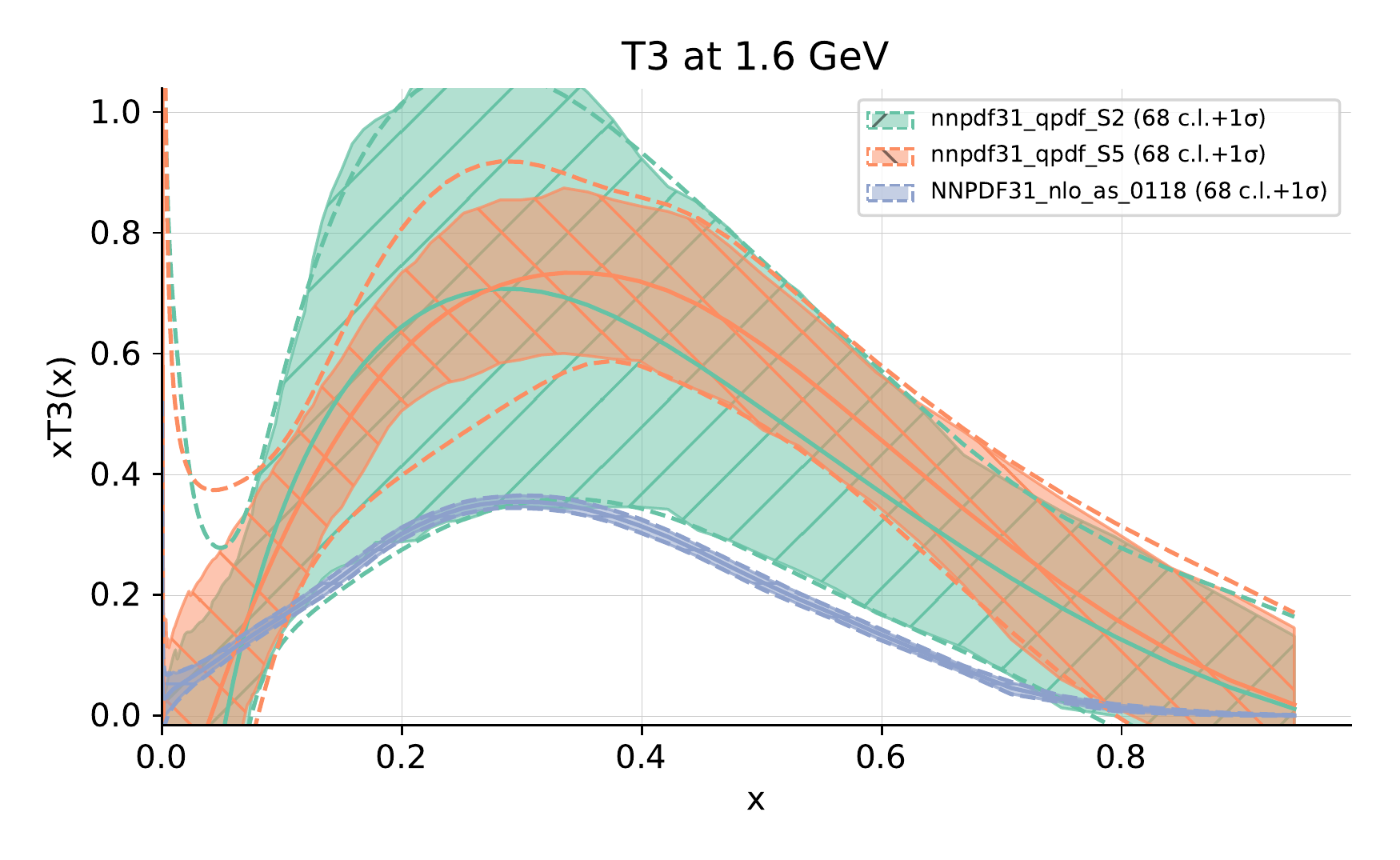}\\
\caption{\label{fig:lattice_data} A comparison of two lattice results (labelled S2 and S5), obtained from an analysis of lattice data with a realistic estimate of systematic uncertainties, with the result of the {\sc NNPDF3.1} global analysis. Results are displayed for the $V_3$ (top) and $T_3$ (bottom) nonsiglet PDF combinations at a scale $\mu=1.6$ GeV. Figure from~\cite{Cichy:2019ebf}.}
\end{figure}

The steps described above can be equally applied to the extraction of PDFs from lattice data, as described and implemented in Refs.~\cite{Ma:2014jla,Ma:2017pxb,Karpie:2019eiq,Cichy:2019ebf}. Since lattice QCD cannot calculate PDFs directly, a factorization theorem is necessary in order to connect some lattice observable to parton distributions, just like how the factorization formula in Eq.~\eqref{eq:DISfact} connects theoretical predictions for DIS experimental data to PDFs. By lattice observable we mean a quantity which in practice can be computed on the lattice: here we will consider the position-space matrix element between proton states of the vector bi-local operator underlying the definition of quasi- and pseudo-PDFs
\begin{align}
\label{eq:Ioffe} 
M_A\left(\nu, z^2\right) &= \langle P |\bar{\psi}(z) \lambda_A \gamma_0 \,   
U(z,0) \psi(0) |P\rangle \, ,
\end{align}
where $\nu = P\cdot z$, $\lambda_A$ denotes the flavour structure and the separation between fields is purely spatial $z = \left(0,0,0,z\right)$. This factorization formula has been worked out up to 1-loop order in perturbation theory, in several independent ways. In the case of quasi-PDFs, the factorization is realized in momentum space, in the limit of large proton momentum, giving
\begin{align}
\label{eq::pdftoqpdf}      
\tilde{f}_A(x , {\mu}^2) = &                                                              
\int_{-1}^{1} \frac{dy}{|y|}\, C_A\left(\frac{x}{y},\frac{\mu}{P_z},\frac{\mu}{\mu'} \right)  f_A(y, {\mu'}^2)
\nonumber \\
&
+ \mathcal{O}\left( \frac{M^2}{P_z^2},\frac{\Lambda^2_{\text{QCD}}}{P_z^2} \right)\, ,                   
\end{align}
$\tilde{f}_A(x,{\mu}^2)$ and $C_A$ being the quasi-PDF and the momentum space matching coefficients respectively. Considering the case of the unpolarized isovector parton distribution, starting from Eq.~\eqref{eq::pdftoqpdf} we can take the real and imaginary part of its inverse Fourier transform, introducing the two following lattice observables
\begin{align}
\label{eq::V3factorization}
\text{Re}\left[M\left(\nu, z^2; \mu\right) \right] 
&= \int_{0}^{1} dx \,\, \mathcal{C}_3^{\text{Re}}\left( \nu x, z, \frac{\mu}{P_z}  \right) V_3\left(x,\mu\right) 
\nonumber\\
&= \mathcal{C}_3^{\text{Re}}\left( z,	   
\frac{\mu}{P_z}  \right) \circledast V_3\left(\mu^2\right)\, , \\
\label{eq::T3factorization}
\text{Im}\left[M\left(\nu, z^2; \mu\right) \right] 
&= \int_{0}^{1} dx \,\, \mathcal{C}_3^{\text{Im}}\left( \nu x, z, \frac{\mu}{P_z}  \right) T_3\left(x,\mu\right)
\nonumber\\
&= \mathcal{C}_3^{\text{Im}}\left( z, \frac{\mu}{P_z}  \right) \circledast T_3\left(\mu^2\right)\, ,  
\end{align}
where $V_3$ and $T_3$ are the nonsinglet distributions 
\begin{align}
& V_3 ( x ) = u\left(x\right) - \bar{u}\left(x\right) -\left[d\left(x\right)-\bar{d}\left(x\right)\right]\, , \\
& T_3 ( x ) = u\left(x\right) + \bar{u}\left(x\right) -\left[d\left(x\right)+\bar{d}\left(x\right)\right]\, .
\end{align}
Eqs.~\eqref{eq::V3factorization} and~\eqref{eq::T3factorization} can then be used to build, together with the corresponding values from the lattice, a $\chi^2$ function, whose minimization would lead to the determination of $V_3$ and $T_3$. In order to define a meaningful $\chi^2$, the statistical and systematical uncertainties over each data point, together with their correlations, have to be known. Once these two pieces of information are provided, a fit to lattice data can be performed using the same fitting framework adopted for experimental data, without any major modification. This exercise was performed in Ref.~\cite{Cichy:2019ebf}, using data for the quasi-PDFs matrix element computed by the ETMC collaboration~\cite{Alexandrou:2018pbm, Alexandrou:2019lfo}, within the {\sc NNPDF} fitting framework. Various assumptions for the systematic uncertainties on the lattice data were considered, corresponding to optimistic, realistic and pessimistic scenarios, respectively. A representative result of this exercise is displayed in Fig.~\ref{fig:lattice_data}, where two lattice results obtained in realistic scenarios (labelled as S2 and S5) are compared to the {\sc NNPDF3.1} global analysis for the $T_3$ and $V_3$ nonsiglet distributions at a scale $\mu=1.6$ GeV. As can be seen from Fig.~\ref{fig:lattice_data}, the lattice result is affected by large uncertainties in comparison to its phenomenological counterpart, and displays a fairly shifted shape. 

The same steps can be repeated for any lattice observable, {\it e.g.} lattice computable hadron matrix elements, like the one in Eq.~(\ref{eq:Ioffe}) but with different parton operators, as long as the observables can be factorized into perturbatively calculable coefficient functions convoluted with the same PDFs~\cite{Ma:2014jla,Ma:2017pxb}. Here, the data of these lattice observables play the same role as experimental data that can be used for extracting PDFs. Although current lattice data cannot cover a kinematic range as wide as experimental data can, lattice calculations could be advantageous for extracting PDFs, or partonic structure in general, of hadrons that are difficult to do experiments with, such as free neutrons, pions~\cite{Sufian:2019bol,Sufian:2020vzb}, and kaons.

\subsubsection{Unpolarized and helicity strange PDFs}
\label{02sec5}

\begin{figure}[!t]
\centering
\includegraphics[width=\linewidth]{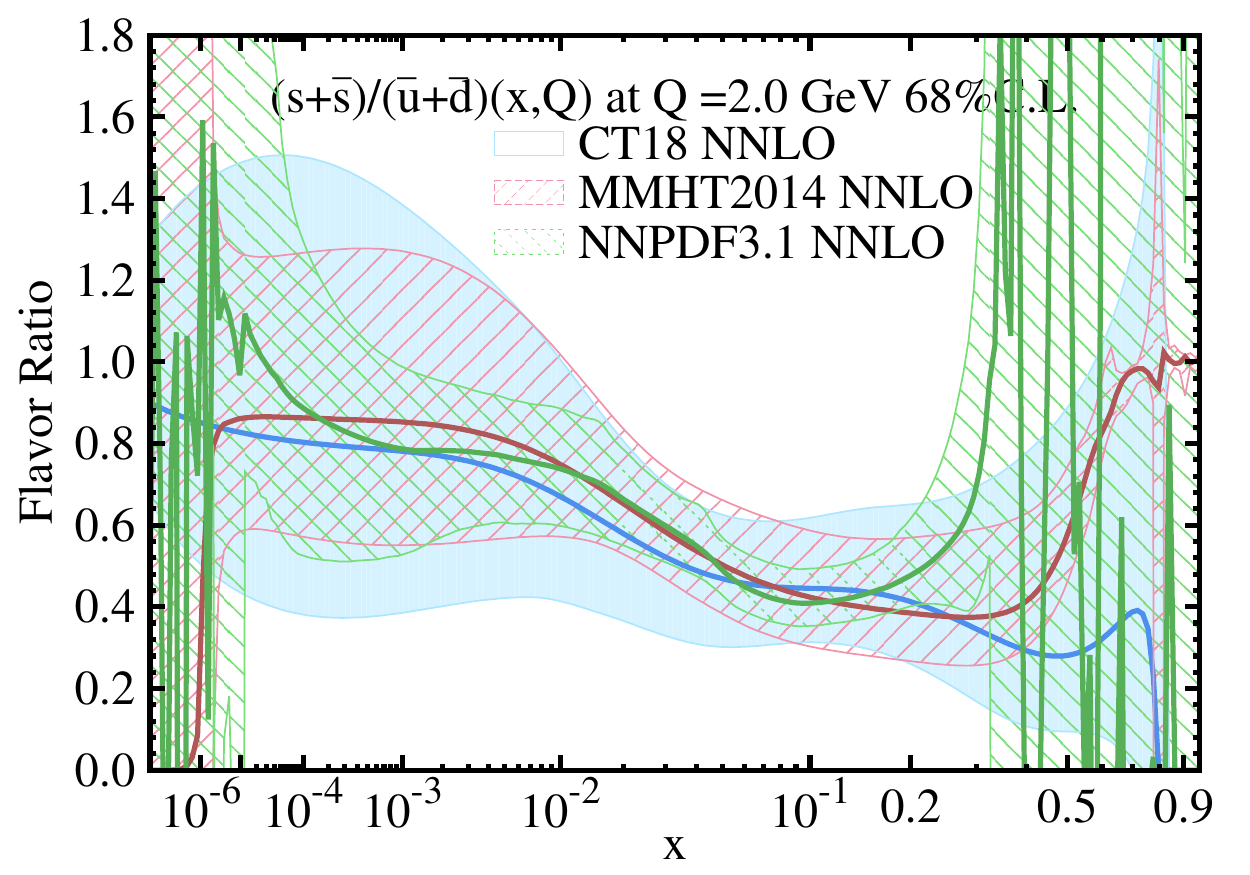}
\caption{\small A comparison of 68\% C.L. uncertainties on the ratio
$\left(s(x,Q)+\bar s(x,Q)\right)/\left(\bar u(x,Q) +\bar d(x,Q)\right)$,
for CT18, MMHT2014 and NNPDF3.1 PDFs at $Q=2$~GeV.
\label{fig:rsall}}
\end{figure}

The strange and antistrange PDFs remain the least known PDFs among all of the unpolarized and helicity distributions. This state of affairs is determined by the lack of experimental data sensitive to this flavor, and is further aggravated by apparent inconsistencies between different data sets. In the unpolarized case, discrepancies have arisen in analyses of inclusive $W^\pm/Z$ production from ATLAS with respect to charged-current neutrino DIS measurements. While the former support a ratio of the strange to non-strange light sea quarks around unity, the latter give a result around one half at a scale of 1.9 GeV$^2$~\cite{Aad:2014xca}. This state of affairs is illustrated in Fig.~\ref{fig:rsall}, where we compare the ratio $\left(s(x,Q)+\bar s(x,Q)\right)/\left(\bar u(x,Q) +\bar d(x,Q)\right)$ for the {\sc CT18}, {\sc MMHT14} and {\sc NNPDF3.1} PDF sets at $Q=2$~GeV. In all of the three PDF sets, the ratio is dominated by neutrino DIS data. The lack of higher-order massive corrections or nuclear uncertainties~\cite{Ball:2018twp} in the description of these data or a sub-optimal estimate of experimental correlations in the ATLAS $W^\pm/Z$ data may be reasons for the discrepancy between the two, which are currently being investigated. The picture is further complicated if measurements of kaon production in SIDIS are added in the QCD analyses, a fact that suppresses the ratio even more~\cite{Sato:2019yez}. 

In the helicity case, the strange distribution is entirely unconstrained unless measurements of SIDIS asymmetries for kaon production are incorporated in the analyses or if SU$_f$(3) constraints from weak baryon decays are assumed. In both cases, the strange helicity PDF is prone to a theoretical bias difficult to quantify, coming respectively from missing uncertainties in the fragmentation of an $s$ quark into a kaon or from the violation of SU$_f$(3) symmetry. 

\begin{figure}[!t]
\centering
\includegraphics[width=0.47\textwidth]{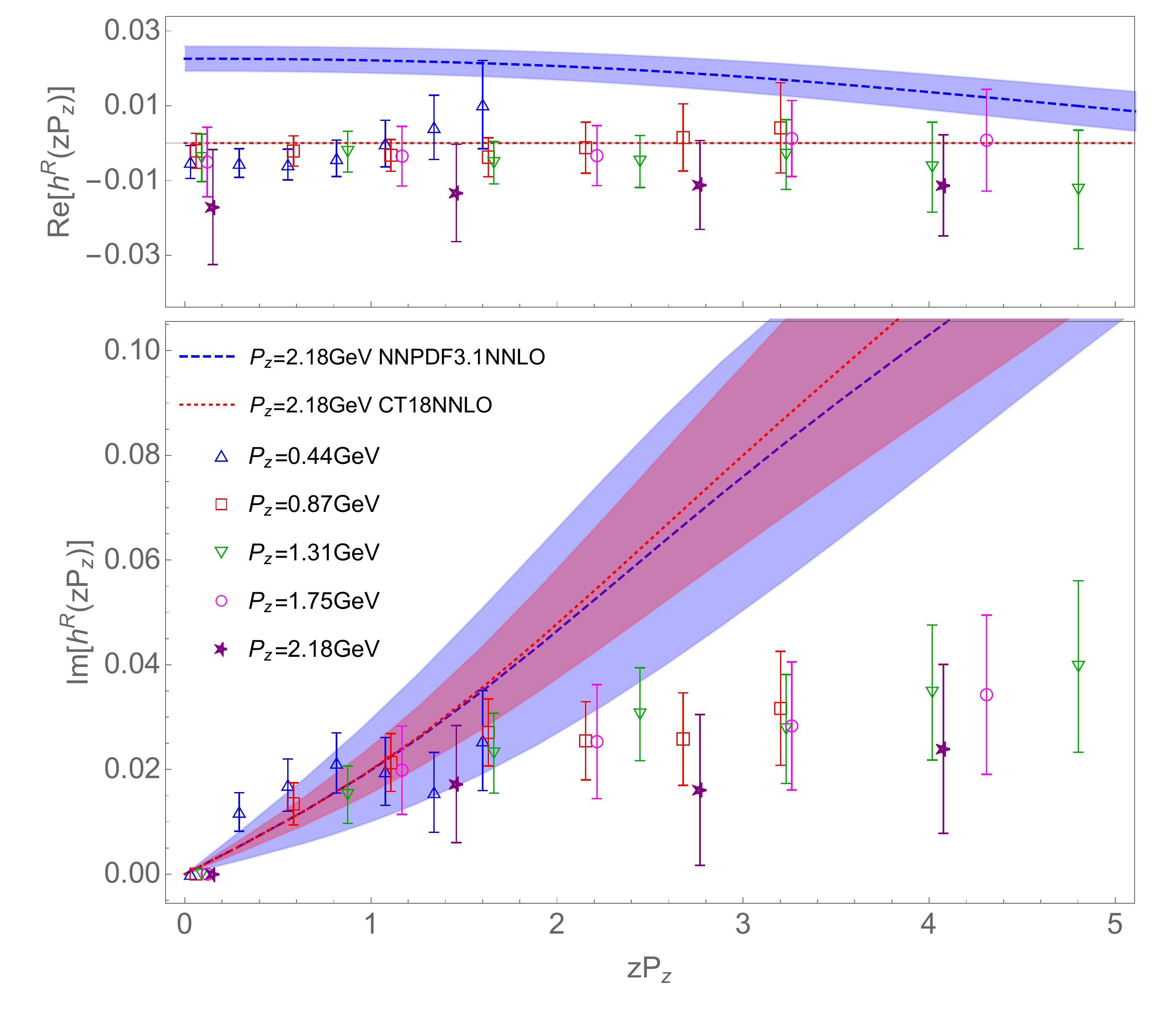}
\caption{The quasi-PDF matrix element renormalized in the RI/MOM scheme at $\mu^R=2.4~\text{GeV}, pz^R=0$ on the lattice compared with the Fourier transform of the NNLO {\sc NNDPF3.1} and {\sc CT18} PDF sets. The upper (lower) plot is the real (imaginary) part. The real part of the CT18 PDF is exactly zero because of the $s(x)=\bar{s}(x)$ assumption entering the fit. The imaginary part of the lattice result grows slower at large $z$, and the real part is consistent with zero at large momentum.}
\label{fig:lamet_quasi_global_msu}
\end{figure}

Exploratory studies using lattice QCD to determine the unpolarized strange PDFs are ongoing. A first lattice calculation of the strange quasi-PDFs~\cite{Zhang:2020dkn} is compared to the Fourier transform of the global fitting results from {\sc NNPDF3.1}~\cite{Ball:2017nwa} and {\sc CT18}~\cite{Hou:2019efy} at NNLO in Fig.~\ref{fig:lamet_quasi_global_msu} at a scale of 4 GeV$^2$. The quasi-PDF matrix elements are extracted from correlators computed on the $a\approx 0.12~\text{fm}$, $24^3\times64$ ensemble with $2+1+1$ flavors of highly improved staggered quarks~\cite{Follana:2006rc} and $M_\pi\approx310~\text{MeV}$ was generated by the MILC collaboration~\cite{Bazavov:2012xda}.  Hypercubic smearing~\cite{Hasenfratz:2001hp} is applied to the configurations. To approach the light-cone distribution, the nucleon operator is constructed with momentum smeared quark sources~\cite{Bali:2016lva} and boosted to momenta up to $P_z=2.18~\text{GeV}$. The matrix elements are renormalized with non-perturbative renormalization factors in the RI/MOM scheme~\cite{Martinelli:1994ty} at $\mu^R=2.4~\text{GeV}, pz^R=0$. Due to the assumption $s(x)=\bar{s}(x)$ in the {\sc CT18} PDF set, the Fourier transformed matrix elements have a vanishing real part. It is observed that the real part of the renormalized quasi-PDF matrix elements are consistent with zero at large momentum. 
The imaginary matrix elements are proportional to the sum of the quark and antiquark distribution; the magnitude is consistently smaller than those from CT18 and NNPDF, possibly due to missing the contributions from other flavor distributions in the matching kernel. A full analysis of lattice-QCD systematics, such as finite-volume effects and discretization, is not yet included, and plans to extend the current calculations are underway.

%% file: 03sec0gpd.tex
\goodbreak
\section{Generalized PDFs (GPD)}
\label{03sec:gpd}
\color{black}

\input{03sec1intro}
\input{03sec2landscape}

\input{03sec3moments}

\input{03sec4machine}
\input{03sec5fits}

%% file: 03sec1intro.tex
\goodbreak
\subsection{Introduction}
\label{03sec1}

More than half a century after the first discovery of the nucleon's internal structure, we still know little about its three-dimensional structure, which is described in terms of the generalized parton distributions (GPDs). 
Interest in GPDs has grown rapidly since they were first introduced in the late 90's \cite{Mueller:1998fv,Ji:1996ek,Radyushkin:1996nd}, because of the unique insight they provide into the spatial distribution of quarks and gluons inside the proton~\cite{Burkardt:2000za}, as well as its mechanical properties, including angular momentum, pressure, and shear forces~\cite{Polyakov:2002yz} (for extensive reviews on the subject see Refs.~\cite{Diehl:2003ny,Belitsky:2005qn,Kumericki:2016ehc}).
While the latter are encoded in the QCD energy-momentum tensor which is canonically probed through gravity, it was realized by Ji~\cite{Ji:1996ek} that its matrix elements between proton states can be described through Mellin moments of GPDs.
GPDs, therefore, bring the energy-momentum tensor matrix elements within experimental grasp through electromagnetic scattering. 
GPDs can be viewed as a hybrid of parton distributions (PDFs), form factors and distribution amplitudes.
Experimentally, they can be accessed in exclusive processes such as deeply virtual Compton scattering or meson electroproduction.
Determining GPDs is an important mission of the US Department of Energy (DOE) drawing global scientific interest.
Experimental collaborations and facilities worldwide have been devoted to searching for these last unknowns of the nucleon, including HERMES at DESY, COMPASS at CERN, GSI in Europe, BELLE and JPAC in Japan, Halls A, B and C at Jefferson Laboratory, and PHENIX and STAR at RHIC (Brookhaven National Laboratory) in the US. 
The study of GPDs and their imaging, or nucleon femtography, has led both the QCD and hadronic physics community worldwide to propose building an Electron-Ion Collider (EIC) 
to further explore these quantities~\cite{Accardi:2012qut}.
In Europe, CERN has plans for the LHeC, adding an electron accelerator to the existing LHC hadron accelerator.
There are also Chinese plans for similar electron-ion colliders.

Analogous to the usual PDFs, GPDs parametrize the quark and gluon correlation functions involving matrix elements of operators at a lightlike separation between the parton fields,
\begin{widetext}
\begin{eqnarray}
\label{eq:def_vec}
W_{\Lambda \Lambda'}^{[\gamma^+]} & =&\frac{1}{2} \int \frac{d y}{2 \pi}^- e^{i k^+ y^-} \langle p', \Lambda'\mid \bar{\psi}\left(-\frac{y}{2} \right) \gamma^+ \, {\cal U}\left(-\frac{y}{2}, \frac{y}{2}\right) \psi\left(\frac{y}{2} \right) \mid p, \Lambda \rangle_{y^+ =y_T=0} \nonumber \\
&  & \hspace{6cm}= \frac{1}{2 P^+} \langle p', \Lambda'\mid \gamma^+ H(x,\xi,t) + \frac{i \sigma^{+j} \Delta_j}{2M} E(x,\xi,t) \mid p, \Lambda \rangle 
\\
\label{eq:def_axi}
\widetilde{W}_{\Lambda \Lambda'}^{[\gamma^+\gamma_5]} & =  &\frac{1}{2} \int \frac{d y}{2 \pi}^- e^{i k^+ y^-} \langle p', \Lambda'\mid \bar{\psi}\left(-\frac{y}{2} \right) \gamma^+ \gamma_5 \, {\cal U}\left(-\frac{y}{2}, \frac{y}{2}\right) \psi\left(\frac{y}{2} \right) \mid p, \Lambda \rangle_{y^+=y_T=0} 
\nonumber \\
&& \hspace{6cm} = \frac{1}{2 P^+} \langle p', \Lambda'\mid \gamma^+ \gamma_5 \widetilde{H}(x,\xi,t) + \frac{ \gamma_5 \Delta^+}{2M} \widetilde{E}(x,\xi,t) \mid p, \Lambda \rangle . \nonumber \\
\end{eqnarray}
\end{widetext}
Two additional kinematic variables enter their definition besides the quark light cone longitudinal momentum fraction, $x=k^+/P^+$ (with $P=(p+p')/2$): the  
light cone component of the longitudinal momentum transfer between the initial and final proton, $\xi=-\Delta^+/(2P^+) \approx x_{Bj}/(2- x_{Bj})>0$ (with $\Delta = p'-p$),  and the transverse component, ${\bf \Delta}_T={\bf p}_T'-{\bf p}_T$; the latter is taken into account through the invariant, $t= \Delta^2= M^2 \xi^2/(1-\xi^2) - \Delta_T^2/(1-\xi^2), t<0$).
At leading (twist-two) level, four quark chirality conserving (chiral even) GPDs, $H$, $E$, $\widetilde{H}$ and $\widetilde{E}$, defined in Eqs.~(\ref{eq:def_vec}) and (\ref{eq:def_axi}), parametrize the quark-proton correlation functions.

Like the PDFs, GPDs are nonperturbative parton-hadron correlation functions.
They are not direct physical observables, since no quark or gluon has been observed in isolation, owing to QCD color confinement.
Lattice QCD cannot calculate GPDs directly, since they are defined in terms of operators with lightlike separation between the parton fields.
GPDs are connected to various physical observables through QCD factorization.
For example, the deeply virtual Compton scattering (DVCS) cross section can be expressed in terms of Compton form factors (CFFs) which are factorized into GPDs convoluted over the variable $x$ with perturbatively calculated short-distance coefficient functions~\cite{Goeke:2001tz,Diehl:2003ny,Belitsky:2005qn,Kumericki:2016ehc}.
In addition to the four chiral even GPDs, we have four more quark-flip/chiral odd GPDs, $H_T$, $E_T$, $\widetilde{H}_T$ and $\widetilde{E}_T$ and four gluon GPDs, $H_g$, $E_g$, $\widetilde{H}_g$ and $\widetilde{E}_g$.
The latter can be singled out in deeply virtual meson-production experiments. 

While several deeply virtual exclusive experiments performed in the past decade are affected by large theoretical and experimental uncertainties on the CFFs (see \cite{Kumericki:2016ehc} for a detailed list), the experimental program of Jefferson Lab at 12 GeV will allow us, in the next decade, to map out the complete helicity structure of the cross section~\cite{Kriesten:2019jep}, making it possible to extract GPDs in the quark sector with sufficiently high precision.
Looking into the future, the upcoming EIC will be instead focused on low $x$, providing a unique probe of the gluon GPDs.

Deeply virtual exclusive processes are measured in coincidence experiments where all the particles in the final state are either directly or indirectly detected.
Extracting GPDs from data involves a much larger number of variables than extracting PDFs.
Furthermore, each experiment can only add a small piece to the picture.
This makes it very difficult, if not impossible, 
to extract GPDs from experiment with the high-precision required to obtain images at the femtoscale (femtography), using traditional global-fitting methods.
More sophisticated analyses are being developed working towards this goal, and a major research effort has been focused on applying machine-learning tools to the extraction of GPDs from data (Sec.~\ref{03sec4}).

%% file: 03sec2landscape.tex
\goodbreak
\subsection{Status of GPD calculations}
\label{03sec2}

A lot of activity and progress has been recently dedicated to extracting information on GPDs from lattice QCD calculations as reported in detail in Sections \ref{sec:moments}, \ref{sec:xdep}.  In addition, several models/parametric forms have been developed that, on one side, facilitate the comparison with experimental data and, on the other, allow us to identify new deeply virtual exclusive observables 
that can be factorized in terms of GPDs. 
The main efforts in this direction include the light-front overlap representation \cite{Chouika:2017itz} used in the PARTON analysis \cite{Berthou:2015oaw}, 
the reggeized diquark model \cite{GonzalezHernandez:2012jv} used in the recent effort in Ref.\cite{Kriesten:2019jep},  conformal moments based models \cite{Mueller:2013caa} used in the analysis of Refs.\cite{Kumericki:2011zc,Kumericki:2015lhb}, as well as parametrizations based on double distributions \cite{Goloskokov:2007nt}.  

To be able to rely on a variety of models is particularly important for the study of GPDs because of the additional constraints that have to be satisfied as compared to PDFs and other one-dimensional quantities. Time reversal and Parity conservation generate discrete symmetries obeyed by GPDs for $\xi \rightarrow -\xi$, $x \rightarrow - x$; Lorentz invariance generates the property of polynomiality, namely that the $x^j$ moments of quark GPDs are even polynomials in $\xi$, and that $x^{j-1}$ moments of gluon GPDs are even polynomials in $\xi$ (see also Section \ref{sec:pseudo}); finally, from the norm of the Hilbert space one obtains positivity constraints, or inequalities between GPDs and the corresponding PDFs \cite{Diehl:2003ny,Belitsky:2005qn,Kumericki:2016ehc}. 
The behavior of GPDs, including their perturbative evolution, under these constraints can be tracked down and monitored using models.  

The wealth of GPD phenomenology combined with its intrinsic difficulty has pushed several groups of theorists and phenomenologists to develop frameworks dedicated to the extraction of GPDs and of the 3D structure of the nucleon from experimental data. In Section \ref{03sec4} we describe platforms that are currently been used for data interpretation.

%% file: 03sec3moments.tex
\goodbreak

 \subsection{Moments of GPD from lattice QCD}
 \label{sec:moments}

A widely-used approach to extract information of GPDs on the lattice is via an operator product expansion. Based on this approach, the resulting local operators can be inserted between hadronic matrix elements, which are calculable in lattice QCD. The Mellin moments, obtained from such matrix elements, provide important information on the structure of the hadron under study, and can be compared against phenomenological fits of experimental data. 
The moments of GPDs are a generalization of the moments of PDF with a nonzero momentum transfer between the initial and final state: form factors and generalized form factors, expressed in terms of the momentum transferred squared, $Q^2$. 
Form factors (FFs) have been studied intensively for many years in lattice QCD due to many decades of scattering experimental results that one can compare with, and recent interests in proton radius puzzles with all possible theoretical calculations.  
On the other hand, there are less interests in the generalized form factors (GFFs) which can be challenging to extract and  reduce the systematic uncertainties. Lack of reliable extraction of GFF experimental data has also put many lattice collaboration to diverge their resources elsewhere. With the exciting results coming out of JLab 12GeV program and other future colliders, one expect more GFFs lattice calculations in the near future.  

\subsubsection{Nucleon electromagnetic form factors}
\begin{figure}[tb]
\centering
\includegraphics[width=0.45\textwidth]{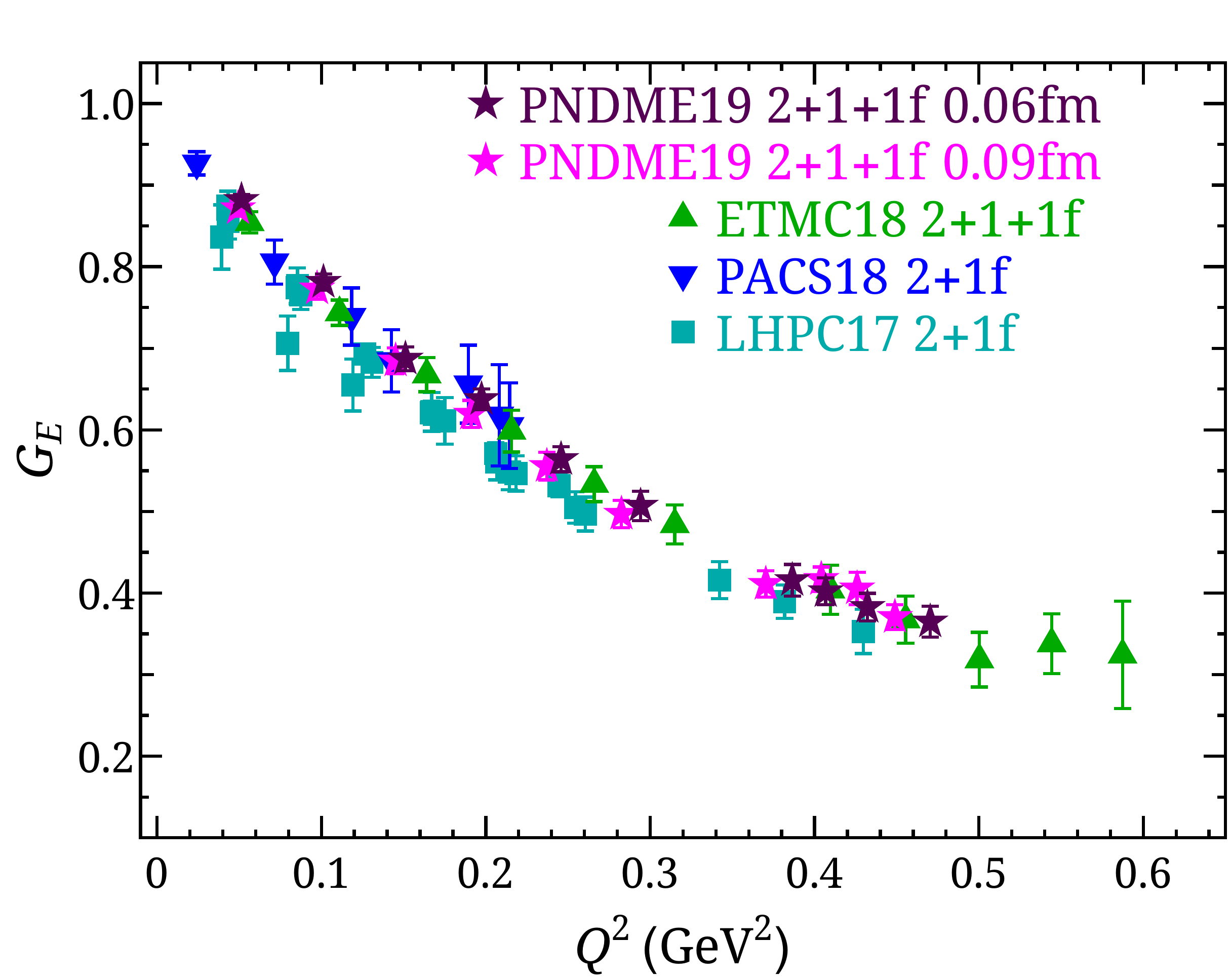} \hfill
\includegraphics[width=0.45\textwidth]{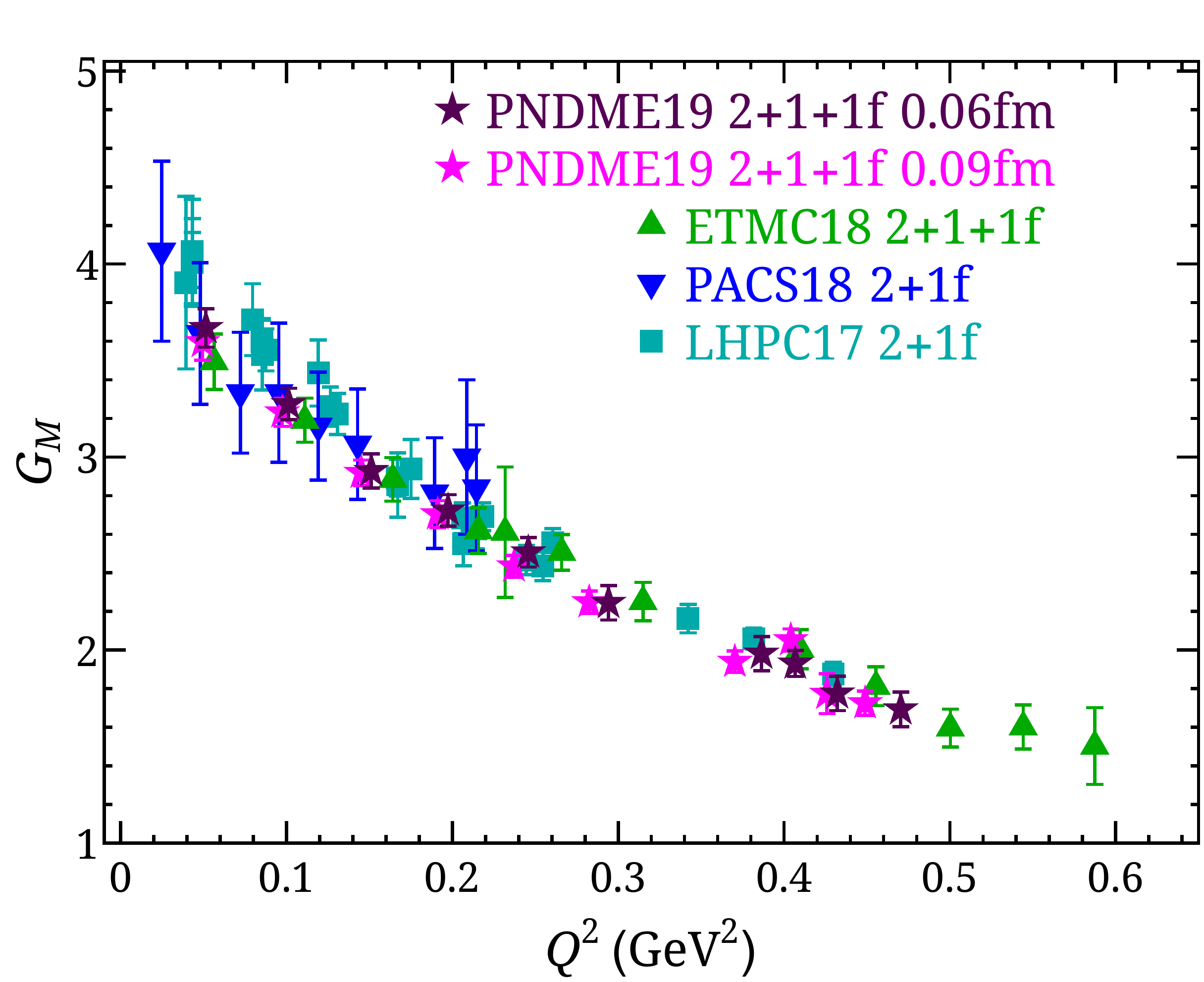}
\caption{\label{fig:EMFF}
Selected nucleon isovector electric (left) and magnetic (right) form factor results from near physical pion mass
as functions of transferred momentum $Q^2$.
The references corresponding to the above works are:
2f ETMC18~\cite{Alexandrou:2017ypw} %
2+1f LHPC14~\cite{Green:2014xba}, %
LHPC17~\cite{Hasan:2017wwt}, %
PACS18~\cite{Shintani:2018ozy}; %
2+1+1f ETMC18~\cite{Alexandrou:2018sjm},
PNDME19~\cite{Jang:2018djx} with 2 lattice spacings of 0.06 and 0.09~fm.
}
\end{figure}

The most commonly calculated form factors are the nucleon isovector electromagnetic form factors, defined by
\begin{align}
   & \langle N(p_f)|V_\mu^+(x)|N(p_i)\rangle = 
    \nonumber \\
    &\bar{u}^{N}\left[
        \gamma_\mu F_1 (q^2) +
        i\sigma_{\mu\nu} \frac{q^\nu}{2M_N} F_2(q^2)
        \right] u_N e^{iq\cdot x},
    \label{eq:ff}
\end{align}
where the isovector current is $V^+_\mu(x)=\overline{u}\gamma_\mu u -\overline{d}\gamma_\mu d$ and
$F_1$ and $F_2$ are Dirac and Pauli form factors, respectively. 
The Sachs electric and magnetic form factors can be obtained from $F_{1,2}$
as $G_E(q^2)=F_1(q^2)+q^2F_2(q^2)/(2 M_N)^2$ and
$G_M(q^2)=F_1(q^2)+F_2(q^2)$, which correspond to the Fourier transforms of nucleon charge and magnetization
density distributions in the Breit frame. 
Inspired by the decades of continuous experimental measurements, there are many calculations at physical pion mass for the electromagnetic form factors 
In Fig.~\ref{fig:EMFF}, only selected results from near-physical pion masses are shown.
PACS has the largest volume among these calculations and is able to probe the smallest $Q^2$. The parameters of the lattice calculations are summarized in Table~\ref{tab:lQCDEMffParams}.
Overall, the lattice-QCD results among different collaborations are in good agreement within 1-2 standard deviations. The magnetic form factors have reasonable agreement with the experimental ones, but the electric form factors, depending on the analysis approach, have a small tension. With more precision experimental electric form factors on their way to resolve the proton radius puzzle, it would be desirable to have lattice form factors at the percent level. To achieve this, one will need to include QED and isospin symmetry breaking in the lattice calculation, which will require significantly more computational resources, as well as new techniques.

\begin{table*}[tbp]
\begin{center}
\hspace*{-0.75cm}
\begin{tabular}{|c|c|c|c|c|c|c|c|}
\hline
Ref. & Sea quarks & Valence quarks & Renormalization &  $N_{\Delta t}$ & a (fm) & $M_\pi$ (MeV) & $M_\pi L$  \\
\hline
ETMC'18~\cite{Alexandrou:2018sjm} & 2f $\&$ 2+1+1f TM &  twisted mass &  RI'-MOM & 3 & 0.080, 0.094 & 130-139~MeV& 3.0--4.0 \\
ETMC'17~\cite{Alexandrou:2017ypw} & 2f TM & twisted mass &  RI'-MOM & 3 & 0.094& 130~MeV& 3.0\\
LHPC'14~\cite{Green:2014xba} & 2+1f clover & clover &  Schr\"odinger functional & 3 & 0.09--0.116 & 149--356 & 3.6--5.0\\
LHPC'17~\cite{Hasan:2017wwt} & 2+1f clover & clover & vector charge & 3 & 0.093 & 135& 4.0\\
PACS'18~\cite{Shintani:2018ozy} & 2+1f clover & clover &   Schr\"odinger functional & 4 & 0.085 & 146 & 8.0\\
PNDME'19~\cite{Jang:2018djx} & 2+1+1f HISQ & clover & RI-MOM &4 &  0.09& 138 & 3.9\\
\hline 
\end{tabular}
\end{center}
\caption{\label{tab:lQCDEMffParams} The lattice parameters used in calculations of the electromagnetic form factors near physical pion mass. $N_{\Delta t}$ indicates the number of source-sink separation used in the lattice calculation to control the ``excited-state'' systematics which can be significant for nucleon structure. 
}
\end{table*}

\begin{figure}[htb!]
 	\centering
    \includegraphics[scale=0.45]{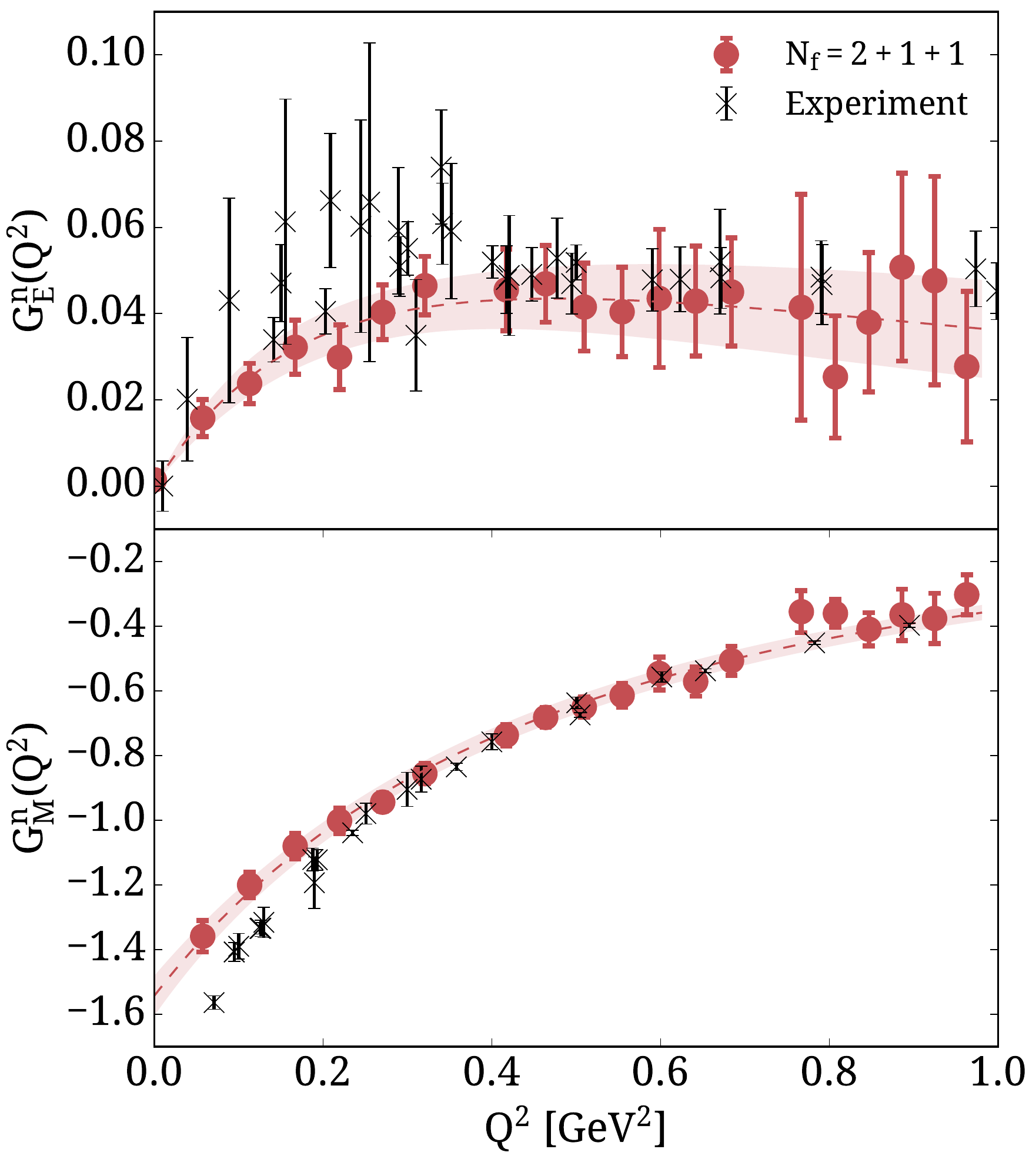}
 	\caption{Neutron electric (top) and magnetic (bottom) form factors as a function of $Q^2$. Lattice data are shown with red circles obtained by ETMC~\cite{Alexandrou:2018sjm}. Experimental data are shown with black stars.}
 	\label{fig:GEM_n}
 \end{figure}

Besides the progress in the calculation of the connected contributions to the electromagnetic form factors, there have been advances in the evaluation of their disconnected (light and strange quark) contributions. The latter result from the self-contractions of the inserted operator, that is, its quark and anti-quark of same flavor contract with each other. Approximately speaking, the operator couples to the hadron via gluons, and are often viewed as sea quark contributions. 
This allows a $u$- and $d$-quark decomposition and the proper extraction of the proton and neutron electromagnetic form factors via:
\begin{eqnarray}
G^p(Q^2) = \frac{1}{2}\left(\frac{1}{3}G^{u+d}(Q^2) + G^{u-d}(Q^2) \right)\,,    \\
G^n(Q^2) = \frac{1}{2}\left(\frac{1}{3}G^{u+d}(Q^2) - G^{u-d}(Q^2) \right)   \,.
\end{eqnarray}
The isoscalar combination in the above equations should be the sum of both connected and disconnected contributions. The electric and magnetic proton form factors including both connected and disconnected diagrams have been calculated by ETMC using twisted mass fermions (TM) ensembles ($N_f=2$ and $N_f=2+1+1$) with different volumes and can be found in Ref.~\cite{Alexandrou:2018sjm}.
Here we show the neutron form factors in Fig.~\ref{fig:GEM_n} for the most accurate data of ETMC at a volume with $L=5.12$ fm (red circles), which are compared against experimental data shown with black crosses. The data are taken from Refs.~\cite{Golak:2000nt,Becker:1999tw,Eden:1994ji,Meyerhoff:1994ev,Passchier:1999cj,Warren:2003ma,Zhu:2001md, Plaster:2005cx, Madey:2003av, Rohe:1999sh, Bermuth:2003qh,Glazier:2004ny,Herberg:1999ud,Schiavilla:2001qe,Ostrick:1999xa}
    for the electric form factor and from
    Refs.~\cite{Anderson:2006jp, Gao:1994ud, Anklin:1994ae, Anklin:1998ae, Kubon:2001rj, Alarcon:2007zza} for the magnetic form factor. There is a very good agreement between lattice and experiment, with exception the small-$Q^2$ for $G_M^n$. Note that the experimental data have, in general, larger uncertainties as compared to the lattice data. We want to emphasize that, such comparison is considered a success for lattice QCD, as it is the first time that the proton and neutron electromagnetic form factors include both connected and disconnected contributions. Such a plot is a representative example of the state-of-the-art of lattice QCD for calculation of form factors.

\begin{figure*}[htb!]
\centering
\includegraphics[width=0.32\textwidth]{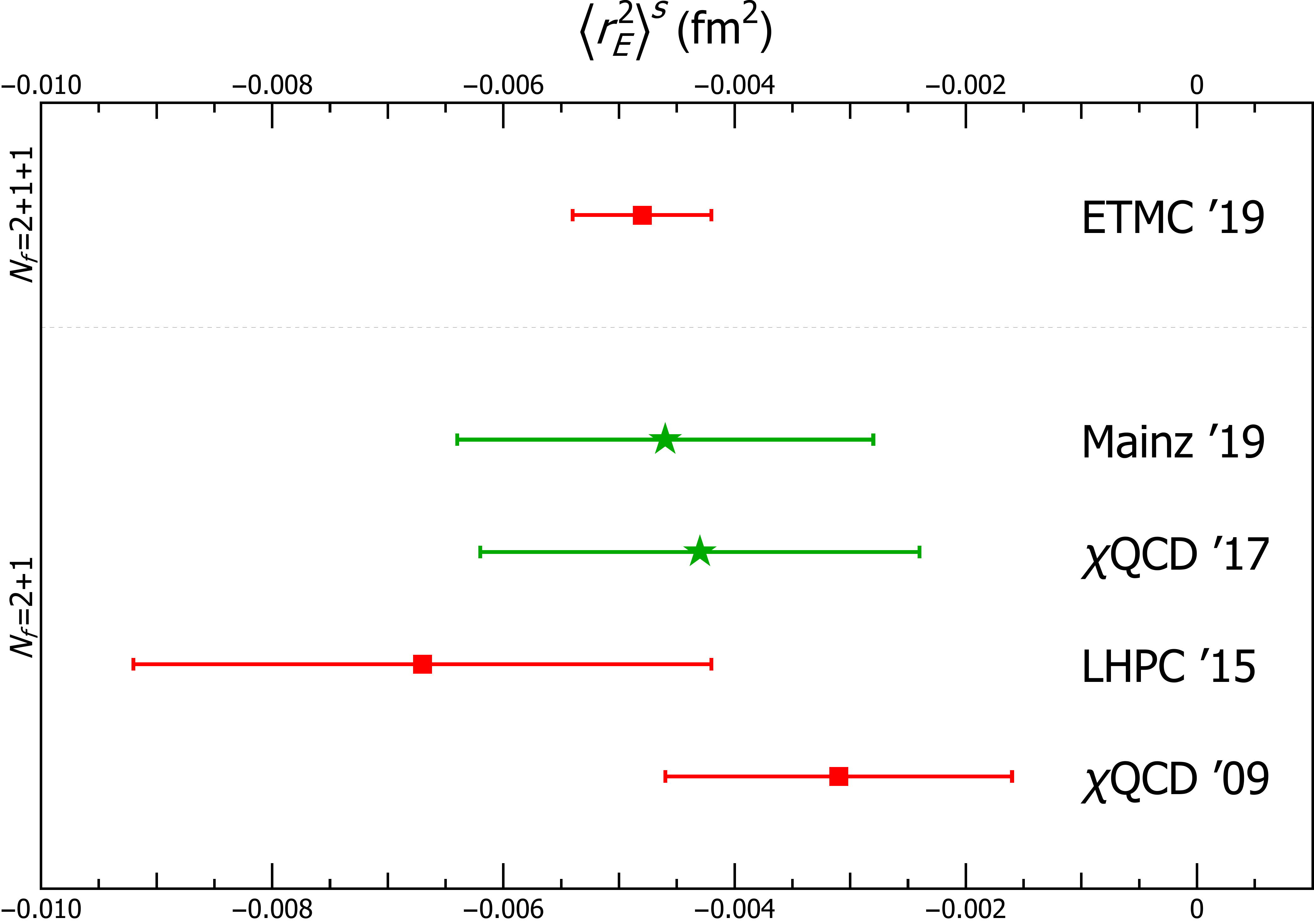}
\includegraphics[width=0.32\textwidth]{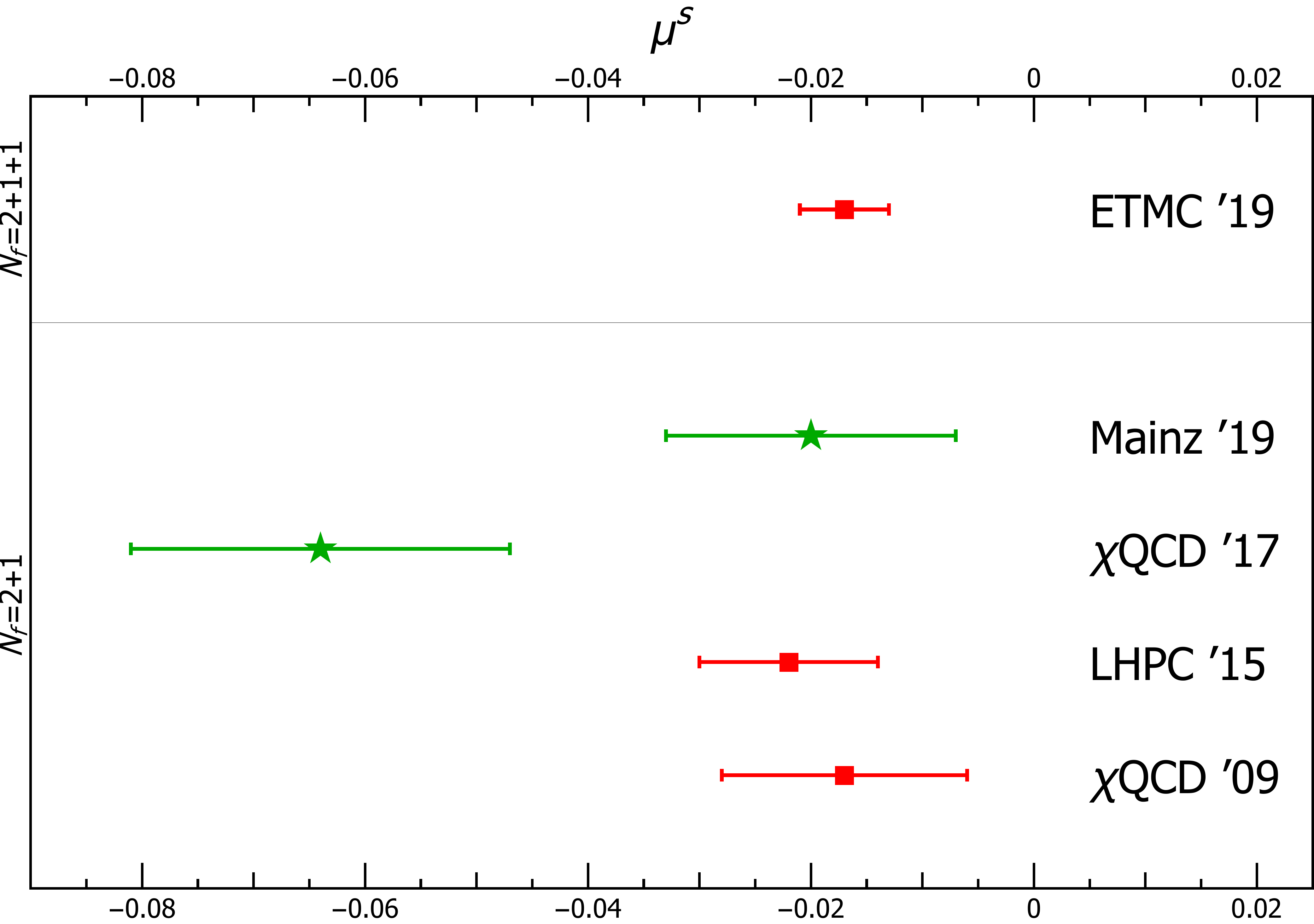}
\includegraphics[width=0.32\textwidth]{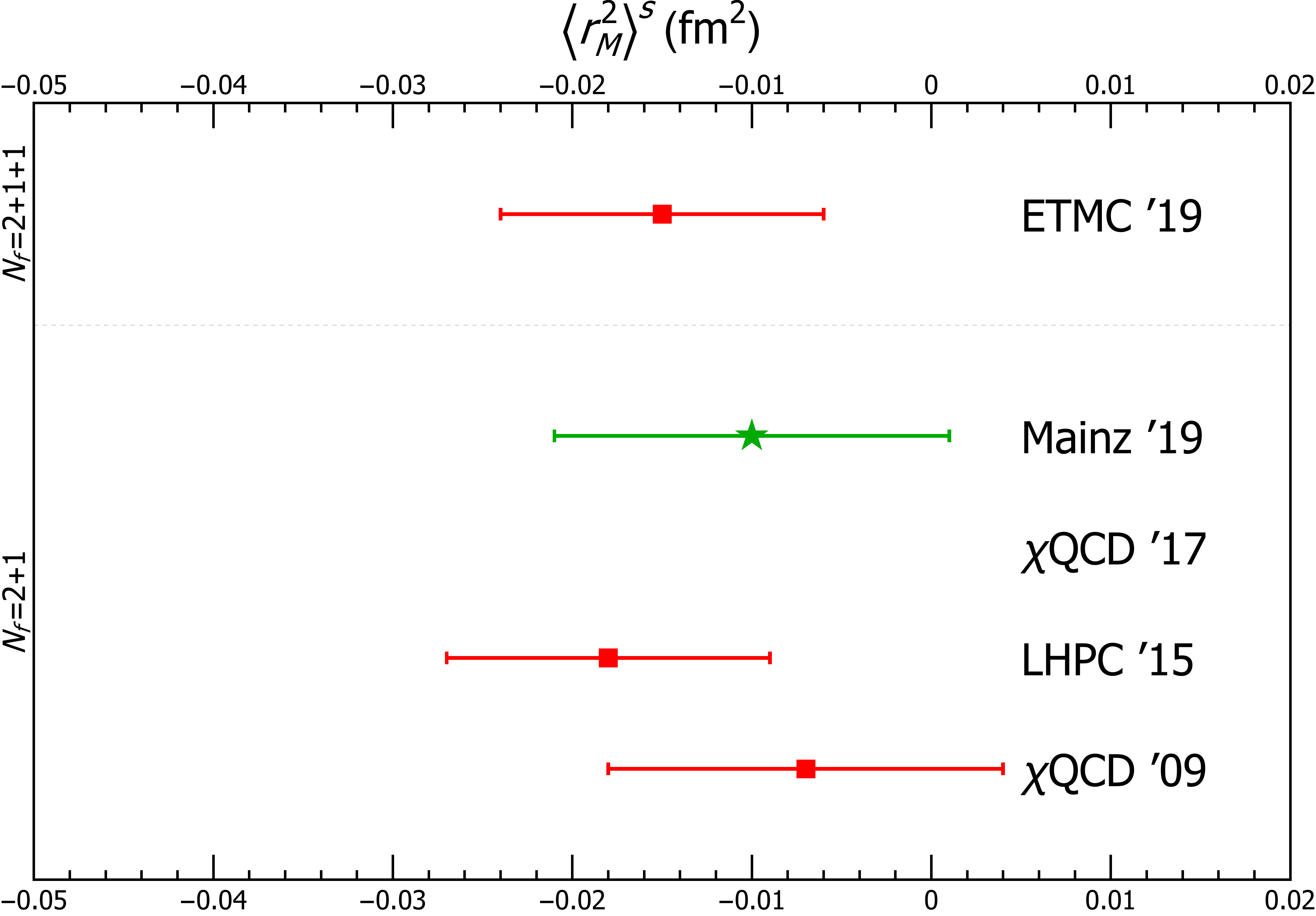}
\caption{Summary of lattice calculations of $\langle r^2_E\rangle^s$ (left), $\mu^s$ (middle) and $\langle r^2_M\rangle^s$ (right) from lattice ensembles with dynamical strange degrees of freedom in the sea. The green symbols show results using multiple ensembles to take the continuum limit $a\rightarrow 0$. Details of the lattice input can be found in Tab.~\ref{tab:rs_mus}.
}
\label{fig:rs_mus}
\end{figure*}

The contributions of strange sea quarks to the nucleon electromagnetic form factors has been of high interest in the last decades. Experimentally, strange electromagnetic form factors can be extracted through the parity-violating asymmetry in the elastic scattering of polarized electrons on unpolarized protons. However, these experimental measurements are difficult and precision determinations of the form factors are not yet available. This is one place where lattice-QCD calculations can provide better results with currently available computational resources. 
A number of collaborations have studied the strange electromagnetic form factors with lattice QCD~\cite{Doi:2009sq,Babich:2010at,Green:2015wqa,Sufian:2016pex,Alexandrou:2018zdf,Djukanovic:2019jtp,Alexandrou:2019olr}; the details of these calculations can be found in Table~\ref{tab:rs_mus}. 
Using the form factors' dependence on transfer-momentum ($Q$), the electric and magnetic radii, as well as the magnetic moment, are obtained via
\begin{equation}
\langle r^2_{E,M}  \rangle^s = -6 \frac{dG^s_{E,M}(Q^2)}{dQ^2}\Big{|}_{Q^2=0}\,,\quad \mu^s\equiv G^s_M(0) .
\end{equation}
We summarize the calculations using dynamical sea quarks in Fig.~\ref{fig:rs_mus}.
Early studies only used only one ensemble with heavy pion masses to reduce computational costs.
There has been great progress made in recent years, including calculations made directly at physical pion mass, such as ETMC's $Nf=2+1+1$ work in Ref.~\cite{Alexandrou:2019olr}. 
$\chi$QCD~\cite{Sufian:2016pex} used four ensembles of valence overlap fermions on $N_f=2+1$ domain wall configurations with multiple volumes, lattice spacings, and pion masses to investigate systematic uncertainties.
The Mainz group~\cite{Djukanovic:2019jtp} used $N_f = 2+ 1$ flavors of ${\cal O}(a)$-improved Wilson fermions, and they also employed an ${\cal O}(a)$-improved vector operator, with lattice spacing as small as 0.05~fm.
Overall, there is agreement among the lattice data, demonstrating that systematic uncertainties are under control.
However, the $\chi$QCD result for the magnetic moment disagrees with the other lattice results, and this tension should be further investigated. 

\begin{table*}[tbp]
\begin{center}
\hspace*{-0.75cm}
\begin{tabular}{|c|c|c|c|c|c|}
\hline
Ref. & Sea quarks & Valence quarks &   a (fm) & $M_\pi$ (MeV) & $M_\pi L$  \\
\hline
$\chi$QCD'09~\cite{Doi:2009sq} &  2+1f clover & clover &  0.12 & 600--840 &  5.97\\ 
BU'10~\cite{Babich:2010at} &  2+1f clover & clover &  0.11 & 416 &  4.79\\ 
LHPC'15 ~\cite{Green:2015wqa} &  2+1f clover & clover &  0.11 & 317 &  5.97\\ 
$\chi$QCD'16~\cite{Sufian:2016pex} & 2+1f DWF& overlap &   0.08--0.14 & 135--400 & 3.89--4.82 \\
ETMC'18~\cite{Alexandrou:2018zdf} & 2f twisted mass & twisted mass & 0.09 & 130 & 2.98 \\
Mainz'19~\cite{Djukanovic:2019jtp} &  2+1f clover & clover  & 0.05--0.09 & 200--360 & 4.14--5.35 \\ 
ETMC'19~\cite{Alexandrou:2019olr} &  2+1+1f twisted mass & twisted mass  & 0.080  & 139 & 3.62 \\
\hline
\end{tabular}
\end{center}
\caption{\label{tab:rs_mus} List of references and details of the lattice-QCD calculations of the nucleon strange form factors.
Here, we omit the calculations using quenched approximation, perturbatively renormalized results, and conference proceedings. 
}
\end{table*}

\subsubsection{Nucleon isovector axial form factors}
There has been also rapidly increasing interest in lattice-QCD nucleon axial form factors ($G_A$) calculations, a key input from the SM for neutrino physics; these form factors can be calculated in terms of the isovector axial current
\begin{eqnarray}\label{eq:cont_axial}
&&\langle N (p_f)| A^+_\mu(x) | N (p_i)\rangle =
\nonumber \\
&&\bar{u}_N
            \left[\gamma_\mu  \gamma_5 G_A(q^2)
             +i q_\mu \gamma_5 {G_P(q^2)} \right]  u_N e^{iq\cdot x}.
\end{eqnarray}
Here, $q=p_f -p_i$ is the momentum transfer between the initial and final state of nucleon; and the
isovector current is $A^+_\mu(x)=\overline{u}\gamma_\mu \gamma_5 u -\overline{d}\gamma_\mu \gamma_5 d$.
$G_A(q^2)$ and $G_P(q^2)$ are known as nucleon axial and induced-pseudoscalar form factors, and
 $\bar{u}_N$ and $u_N$ are the associated nucleon spinors.
In the limit $|{\vec q}| \rightarrow 0$, $G_A(q^2=0)$ should recover the nucleon axial charge $g_A = -1.2723(23)$, which is well determined from neutron $\beta$-decay experiments~\cite{Tanabashi:2018oca}.
Contrariwise, the axial-charge radius squared $r_A^2$ is less known; once $G_A(q^2)$ is obtained,
one can obtain the radius via
\begin{equation}
    r_A^2 \equiv \frac{6}{G_A(0)} \left. \frac{dG_A}{dq^2}\right|_{q^2=0}.
\end{equation}

\begin{figure*}[ht!]
\centering
\includegraphics[scale=0.32]{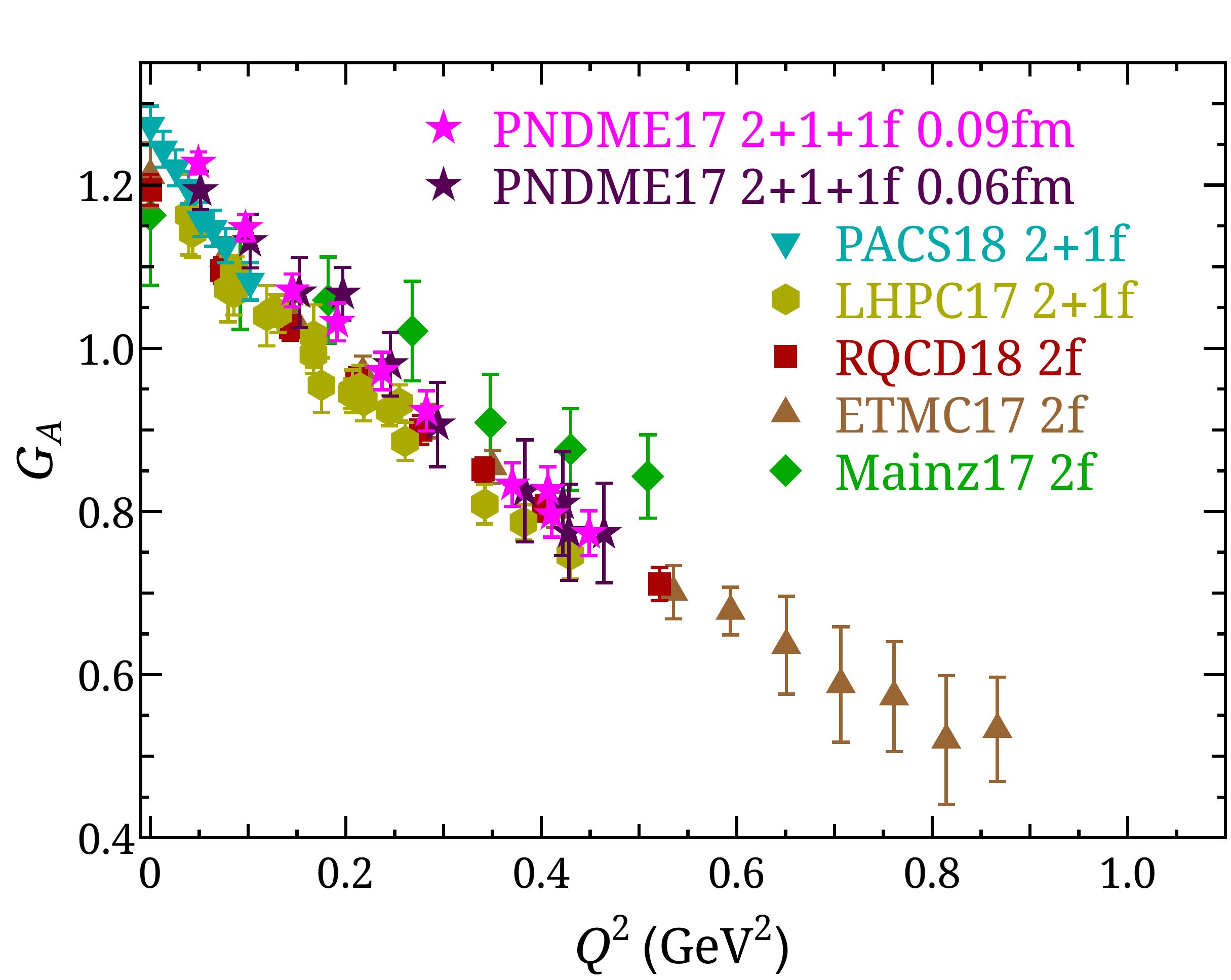} \hfil
\includegraphics[scale=0.32]{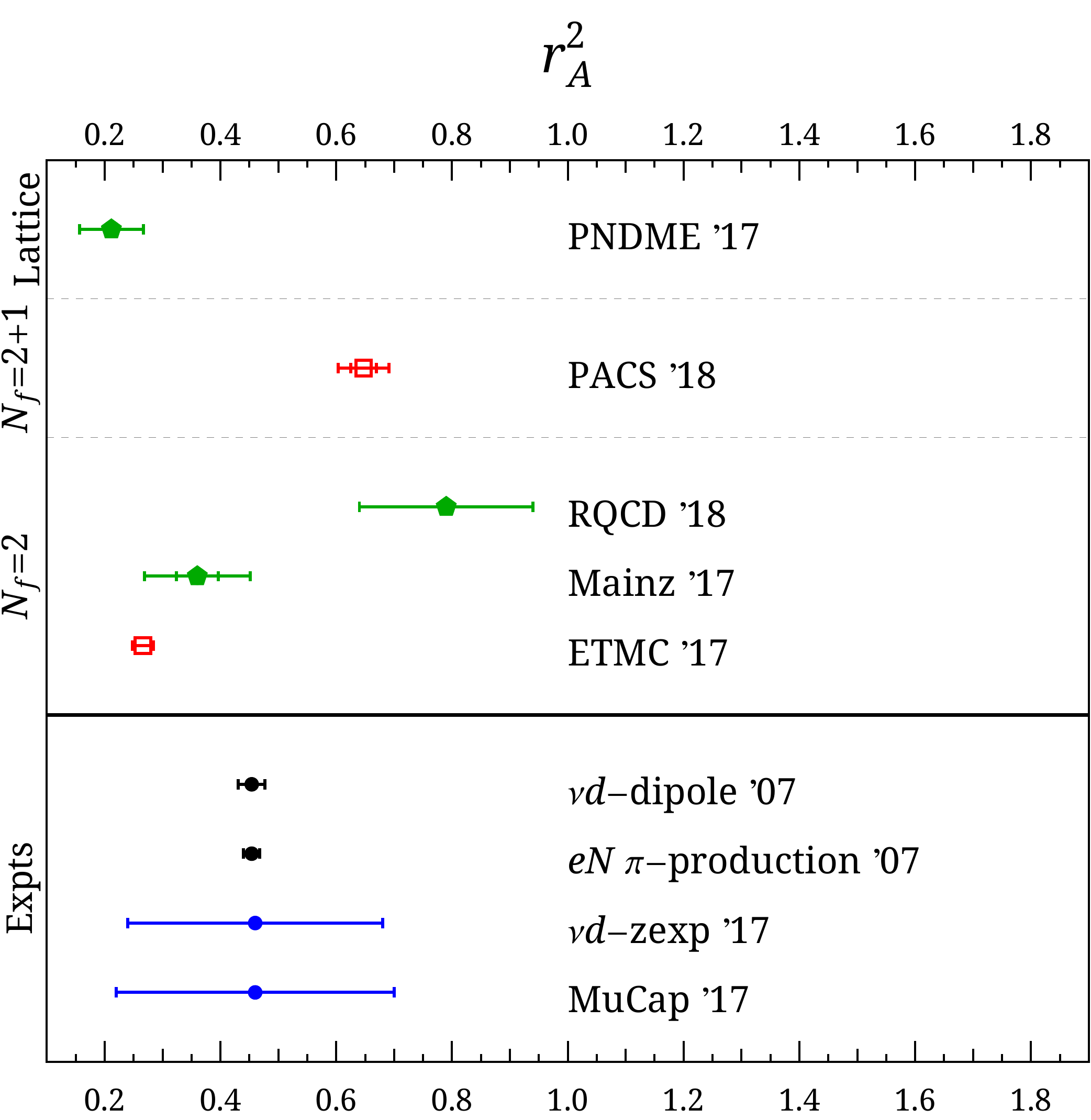}
\caption{\label{fig:axialFF}
Left: nucleon isovector axial form factor as functions of $q^2$. 
Right: summary of $r_A^2$ from calculations on ensembles near physical pion mass~\cite{Capitani:2017qpc,Alexandrou:2017hac,Bali:2018qus,Hasan:2017wwt,Shintani:2018ozy,Rajan:2017lxk}, together with non-lattice determinations.
The color code for $r_A^2$ is adapted from the Flavor Lattice Averaging Group~\cite{Aoki:2016frl}, as specified in the
Appendix of Ref.~\cite{Bhattacharya:2015wna}.
The legends and references for $r_A^2$ are as follows: $\nu d$~(dipole) and $e N \to e N' \pi$~\cite{Bodek:2007ym}, $\nu d$ ($z$ expansion)~\cite{Meyer:2016oeg}, ``MuCap''~\cite{Hill:2017wgb}.
}
\end{figure*}
Although the first lattice-QCD calculation of the isovector nucleon axial form factors can be traced back to the early '90s, there has been significant progress and improvement by the global lattice community. One of the systematics, excited-state contamination, was not consistently addressed 10 years ago, which resulted in a lower nucleon axial coupling $g_A$ and the wrong form factors. Removing this systematic has become essential for any lattice-QCD nucleon calculation.
Another exciting breakthrough in the past decade is the increasing number of lattice nucleon calculations at the physical pion mass, mostly due to recent advances in both algorithms and a worldwide investment in pursuing the first exascale computing machine.
Many calculations now comes with high statistics (${\cal O}(100k)$ measurements) and some with multiple lattice spacings and volumes to control lattice artifacts. Such programs would have been impossible 5 years ago.  Figure~\ref{fig:axialFF} gives a summary of the recent nucleon axial form factors done near the physical pion mass, and the status of lattice-QCD calculations and examples of the phenomenological determinations of $r_A^2$ are shown on the right-hand side of Fig.~\ref{fig:axialFF}. The parameters for each ensemble employed are given in Table~\ref{tab:lQCDAffParams}.
The analysis with the $z$ expansion~\cite{Meyer:2016oeg} eliminates the uncertainty estimates of determinations predicated on the dipole form. The model-independent results (red; between the horizontal lines) illustrate the best estimate of $r_A^2$ without such strong assumptions.

\begin{table*}[tb]
\begin{center}
\hspace*{-0.75cm}
\begin{tabular}{|c|c|c|c|c|c|c|c|}
\hline
Ref. & Sea quarks & Valence quarks & Renormalization &  $N_{\Delta t}$ & a (fm) & $M_\pi$ (MeV) & $M_\pi L$  \\
\hline
ETMC '17~\cite{Alexandrou:2017hac} & 2f twisted mass &  twisted mass &  RI'-MOM & 3 & 0.094& 130~MeV& 3.0\\
LHPC '17~\cite{Hasan:2017wwt} & 2+1f clover & clover &  RI'-MOM & 3 & 0.093& 135~MeV& 4.0\\
Mainz '17~\cite{Capitani:2017qpc} & 2f clover & clover &  Schr\"odinger functional & 4--6 & 0.05--0.08 & 193-- 73 & 4.0--6.0\\
PACS '18~\cite{Shintani:2018ozy} & 2+1f clover & clover & Schr\"odinger functional & 4 & 0.085& 146& 8.0\\
PNDME '17~\cite{Rajan:2017lxk} & 2+1+1f HISQ & clover &  RI'-MOM & 3--4 &   0.06--0.12 & 130--320 & 3.7--5.5\\
RQCD '18~\cite{Bali:2018qus} & 2f  clover & clover &  RI'-MOM & 2--6 & 0.06--0.08 & 150--490 & 3.4--6.7\\
\hline
\end{tabular}
\end{center}
\caption{\label{tab:lQCDAffParams} The lattice parameters used in the near physical pion mass axial form-factor calculations }
\end{table*}

\subsubsection{Nucleon isovector generalized form factors}
There has been lattice-QCD efforts in moments of GPDs calculations in the past few decades, 
and in recent years, there has been emerging direct physical pion mass calculation. 
A common lattice approach is to use the operator product expansion to local operators with  
the nucleon matrix. 
Most calculations focusing on the leading-twist operators, computing 
isovector nucleon matrix elements of the one-derivative operators to obtained the generalized form factors ($A_{20}$, $B_{20}$ , $C_{20}$, $\tilde A_{20}$, $\tilde B_{20}$,$\tilde C_{20}$) though:
\begin{widetext}
  \begin{align}
    \langle N(p^\prime,s^\prime)| \mathcal{O}_{V}^{\mu\nu}| N(p,s)\rangle =
    \bar u_N(p^\prime,s^\prime)\frac{1}{2}\Bigl[& A_{20}(q^2)\, \gamma^{\{ \mu}P^{\nu\}} +B_{20}(q^2)\, \frac{i\sigma^{\{\mu\alpha}q_{\alpha}P^{\nu\}}}{2m_N}+C_{20}(q^2)\, \frac{1}{m_N}q^{\{\mu}q^{\nu\}} \Bigr] u_N(p,s), \nonumber \\
    \langle N(p^\prime,s^\prime)| \mathcal{O}_{A}^{\mu\nu}| N(p,s)\rangle =
    \bar u_N(p^\prime,s^\prime)\frac{i}{2}\Bigl[& \tilde A_{20}(q^2)\, \gamma^{\{\mu}P^{\nu\}}\gamma^5 + \tilde B_{20}(q^2)\,\frac{q^{\{\mu}P^{\nu\}}}{2m_N}\gamma^5\Bigr]u_N(p,s),\,%
  \end{align}
\end{widetext}
where $u_N$ are nucleon spinors, $q=p'-p$ is the momentum transfer,
$P=(p'+p)/2$, $m_N$ is the nucleon mass, and operators 
\begin{align}
  \mathcal{O}_{V}^{\mu\nu} =& \bar{\psi}\gamma^{\{\mu}\overleftrightarrow{D}^{\nu\}}\frac{\tau^3}{2}\psi,\nonumber\\
  \mathcal{O}_{A}^{\mu\nu} =& \bar{\psi}\gamma_5\gamma^{\{\mu}\overleftrightarrow{D}^{\nu\}}\frac{\tau^3}{2}\psi,\,\textrm{and}%
\end{align}
where $\psi$ ($\bar{\psi}$) are (anti-)light quark flavor, 
$\tau^3$ as Pauli matrix, 
and the curly (square) brackets denote
symmetrization (antisymmetrization) 
\begin{equation}
  \overleftrightarrow{D}_\mu = \frac{1}{2}(\overrightarrow{D}_\mu - \overleftarrow{D}_\mu),\, D_\mu = \frac{1}{2}(\nabla_\mu + \nabla_\mu^*)
\end{equation}
with $\nabla_\mu$  ($\nabla_\mu^*$) denoting the forward (backward) 
derivatives on the lattice. For space limitations, we do not include the decomposition of the tensor current ($\mathcal{O}_{T}^{\mu\nu\rho}$), which leads to the GFFs $A_{T20}$, $B_{T20}$, $C_{T20}$, as we do not present any results here.

These nucleon matrix elements
can be expanded in terms of generalized form factors (GFFs), which are
Lorentz invariant functions of the momentum transfer squared. At zero
momentum transfer, these nucleon matrix elements yield the second Mellin
moments of the unpolarized, helicity and transversity PDFs. There are very limited calculations on the generalized form factors, as can be seen in Table~\ref{tab:lQCDGFFsParams}. In this summary we exclude calculations using quenched approximation, perturbatively renormalized results, as well as conference 
proceedings. In Fig.~ref{fig:LatGFF} we give the results obtain with simulations at the physical point by ETMC using three ensembles~\cite{Alexandrou:2019ali}, and the near physical point calculation or RQCD~\cite{Bali:2018zgl}.  We show the unpolarized GFFs ($A_{20},\,B_{20}$) and the linearly polarized GFFs ($\tilde A_{20},\,\tilde B_{20}$). The unpolarized GFF $C_{20}$ is found consistent with zero and, thus, not shown here. It is interesting to observe that the two data sets for $A_{20}$ in the ETMC calculation exhibit some tension. This is an indication of systematic uncertainties. Given that the blue points correspond to finer lattice spacing, and larger volume and larger $m_\pi L$ value, we expect that the blue points have suppressed systematic uncertainties. Both ensembles of ETMC lead to compatible results for the remaining GFFs. The comparison between the $N_f=2$ ETMC data and $N_f=2$ RQCD data reveals agreement for $A_{20}$,  $B_{20}$ and $\tilde B_{20}$. However, the RQCD data have a different slope than the ETMC data, which is attributed to the different analysis methods, and systematic uncertainties.

\begin{table*}[tbp]
\begin{center}
\hspace*{-0.75cm}
\begin{tabular}{|c|c|c|c|c|c|c|c|}
\hline
Ref. & Sea quarks & Valence quarks & Renormalization &  $N_{\Delta t}$  & a (fm) & $M_\pi$ (MeV) & $M_\pi L$  \\
\hline
  ETMC '19 \cite{Alexandrou:2019ali} &  2f $\&$ 2+1+1f TM &  twisted mass &  RI'-MOM & 3 & 0.080, 0.094& 130--139~MeV& 3.0--4.0 \\
  ETMC '13 \cite{Alexandrou:2013joa} &  2+1+1f twisted mass & twisted mass & RI'-MOM  & 1  & 0.06--0.08 & 210,370 & 3.4--5.0 \\ 
  ETMC '11 \cite{Alexandrou:2011nr} &  2f twisted mass & twisted mass & RI'-MOM  & 1 & 0.056--0.089  & 262--470 & 3.3--5.3\\ 
 LHPC '10 \cite{Bratt:2010jn} &  2+1f Astad & domain wall  & RI'-MOM  & 1 & 0.124 & 293--596 & 3.7--7.5\\
  LHPC '07 \cite{Hagler:2007xi} &  2+1f staggered & domain wall  & RI'-MOM  & 1 & 0.124 & 350--760 & 4.4--9.6\\
    RQCD '19 \cite{Bali:2018zgl} &  2f clover & clover & RI'-MOM & 1--6 & 0.060--0.081  & 150--490 & 2.8--6.7\\

\hline
\end{tabular}
\end{center}
\caption{\label{tab:lQCDGFFsParams} List of references and details of the lattice-QCD moments of generalized form factors calculations.
Here we omit the calculations using quenched approximation, perturbatively renormalized results, and conference proceedings. 
}
\end{table*}

\begin{figure*}[tb]
\centering
\includegraphics[width=0.42\textwidth]{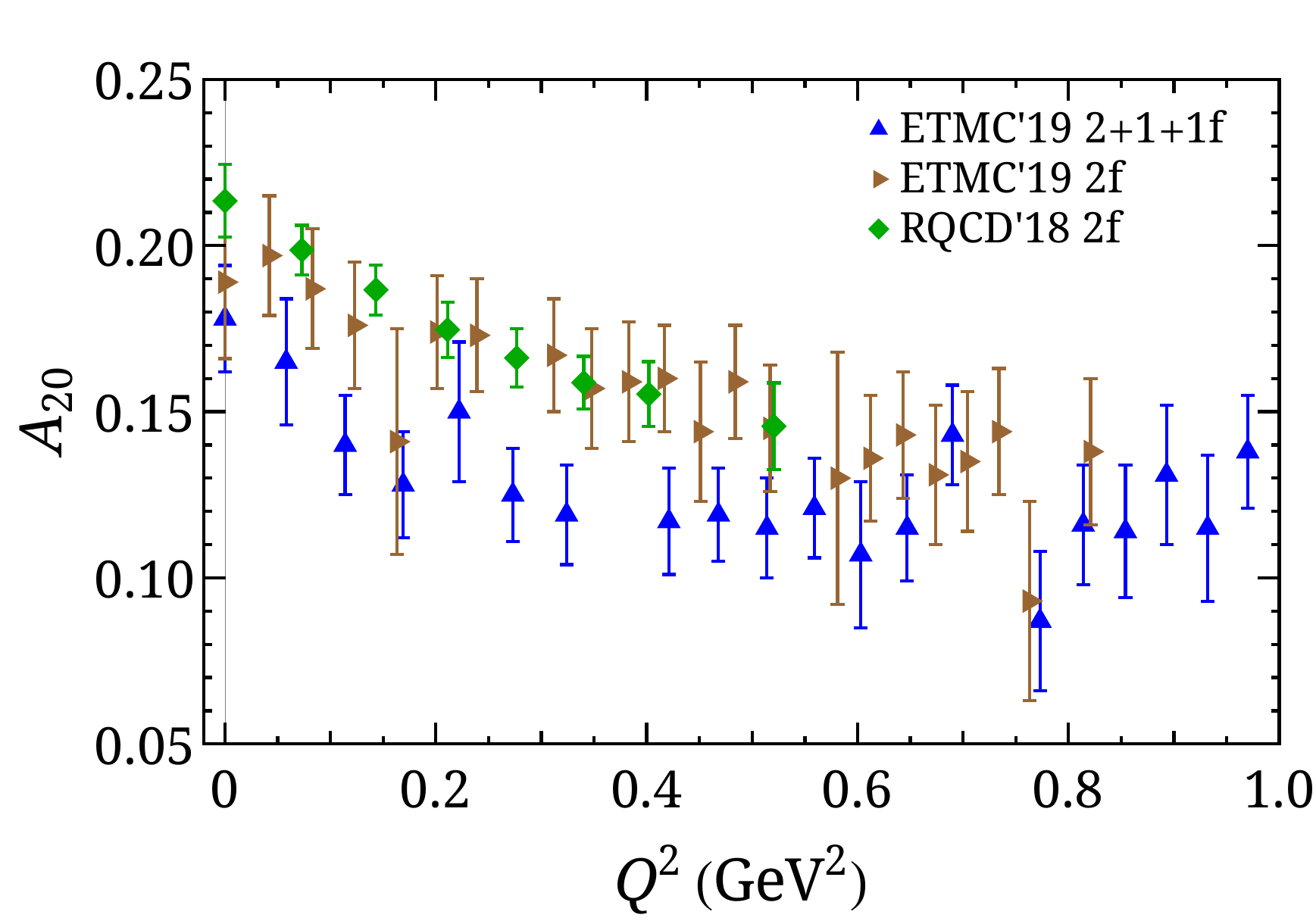} \hfil
\includegraphics[width=0.42\textwidth]{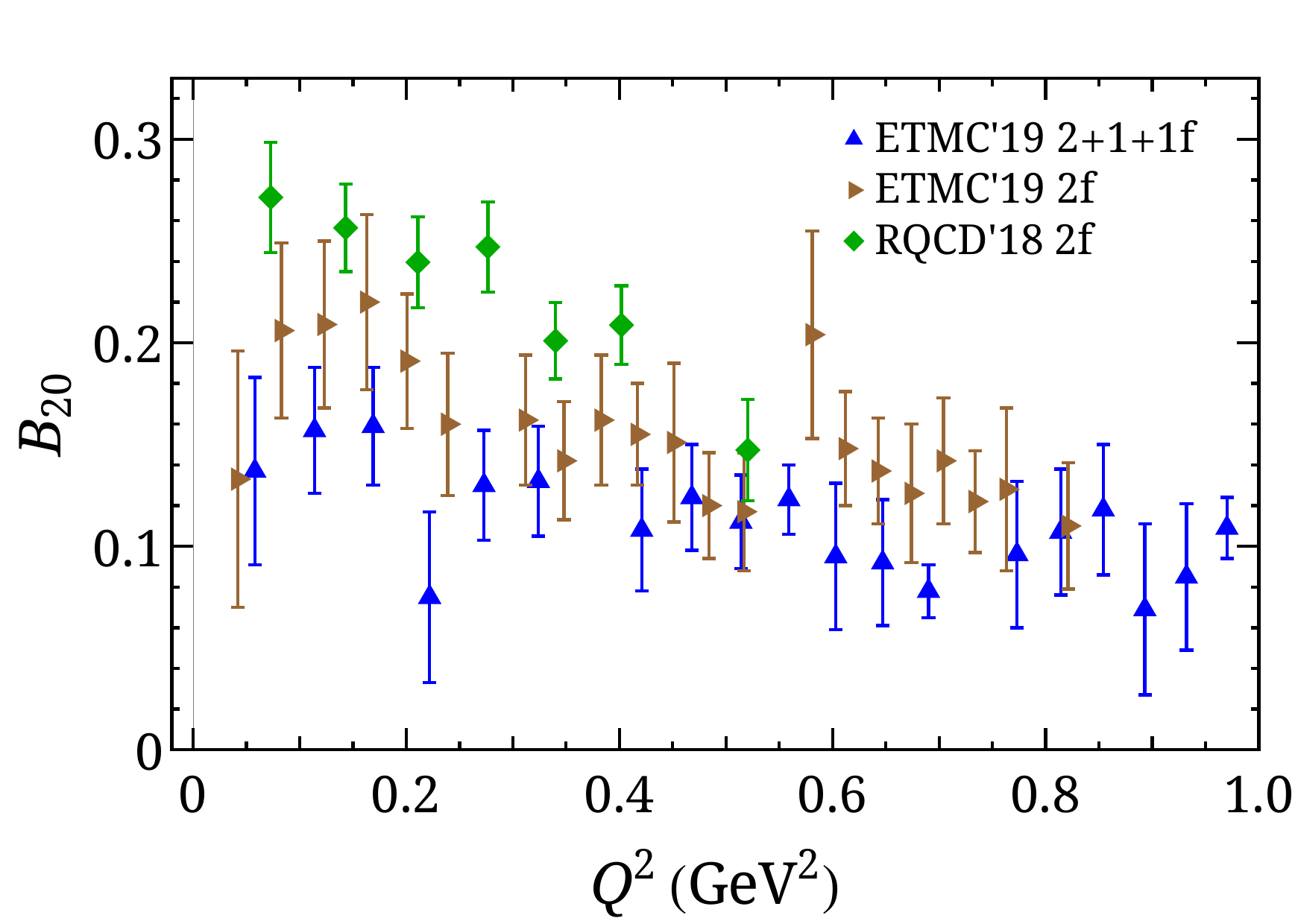}
\includegraphics[width=0.42\textwidth]{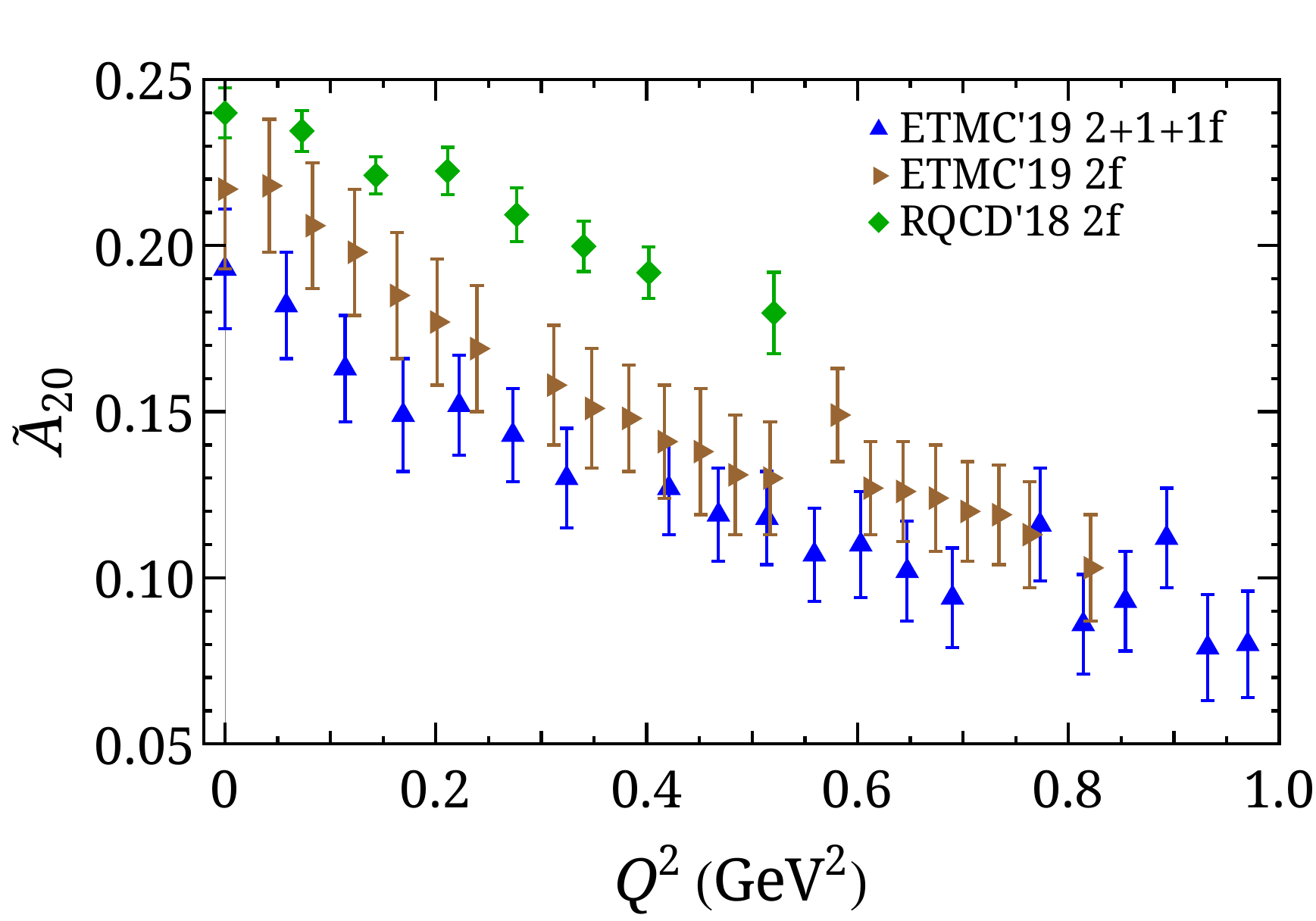} \hfil
\includegraphics[width=0.42\textwidth]{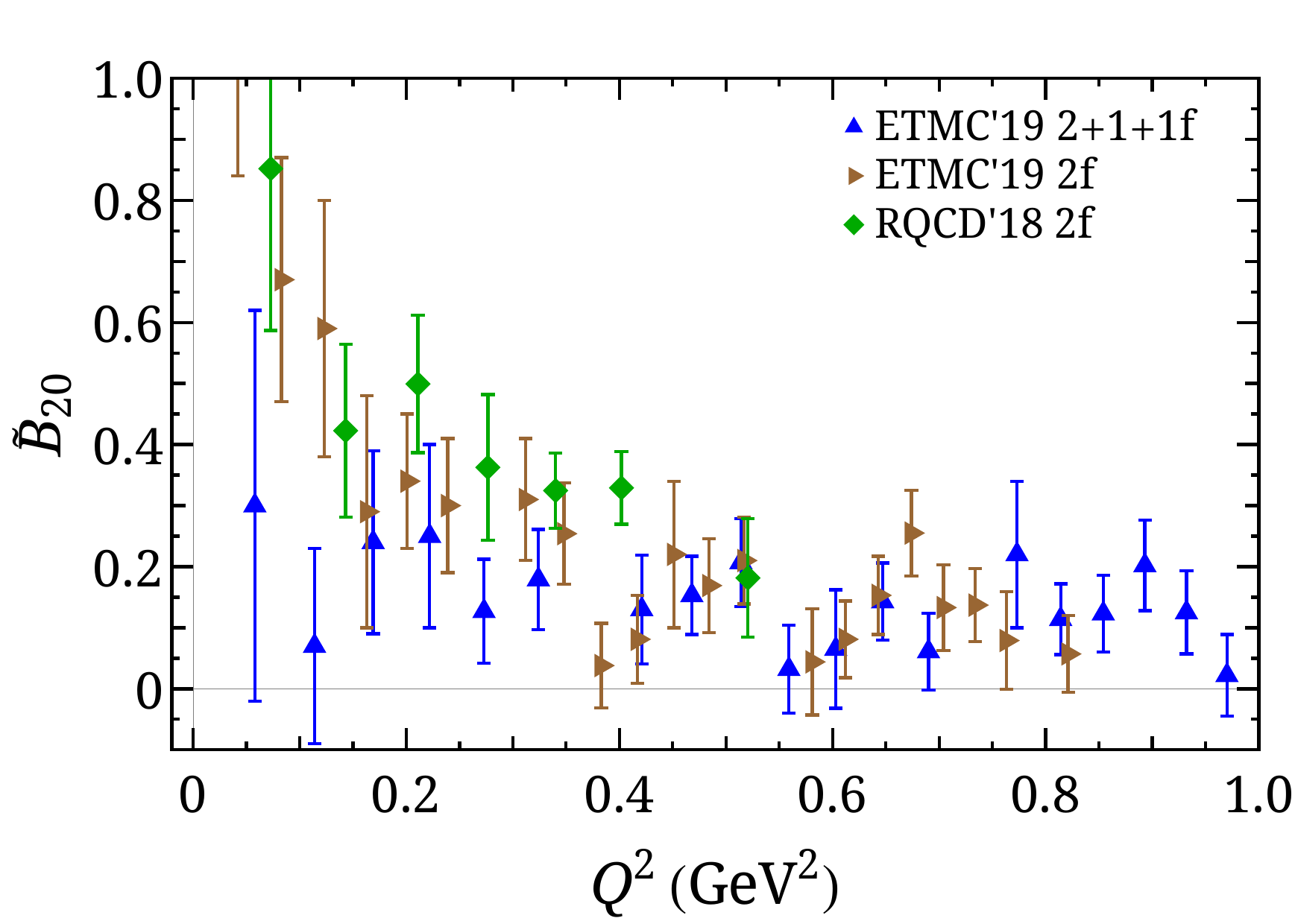}
\caption{\label{fig:LatGFF}
Unpolarized (left) and linearly polarized (right) nucleon isovector GFFs from near physical pion mass
as functions of transferred momentum $Q^2$.
The references corresponding to the above works are: 
2+1+1f ETMC19~\cite{Alexandrou:2019ali},
2f ETMC19~\cite{Alexandrou:2019ali} (only the larger-volume results is quoted here), 
2f RQCD19~\cite{Bali:2018zgl}. 
}
\end{figure*}

\subsection{x-dependent GPDs from lattice QCD}
\label{sec:xdep}

\vspace{0.3cm}
Information on GPDs from lattice QCD has been available via their form factors and generalized form factors, using Operator Product Expansion (OPE). As in the case of PDFs, such information is limited due to suppression of the signal as the order of the Mellin moments increases, but also as the momentum transfer between the initial and final state increases. 

Significant progress has been made towards new methods to access the $x$- and $t$-dependence of GPDs ($t=-Q^2$), which is driven by the advances in the calculations of PDFs presented in Sec.~\ref{02sec3}. Here we present the methodology for obtaining GPDs from the lattice using the quasi- and pseudo-distributions approach, as well as first results for the nucleon and pion GPDs.

\subsubsection{GPDs using the quasi-distribution approach}
In the quasi-distribution approach, GPDs
can be extracted in the same way as extracting PDFs, except that one needs to take into account the off-forward kinematics. For the unpolarized quark GPDs in the nucleon defined by Eq.~(\ref{eq:def_vec}) for the unpolarized case. On the lattice we can calculate the so-called quasi-GPDs
\begin{align}\label{eq:quasiGPD}
&\tilde F(x,\tilde{\xi},t,{\bar P}_3)=\frac{{\bar P}_3}{{\bar P}^0}\int \frac{dz}{4\pi}e^{i x z {\bar P}_3 }\langle P'|{\tilde O}_{\gamma^0}(z)|P\rangle\nn\\
&=\frac{\bar{u}(P')}{2{\bar P}^0}\bigg\{\widetilde H(x,\tilde\xi,t,{\bar P}_3)\gamma^0
+\widetilde E(x,\tilde\xi,t,{\bar P}_3) \frac{i\sigma^{0\mu}\Delta_\mu}{2M}\bigg\}u(P)\,,
\end{align}
as Eq.~(\ref{eq:def_vec}) is not accessible on a Euclidean lattice. The skewness is replaced by the quasi-skewness
\begin{align}
\tilde{\xi}=-\frac{P'_3-P_3}{P'_3+P_3}=-\frac{\Delta_3}{2{\bar P}_3} =  \xi +\mathcal{O}\left( \frac{M^2}{({\bar P}_3)^2}, \frac{t}{({\bar P}_3)^2} \right)
\end{align}
differs from the light-cone skewness $\xi$ by power suppressed corrections. It can be replaced by $\tilde\xi$ with the difference being attributed to generic power corrections. Then the quasi-GPDs can be factorized into the GPDs as~\cite{Ji:2015qla,Liu:2019urm}
\begin{align}
	&\tilde F(y,\xi,t,{\bar P}_3,\mu) \nn\\
	&\hspace{2em}=  \int_{-1}^1 \frac{dx}{|\xi|} C\left(\frac{y}{\xi}, \frac{x}{\xi}, \frac{\mu}{\xi {\bar P}_3 }\right)F(x,\xi,t,\mu) +\ldots,
\end{align}
where $\ldots$ denotes power corrections of the form $\mathcal O(M^2/{\bar P}_3^2,t/{\bar P}_3^2,\Lambda_{\rm QCD}^2/({y^2{\bar P}_3^2}))$.

Prior to lattice calculations, quasi-GPDs have been studied in models, which provides a qualitative qualitative understanding of GPDs. In Ref.~\cite{Bhattacharya:2018zxi} the quasi-GPDs are studied in the scalar diquark spectator model. Among other, some robust features of quasi-GPDs were revealed concerning the behavior at large $x$, the behavior around the cross-over points $x = \pm \xi$, as well as their behavior in the ERBL region. In the follow-up work presented in Ref.\cite{Bhattacharya:2019cme} the Authors findings include two model-independent results that are interesting for lattice calculations. This regards a discussion of the moments of quasi-GPDs (and quasi-PDFs), and the behavior of quasi-GPDs under $\xi \to -\xi$.  While for light-cone GPDs the $\xi$-symmetry has been of academic interest, for the quasi-GPDs this point could potentially be of practical relevance to decrease statistical uncertainties in lattice calculations.

In lattice QCD, there are several challenges in calculating quasi-GPDs. In contrast with the PDFs, the extraction of GPDs is more challenging because they require momentum transfer, $Q^2$, between the initial (source) and final (sink) states. Another complication is the fact that the GPDs are defined in the Breit frame, in which the momentum transfer is equally distributed to the initial and final states.

Working in the Breit frame has no conceptual difficulty, however, it increases the computational cost, as separate calculations are necessary for each value of the momentum transfer. The bare matrix element
\begin{equation}
N(P3{+}Q/2)|\bar\psi\left(z\right)\Gamma W(0,z)\psi\left(0\right)|N(P_3{-}Q/2)\rangle
\end{equation}
is calculated using the techniques developed for PDFs, and the use of momentum smearing method~\cite{Bali:2016lva},

Another factor that increases the computational cost is the need of an optimized ratio of 3pt- and 2pt-functions, that is:
\begin{eqnarray}
&&\hspace*{-0.25cm}R_{\cal O}(\Gamma;\vec{p}',\vec{p}; t_s, t_{in})=\frac{C^{3pt}(\Gamma;\vec{p}',\vec{p}; t_s, t_{in})}{C^{2pt}(\Gamma_0;\vec{p}', t_{in})}\times \nonumber\\
&&\hspace*{-0.25cm}\sqrt{\frac{C^{2pt}(\Gamma_0;\vec{p},t_s{-}t_{in})\,C^{2pt}(\Gamma_0;\vec{p}', t_{in})\,C^{2pt}(\Gamma_0;\vec{p}', t_s)}{C^{2pt}(\Gamma_0;\vec{p}',t_s{-}t_{in})\,C^{2pt}(\Gamma_0;\vec{p}, t_{ins})\,C^{2pt}(\Gamma_0;\vec{p}, t_s)}}\,. 
\end{eqnarray}
As can be seen, one needs to calculate the 2pt-functions with momentum boost equal to the source momentum, as well as, the sink momentum. The momentum smearing method is applied on the 2pt-functions, with the exponent of the additional phase, being in parallel to the direction of the momentum boost. In addition to the above, different projectors are needed (for baryons) to disentangle the GPDs.

The aforementioned matrix elements decompose into two form factors, which after the Fourier transform and the matching lead to the light-cone GPDs $H,\,E$ for the unpolarized case, and $\widetilde{H},\,\widetilde{E}$ for the helicity case. Therefore one can extract their $x$ dependence at each value of $P_3$ and momentum transfer. For GPDs, so-called ERBL region is defined for $\xi\ne0$, via $|x|<\xi$. As discussed in Ref.~\cite{Liu:2019urm}, the matching for $\xi=0$ is the same as the one for PDFs, while $\xi\ne0$ requires different matching, which is derived in Ref.~\cite{Liu:2019urm} using an RI-type scheme. 

\begin{figure}[tb!]
 	\centering
    \includegraphics[width=0.45\textwidth]{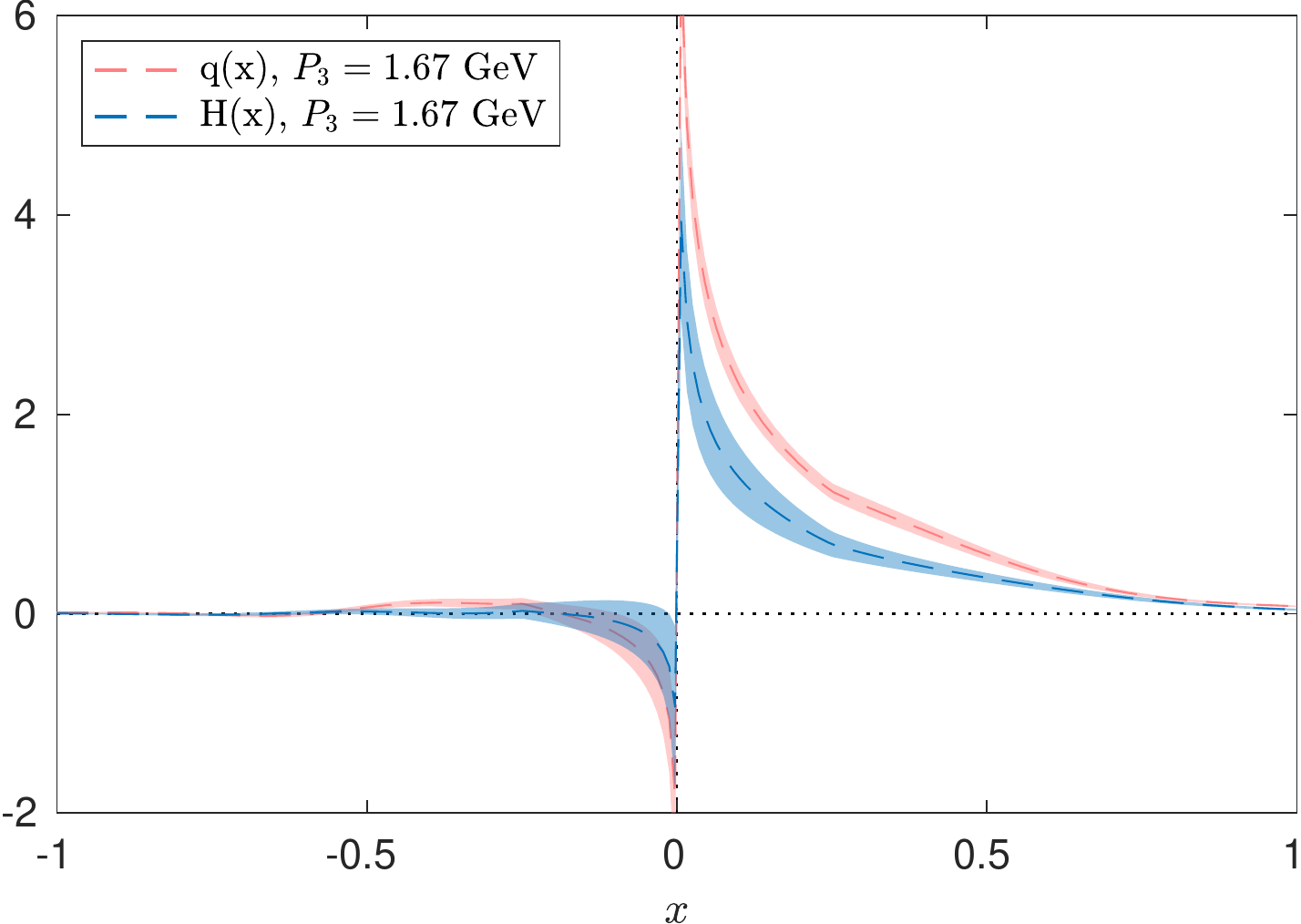} 
 	\caption{$H$-GPD for $P_3=1.67$ GeV, $Q^2=0.69$ GeV$^2$, and $\xi=0$, obtained by the ETM Collaboration.}
 	\label{fig:H_GPD}
 \end{figure}

\begin{figure}[tb!]
 	\centering
    \includegraphics[width=0.45\textwidth]{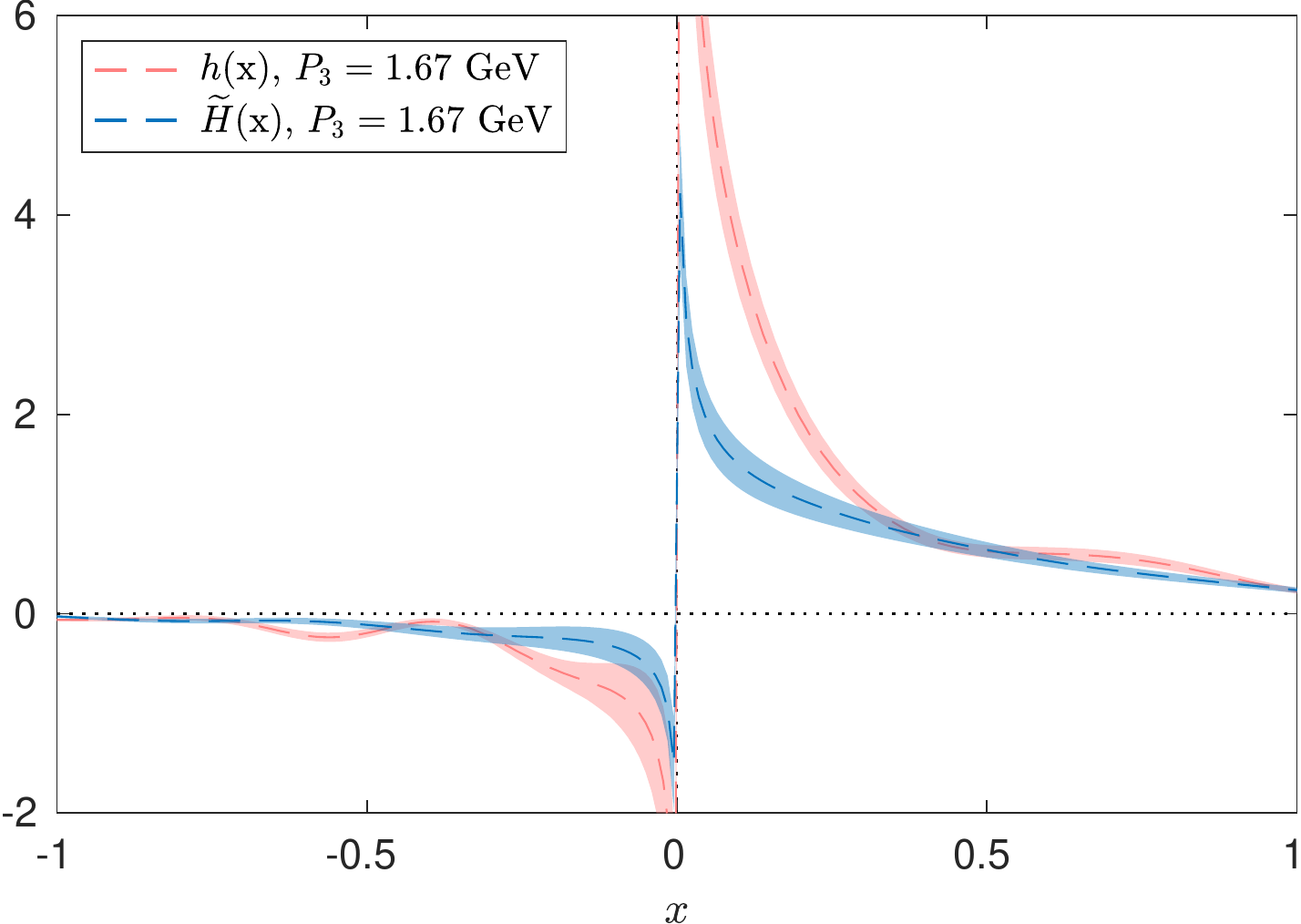}
 	\caption{$\widetilde{H}$-GPD for $P_3=1.67$ GeV, $Q^2=0.69$ GeV$^2$, and $\xi=0$, obtained by the ETM Collaboration.}
 	\label{fig:Htilde_GPD}
 \end{figure}

To date there has been one calculation for the nucleon GPDs by ETMC, and one for the pion GPDs~\cite{Chen:2019lcm}. The former uses one ensemble of $N_f=2+1+1$ gauge configurations at a pion mass of 260 MeV. Part of these results have been presented in Ref.~\cite{Alexandrou:2019dax}. This work presents results of the unpolarized and helicity GPDs, for both $\xi=0$ and $\xi\ne0$. Here we give results for the GPDs at $\xi=0$. Note that $\widetilde{E}$-GPD cannot be accessed in this case, as its kinematic factor becomes zero. Fig.~\ref{fig:H_GPD} compares the $H$-GPD at $Q^2=0.69$ GeV$^2$ with the unpolarized PDF, in order to see the effect of nonzero $Q^2$. The momentum boost is 1.67 GeV$^2$ for both cases. One can see that the $H$-GPD is suppressed as compared to the PDF, especially in all regions of $x$.
A similar comparison can be seen in Fig.~\ref{fig:Htilde_GPD} for the $\widetilde{H}$-GPD and the helicity PDF, for $P_3=1.67$ GeV. Similar to the case of the $H$-GPD, the introduction of the momentum transfer suppresses the GPD, as expected from the behavior of the form factors and generalized form factors. For the intermediate- to large-$x$ region we find that the $H$-GPD is compatible with the helicity PDF.

\begin{figure}[tb]
\includegraphics[width=0.49\textwidth]{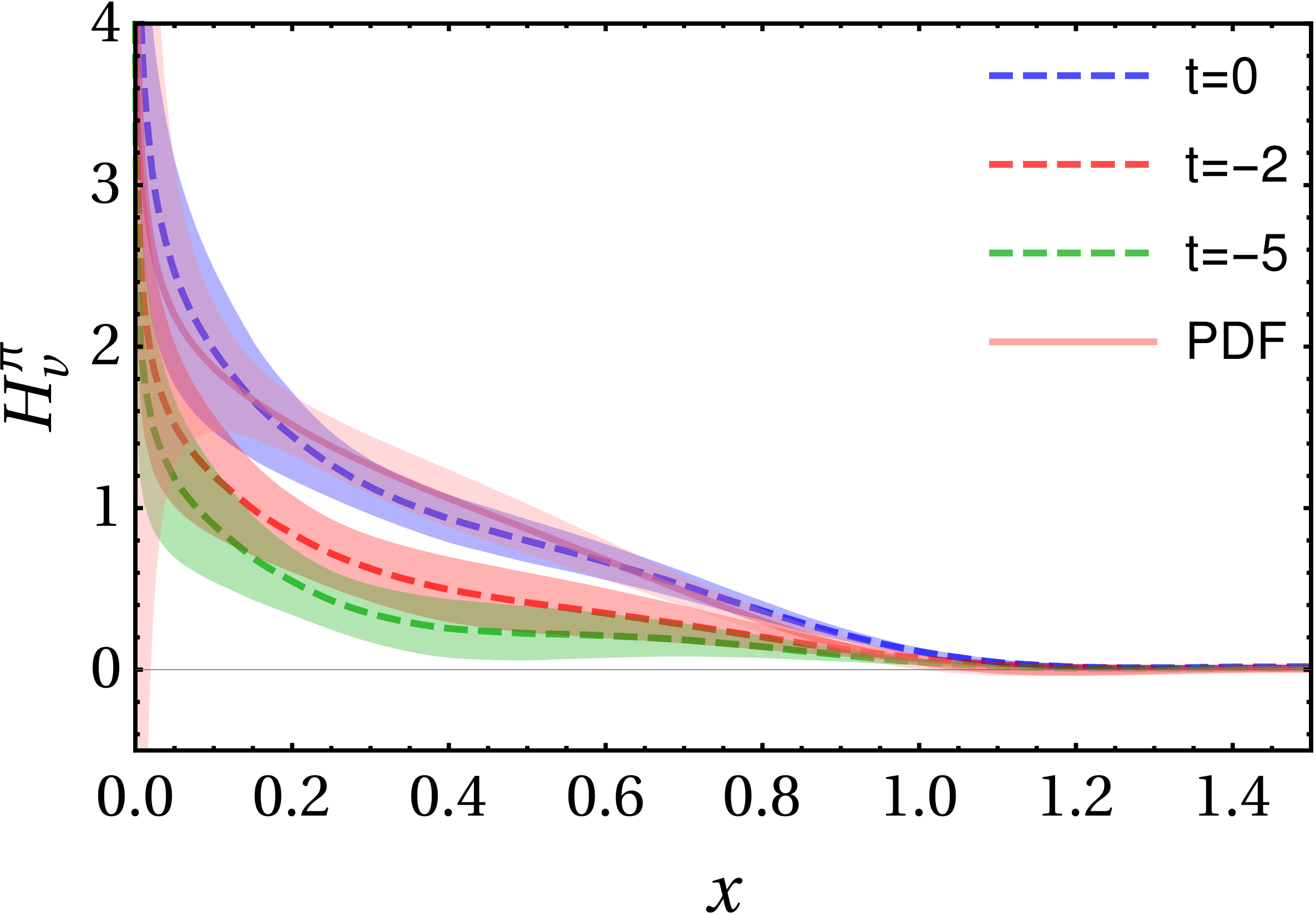}
\caption{
The zero-skewness pion valence quark GPD {$H^{\pi^{+}}_{v}(x,\xi=0,t,\mu=4\text{GeV})$
for $t=\{0, -2, -5\}(2\pi/L)^2$} after one-loop matching  and the meson-mass correction. {``PDF" denotes the pion PDF result in Ref.~\cite{Chen:2018fwa}}.}
\label{fig:xqvpi}
\end{figure}

 In~\cite{Chen:2019lcm}, the pion valence quark GPD at zero skewness was calculated using clover valence fermions on an ensemble of gauge configurations with $2+1+1$ flavors (degenerate up/down, strange and charm) of highly improved staggered quarks (HISQ) with lattice spacing $a \approx 0.12$~fm, box size $L \approx 3$~fm and pion mass $m_\pi \approx 310$~MeV. The result is shown in Fig.~\ref{fig:xqvpi}. It turns out that, with current uncertainties, the result does not show a clear preference among different model assumptions about the kinematic dependence of the
GPD. To disentangle different models, further studies using higher-statistics data will be crucial.

\subsubsection{GPDs using pseudo-distribution approach}

The generalization of the pseudo-distributions approach to include GPDs was developed recently~\cite{Radyushkin:2019owq}.
For lattice applications, and the discussion presented here we choose  $z=z_3$ without loss of generality.  Decomposing the hadron momenta $p_1$ and $p_2$ into
\mbox{$p_1=\{E_1,\Delta_{1,\perp},P_1\}$}  and 
\mbox{$p_2=\{E_2,  \Delta_{2,\perp},P_2\}$,}  
one deals with two Ioffe-time invariants $\nu_1= P_1 z_3$ and $\nu_2 =P_2 z_3$. 

The choice of operator $\gamma^0$ (instead of the alternative $\gamma^3$)  eliminates the $z^0$ and $\Delta_\perp^0$ parts from the parametrization
  \begin{align}
  \langle p_2 | \bar   \psi(-z_3/2) \gamma^0 \ldots  \psi (z_3/2)|p_1 \rangle = 2 {\cal P}^{0}   {M} (\nu_1,\nu_2,t;z_3^2)
\,  
 \
 \label{Mnn}
\end{align} 
defining  the {\it double}  Ioffe-time pseudodistribution  ${M} (\nu_1,\nu_2,t;z_3^2)$. After denoting 
  \mbox{$\nu=(\nu_1+\nu_2)/2
   $,}  the latter 
 converts into  the {\it generalized Ioffe-time pseudodistribution} (pseudo-GITD)
$ {\cal M} (\nu, \xi ,t;z_3^2)$.  To remove link-related UV divergences, one may introduce 
 the {\it reduced}  pseudo-GITD   \cite{Radyushkin:2019owq}
   \begin{align}
{\mathfrak M} (\nu, \xi,t; z_3^2) \equiv \frac{
{\cal M} (\nu, \xi, t; z_3^2)}{ {\cal M} (0,0,0; z_3^2)} \  .
 \label{redITDGPD}
\end{align}
For small $z_3^2$, it may be expressed  (see Ref.   \cite{Radyushkin:2019owq}) in terms of  the  light-cone ITD 
  \begin{align}
  & {\cal I} (\nu, \xi,t;\mu^2)
 \equiv  \,  \int_{-1}^1 d x \, e^{i x \nu } \, 
    { H} \left (x,\xi,t;\mu^2\right ) 
\,  
 \
 \label{MH}
\end{align}  
 by a perturbative  matching relation that  has  the  usual  pQCD factorization  form 
  \begin{align} 
{\mathfrak  M } (\nu, \xi, t; z_3^2 )=  \int_{-1}^1  {d w} \, C (w,\xi,  z_3^2 \mu ^2) \,& {\cal I} (w \nu,\xi,t ; \mu^2)  
\nonumber \\&
 + {\cal O} (z_3^2) 
  \  .
\label{IOPE}
\end{align}

\subsubsection{Extraction of GPDs from data using pseudo-distribution approach}
\label{sec:pseudo}
   To extract  GPDs, it is proposed   to model   ${\cal I} (w \nu,\xi,t ; \mu^2)  $ from  some parametrization 
for $H(x,\xi,t;\mu^2)$  \cite{Radyushkin:2019owq},  and    fit its  parameters  using  lattice data on ${\mathfrak M} (\nu, \xi,t; z_3^2) $.
 The {\it polynomiality} property \cite{Mueller:1998fv,Ji:1996ek,Radyushkin:1997ki} of 
   GPDs may be 
efficiently     taken into account by  using  the {\it double distribution Ansatz} \cite{Radyushkin:1998es} 
in the GPD modeling. 

An equivalent   strategy %
  is to convert (\ref{IOPE})  into a  kernel relation  \cite{Radyushkin:2019owq,Cichy:2019ebf,Radyushkin:2019mye}
\begin{align} 
{\mathfrak  M } (\nu, \xi, t; z_3^2 ) =  \int_{-1}^1  {d x} \, R (x \nu, \xi, z_3^2 \mu ^2) \,& {H} (x,\xi,t; \mu^2)  
\nonumber \\&
 + {\cal O} (z_3^2) 
  \  
\label{ker}
\end{align}
 obtained by  writing 
 $  {\mathcal I}(w\nu,\xi,t,\mu^2) $  in terms of $H(x,\xi,t;\mu^2)$ using  Eq. (\ref{MH}).
Note that  Eq. (\ref{ker}) directly relates the light-cone GPD $H(x,\xi,t;\mu^2)$  with the 
 lattice data on ${\mathfrak M} (\nu, \xi,t; z_3^2)$ through a perturbatively calculable 
 kernel $ R (x \nu, \xi, z_3^2 \mu ^2)$. Hence,  no intermediaries 
(like quasi-PDFs) are needed in this approach.

\begin{figure}[tb]
\begin{center}
\includegraphics[width=0.49\textwidth]{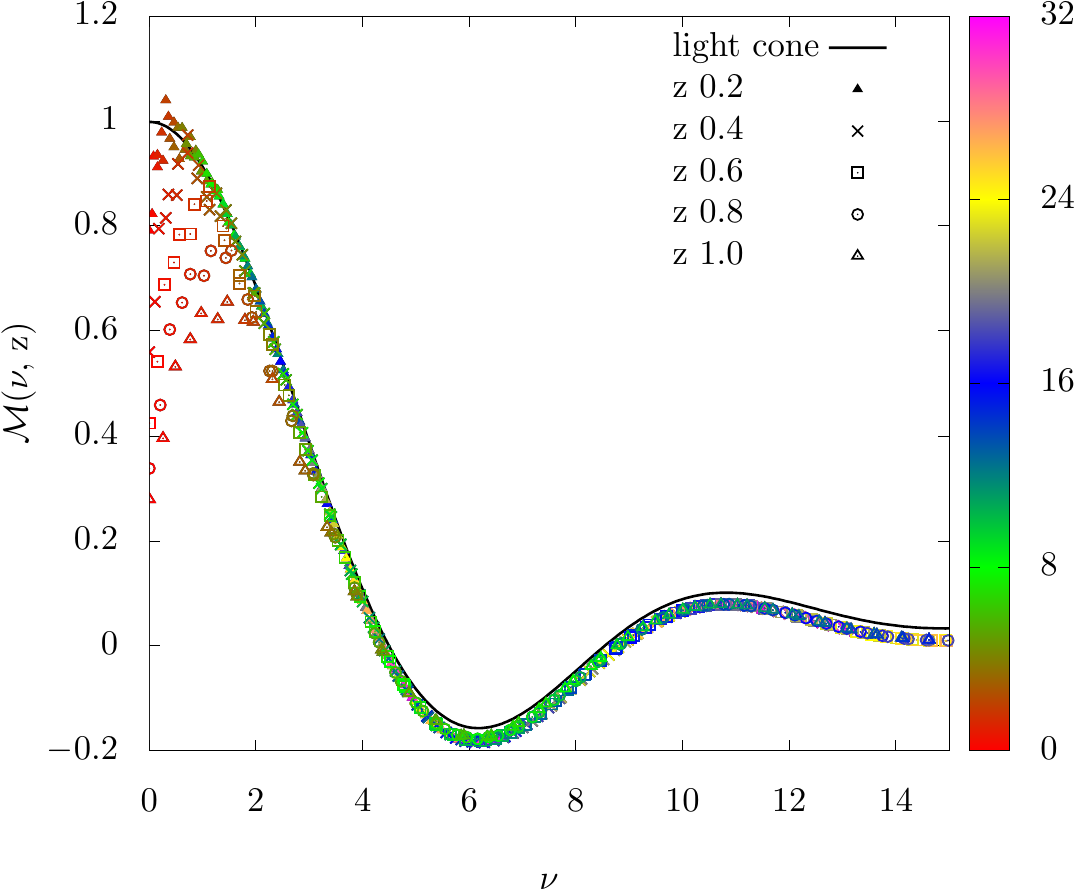}
\end{center}
\caption{
The proton  unpolarized u quark pseudo distribution normalized to one is shown for different values of quark field separation z in GeV$^{-1}$. $P_3$ is tagged from 0 to 32 GeV according to color. We see that at small Ioffe time $\nu$, for lower values of $P_3$, the pseudo pdf deviates substantially from the expected Ioffe time behavior of the light cone parton distribution. The momentum space GPD used to evaluate the curves in this figure was obtained using the reggeized diquark model from Ref.~\cite{GonzalezHernandez:2012jv}. \label{fig:pseudo_regge}
}
\end{figure}

To aid a quantitative analysis one can draw information from models. Building a relatively simple model to evaluate GPDs    can help understanding some of the essential features of the pseudo-pdfs approach. For example, the reggeized diquark model \cite{GonzalezHernandez:2012jv}  can be used to have a better understanding of the ``off-the-light-cone" behavior of the PDFs/GPDs, and to put more stringent constraints on the onset of perturbative behavior in coordinate space.
In models one has the flexibility to perform calculations that mimic the equal time constraint on the lattice. Instead of obtaining the usual light cone PDF, $f(x)$, using this procedure, one obtains the off-the-light-cone PDFs which are a function of $k_3$ and $P_3$. Upon performing the Fourier transform in $k_3$ one obtains the Ioffe time distributions which are now a function of $z$ and $P_3$ or, equivalently, the two Lorentz scalars $P\cdot z$ and $z^2$. One expects that as $z^2$ tends to zero and $P_3$ increases the pseudo Ioffe time distribution approaches the light cone Ioffe time distribution. This is illustrated in Figure \ref{fig:pseudo_regge}. 

Moving off forward, a similar exercise can be repeated for the GPDs. The presence of an additional vector $\Delta$ introduces, in coordinate space, the dependence on another Lorentz scalar $\Delta\cdot z$. For zero skewness, this quantity is zero for GPDs on the lightcone. For pseudo GPDs, however, this is no longer true. Hence, in addition to $z^2$, in the case of PDFs, which captures the ``off the light-coneness" of pseudo PDFs, in the case of GPDs, one also sees the dependence on $\Delta \cdot z$ for pseudo GPDs at zero skewness. Model calculations of pseudo GPDs at zero skewness reveal that the dependence on $\Delta \cdot z$ is more pronounced at larger $t$.

\subsubsection{Evaluation of GPDs from their Mellin Moments and Ioffe Time Behavior}

Complementary to the effort of calculating non-local operators on the lattice is the evaluation of Mellin moments of PDFs. While having their own set of complications in the evaluation of higher moments, they provide valuable input for a completely independent way of obtaining the PDFs. In the target rest frame, Ioffe time $\nu\equiv P\cdot z$ quantifies the distance along the light-cone that the quark fields describing the PDF are separated by. In this sense, it is a natural candidate for separating the short distance from the long distance physics. Using the Mellin moments, one can map out the Ioffe time behavior of the PDF for smaller values of Ioffe time. Having more moments allows one to describe the Ioffe time behavior of the PDF for higher and higher values of $\nu$. The large Ioffe time behavior, on the other hand, is described by a term of the form $\nu^{-\alpha}$ which essentially captures the small $x$ physics. 

\begin{figure}[tb]
\begin{center}
\includegraphics[width=0.49\textwidth]{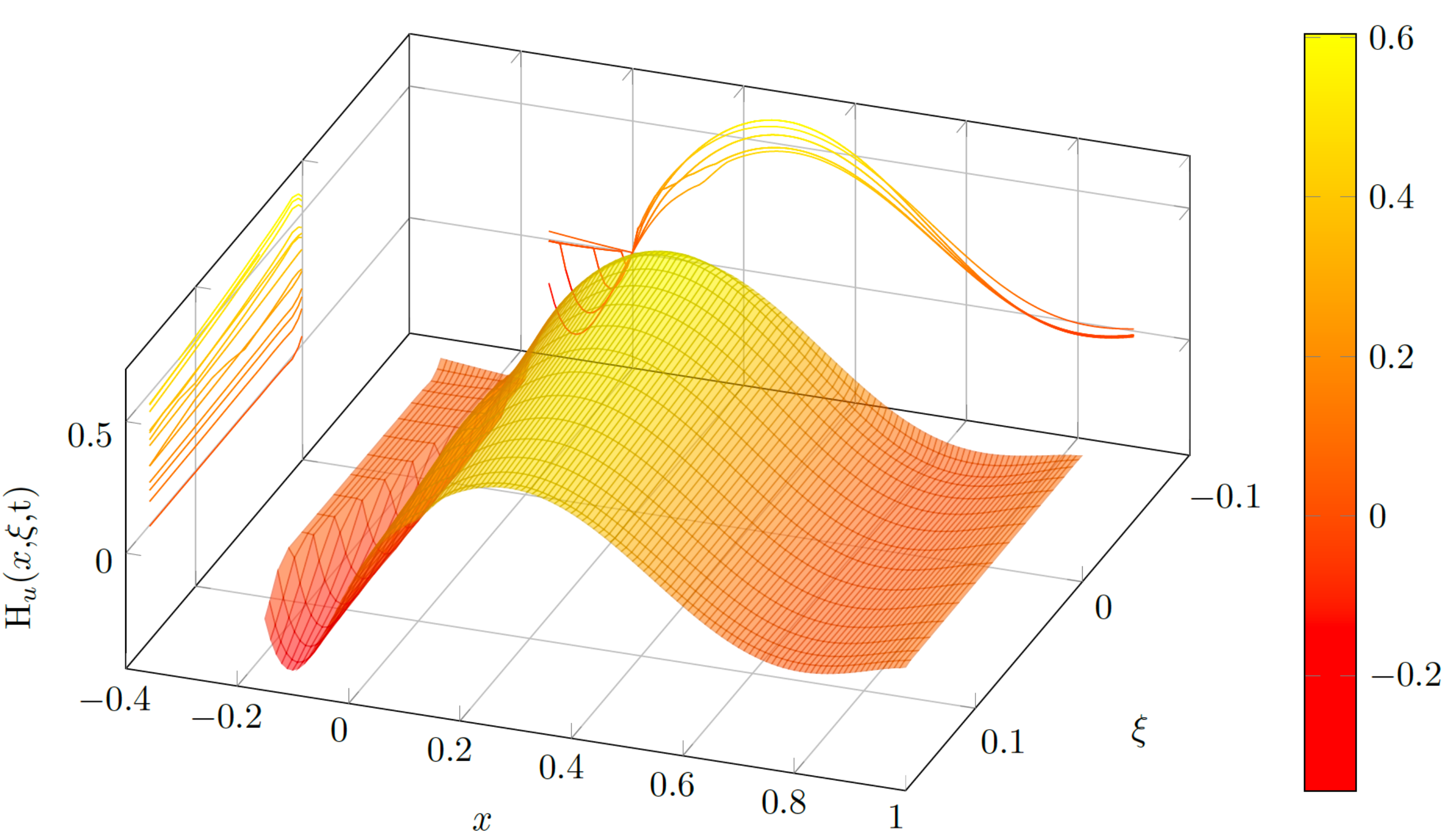}
\end{center}
\caption{\label{fig:reconstrGPD}The GPD $H$ for the $u$ quark at $t=-0.1 GeV^2$ reconstructed using the Generalized Form Factors calculated on the lattice. 
}
\end{figure}

In the case of GPDs, polynomiality governs the behavior of the Mellin moments for non-zero skewness. The calculation of Generalized Form Factors on the lattice allows one to reconstruct the GPD in Ioffe time space for a given skewness similar to the case of PDFs as described above. By varying $\xi$, one obtains the $x$ and $\xi$ dependence of the GPD for a given $t$. In Fig.~\ref{fig:reconstrGPD} we show the GPD $H$ for the $u$ quark evaluated by Fourier transformation from coordinate space. The Ioffe time dependence of the GPD was evaluated using the first three Mellin moments calculations from the lattice (see Section \ref{sec:moments}), and assuming Regge behavior for the asymptotic $z^-$ dependence. 

%% file: 03sec4machine.tex
\goodbreak
\subsection{Machine Learning for GPDs}
\label{03sec4}

Deeply virtual exclusive processes are measured in coincidence experiments where all the particles in the final state are either directly or indirectly detected. Therefore, extracting GPDs from data involves a much larger number of variables than in inclusive deep inelastic scattering. 
Furthermore,  each experiment can only add a small piece to the picture characterized by a specific polarization configuration.
This makes our problem virtually impossible to solve with traditional methods: for high precision femtography which is required to obtain proton images, we need to develop  more sophisticated analyses. 
Working towards this goal, recent efforts have focused on developing Machine Learning (ML) tools for the extraction of GPDs from data.

Three different groups have been attacking the problem.

\subsubsection{The PARTONS framework}
\begin{figure}[tb]
  \centering
  \includegraphics[width=0.35\textwidth]{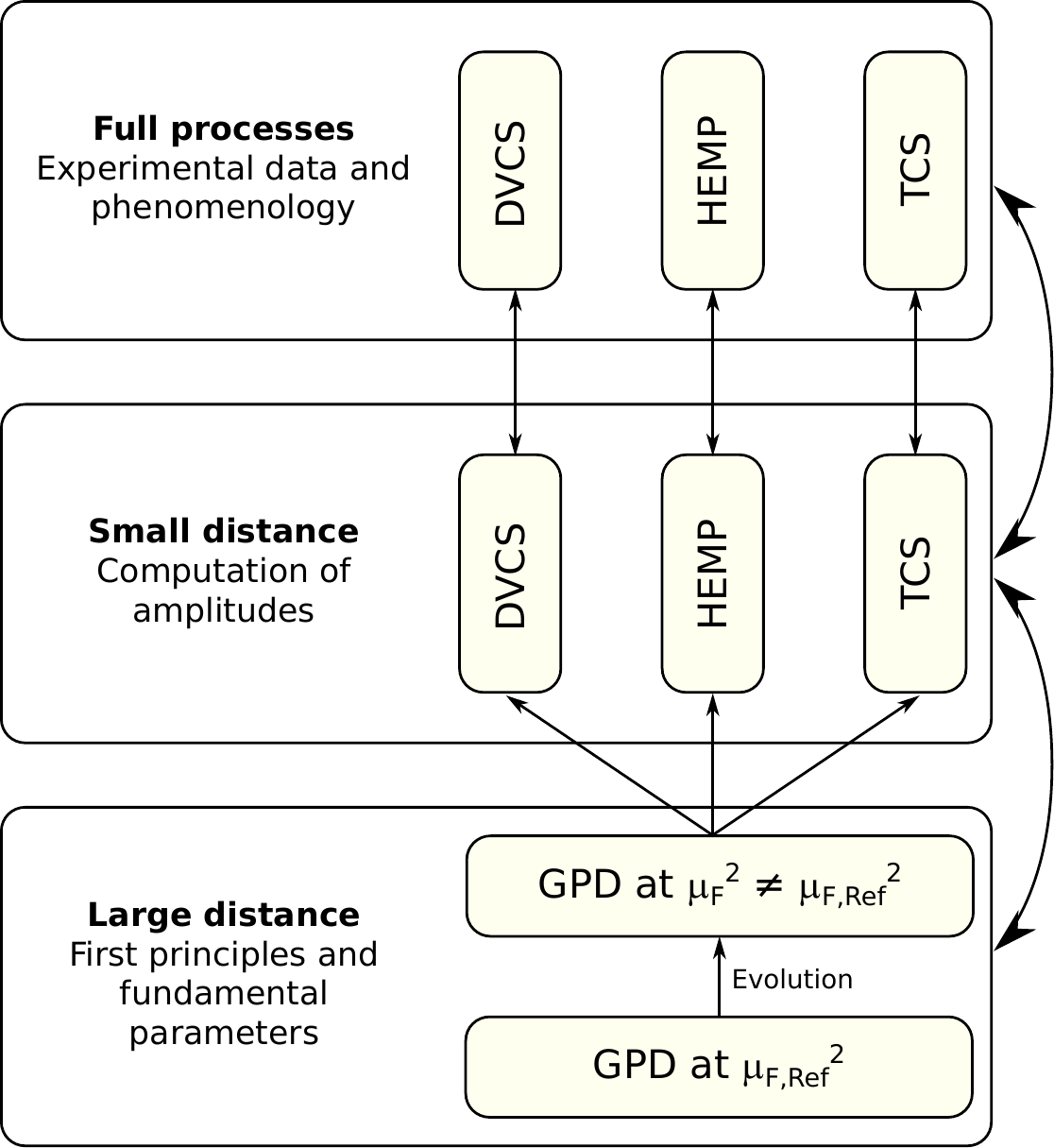}
  \caption{The computation of an observable in terms of GPDs is generically layered in three basic steps: description of the hadron structure with nonperturbative quantities, computation of coefficient functions, and evaluation of cross sections.}
  \label{fig:Parton_Structure}
\end{figure}

PARTONS (PARtonic Tomography of Nucleon Software) is a platform that handles multiple experimental channels where most of the tasks are automated, allowing the users to set physics assumptions at various point of the computation~\cite{Berthou:2015oaw}. 
This is achieved by splitting experimental channels into three pieces (see fig. \ref{fig:Parton_Structure}), accordingly to the framework of collinear factorization. The large distance part contains the physics related to GPDs themselves. Various models \cite{Goloskokov:2007nt,Mezrag:2013mya} are provided, together with leading-order evolution routines \cite{Vinnikov:2006xw}. The second step consists in the computations of amplitudes from GPDs. Next-to-leading order computations of Compton Form Factors (CFFs) are also available. Finally, CFFs are connected to observables. 
PARTONS has been used in recent extractions of CFF from experimental data, using both functional parametrizations \cite{Moutarde:2018kwr} and Artificial Neural Networks (ANNs) \cite{Moutarde:2019tqa}. 
This analysis focuses on the suppression of any model dependence, and on a careful estimation of uncertainties. This was achieved by using separate neural networks for the real and imaginary parts of each CFF, and by avoiding any pre-factors driving the functional forms of CFFs. The replica method was used for the propagation of uncertainties, while the early-stopping regularisation method was utilised to provide the anticipated growth of uncertainties in the unconstrained kinematic domains. The networks were trained to nearly all available DVCS data with the genetic algorithm.

Work within the PARTONS framework 
has been also recently extended to estimate time-like Compton scattering (TCS) observables taking advantage of the complementarity between 
he Leading Order (LO) and Next to Leading Order (NLO) descriptions \cite{Grocholski:2019pqj}; 
for a model free extraction of the subtraction constant entering the dispersion relations approach to GPDs \cite{Diehl:2007jb} and its impact on the energy momentum tensor matrix elements \cite{Polyakov:2002yz}.
It is also currently used to model light nuclei GPDs, and to develop a DVCS event generator for the future EIC. This software is free and open-source, released on GPLv3. The code and its documentation can be found on the PARTONS website\footnote{http://partons.cea.fr}.
\subsubsection{Stepwise Regression Method}
In a series of papers~\cite{Kumericki:2011zc,Kumericki:2013br,Kumericki:2015lhb,Kumericki:2019mgk}, the authors propose to use the stepwise regression method in conjunction with an ANN analysis to address the
question of which of the four leading order chiral-even CFFs, ${\cal H}$, ${\cal E}$, $\widetilde{\cal H}$, $\widetilde{\cal E}$ can be
reliably extracted from exclusive photon electroproduction data.  
In stepwise regression, the number of real and imaginary components of the CFFs is gradually increased
and all combinations are tried, until there is no statistically significant improvement in the description of the data. It was found \cite{Kumericki:2019mgk} that the DVCS data available so far are only sensitive to $\Im m {\cal H}$, $\Re e {\cal E}$ and $\Im m \widetilde{H}$. This method was used recently to   extract the energy momentum tensor pressure form factor from DVCS data. A different conclusion from the one reported in \cite{Burkert:2018bqq} was found using the same set of experimental data \cite{Kumericki:2019ddg}.    

\subsubsection{Femtography effort at Jefferson Lab}
An ongoing effort at Jefferson Lab follows the footpath of a successful pilot project launched in the Summer 2019 where the physics community was asked to address femtography within collaborative projects. These projects are now a reality: computer scientists, data scientists, mathematicians, visualization experts are collaborating to build a platform addressing the computational and theoretical challenges in mapping out the proton's 3D structure. 
The existent platform will allow us to refine our focus and to take further steps into providing both machine learning based and virtual reality shareable tools for the physics community. 

%% file: 03sec5fits.tex
\goodbreak
\subsection{Towards global fits for GPDs}
\label{03sec5}
The cleanest probe of GPDs is DVCS, which is measured in exclusive production of a hard photon in lepton-hadron scattering, $e p \rightarrow e' p' \gamma$. The virtual photon four-momentum squared, $Q^2= -(k_e-k_e')^2$, with $k_e$ ($k'_e$) being the initial (final) electron four-momentum, provides a hard scale in the one photon exchange approximation. The presence of a hard scale allows us to single out the perturbative, short distance reaction from the non-perturbative, long distance
matrix elements according to the factorization property of QCD (proofs of factorization for deeply virtual exclusive processes can be found in  \cite{Collins:1996fb,Ji:1997nk,Ji:1998xh}).  Since the DVCS observable is not sensitive to flavor of the parton participating in the hard collision, probed proton GPD is written in terms of a sum of all quark flavor GPDs as,
\begin{eqnarray}
H & = & \sum_q e_q^2 \, H_q 
\end{eqnarray}
$e_q$ being the quark charge. Gluon GPDs can also contribute to DVCS cross section at high orders.  The neutron GPD can be obtained using isospin symmetry.  

The exclusive hard photon can also be produced via the 
Bethe-Heitler (BH) scattering,  
where the photon is emitted from the electron. The BH process provides a means towards a cleaner extraction of the QCD matrix elements since these appear as linear combinations in the interference contribution to the cross section between the BH and DVCS scattering amplitudes, at variance with their bilinear counterparts in the pure DVCS contribution.

Experimental information on GPDs is also obtained from %
related channels, namely exclusive photon proton scattering, or timelike Compton scattering, exclusive Drell-Yan processes, and exclusive meson production. GPDs enter the cross section encoded in Compton Form Factors (CFFs) which, using QCD factorization,  are defined as convolutions over the parton light cone momentum fraction $x$, of GPDs and QCD Wilson coefficient functions \cite{Goeke:2001tz,Diehl:2003ny,Belitsky:2005qn,Kumericki:2016ehc}.
At leading order one has, in the chiral-even sector,
\begin{eqnarray}
\label{eq:cffe}
\mathcal{F}_q(\xi,t)  & = & {\cal C} \left( C^+ \, F_q \right) \equiv \int_{-1}^{1} dx \,   C^+(x,\xi) F_q(x,\xi,t), 
\nonumber \\ \\
\label{eq:cffo}
\mathcal{\widetilde{F}}(\xi,t) & = & {\cal C} \left( \, C^- \, \widetilde{F}_q \right) \equiv \int_{-1}^{1} dx   C^-(x,\xi) \, \widetilde{F}_q(x,\xi,t) \nonumber \\
.%
\label{CFF}
\end{eqnarray}
where ${\cal F}_q$ = (${\cal H}_q$, ${\cal E}_q$), and $\widetilde{\cal F }_q$= ($\widetilde{\cal H}_q$ $\widetilde{\cal E}_q$), respectively, 
with leading order coefficients functions given by,
 \begin{equation}
 { C^\pm(x,\xi) = \frac{1}{x-\xi - i \epsilon} \mp \frac{1}{x+\xi - i \epsilon} . } %
 \end{equation}
GPDs observe crossing symmetry relations with respect to $x \rightarrow -x$, which allow us to introduce valence (symmetric) and quark singlet (anti-symmetric) distributions. 

Traditional QCD global fits to extract GPDs from data are particularly challenging because of the complexity of the higher order Wilson coefficient functions (replacing $C^{\pm}$ in Eqs.~(\ref{eq:cffe}) and (\ref{eq:cffo})), and of the fact that the relative momentum fraction, $x$, of GPDs is not directly accessible from experiment but only indirectly relevant through the $\xi$-dependence of CFFs. In particular, it is very difficult to extract GPDs with an accurate flavor separation and determination of the scale dependence, with a similar accuracy level to what has been  accomplished for PDFs. With the present kinematic coverage of experiments worldwide, at the best of our knowledge, it has not been possible to perform global fits of GPDs that span from the valence region measured in fixed target experiments, to the small-$x$ which is accessible at high energy colliders. 
What makes such a global analysis very demanding is first of all, as mentioned above, that GPDs need to fulfill a set of properties derived from the symmetries of QCD, such as positivity and polynomiality. Fulfilling both at the same time has been a long issue for GPD phenomenologists. Recently, new modeling techniques have been introduced that might solve this longstanding issue \cite{Chouika:2017dhe}, but have not been applied in practical fits. 
Another important difference with PDFs is that the CFFs from which GPDs are extracted, enter the cross section in bilinear forms. Therefore,  
 GPDs corresponding to different polarization configurations all contribute to specific helicity states of the various observables: for example, in DVCS, the unpolarized structure function, $F_{UU,T}$ \cite{Kriesten:2019jep}, contains both the vector, $H,E$, and axial-vector, $\widetilde{H},\widetilde{E}$, GPDs (the other polarization configurations behave similarly).  
As a result, in order to extract information from experiment, many more observables need to be analyzed simultaneously. This is what motivated, for instance, the first attempts to global fits in \cite{Kumericki:2015lhb,Kumericki:2019ddg}.
 To determine the various observables that are necessary to extract GPDs from data, current experimental programs at Jefferson Lab, COMPASS at CERN and J-PARC are set to measure DVCS and the crossed channel reactions to DVCS, namely both TCS, $\gamma p \rightarrow p' \mu^+ \mu^-$ \cite{Berger:2001xd,Moutarde:2013qs,Boer:2015fwa}, and the exclusive Drell-Yan process, $\pi^{\pm} p \rightarrow p' \mu^+ \mu^-$ \cite{Sawada:2016mao}.
 On the other side,  GPDs can also be extracted from deeply virtual meson production %
 although this requires a clear understanding of the meson  distribution amplitudes. 

NLO corrections in $\alpha_S$ to CFFs and evolution of GPDs are also challenging in a global fit. Gluon GPDs are, in fact, expected to bring a significant contribution to both DVCS and TCS CFFs \cite{Moutarde:2013qs}. 
In the perspective of the upcoming EIC, NLO studies will become more and more pressing. Furthermore, many available data present a virtuality $Q^2$ of few GeV$^2$, {\it i.e.} of the same order of magnitude as the square of the proton mass. Therefore, target mass corrections, but also finite $t$ corrections \cite{Braun:2012hq} need to be taken into account. They could be responsible for half of the signal measured at some kinematical bins at JLab \cite{Defurne:2015kxq}, and may, therefore, be an essential piece to connect the lower $Q^2$ data (few GeV$^2$) with higher ones at the EIC.

The four additional GPDs appearing in the chiral-odd sector, $H_T, E_T, \widetilde{H}_T, \widetilde{E}_T$,  can be measured in processes that allow for quark helicity flip at the amplitude level, for instance in $\pi^o$ and $\eta$ photoproduction \cite{Ahmad:2008hp,Goloskokov:2011rd,Goldstein:2013gra} ($\pi^o$ electroproduction also constitutes the main background process for DVCS). It should also be noticed that factorization has not been proven for transversely polarized virtual photon exchange which dominates the process in this case, and that the quark flip process involves a beyond leading-twist pion lightfront wave function.
Exclusive $\pi^0$ production data seem to indicate that standard colinear factorization might not be working at JLab kinematics \cite{Defurne:2016eiy}. A similar situation has been found for the $Q^2$ dependence of recent DVCS data \cite{Georges:2018kyi}. Future experiments will help clarifying the issue of the scale dependence in both processes.
 
Chiral-odd gluon GPDs, contribute to DVCS at NLO, generating a $\cos 2\phi$ modulation in the cross section. Although a hint of this modulations is consistent with the data in Ref.\cite{Defurne:2017paw}, the experimental precision remains too low to disentangle it from possible higher-twist contributions.

In summary, at a time when 
the US EIC project is getting off the ground, GPD phenomenologists are facing many challenges, both  theoretical and in setting up an appropriate computational framework. The solutions that will be brought forward to the various issues mentioned in this report will certainly impact our understanding of the future EIC measurements, but first and foremost they will be key to interpreting the data that are already being taken at fixed target facilities. A better control of the cross section structure for DVCS and its crossed channels, TCS and exclusive Drell Yan, of higher-twist terms and their factorization, of kinematic and target mass corrections and NLO evolution, will allow for significant steps towards mapping out the 3D structure of the nucleon.

%% file: 04sec0tmd.tex
\goodbreak
\section{Transverse Momentum Dependent PDFs (TMD PDFs)}
\label{04sec:tmd}
\input{04sec1intro}

\input{04sec2quasi}

\input{04sec3fits}

%% file: 04sec1intro.tex
\goodbreak
\subsection{Introduction to TMD PDFs}
\label{04sec1}

The TMD PDF concept, in the most basic sense, directly extends the ordinary PDF concept to incorporate dependence on the parton's transverse momentum.
That is, instead of having a number density $f(x)$ of partons per unit of lightcone momentum fraction $x$, one has a density $f(x,{\bf k}_T)$ of partons per unit of both the momentum fraction $x$ \emph{and} the (small) transverse momentum components.
A similar relation exists between standard fragmentation functions and TMD fragmentation functions.
Standard factorization theorems involve collinear PDFs (and fragmentation functions) only, whereas situations that use TMD PDFs require a different kind of ``TMD''  factorization theorem~\cite{Collins:1984kg}.

Collinear factorization relies on processes being rather inclusive, such that the details of small variations in parton transverse momentum are not important.
Transverse-momentum--dependent PDFs (TMD PDFs), by contrast, arise in less inclusive processes, particularly where low, and especially nonperturbative, transverse momentum transfer is important.
There are a number of processes where a complete QCD treatment calls for the TMD PDF concept.
These include especially cross sections differential in a measured transverse momentum --- so-called multiscale processes, where there is an overall physical hard scale (usually labeled $Q$) and a transverse momentum which may also be hard in some regions but is typically very different from $Q$.
Drell-Yan scattering is a classic example of this.
There, the overall hard scale is set by the invariant energy $Q$ of the produced dilepton pair, while the total transverse momentum of the pair $\qt $ provides the transverse momentum dependence.
For a cross section integrated over $\qt $, collinear factorization is relevant, whereas cross sections differential in $\qt $ require a form of TMD factorization, especially in the region $q_T \ll Q$ (we denote $|\qt | \equiv q_T $ here and in the following). More on this example follows below.
The information about transverse degrees of freedom is essential in many non-inclusive collider observables~\cite{Angeles-Martinez:2015sea}.
The sensitivity to a nonperturbative intrinsic transverse momentum is especially exciting from the perspective of hadron structure also because the intrinsic transverse momentum may couple in nontrivial ways with spin degrees of freedom.
Early classic work on TMD structures~\cite{Mulders:1995dh} enumerated the large number of TMD PDF structures that can arise once a transverse momentum degree of freedom is allowed for the parton. 

The TMD PDF concept also arises in many other areas of QCD physics.
Another approach to determine the transverse momentum dependence, valid at low $x$, can be developed within the color-glass condensate~\cite{Gelis:2010nm} framework, an effective model formulated within QCD.
A special limit, so-called improved transverse-momentum--dependent~\cite{Kotko:2015ura,Altinoluk:2019fui} factorization, addresses the situation where one of the colliding hadrons is dilute and is characterized by collinear PDFs, and the other one (the target) is dense and is parametrized by a set of gluon densities dependent on longitudinal and transverse momentum.
The processes where this factorization applies are limited to forward production.
There is also a Monte Carlo approach to obtain TMD distributions called Parton Branching TMD\cite{Hautmann:2017fcj,Martinez:2018jxt}. It is essentially based on DGLAP dynamics.

In this section, we elaborate on the TMD PDF concept, giving a brief introduction to situations where TMD PDFs appear instead of collinear PDFs, leaving technical details of their definition to Sec.~\ref{04sec1.5}.

It is instructive to contrast TMD PDFs with the collinear PDFs studied in Sec.~\ref{02sec:pdf}, and so we consider the specific case of the Drell-Yan process, the production of a lepton pair with momentum $q^\mu$ through the process $p p \to Z/\gamma* \to \ell^+ \ell^-$ at a hadron collider, while noting that a similar situation also arises in semi-inclusive DIS and electron-positron annihilation such as will be studied at a future EIC~\cite{Accardi:2012qut}.

First, consider the case that one is only interested in measuring the invariant mass $Q = \sqrt{q^2}$ of the lepton pair, but not its transverse momentum $\qt$.
This is described by collinear factorization, which allows one to factorize the cross section as
\begin{align} \label{eq:DY_fact_coll}
 \frac{d \sigma}{d Q^2} &
 = \sum_{i,j} \int_0^1 d \xi_a d \xi_b \, f_{i/P_a}\!(\xi_a) f_{j/P_b}\!(\xi_b) \frac{d \hat\sigma_{ij}(\xi_a,\xi_b)}{d Q^2}
 \nn\\&\quad
 \times \left[1 + \cO\left(\frac{\lqcd}{Q}\right)\right]
\,.\end{align}
Here, $f_{i/P}(\xi)$ is the PDF for a parton of type $i$ carrying the fraction $\xi$ of the momentum of its parent hadron $P$, we sum over all possible flavors $i$ and $j$, and $\hat\sigma_{ij}$ is the partonic cross section for the Drell-Yan process.
In an observable like \eqref{eq:DY_fact_coll}, that has been averaged over all allowed transverse momenta $\qt$ of the lepton pair, it is reasonable to assume that the intrinsic transverse motion of partons in the proton is numerically not relevant, and thus the PDFs in \eqref{eq:DY_fact_coll} are functions of longitudinal momentum components only.
Note that \eqref{eq:DY_fact_coll} receives corrections in $\lqcd/Q$, which are typically negligible at high-energy experiments, but may become important at low-energy colliders.

The aim of TMD factorization is to generalize Eq.~\eqref{eq:DY_fact_coll} to also measure the transverse momentum $\qt$ of the produced lepton pair.
For very large $\qt$, it is again reasonable to assume that the intrinsic motion of partons inside the hadrons is small, and collinear factorization applies:
\begin{align} \label{eq:DY_fact_coll_large_qT}
 q_T \sim Q &\gg \lqcd:
 \\ \nn
 \frac{d \sigma}{d Q^2 d^2\qt } &
 = \sum_{i,j} \int_0^1 d \xi_a d \xi_b \, f_{i/P_a}\!(\xi_a) f_{j/P_b}\!(\xi_b) \frac{d \hat\sigma_{ij}(\xi_a,\xi_b)}{d Q^2 d^2\qt }
 \\ \nn & \quad
 \times \left[1 + \cO\left(\frac{\lqcd}{Q}, \frac{\lqcd}{q_T}\right)\right]
\,.\end{align}
In this case, $\qt$ is dominantly generated by the hard partonic process $\hat\sigma_{ij}$, which is perturbatively calculable, while the only nonperturbative input to \eqref{eq:DY_fact_coll_large_qT} is through the standard PDFs.
Compared to \eqref{eq:DY_fact_coll}, \eqref{eq:DY_fact_coll_large_qT} also receives corrections in $\lqcd/q_T$, signaling that it breaks down for small $q_T$.
Indeed, the situation becomes drastically different if one considers transverse momenta much smaller than the invariant mass of the lepton pair, $q_T \ll Q$.
This is the realm of \emph{TMD factorization}, where one has
\begin{align} \label{eq:DY_fact_coll_small_qT}
 q_T \ll Q:&
 \\\nn
 \frac{d \sigma}{d Q^2 d^2\qt} &
 = \sum_{i,j} H_{ij}(Q) \int_0^1 d \xi_a d \xi_b \, \int d^2\bt \, e^{i \bt \cdot \qt}
 \\\nn &
 \hspace{2.5cm}\times
 f_{i/P}(\xi_a, \bt)  f_{j/P}(\xi_b, \bt)
 \\ \nn & \quad
 \times \left[1 + \cO\left(\frac{\lqcd}{Q}, \frac{q_T}{Q}\right)\right]
\,.\end{align}
Here, the hard factor $H_{ij}(Q)$ describes the underlying partonic process $q_i q_j \to Z^/\gamma^* \to \ell^+ \ell^-$,
and the $f_{i/P}(\xi,\bt)$ are TMD PDFs in Fourier space, describing the probability to find a parton of type $i$ with a longitudinal momentum fraction $\xi$ and transverse momentum conjugate to\footnote{The unfortunate overlapping use of the notation $\bt$ for two distinct objects in the TMD PDF and GPD literatures should be noted. In the GPD literature, $\bt$ denotes the impact parameter of the struck quark with respect to the center of momentum of the hadron; it is Fourier conjugate to the momentum transfer $\Delta_{T}$.
In the TMD PDF literature, $\bt $ denotes the Fourier conjugate to the transverse momentum $\kt$ of the struck quark, corresponding to a {\em relative} separation in the relevant quark bilinear operator, cf.~the definition (\ref{eq:tmdpdf_def}) below.
Put succinctly, the relation between the conventions in the two literatures is akin to the relation between center-of-mass and relative coordinates. In the present section, the impact parameter will instead be denoted by $\rt$, cf.~the discussion at the end of section \ref{04sec2}.}  $\bt$.
Upon taking the inverse Fourier transform in \eqref{eq:DY_fact_coll_small_qT}, one then recovers the $\qt$ distribution of the lepton pair.
\eqref{eq:DY_fact_coll_small_qT} is valid both in the nonperturbative regime, where $q_T \sim \lqcd \ll Q$ and the TMD PDFs are intrinsically nonperturbative objects, and in the perturbative regime where $\lqcd \ll q_T \ll Q$ and the TMD PDFs can be perturbatively calculated.
It also receives corrections in $q_T/Q$, which in the literature are often included as the so-called $Y$ term.
These corrections are important to smoothly transition between the regimes described by \eqref{eq:DY_fact_coll_large_qT} and \eqref{eq:DY_fact_coll_small_qT}.

\goodbreak
\subsection{Definitions of TMD PDFs}
\label{04sec1.5}

Setting up consistent definitions for TMD PDFs, useful beyond the most basic parton model approaches, involves a significant number of 
subtleties beyond what are encountered directly in 
the more standard collinear factorization. The history of these efforts is long~\cite{Collins:2003fm} and we will not be 
able to cover all aspects here. See, however, 
reviews in \cite{Rogers:2015sqa,Diehl:2015uka} for somewhat 
broader summaries that highlight some of the more subtle aspects.

In this section, we provide more details on the definition of TMD PDFs.

\textbf{Classification of TMD PDFs.}
First, we note that TMD PDFs can be classified according to the scheme used for the hard factor $H_{ij}$ in \eqref{eq:DY_fact_coll_small_qT}.
In the original formulation by Collins, Soper and Sterman, often referred to as CSS1, one has $H_{ij} = 1$ such that the hard factor is effectively absorbed into the TMD PDFs~\cite{Collins:1984kg}.
In its revised formulation (CSS2) due to Collins~\cite{Collins:2011zzd}, see also~\cite{Aybat:2011zv}, $H_{ij}$ is defined as the form factor of the Drell-Yan process in the $\overline{\text{MS}}$ scheme. This scheme choice is also adopted by formulations of TMD factorization employing soft-collinear effective theory (SCET)~\cite{Becher:2010tm, Echevarria:2012js, Chiu:2012ir, Li:2016axz}.
Lastly, TMD PDFs as introduced by Ji, Ma and Yuan (JMY)~\cite{Ji:2004wu} employ a scheme where $H_{ij}$ depends not only on the $\overline{\text{MS}}$-renormalization scale $\mu$, but also on an additional parameter $\rho$ related to the regulation of rapidity divergences, which we discuss below.
All of these formulations are perturbatively related to each other, and explicit relations are given in~\cite{Prokudin:2015ysa, Collins:2017oxh}.

\textbf{Definitions of TMD PDFs.}
Consider a hadron $p$ with momentum $P^\mu = (\sqrt{M^2 + P_z^2}, 0, 0, P^z)$.
It is useful to introduce a basis of lightlike vectors
\begin{align}
 n^\mu = \frac{1}{\sqrt2}(1,0,0,1) \,,\quad \bn^\mu = \frac{1}{\sqrt2}(1,0,0,-1)
\,,\end{align}
such that $p$ is moving close to the $n$ direction.
The TMD PDF can then be defined as
\begin{align} \label{eq:tmdpdf_1}
 f_{i/P}(x, \bt, \mu , \zeta) &
 =  \lim_{\substack{\epsilon\to 0 \\ \tau\to 0}} Z_\text{uv}^i(\mu,\zeta,\epsilon) \,
    \frac{f_{i/P}^{0\,(u)}\bigl(x, \bt, \epsilon, \tau, x P^+ \bigr)}{S^{0\,\rm sub}_{n \bn}(\bt,\epsilon,\tau)}
 \nn\\&\hspace{1.2cm} \times
    \sqrt{S_{n \bn}^{0}(\bt,\epsilon,\tau)}
\,.\end{align}
Here, $i$ denotes the parton flavor, $x$ is its longitudinal momentum fraction, $\bt$ is Fourier-conjugate to the transverse momentum of the struck quark, $\mu$ denotes the $\overline{\text{MS}}$-renormalization, and $\zeta$ is the so-called Collins-Soper scale~\cite{Collins:1981va}.
On the right-hand side, $\epsilon$ is the UV regularization parameter, and UV divergences in $1/\epsilon$ are absorbed by the $\overline{\text{MS}}$-counterterm $Z_\text{uv}^i$.
$f_{i/P}^{0\,(u)}$ is the bare unsubtracted TMD PDF, $S_{n \bn}^0$ is the bare soft function, and the factor $S^{0\,\rm sub}$ removes overlap between $f_{i/P}^{0\,(u)}$ and $S_{n \bn}^0$.%
\footnote{In the SCET-based literature, the combination $f_{i/P}^{0\,(u)} / S^{0\,\text{sub}}_{n \bn}$
is often referred to as the beam function.}
All these functions are plagued by additional so-called rapidity divergences~\cite{Collins:1981uk,Collins:1992tv,Collins:2008ht,Becher:2010tm,GarciaEchevarria:2011rb,Chiu:2012ir},
which are regulated by $\tau$.
There are many different rapidity regulators $\tau$ employed in the literature,
leading to different results for the individual ingredients on the right-hand side of \eqref{eq:tmdpdf_1}, but the final TMD PDF is well-defined and unique up to choice of scheme for $H_{ij}$ as discussed above.

\begin{figure}[tbp]
 \includegraphics[width=0.35\textwidth]{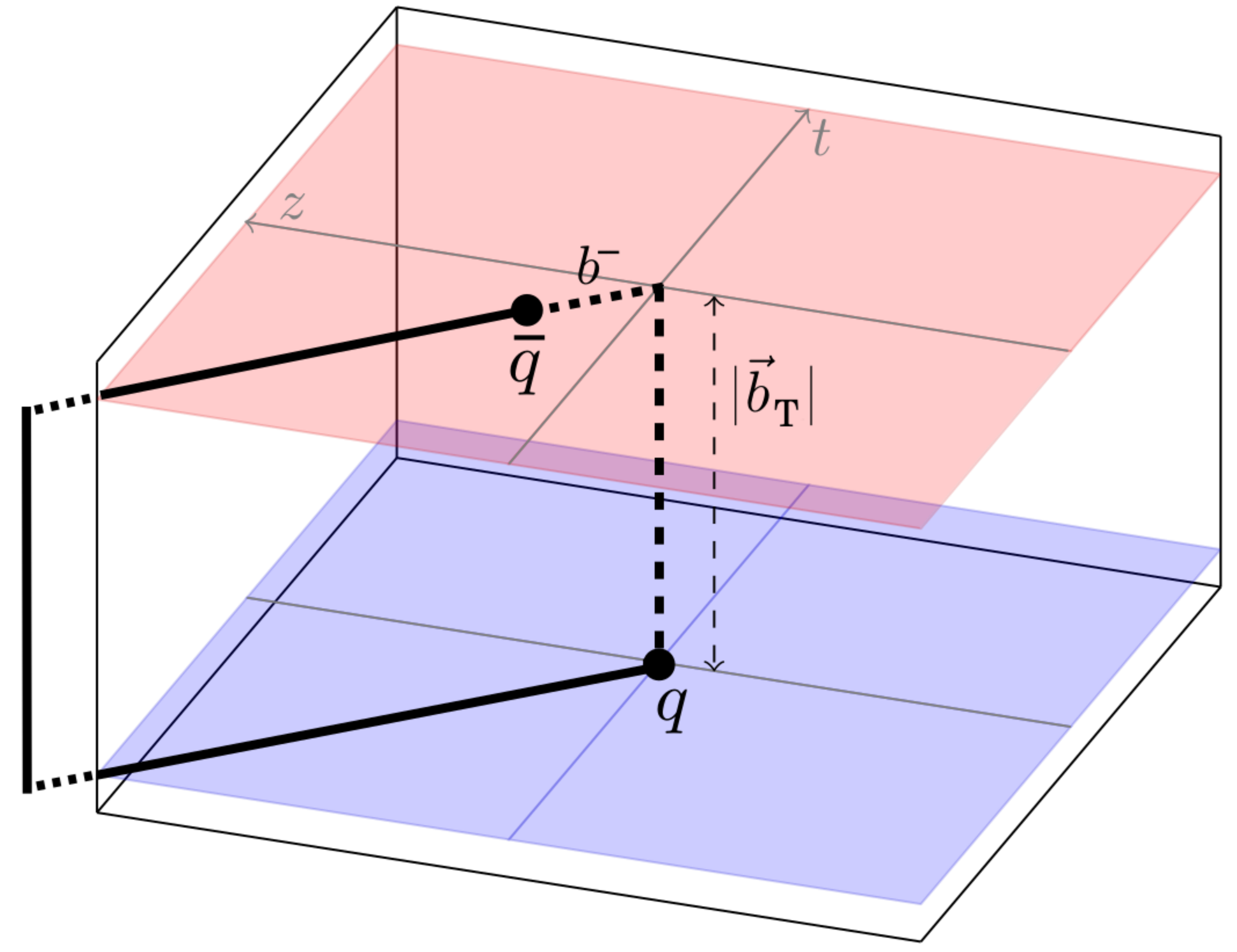}
 \\
 \includegraphics[width=0.35\textwidth]{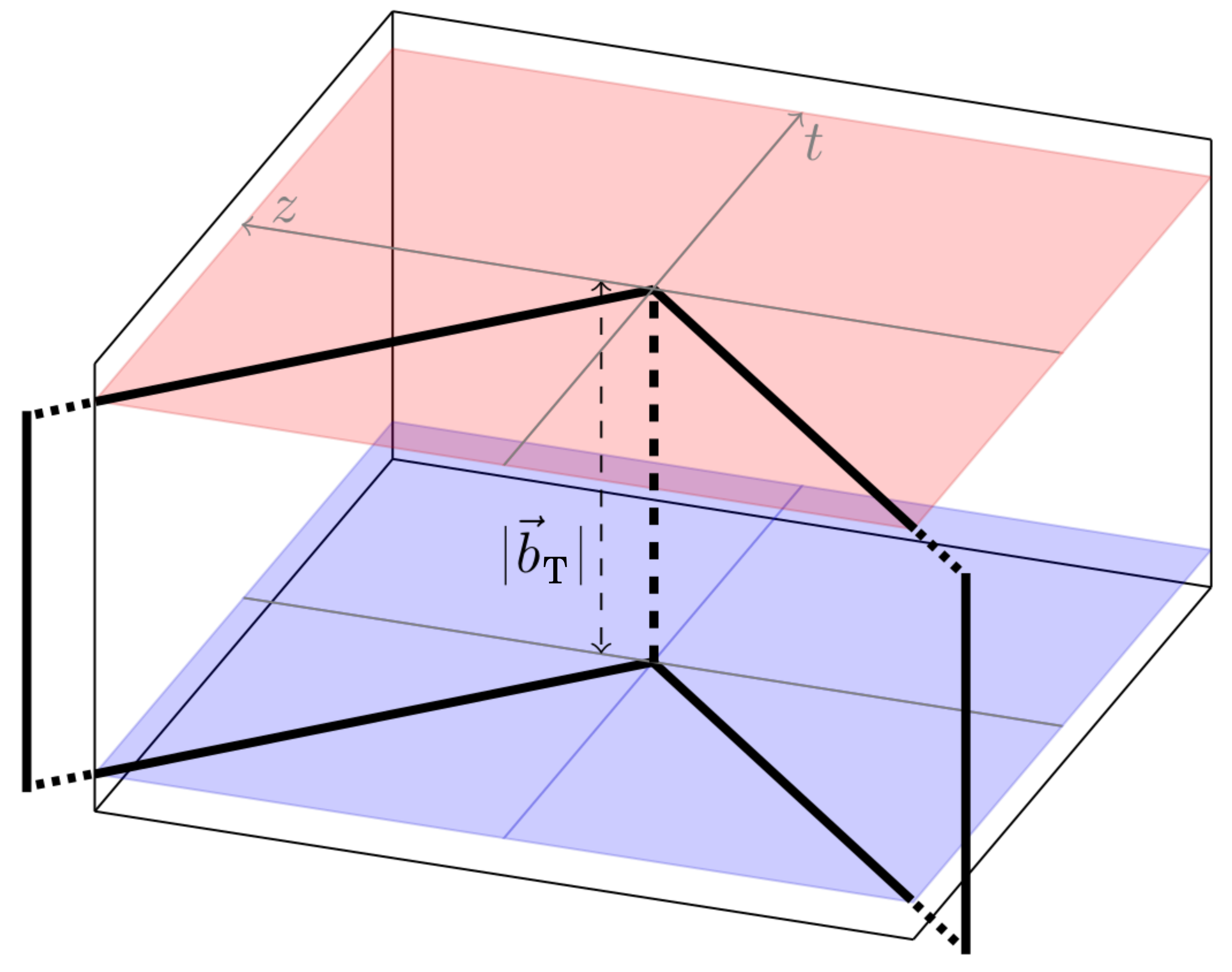}
 \caption{Illustration of the Wilson-line structures of the unsubtracted TMD PDF (top) and soft function (bottom), as defined in Eq.~\eqref{eq:tmdpdf_def}.
 Figure taken from~\cite{Ebert:2019okf}, with slight adjustment of the $b$-component labeling to match the notation used in the present work. 
 }
 \label{fig:wilsonlines}
\end{figure}

The unsubtracted TMD PDF $f_{i/P}^{0\,(u)}$ is defined as a hadron matrix element, and as such is sensitive to both $x$ and $\bt$, while the soft factor $S_{n \bn}^0$ is a vacuum matrix element only sensitive to $\bt$. Their definitions read
\begin{widetext}
\begin{align} \label{eq:tmdpdf_def}
 f_{i/P}^{0\,(u)}(x,\bt,\epsilon,\tau,x P^+) &= \int\frac{d b^-}{2\pi} e^{-i b^- (x P^+)}
 \Bigl< p(P) \Bigr|  \Bigl[ \bar q(b^\mu) W_{\bn}(b^\mu; -\infty, 0) \frac{\gamma^+}{2} W_{n_\perp}^{\dagger} (-\infty \bn, \bt,0)
 W_{\bn}^\dagger(0; -\infty, 0) q(0) \Bigr]_\tau \Bigl|p(P) \Bigr>
,\nn\\
 S^{0}_{n \bn}(\bt,\epsilon,\tau) &= \frac{1}{N_c} \bigl< 0 \bigr| \mbox{Tr} \bigl[
   W_{n}^\dagger(\bt;-\infty,0) W_{\bn}(\bt;-\infty,0) W_{n_\perp}^{\dagger }(-\infty \bn; \bt,0)
   \nn\\&\hspace{1.5cm}\times
   W_{\bn}^\dagger(0;-\infty,0) W_{n}(0;-\infty,0) W_{n_\perp}(-\infty n; \bt,0)
 \bigr]_\tau \bigl|0 \bigr>
\,.\end{align}
\end{widetext}
with Tr denoting the color trace.
Here, $b^\mu = b^- \bn^\mu + b_T^\mu$, lightcone coordinates are defined as $x^\pm = (x^0 \pm x^3)/\sqrt{2}$,
and the $W_n$ are path-ordered Wilson lines defined as
\begin{align} \label{eq:Wilson_lines}
 W_n(x; a,b) &= P \exp\biggl[ -i g_s \int_a^b d s \, n{\cdot}A(x^\mu + s n^\mu) \biggr]
\,.\end{align}
The paths of the Wilson lines in \eqref{eq:tmdpdf_def} are illustrated in Fig.~\ref{fig:wilsonlines}.
They form a closed path between the quark fields in the TMD PDF and a closed loop in the soft function, thereby rendering these functions gauge invariant.

In \eqref{eq:tmdpdf_def}, the regulator $\tau$ is not specified, but is typically implemented to modify the Wilson line structure.
The regulator most closely connected to the lattice studies presented in section \ref{04sec2} is that of Collins, which is implemented by taking the lightlike reference vectors $n$ and $\bn$ off the lightcone as~\cite{Collins:2011zzd}
\begin{align} \label{eq:Collins_rap}
 n^\mu &\quad\to\quad n_{y_A}^\mu \equiv n^\mu - e^{-2 {y_A}} \bn^\mu%
\,,\nn\\
 \bn^\mu &\quad\to\quad \bn_{y_B}^\mu \equiv \bn^\mu - e^{+2 {y_B}} n^\mu%
\,.\end{align}
The regulator is removed through the limit $y_A\to\infty, y_B\to-\infty$.
Concretely, in this regulator scheme the TMD PDF is obtained as
\begin{align} \label{eq:tmdpdf_collins}
 f_{i/P}(x, \bt, \mu, \zeta)
 &= \lim_{y_B\to-\infty} Z_\text{uv}
   \frac{f_{i/P}^{0\,(u)}(x,\bt,\epsilon,y_B,x P^+)}{\sqrt{S^{0}_{n_{y_n} \bn_{y_B}}(\bt,\epsilon)}}
\,.\end{align}
Here, $y_n$ is a free parameter appearing only in the Wilson lines of the soft function, and the Collins-Soper (CS) scale is defined as $\zeta = (x P^+ e^{-y_n})^2$.
In the TMD PDF for a proton moving along the $\bn$ direction, the roles of $y_A$ and $y_B$ will be reversed.

\textbf{Evolution equations.}
The TMD PDF defined in the $\overline{\text{MS}}$-scheme depends on two scales, namely the $\overline{\text{MS}}$-renormalization scale $\mu$ and the Collins-Soper scale $\zeta$. The all-order form of its evolution is given by
\begin{align} \label{eq:gamma_mu}
 &\frac{d \ln f_{i/P}(x, \bt, \mu, \zeta)}{d \ln\mu}
 = \gamma^i_\mu(\mu,\zeta)
 \nn\\&
 = \Gamma_\text{cusp}^i[\as(\mu)] \ln\frac{\mu^2}{\zeta} + \gamma_\mu^i[\as(\mu)]
\quad , \\ \label{eq:gamma_zeta}
 &\frac{d \ln f_{i/P}(x, \bt, \mu, \zeta)}{d\ln\zeta}
 = \frac{1}{2} \gamma^i_\zeta(\mu,b_T)
 \nn\\&
 = -2 \int_{1/b_T}^\mu \frac{d\mu'}{\mu'} \Gamma_\text{cusp}^i[\as(\mu')] + \gamma_\zeta^i[\as(1/b_T)]
\quad .
\end{align}
Here, $\Gamma_\text{cusp}^i$ is the so-called cusp anomalous dimension, and the $\gamma^i[\as]$ denote the non-cusp anomalous dimensions.
Note, however, that the CS kernel $\gamma_\zeta$,
which in the literature is also often referred to as $K$, becomes nonperturbative
when $q_T \sim b_T^{-1} \sim \lqcd$, independent of the choice of $\mu$.
In perturbation theory, $\gamma_\zeta$ is known at three loops~\cite{Li:2016ctv,Vladimirov:2016dll}.

A solution to Eqs.~\eqref{eq:gamma_mu} and \eqref{eq:gamma_zeta} is given by
\begin{align} \label{eq:TMD_evolution}
 & f_{i/P}(x, \bt, \mu, \zeta) = f_{i/P}(x, \bt, \mu_0, \zeta_0)
 \\ \nn & \quad \times
 \exp\biggl[ \int_{\mu_0}^\mu \frac{d\mu'}{\mu'} \gamma_\mu^i(\mu',\zeta_0) \biggr]
 \exp\biggl[ \frac12 \gamma_\zeta^i(\mu,b_T) \ln\frac{\zeta}{\zeta_0} \biggr]
\,.\end{align}
Note that the Collins-Soper evolution in $\zeta$ governs the energy dependence of the TMD PDF, as $\zeta \propto (x P^+)^2$ corresponds to the momentum of the struck quark. Thus, since $\gamma_\zeta$ itself becomes nonperturbative, TMD PDFs at different energies are related through a nonperturbative evolution, and hence calculating $\gamma_\zeta$ from lattice QCD is of major interest to TMD physics.

\textbf{Spin-dependent TMD PDFs.}
For simplicity, we have so far restricted ourselves to unpolarized processes. However, due to the vector nature of $\qt$, or equivalently $\bt$, TMD PDFs are sensitive to both the polarization of the hadron and the spin of the struck parton.
This is especially useful and can be exploited 
when nucleon or hadron structure is the target topic of interest.

In the language and notation of Mulders, Tangerman, and collaborators~\cite{Tangerman:1994eh,Mulders:1995dh,Bomhof:2007xt}, the leading TMD PDFs arise through the following decomposition,
\begin{align}
    &\Phi(x,{\bf k}_T,P,S) = \frac{1}{2} \left\{
    f_1(x,k_T^2) \sla{P} + \frac{1}{2} h_{1T}(x,k_T^2) \gamma_5 \left[ \sla{S}_T,\sla{P} \right] \right. \nonumber \\
    &{} \left. + S_L g_{1L}(x,k_T^2) \gamma_5 \sla{P} + \frac{{\bf k}_T \cdot {\bf S}_T}{M} g_{1T}(x,k_T^2) \gamma_5 \sla{P} 
    \right. \nonumber \\
    &{} \left. + S_L h_{1L}^\perp(x,k_T^2) \gamma_5 \frac{\left[ \sla{k}_T, \sla{P} \right]}{2M} + \frac{{\bf k}_T \cdot {\bf S}_T}{M} h_{1T}^\perp(x,k_T^2) \gamma_5 \frac{\left[ \sla{k}_T, \sla{P} \right]}{2M}
    \right. \nonumber \\
    &{} \left. +  i h_{1}^\perp(x,k_T^2) \frac{\left[ \sla{k}_T, \sla{P} \right]}{2M}- \frac{\epsilon_T^{k_T S_T}}{M} f_{1T}^\perp(x,k_T^2) \sla{P}
    \right\} \, , \label{eq:phi_decomp}
\end{align}
where $\Phi$ is, up to normalization conventions, the momentum space version of the unsubtracted TMD PDF definition on the first line of Eq.~\eqref{eq:tmdpdf_def}, but with a general Dirac structure rather than the trace over $\Gamma$. Equation~\eqref{eq:phi_decomp} displays the eight different TMD PDF functions that parametrize the range of possible combinations 
of correlations between parton polarization, hadron polarization, and transverse momentum. 
Their contribution to, for example, the SIDIS 
cross section can be seen directly in 
\cite{Bacchetta:2006tn}. 
By contrast, only three correlation functions survive the reduction to the collinear factorization case.

Several of the polarization-dependent TMD correlation functions have attracted particular 
attention for having novel properties unique to QCD with TMD factorization. For example, the 
Sivers function $f_{1T}^\perp(x,k_T^2)$ vanishes 
in naive definitions of the TMD PDF by PT invariance~\cite{Collins:1992kk}. A nonzero 
contribution is only made possible by the nontrivial 
Wilson line in Fig.~\ref{fig:wilsonlines}~\cite{Brodsky:2002cx}. The Wilson line introduces a pattern of non-universality in the form of a sign change of the Sivers TMD function between Drell-Yan-like processes and SIDIS-like processes~\cite{Collins:2002kn}.

\textbf{Large Transverse Momentum.}
The TMD factorization treatment discussed above 
is derived for the limit that parton transverse momentum is much smaller than the overall hard
scale of the problem. Approximations made in this 
limit are no longer valid when $q_T$ becomes of 
order or larger than the hard scale $Q$. There, a purely fixed-order collinear perturbative treatment that takes into account nonfactorizable transverse momentum behavior is the appropriate approach. Understanding the transition between these two different types of 
transverse-momentum 
dependence is important for understanding the TMD PDFs in particular, especially at moderate $Q$, where there is higher sensitivity to nonperturbative transverse momentum but a less-clear demarcation between large and small transverse momentum.

Schematically, the contributions from small 
and large transverse momentum are combined 
in the following additive way:
\begin{equation}
\frac{d \sigma}{d Q^2 d^2\qt}  = W + Y \, , 
\label{eq:WpY}
\end{equation}
where the ``$W$-term'' is the factorization formula 
involving TMD PDFs, Eq.~\eqref{eq:DY_fact_coll_small_qT}, while 
the second ``$Y$-term'' is the correction for 
large transverse momentum, perturbatively calculable using collinear factorization techniques. Achieving a robust matching between 
the two regions is critical for establishing that 
each region is under theoretical control. A current challenge is that calculations done with existing collinear parton distribution and fragmentation functions produce significant tension with data in the $Y$-term region of transverse momentum~\cite{Gonzalez-Hernandez:2018ipj,Wang:2019bvb,Bacchetta:2019tcu,Moffat:2019pci}. Work is needed to understand and resolve this tension.

One consequence of the large transverse momentum 
contribution is that the relationship between the 
TMD and collinear versions of factorization is not a simple integration over transverse momentum. That is, from Eq.~\eqref{eq:WpY},
\begin{equation}
\int d^{2}\qt \frac{d \sigma}{d Q^2 d^2\qt}  
\neq \int d^{2}\qt \, W \, .
\end{equation}
Ultra-violet divergences in the transverse momentum integrals of TMD functions are one symptom of the absence of a $Y$ term in the treatment. The sensitivity to the ultraviolet transverse momentum can be nontrivial~\cite{Qiu:2020oqr}. More work in the direction of quantifying and understanding these ultraviolet effects is needed.

When the large transverse momentum $q_T$ becomes even larger than $Q$, both observed momentum scales are in the perturbative regime. In the case of the Drell-Yan process, since the transverse momentum of the colliding partons is sufficiently smaller than both $Q$ and $q_T$, the Drell-Yan cross section can be factorized in terms of PDFs convoluted with a perturbatively calculable short-distance hard part. However, due to the large difference of the two observed momentum scales, $q_T \gg Q$, the large logarithms $\log(q_T /Q)$ of the short-distance hard part need to be resummed, and can be systematically subsumed into an effective fragmentation function for a parton to fragment into a Drell-Yan lepton pair \cite{Berger:2001wr,Fai:2003zc}.

%% file: 04sec2quasi.tex
\goodbreak
\subsection{TMD observables from lattice QCD}
\label{04sec2}

An ongoing program of evaluating transverse-momentum--dependent observables in hadrons within lattice QCD has been reported in Refs.~\cite{Hagler:2009mb,Musch:2010ka,Musch:2011er,Engelhardt:2015xja,Yoon:2017qzo,Engelhardt:2017miy,Engelhardt:2018zma}.
These studies are based on calculating hadron matrix elements of the type
\begin{equation}
\widetilde{\Phi}^{[\Gamma ]}
\equiv \frac{1}{2} \langle P^\prime ,S^\prime | \bar{q} (-b/2) \Gamma \,
{\cal U} [-b/2,b/2] q(b/2) |P,S\rangle
\label{eq:matelm} \ ;
\end{equation}
$\widetilde{\Phi}$ is a version of the unsubtracted TMD PDF defined in the first line of (\ref{eq:tmdpdf_def}) cast purely in terms of spacetime separations, i.e., even the dependence on momentum fraction $x$ has been replaced by a dependence on the Fourier-conjugate longitudinal separation $b^-$ (cf.~also the complementary definition (\ref{eq:phi_decomp}), which instead is cast purely in momentum space).
Several generalizations and adjustments are made to arrive at a formulation suitable for lattice calculations. An arbitrary Dirac structure $\Gamma$ is allowed for, and the states can also carry definite spin in addition to momentum. Also off-forward matrix elements, $P^\prime \neq P$, are of interest, as will be detailed further below. In a concrete lattice calculation, the staple-shaped gauge connection between the
quark operators $\bar{q}$, $q$, summarized here by ${\cal U}$, has finite extent; in the following, the vector $v$ specifies the direction of the staple legs, with their length scaled by the
parameter $\eta$. For $\eta =0$, the path becomes a straight link between the quark operators. Standard TMD observables are obtained by extrapolating the obtained data to $\eta \rightarrow \infty$.

As discussed further above, unsubstracted matrix elements of the type (\ref{eq:matelm}) contain divergences, which have to be absorbed into corresponding multiplicative soft factors. The TMD observables considered in the following are, however, appropriate ratios in which the soft factors cancel. As a result, the soft factors do not need to be specified in detail for present purposes.
To regulate rapidity divergences, the staple direction $v$ is taken off the lightcone into the spacelike domain~\cite{Collins:2011zzd}. This scheme is particularly suited for the connection to lattice QCD, as will become clear presently. A useful parameter characterizing how close
$v$ is to the lightcone is the Collins-Soper type parameter
$\hat{\zeta} = v\cdot P /(|v||P|)$, in terms of which the lightcone
is approached for $\hat{\zeta} \rightarrow \infty$.

The application of standard lattice QCD methods to evaluate (\ref{eq:matelm})
requires the operator in (\ref{eq:matelm}) to exist at a single time; given
that $b$ and $v$ are spacelike (the latter by virtue of the rapidity regulator scheme), there is no obstacle to boosting the problem
to a Lorentz frame in which this is the case. The transformation of the
results back to the original frame is facilitated by a decomposition of
(\ref{eq:matelm}) into Lorentz invariants that parallels the decomposition (\ref{eq:phi_decomp}) of the momentum space correlator into TMD PDFs. For example, in the
$\Gamma = \gamma^+$ channel, for a proton~\cite{Musch:2011er},
\begin{equation}
\frac{1}{2P^+} \widetilde{\Phi}^{[\gamma^+ ]} =
\widetilde{A}_{2B} + im_N \epsilon_{ij} b_i S_j \widetilde{A}_{12B}
\end{equation}
The invariants $\widetilde{A}_{iB}$ essentially correspond to
Fourier-transformed TMD PDFs. Through them, one can finally define
observables such as the generalized Sivers shift~\cite{Musch:2011er},
\begin{multline}
 \langle k_T \rangle_{TU}
(b^2 ,b\cdot P,\hat{\zeta} ,\eta v\cdot P,\ldots ) =  \\
 -m_N \frac{\widetilde{A}_{12B}
(b^2 ,b\cdot P,\hat{\zeta} ,\eta v\cdot P,\ldots )}{\widetilde{A}_{2B}
(b^2 ,b\cdot P,\hat{\zeta} ,\eta v\cdot P,\ldots )} \ .
\label{eq:sivrat}
\end{multline}
In the $b_T \rightarrow 0$ limit, (\ref{eq:sivrat}) formally represents the
average transverse momentum $k_T$ of unpolarized (``$U$'') quarks orthogonal
to the transverse (``$T$'') spin of the proton, normalized to the
corresponding number of valence quarks. Note that any multiplicative soft
factors renormalizing the $\widetilde{A}_{iB}$ are canceled by forming this type of ratio.

\begin{figure*}[htb]
\centering
\includegraphics[width=0.49\textwidth]{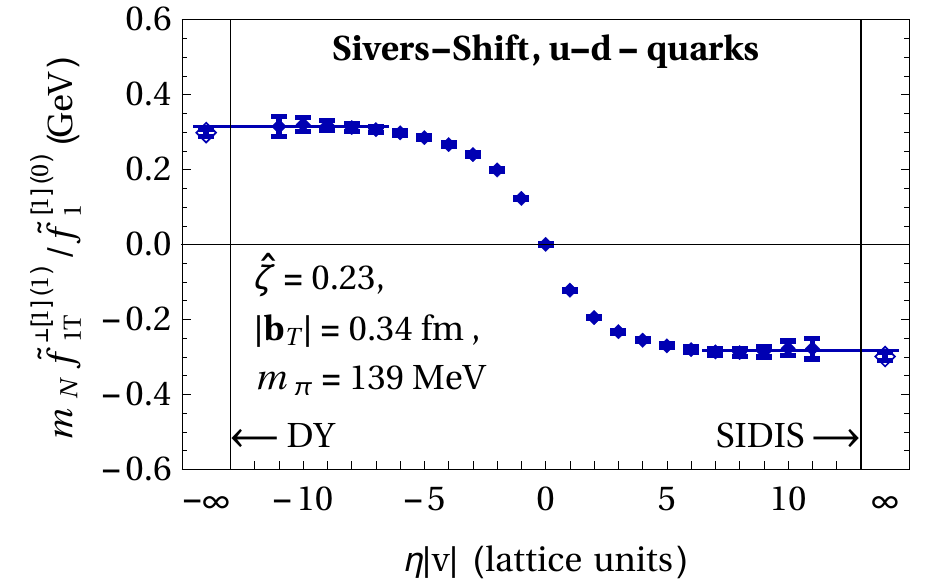}\hfil
\includegraphics[width=0.475\textwidth]{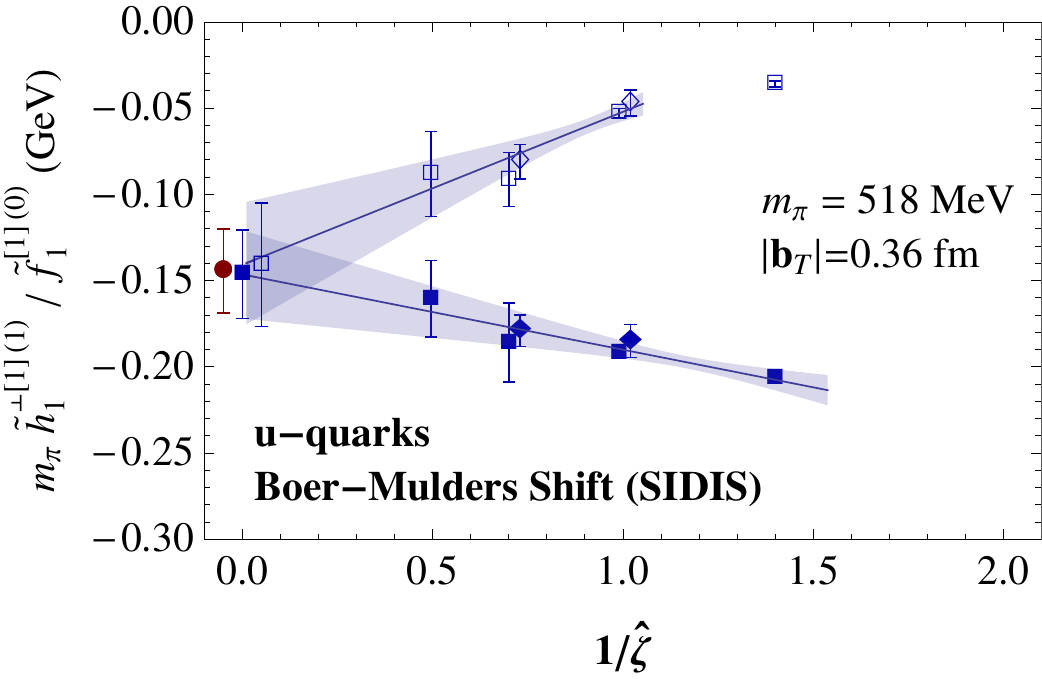}
\caption{\small Left: Proton Sivers shift as a function of staple
length for fixed $b_T$ and $\hat{\zeta}$; $\eta \rightarrow \infty$
defines the SIDIS limit. Right: Extrapolation of the
SIDIS-limit data for the pion Boer-Mulders shift to large $\hat{\zeta}$
at fixed $b_T$~\cite{Engelhardt:2015xja}. Open symbols represent a partial
contribution that dominates at large $\hat{\zeta}$, providing further
insight into the approach to the asymptotic regime.}
\label{fig:one}
\end{figure*}

Lattice calculations to date have predominantly focused on the special case $b\cdot P=0$, corresponding to evaluating integrals of the TMD PDFs in
question over the longitudinal quark momentum fraction $x$, and thus concentrating purely on the transverse-momentum dependence. Some example results are exhibited below. Generalizing these calculations to a scan of the $(b\cdot P)$-dependence allows one to access also the dependence on $x$, which is Fourier conjugate to $b\cdot P$. Note that, since soft factors do not depend on $b\cdot P$, they can be factored outside the corresponding Fourier transforms and still canceled in ratios. Cast in Lorentz-invariant form, the lattice geometries relevant for TMD observables must obey the relation~\cite{Musch:2011er}
\begin{equation}
\frac{v\cdot b}{v\cdot P} = \frac{b\cdot P}{m_N^2} \left( 1-\sqrt{1+1/\hat{\zeta}^2 } \right) \ ,
\end{equation}
which forces one to use general off-axis directions on the lattice, significantly complicating the analysis.
Data from a preliminary scan of the $(b\cdot P)$-dependence indicate that it is feasible to obtain also the $x$-dependence of TMD ratios such as
(\ref{eq:sivrat}). A corresponding comprehensive data production effort
is currently in progress. A complementary approach to the $x$-dependence of TMD PDFs based on quasi-TMD PDFs is presented in detail in the next section.

\begin{figure*}[tb]
\centering
\includegraphics[width=0.95\textwidth]{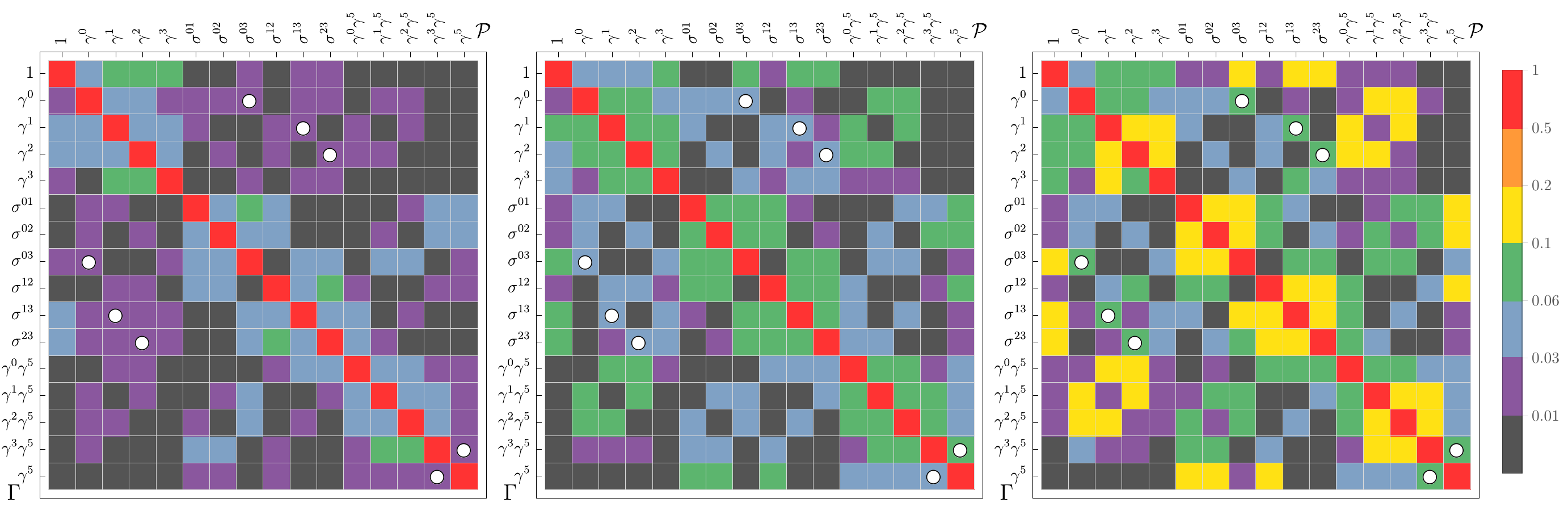}
\caption{Mixing pattern in the RI'/MOM scheme for quark bilinear operators constructed using improved Wilson fermions with staple-shaped gauge connections. The quark operator separation $b$ is purely transverse, with $b_T/a=3,7,11$ from left to right, where $a=0.06$~fm denotes the lattice spacing. The staple length is given by $\eta /a =14$. Colors indicate mixing strengths. White circles indicate mixings already obtained in one-loop lattice perturbation theory~\cite{Constantinou:2019vyb}.}
\label{fig:mixing}
\end{figure*}

\begin{figure*}[tb]
\centering
\includegraphics[width=0.48\textwidth]{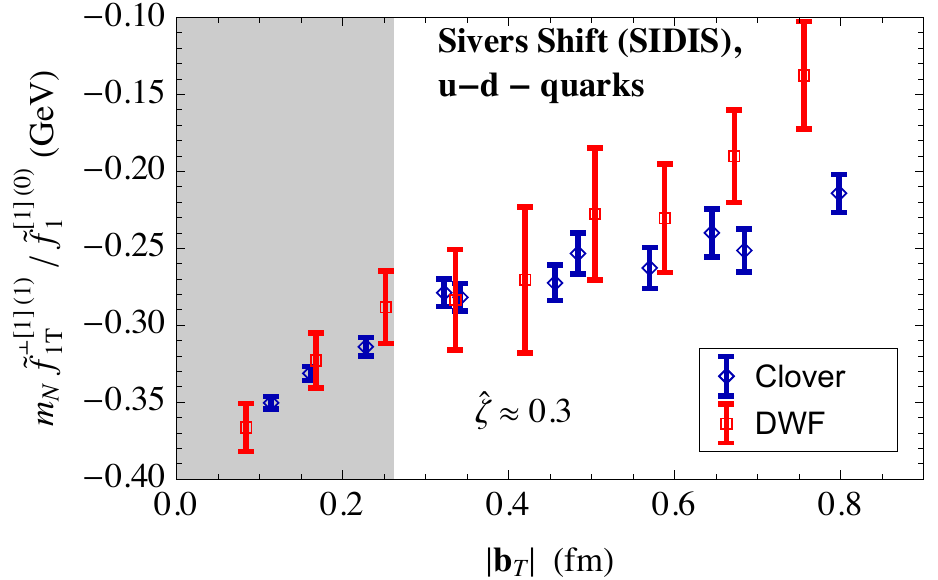}\hfil
\includegraphics[width=0.45\textwidth]{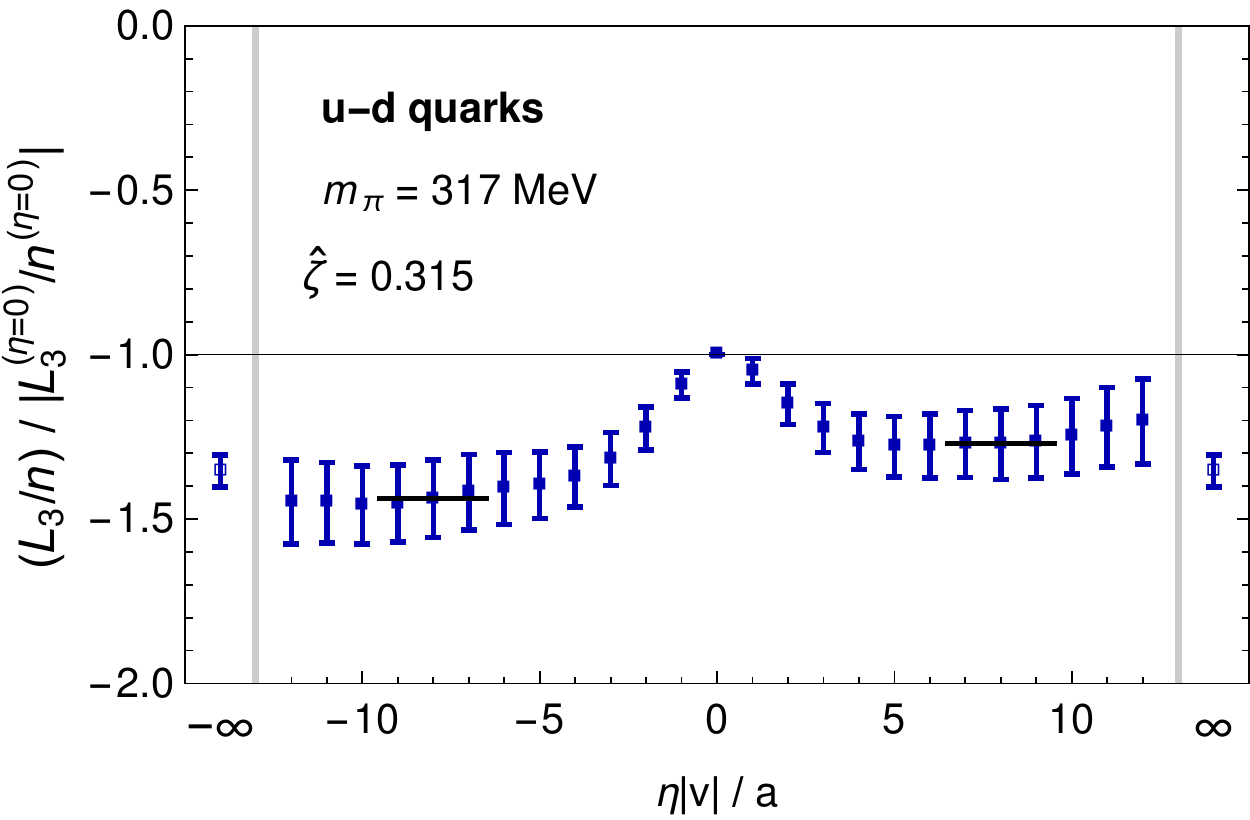}
\caption{\small Left: Comparison between SIDIS-limit data for
the proton Sivers shift obtained for two distinct lattice discretizations,
as a function of $b_T$ at fixed $\hat{\zeta}$~\cite{Yoon:2017qzo}.
The data are compatible within uncertainties, suggesting that no
significant violations of multiplicative renormalization are present.
Right: Longitudinal quark OAM in the proton $L_3$ as a function of
staple length at fixed $\hat{\zeta}$~\cite{Engelhardt:2018zma}. The
limit $\eta =0$ yields Ji OAM, $\eta \rightarrow \pm \infty$
Jaffe-Manohar OAM. The ratio of $L_3$ to the number of valence quarks
$n$ is evaluated to cancel multiplicative renormalizations, analogous
to Eq.~(\ref{eq:sivrat}). Data are shown in units of the absolute value
of Ji OAM.}
\label{fig:two}
\end{figure*}

In a concrete lattice calculation, one
must extrapolate the data to the $\eta \rightarrow \infty$ limit in order to make contact with standard TMD PDFs. This
extrapolation is typically under good control, as shown in
Fig.~\ref{fig:one} (left), exhibiting first data recently obtained
directly at the physical pion mass. Moreover, the lattice
data need to be extrapolated to the regime of large $\hat{\zeta}$.
This presents a considerable challenge, since it requires data at
sufficiently high hadron momenta. Figure~\ref{fig:one} (right) displays
results of a corresponding dedicated study of the Boer-Mulders shift in
a pion~\cite{Engelhardt:2015xja}; the Boer-Mulders shift is a counterpart
to the Sivers shift in which the hadron is unpolarized, but the quark is
transversely polarized. A further question to be faced is whether the
multiplicative renormalization pattern of continuum TMD PDFs carries over to
the lattice formulation. This issue was explored in Ref.~\cite{Yoon:2017qzo}
by varying the discretization scheme and testing for violations of a
multiplicative renormalization pattern; a sample comparison is shown in
Fig.~\ref{fig:two} (left), corroborating that soft factors cancel in the
Sivers shift ratio to the accuracy accessible in the calculation. Note
that, for selected Dirac structures and discretization schemes, operator
mixing has been shown to occur for quark bilinear operators with staple-shaped gauge connections~\cite{Constantinou:2019vyb,Shanahan:2019zcq,Green:2020xco},
invalidating simple multiplicative renormalization.
Fig.~\ref{fig:mixing} illustrates the mixing pattern in the RI'/MOM scheme for quark bilinear operators constructed using improved Wilson fermions with staple-shaped gauge connections in the case of a purely transverse quark operator separation $b$~\cite{Shanahan:2019zcq}.
 
Finally, by extending calculations of (\ref{eq:matelm}) to nonzero transverse
momentum transfer $\Delta_T =P^\prime -P$, one can correlate
quark transverse momentum with position; $\Delta_T$ is Fourier
conjugate to the quark impact parameter $\rt$. This allows one to
directly access longitudinal quark orbital angular momentum (OAM),
$\langle \rt \times \kt \rangle$. The choice of gauge link ${\cal U}$
corresponds to different decompositions of proton spin. A staple link
extending to infinity, such as used in standard TMD PDF studies, yields
Jaffe-Manohar OAM, whereas the $\eta =0$ limit yields Ji OAM~\cite{Hatta:2011ku,Ji:2012sj,Burkardt:2012sd,Rajan:2016tlg,Raja:2017xlo}.
By varying $\eta$ within a lattice calculation, a continuous,
gauge-invariant interpolation between the two is obtained~\cite{Engelhardt:2017miy}. Fig.~\ref{fig:two} (right) displays corresponding
results reported in Ref.~\cite{Engelhardt:2018zma}. Jaffe-Manohar orbital angular
momentum is enhanced in magnitude compared to its Ji counterpart.

\subsection{Quasi-TMD PDFs}
An alternative approach to calculating TMD PDFs from lattice QCD using so-called
quasi-TMD PDFs~\cite{Ji:2013dva,Ji:2014gla} has been been explored recently
by several groups~\cite{Ji:2014hxa, Ji:2018hvs, Ebert:2018gzl, Ebert:2019okf,Ebert:2019tvc, Ji:2019sxk, Ji:2019ewn, Vladimirov:2020ofp, Ji:2020ect, Ebert:2020gxr}.
Here, the key idea is to construct an equal-time correlator by replacing lightlike
separations in \eqref{eq:tmdpdf_def} by a spacelike separation along the $z$ direction alone.
Under a Lorentz boost, these spacelike directions approach the lightcone,
with the mismatch being accounted for by a perturbative matching.

Following the notation of Ref.~\cite{Ebert:2019okf},
we define the quasi-TMD PDF analogous to \eqref{eq:tmdpdf_1} as
\begin{multline} \label{eq:qtmdpdf_1}
 \tilde f_{i/P}(x, \bt, \mu , P^z) 
 =  \lim_{\substack{a\to 0 \\ \eta\to\infty}} \tilde Z_\text{uv}^i(\mu,a) \\
    \times \tilde f_{i/P}^{0\,(u)}\bigl(x, \bt, a, \eta, P^z \bigr)
    \tilde\Delta_S^q(\bt, a, \eta)
\,.\end{multline}
Here, the lattice spacing $a$ acts as UV regulator, $\eta$ limits the length of
the Wilson lines, and $\tilde\Delta_S^q$ is a quasi-soft factor required to cancel
divergences as $\eta/b_T\to\infty$. While it has been shown in Ref.~\cite{Ebert:2019okf}
that there is no straightforward construction of a quasi-soft factor
directly related to the soft functions appearing in \eqref{eq:tmdpdf_1}, Ref.~\cite{Ji:2019sxk}
recently proposed to calculate it within lattice QCD, employing a formulation based on heavy-quark effective theory. Importantly, this factor cancels in ratios
of quasi-TMD PDFs, and hence, similarly to the method discussed in \ref{04sec2},
a determination of $\tilde\Delta_S^q$ is not needed to access such ratios.

The unsubtracted quasi-TMD PDF is defined as
\begin{multline} \label{eq:qtmdpdf_def}
 \tilde f_{i/P}^{0\,(u)}(x, \bt, a, \eta, P^z)
 = \int\frac{d b^z}{2\pi} e^{i b^z (x P^z)} \, N_{\tilde\Gamma} \\
 \times
 \, \bigl<\!P \big|\bar q(b/2) \, {\cal \tilde U}
 \tfrac{\tilde\Gamma}{2} q(-b/2) \big| P \bigr>
\,,\end{multline}
where $b^\mu = (0,\bt,b^z)$, and the Wilson line path $\cal \tilde U$
is chosen such that it connects $b/2 \to (0,\bt/2,\eta) \to (0,-\bt/2,\eta) \to -b/2$.
For unpolarized TMD PDFs, the Dirac structure can be chosen as $\tilde\Gamma = \gamma^0, \gamma^3$,
with the normalization factor $N_{\gamma^0} = 1, N_{\gamma^3} = P^z/P^0$.

For polarized quarks and protons, one can generalize \eqref{eq:qtmdpdf_def}
using different Dirac structures $\tilde\Gamma$. However, the relation
between quasi-TMD PDFs and TMD PDFs has been argued to be
spin-independent~\cite{Vladimirov:2020ofp,Ebert:2020gxr}, and hence the following results
also apply to the polarized case.

The key relation between quasi-TMD PDFs and TMD PDFs is~\cite{Ebert:2019okf, Ji:2019ewn, Ji:2020ect, Vladimirov:2020ofp}
\begin{multline} \label{eq:qtmd_tmd}
 \tilde f_{\text{ns}}(x, \bt, \mu, P^z)
 = C_\text{ns}(\mu,x P^z) g_q^S(b_T,\mu) \\
   \times \exp\biggl[\frac12 \gamma_\zeta^q(\mu,b_T) \ln\frac{(2xP^z)^2}{\zeta} \biggr]
   f_{\text{ns}}(x, \bt, \mu, \zeta)
\,,\end{multline}
which holds up to corrections in $b_T/\eta$, $1/(b_T P^z)$ and $1/(P^z \eta)$.
In \eqref{eq:qtmd_tmd}, the quasi-TMD PDF $\tilde f_\text{ns}$ in the nonsinglet $\text{ns}=u-d$ channel
is related to the TMD PDF $f_\text{ns}$ through a perturbative kernel $C_\text{ns}$,
which is known at one loop~\cite{Ebert:2019okf}.
It also involves a nonperturbative factor $g_q^S$ because the quasi-TMD PDF was not defined
with the physical soft function. In the approach of Ref.~\cite{Ji:2019sxk}, it corresponds
to the reduced soft factor, $g_q^S = \sqrt{S_r}$, see also Ref.~\cite{Ji:2020ect}.
Once calculations of this reduced soft factor become available,
quasi-TMD PDFs can be formulated without this nonperturbative factor~\cite{Ji:2019sxk,Ji:2019ewn}.
Lastly, \eqref{eq:qtmd_tmd} also involves the nonperturbative CS evolution kernel, see \eqref{eq:gamma_zeta}, which relates the hadron energies $P^z$ and $\zeta$.

The Collins-Soper kernel $\gamma_\zeta$ is required to relate TMD PDFs at different hadron energies,
and thus its nonperturbative determination from lattice QCD is of key interest.
Based on \eqref{eq:qtmd_tmd}, Refs.~\cite{Ebert:2018gzl,Ebert:2019okf} proposed to determine it
from ratios of quasi-TMD PDFs at different momenta $P^z_1 \ne P^z_2$,
\begin{align} \label{eq:gamma_zeta_lattice}
 \gamma^q_\zeta(\mu, b_T) &
 = \frac{1}{\ln(P^z_1/P^z_2)}
  \ln \frac{C_\text{ns}(\mu,x P_2^z)\, \tilde f_\text{ns}(x, \bt, \mu, P^z_1)}
           {C_\text{ns}(\mu,x P_1^z)\, \tilde f_\text{ns}(x, \bt, \mu, P^z_2)}
\,,\end{align}
see also Ref.~\cite{Vladimirov:2020ofp} for a related proposal.
In the ratio in \eqref{eq:gamma_zeta_lattice}, the quasi-soft factor
$\tilde \Delta_S^q$ cancels, and thus the Collins-Soper kernel can be obtained from a ratio
of unsubtracted (but renormalized) quasi-TMD PDFs alone.
Another key advantage is the independence of \eqref{eq:gamma_zeta_lattice}
on the hadron state, and hence it can be calculated in a pion state rather than a proton state.

Without $\tilde \Delta_S^q$, both numerator and denominator in \eqref{eq:gamma_zeta_lattice}
suffer from divergences associated with Wilson-line self energies, that is, divergences in $\eta/b_T$.
These divergences cancel in the ratio, but in practice it can be numerically more reliable
to enforce this cancellation separately in numerator and denominator.
In Ref.~\cite{Ebert:2019tvc}, it was suggested to insert a common renormalization factor
determined nonperturbatively in the RI'/MOM scheme on the lattice,
and its conversion to the $\overline{\text{MS}}$ scheme was calculated at one loop.
The first lattice studies of this renormalization factor were explored in Ref.~\cite{Shanahan:2019zcq},
revealing significant operator mixing on the lattice, cf.~Fig.~\ref{fig:mixing}, which was treated
by diagonalizing the renormalization and matching matrices for the RI'/MOM scheme.

The feasibility of the above method of determining the Collins-Soper kernel was demonstrated in Ref.~\cite{Shanahan:2020zxr}
using a lattice of size %
$L^3 \times T = (2\text{ fm})^3 \times 4\text{ fm}$ and lattice spacing
$a=0.06\text{ fm}$  with a heavy pion mass $m_\pi = 1.2\text{ GeV}$.
Taking advantage of \eqref{eq:gamma_zeta_lattice} being independent of the hadron state,
the Collins-Soper kernel for $n_f = 0$ quark flavors was extracted from a quenched
calculation of quasi TMD PDFs in a pion state with momenta $P^z \in \{1.29, 1.94, 2.58\}\text{ GeV}$.
The study found that the extrapolation of the position-space matrix element to $b^z\to\infty$
poses a key challenge in the Fourier transform in \eqref{eq:qtmdpdf_def}.
Two different fits to the lattice data were performed using Bernstein and Hermite polynomials,
and strong sensitivity to the finite size of the lattice was observed 
by comparing the extrapolations of the two polynomial fits to $b^z > L/2$.
Additional systematic effects from the power corrections mentioned below \eqref{eq:qtmd_tmd}
could not be resolved with the limited data from the employed lattice simulation.
Nevertheless, promising results were obtained, demonstrating that lattice results for the CS kernel
in the nonperturbative region up to $b_T \sim 1\text{ fm}$ are tractable
provided that future calculations are performed on significantly larger lattice volumes.

\begin{figure}[tb]
\centering
\includegraphics[width=0.45\textwidth]{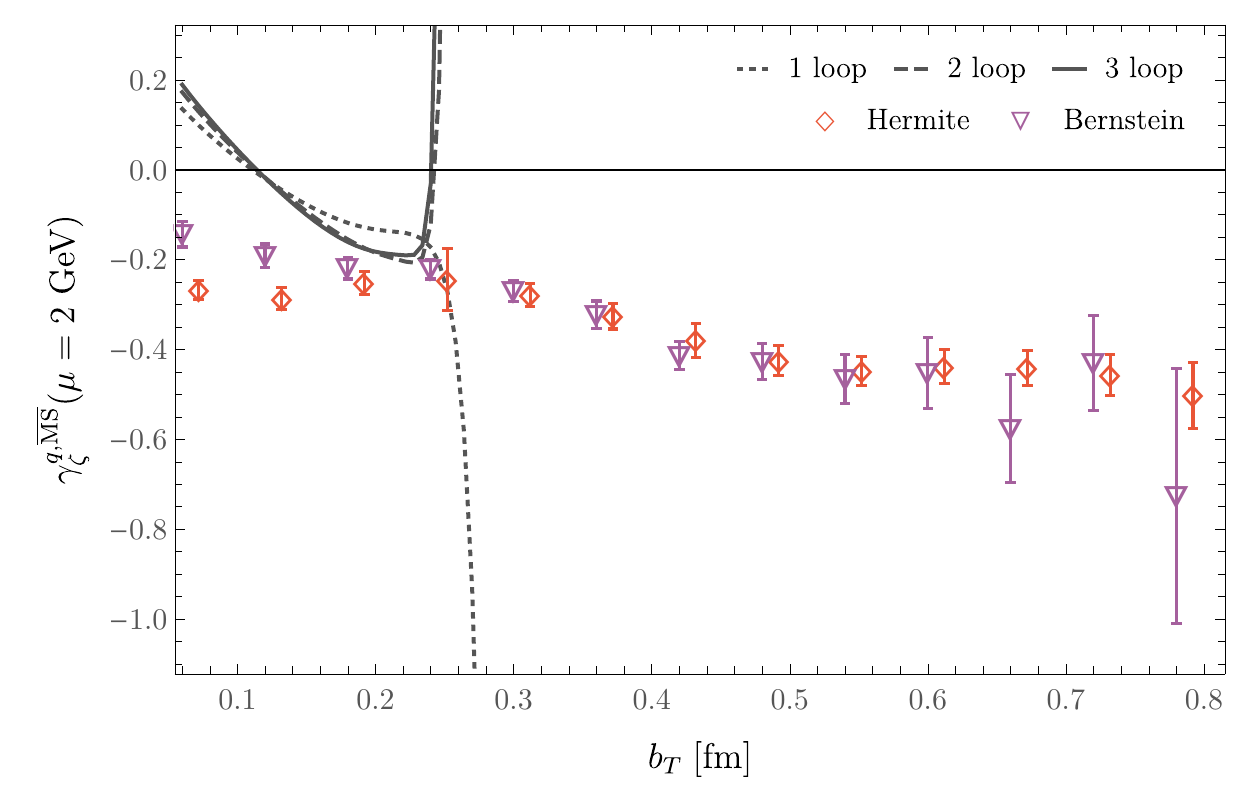}
\caption{Results of Ref.~\cite{Shanahan:2020zxr} for the $b_T$-dependence of Collins-Soper evolution.
The red and purple data exhibit results using two different interpolations of the unsubtracted quasi-TMD PDF.
The dashed and solid lines indicate perturbative results for the same quantity, which break down at the Landau pole at $b_T \approx 0.25~\text{fm}$.
Both in the small-$b_T $ limit as well as in the large-$b_T $ region there are additional systematic uncertainties owing to power corrections that have not been estimated in the figure.
}
\label{fig:CS}
\end{figure}

Figure \ref{fig:CS} shows the result of Ref.~\cite{Shanahan:2020zxr} for the $b_T$-dependence of the CS kernel, comparing their two extractions using Bernstein and Hermite polynomials to perturbative predictions of the same quantity.
The perturbative results diverge at $b_T \approx 0.25$~fm due to the Landau pole of the strong coupling, while the lattice calculation yields encouraging results for all values of $b_T$, except for very small $b_T$ where $1/(b_T P^z)$ power corrections become important.

It will also be interesting to consider other ratios of TMD PDFs obtained from lattice QCD, for example ratios of (quasi-)TMD PDFs in different hadron states,
or ratios of spin-dependent (quasi-)TMD PDFs. Such ratios of quasi-TMD PDFs
are independent of the soft factor, and, when evaluated at equal momenta,
are furthermore independent of the CS evolution in \eqref{eq:qtmd_tmd}; they can
thus be related perturbatively to corresponding ratios of TMD PDFs.
Of particular value will be lattice results of sufficient quality, and in the relevant kinematical regimes, to be useful as auxiliary input to phenomenological fits of TMD PDFs, discussed in the next section.

%% file: 04sec3fits.tex
\goodbreak
\subsection{Status of global TMD fits} 
\label{04sec3}
TMD PDFs carry rich information on parton confined motion inside a bound hadron, as well as on correlations between the motion of quarks and gluons and the direction of hadron spin.  
Their information can be extracted from data on QCD factorizable hadronic cross sections or lattice observables (such as quasi- or pseudo-TMD PDFs), and their properties could also be studied in terms of model calculations. 
The limited kinematic coverage for the differential cross sections
provided by experimental data leads to a challenging inverse problem; it is unlikely, if not impossible, that one can completely fix the TMD PDFs as continuous functions over the full range of kinematics. 
To pin down TMD PDFs to the best possible accuracy, we need data from multiple observables, experimentally measured or calculated within Lattice QCD, that are related to the same universal set of TMD PDFs, covering a wide kinematic regime, and to perform QCD global analyses and fits, similar to what is done to extract the PDFs as described in Sec.~\ref{02sec4}.

However, there are fundamental differences between global fits to extract PDFs and the global fits for extracting TMD PDFs.  With the linear DGLAP evolution equations, PDFs are uniquely determined once we have a set of input PDFs as a function of $x$ at a lowest possible input hard scale $Q_0\sim{\rm GeV}$, since all DGLAP evolution kernels are perturbatively calculable for scales larger than $Q_0$.  
Then, PDFs at $Q>Q_0$ {\it can} be generated by the DGLAP evolution, which is the predictive power of QCD dynamics and its factorization formalism.  

For TMD PDFs with two momentum scales, $Q$ and $k_T$ (or its Fourier conjugate $b_T $), QCD evolution involves two coupled evolution equations and covers a two-dimensional phase space, e.g., $(Q,b_T )$, where $b_T  \in [0,\infty)$ \cite{Collins:1984kg}. As a consequence, the path used in solving these two coupled equations is not unique, which could affect the size of higher order corrections, leading to an additional scheme dependence of the TMD PDFs \cite{Scimemi:2019cmh}. However, the fundamental and most important difference from DGLAP evolution is that evolution kernels for evolving TMD PDFs from an input scale $Q_0$ to any higher observed scale $Q$, referred as the Collins-Soper kernels (cf.~Sec.~\ref{04sec1.5}), depend on the value of $b_T $, and are not perturbatively calculable for the region where $b_T > 1/Q_0$.  That is, if the Fourier transform to obtain momentum space $k_T$-dependent TMD PDFs, which are needed for evaluating momentum-space cross sections, is sensitive to the large $b_T $-region of the evolved $b_T $-distributions, QCD perturbation theory and its factorization formalism {\it cannot} uniquely generate TMD PDFs at a large scale $Q$ from input TMD PDFs at $Q_0$ as a function of $x$ and $k_T$ (or $b_T $ in Fourier space) \cite{Collins:1984kg}. 

In high energy collisions, a parton shower develops following the breaking up of the colliding hadron(s), and the characteristics of the shower depend on the hard scale of the collision $Q$ and the available phase space for the shower, which is sensitive to probed parton momentum fraction $x$ \cite{Grewal:2020hoc}.  Consequently, the probed active parton's $k_T$ in hard collisions, described by the measured TMD PDFs, is not the same as the intrinsic transverse momentum $k_{T_0}$ of the same parton inside a bound hadron.  The difference is encoded in the QCD evolution of TMD PDFs. For physical observables whose momentum transfer $Q$ is large and active parton $x$ is effectively small, such as the transverse momentum $\qt$-dependence of $W/Z$ or Higgs production at collider energies, QCD evolution (or the effect of the shower) is dominated by perturbatively calculable large logarithms, or equivalently, the $\bt$-dependence of the relevant TMD PDFs in Fourier space is completely dominated by the perturbative small-$b_T $ region \cite{Qiu:2000hf,Grewal:2020hoc}.  In this case, non-perturbative evolution kernels from the large-$b_T $ region are effectively irrelevant, the extracted $k_T$-dependence of TMD PDFs for these observables is perturbatively generated (by the calculable part of the shower), and the only relevant non-perturbative information is given by normal collinear PDFs.  Physically, for such observables and the corresponding TMD PDFs, the information on non-perturbative partonic motion (or intrinsic $k_{T_0}$-dependence) inside a bound hadron was diluted by the tremendous shower developed during the collision (or the effective $\langle \mathbf{k_{T_0} }^{2} \rangle \ll \langle \qt^2\rangle$). In this case, QCD perturbation theory and its TMD factorization formalism should have excellent predictive power similar to that of collinear factorization \cite{Qiu:2000ga,Berger:2002ut,Grewal:2020hoc}.  QCD global analyses and fits of data on this kind of observable at high energy colliders should provide excellent precision tests of QCD factorization and resummation of large logarithms, and additional opportunities or channels to explore new physics; they should be pursued for LHC physics.

On the other hand, for extracting information on the motion of quarks and gluons inside a bound hadron, a fundamental emergent QCD phenomenon, we need to probe TMD PDFs through scattering processes with relatively less dilution from partonic showers, and observables with a relatively smaller momentum transfer $Q$. In this case, QCD evolution (or the connection between the measured $k_T$-dependence at the hard collision scale $Q$ and the intrinsic $k_{T_0}$ dependence inside a bound hadron) is very much sensitive to the non-perturbative large $b_T $ region \cite{Qiu:2000hf,Scimemi:2019cmh,Grewal:2020hoc}.  Any global analyses of TMD PDFs and fits of data from this kinematic regime have to clearly define how the non-perturbative evolution in the large $b_T $-regime is treated \cite{Collins:2014jpa}.  

There have been three types of approaches to treating the non-perturbative evolution in the large-$b_T $ regime.  As originally proposed in the pioneering paper on TMD physics (or physics of two-scale observables) \cite{Collins:1984kg}, the evolution is completely taken care of by perturbative Collins-Soper evolution kernels in the small $b_T $-regime by introducing a ``$b^*$-prescription",
\begin{equation}
f_{i/P}(x,\bt,Q) \equiv f_{i/P}(x,\bt^*,Q)\, F^{\rm NP}(x,\bt,Q)\, ,
\label{eq:b-star}
\end{equation}
where $\bt^* \equiv \bt/\sqrt{1+(b_T /b_{T_{\max}})^2} $, such that $b_T^* < b_{T_{\max}}$ for $b_T \in[0,\infty)$, and $F^{\rm NP}(x,\bt,Q)$ is a nonperturbative function to be fitted through data from global fits \cite{Davies:1984sp,Meng:1995yn,Landry:2002ix,Collins:2014jpa}.  The second approach is to keep the perturbatively evolved TMD PDFs in the small-$b_T $ region as they are, and introduce a model evolution kernel for the large-$b_T $ region whose parameters are fitted by experimental data \cite{Qiu:2000hf,Scimemi:2019cmh}.  The third approach is to calculate these non-perturbative evolution kernels within lattice QCD, as discussed in the previous subsection, an approach which needs to be explored further.
In view of the reach of lattice TMD calculations performed to date, it seems plausible to anticipate that, in the medium term, obtaining non-perturbative evolution kernels with an accuracy in the range of 10-20\% up to $b_T \sim 1\, \mbox{fm} $ at hadron momenta up to about $2\, \mbox{GeV} $ at the physical pion mass will be possible. The restriction in the hadron momentum is likely to remain a significant source of systematic uncertainty, engendering higher-twist power corrections that must be brought under numerical control. Also effects of the finite size of the lattice are still only rudimentarily understood. Nonetheless, such input could provide valuable additional information for global fits.

Phenomenologically, most efforts to extract TMD PDFs from existing data are concentrated on the unpolarized $f_1 (x,k_T^2)$ in Eq.~\eqref{eq:phi_decomp}. In recent years, there has been tremendous progress both in the accuracy of the theoretical framework and in the dimension of the data set included in the fit. Extractions of $f_1 (x,k_T^2)$ from SIDIS data were performed in the so-called extended parton model~\cite{Signori:2013mda,Anselmino:2013lza} (the latter being the only analysis so far with an explicit flavor dependence in the fitting parameters of the functional form), and were followed by extractions in the appropriate TMD framework but limited to Drell-Yan and $Z$-boson production data. In these cases, the description of the TMD evolution formula~\eqref{eq:TMD_evolution} was continuously improved, moving from the NLL level~\cite{Echevarria:2014xaa,Su:2014wpa,Bacchetta:2017gcc} through the NNLL one~\cite{DAlesio:2014mrz,Scimemi:2017etj,Bertone:2019nxa}, up to the NNNLL level~\cite{Bacchetta:2019sam,Scimemi:2019cmh} that matches the accuracy of standard phenomenology at the LHC (see Table 1 and Sec. 2.3 of Ref.~\cite{Bacchetta:2019sam} for more details on the perturbative expansion). In the most recent analyses~\cite{Scimemi:2017etj,Bertone:2019nxa,Bacchetta:2019sam,Scimemi:2019cmh}, very precise data from the LHC (in particular, from ATLAS) were included in the fit and turned out to have a large impact on the behaviour of $f_1 (x,k_T^2)$ at small $x$. Conversely, the impact of a flavor-dependent TMD (as obtained from the low-energy analysis of Ref.~\cite{Signori:2013mda}) on the extraction of Standard Model parameters at the LHC was explored in Ref.~\cite{Bacchetta:2018lna} using the template-fit technique, reaching the conclusion that flavor sensitivity in the intrinsic quark transverse momentum might induce an additional uncertainty in the extraction of the $W$ boson mass that is comparable to the error correlated to PDF uncertainties. 

The first extraction of $f_1 (x,k_T^2)$ in the TMD framework from a global fit of both SIDIS data and measurements of the Drell-Yan and $Z$-boson production processes can be found in Ref.~\cite{Bacchetta:2017gcc}. The approach is based on the description at NLL of the perturbative Collins-Soper evolution kernel plus a modified ``$b^\ast$-prescription''. The fit includes 8059 data points with a quality of $\chi^2$/d.o.f. = 1.55 $\pm$ 0.05 with only 11 parameters. The results confirm the universality of TMD PDFs but show a marked anti-correlation between TMD PDFs and TMD FFs, calling for an independent extraction of transverse-momentum dependent fragmentation functions from $e^+$-$e^-$ annihilation data (which is still missing). The same finding is obtained in the more recent global fit of Ref.~\cite{Scimemi:2019cmh}, where the perturbative accuracy of the TMD at small $b_T$ is pushed to the NNNLO level and the nonperturbative part of the evolution at large $b_T$ is fitted to data. The analysis of Ref.~\cite{Scimemi:2019cmh} includes recent very precise LHC data, although the total number of analyzed data points (1039) falls short of the hitherto largest set of 8059 analyzed in Ref.~\cite{Bacchetta:2017gcc}. 

\begin{figure}[tb]
\centering
\includegraphics[width=0.45\textwidth]{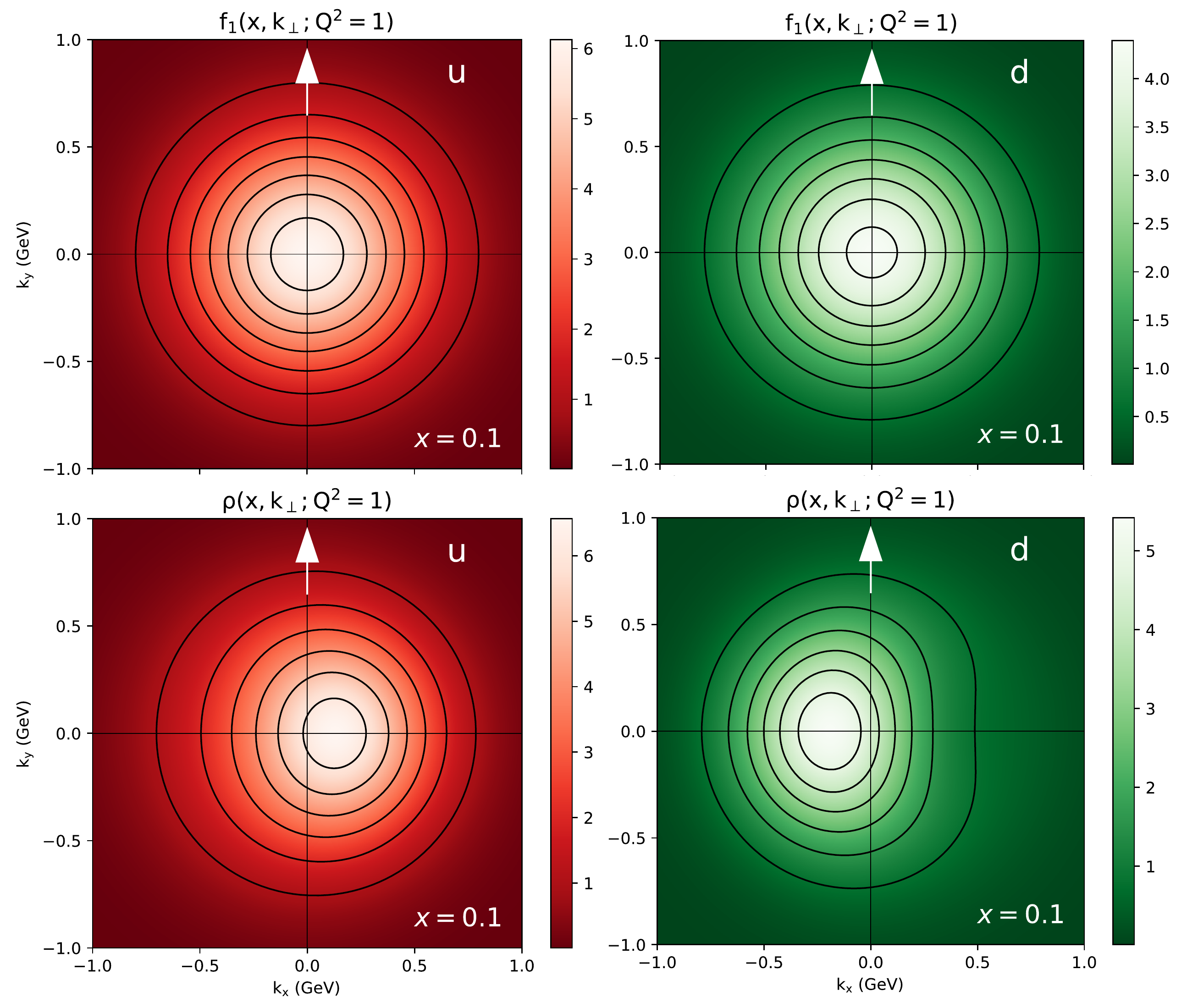}
\caption{Upper panels: quark density $f_{q/N} = f_1 (x=0.1, k_T^2)$ at $Q^2=1$ GeV$^2$ for an unpolarized proton moving ideally towards the reader; left panel for up quark, right panel for down quark. Lower panels: quark density $f_{q/N^\uparrow} = f_1 (x=0.1, k_T^2) - f_{1T}^{\perp q} (x=0.1, k_T^2) k_x/M$ in the same conditions but for a proton transversely polarized along the $\hat{y}$ direction (see text). 
}
\label{fig:Siversdensity}
\end{figure}

As for polarized TMD PDFs in Eq.~\eqref{eq:phi_decomp}, the best known is the Sivers function $f_{1T}^{\perp} (x,k_T^2)$. It describes how the $k_T$ distribution of an unpolarized quark is distorted in a transversely polarized hadron~\cite{Sivers:1990cc}. In fact, for a nucleon with mass $M$ moving along the $\hat{z}$ direction and transversely polarized along the $\hat{y}$ direction, the probability density of a quark with flavor $q$ is given, at a certain (understood) scale $Q$, by
\begin{equation}
f_{q/N^\uparrow} (x, k_x, k_y) = f_1^q (x, k_T^2) - f_{1T}^{\perp q} (x, k_T^2) \, \frac{k_x}{M} \; ,
\label{eq:Siversdensity}
\end{equation}
cf.~Eq.~(\ref{eq:phi_decomp}).
If the nucleon were unpolarized, the $f_{q/N}$ would be determined only by the unpolarized TMD PDF $f_1$ and the density would be perfectly symmetric around the $\hat{z}$ direction. The transverse polarization of the nucleon induces a distortion of the density along the $\hat{x}$ direction through the Sivers TMD PDF $f_{1T}^{\perp}$. Evidently, $f_{1T}^{\perp}$ describes a spin-orbit effect at the partonic level. The Sivers function is representative of the class of na\"ive T-odd TMDs, namely of those TMDs that are not constrained by T-reversal invariance~\cite{Boer:1997nt}. Their universality is broken but in a calculable way. For example, the Sivers function extracted in a Drell-Yan process with a transversely polarized proton should turn out opposite to the one that is extracted in SIDIS. The $f_{1T}^{\perp} \vert_{\mathrm{DY}} = - f_{1T}^{\perp} \vert_{\mathrm{SIDIS}}$ prediction is based on very general assumptions, and it represents a fundamental test of QCD~\cite{Collins:2002kn}. Therefore, it is the subject of intense experimental investigation. Preliminary results hint to statistically favor the prediction~\cite{Adamczyk:2015gyk,Aghasyan:2017jop,Anselmino:2016uie} although more precise data are needed to draw a sharp conclusion. Many parametrizations of the Sivers function are available (for example, see Refs.~\cite{Vogelsang:2005cs,Collins:2005ie,Bacchetta:2011gx,Anselmino:2012aa,Aybat:2011ta,Sun:2013hua,Boer:2013zca}). In Ref.~\cite{Bacchetta:2020gko}, the density $f_{q/N^\uparrow}$ of Eq.~\eqref{eq:Siversdensity} is reconstructed by combining the extraction of the unpolarized $f_1$ from Ref.~\cite{Bacchetta:2017gcc} with the extraction of the Sivers $f_{1T}^{\perp}$ from SIDIS data in the same approach. In this way, the quark density is reconstructed in a consistent way from real experimental data for nucleons with or without transverse polarization. In Fig.~\ref{fig:Siversdensity}, $f_{q/N^\uparrow}$ is represented at $x=0.1$ and $Q^2=1$ GeV$^2$ for a proton ideally moving towards the reader. The upper panels correspond to the symmetric situation of an unpolarized proton. The lower panels show the distortion along the $\hat{x}$ direction induced by the transverse polarization along the $\hat{y}$ direction, which turns out opposite for the up quark (left panel) and the down quark (right panel). The tomography depicted in Fig.~\ref{fig:Siversdensity} gives a realistic estimate of the non-trivial correlation between the motion of quarks and the spin direction of the parent proton.

As for the transversity TMD PDF $h_1 (x,k_T^2)$ in Eq.~\eqref{eq:phi_decomp}, we refer to Sec.~\ref{sec:02sec1h1}. Very recently, a global fit of all single-spin asymmetries involving polarized TMDs in SIDIS, Drell-Yan, $e^+$-$e^-$ annihilation, and hadron-hadron collision processes has been presented in Ref.~\cite{Cammarota:2020qcw}, extracting a universal set of TMD functions including the Sivers and the transversity distributions. However, the analysis was not performed in the TMD framework, thus neglecting the effects of TMD evolution which might be important when connecting experimental data collected at very different scales. 

In summary, it has now become possible to
 build the first accurate tomography in momentum space of quarks inside (polarized) nucleons, and we can describe how this picture evolves with the hard scale $Q$ and how it changes with the quark longitudinal momentum fraction $x$. However, we have no similar results for the gluon TMD PDFs. Several studies have explored useful channels at RHIC and the LHC with $p-p$ collisions leading to $J/\psi+X$, or $J/\psi+\gamma+X$, or $\eta_c+X$ final states~\cite{Dunnen:2014eta,Boer:2016fqd,DAlesio:2017rzj,DAlesio:2019gnu}. At the EIC, it would be possible to consider also SIDIS processes like $e-p$ collisions leading to $J/\psi+X$, $h_1+h_2+X$, jet+jet+$X$, $J/\psi$+jet+$X$ final states~\cite{Mukherjee:2016qxa,Boer:2016fqd,Rajesh:2018qks,Bacchetta:2018ivt}.

With more and higher quality data becoming available from Jefferson Lab at 12GeV, COMPASS at CERN, and the future EIC, and a better understanding of non-perturbative evolution at large $b_T $ from lattice QCD calculations, the community should be able to perform much more accurate global fits of TMD PDFs to learn about the ``true'' confined partonic motion in hadrons, isolating it from the contamination from parton showers triggered by high energy collisions.
In addition to the specific issue of the Collins-Soper evolution kernel highlighted above, lattice QCD can provide further auxiliary input for future global fits. TMD observables at separations up to $b_T \sim 1\, \mbox{fm} $ at hadron momenta up to about $2\, \mbox{GeV} $ will be accessible at the physical pion mass in the medium term, where it can be expected that moments in the momentum fraction $x$ will continue to be obtained with better precision than the $x$-dependent quantities, even as determinations of the latter are developed.
Used to anchor phenomenological fits, such moments may represent the most accessible avenue for lattice QCD to contribute to global fit efforts.

%% file: 05sec0outlook.tex
\goodbreak
\section{Outlook, Challenges \& Conclusions}
\label{05sec:outlook}
\subsection{Outlook}

The study of the distributions of partons into hadrons is an active interdisciplinary research 
field lying at the crossroads of high-energy, hadronic and nuclear physics.
In this report, we have provided a succinct review of recent
developments concerning PDFs, GPDs and TMD PDFs from both nonperturbative lattice QCD and
perturbative global analysis.
Throughout, we have highlighted the synergy between these
two distinct approaches and provided examples of where each can
reinforce the other.

From the perspective of QCD global analysis, fits of PDFs, GPDs and TMD PDFs 
are often limited by kinematical regions which have either
been weakly probed experimentally or are inaccessible to contemporary
measurements, such as (very) high or (very) low $x$, in the case
of PDFs. Similarly, precise extractions of the flavor dependence
of the PDFs and related quantities are frequently hindered by
the sparsity of data with direct sensitivity.  For instance,
in studies of unpolarized PDFs, determinations of the nucleon's
strangeness content commonly require final-state tagging of charm
quarks in dimuon production off nuclei or other semi-inclusive
processes for which various ambiguities exist regarding the treatment of
nuclear medium effects, hadronization, and other issues.
Due to these complexities, lattice-QCD data on PDF moments or
quasi-/pseudo-distributions can already have a valuable impact
as fitted data or theoretical priors in global PDF fits.  In contrast,
contemporary data informing the $u$- and $d$-quark distributions
are sufficiently constraining that lattice calculations on the
lowest PDFs moments do not yet possess the needed precision to
be quite as informative in a global analysis.  All the same,
technical improvements in the evaluation of PDF moments, the extension
of lattice calculations to higher moments, and supplementation
of this information with quasi- and pseudo-PDFs will gain pace
in coming years.  These developments suggest the possibility of
future global PDF fits in which extractions of unpolarized
PDFs are augmented by an array of lattice calculations that
provide enhanced precision.

Strategies and theory developments for this approach can be
guided by lattice calculations and QCD analyses for the comparatively
less constrained partonic distributions, such as the helicity and transversity
collinear distributions, GPDs, and TMD PDFs..
The impact of lattice calculations in QCD global fits can be reciprocally
enhanced by the phenomenological determinations themselves, which,
by their precision, can provide benchmarks for many standard lattice
observables.
We therefore envision a PDF-Lattice synergy deriving from the ability of
QCD global fitters to drive improvements in lattice calculations with benchmarks
informed by high-energy data, while the lattice provides informative
constraints in kinematical regions that are otherwise challenging to
constrain empirically.

As an essential ingredient for this relationship, both communities
must establish a common basis for comparing results from lattice
QCD and global fits --- a challenging undertaking given the complex
contemporary landscape of lattice calculations and global fits,
which involve a patchwork of theoretical settings, systematic
assumptions, and, in the case of QCD analyses, empirical data
sets.
This is a primary goal of this report, which, in addition
to numerous  updates, aimed to extend the comparative PDF-lattice
basis from the arena of collinear distributions developed in
the 2017 PDFLattice white paper to multidimensional distributions for which
lattice information can more immediately drive global fits.
The study of 1\nobreakdash-dimensional collinear unpolarized PDFs is a very mature field and benefits
from decades of intense experimental and theoretical accomplishments.
By contrast, study of the collinear helicity and transversity PDFs and of the
3\nobreakdash-dimensional hadronic structure represents
a comparatively new field which offers a wealth of opportunities for
new measurements, tools, and ideas.

The TMD PDFs and GPDs discussed in this report are examples of how we can
augment our current collinear PDFs to more fully map out the hadronic
structure. These generalized distributions provide a special
opportunity with new facilities on the horizon such as the
HL\nobreakdash-LHC, LHeC, and EIC.
For example, the EIC is well suited to perform hadron tomography measurements, given its
high-luminosity coverage of the low $Q^2$   quark-hadron
transition region in the kinematical parameter space.
These hadron tomography measurements can be crucial in unfolding the nucleon’s collinear and
transverse structure at scales adjacent to the nucleon mass, and for constraining the quantities accessible in next-generation lattice-QCD calculations.

\subsection{Challenges}

Looking ahead to the next decade, what will we need to accomplish?.

\begin{itemize}

    \item 
For the unpolarized proton PDFs, the goal is to continue reducing uncertainties so 
these are no longer a limiting factor for many precision measurements. 
These analyses are performed at NNLO as standard. 
It will take a concerted effort on both the experimental and theoretical fronts 
to move these analyses to the next desired precision. 

\pagebreak[4]
   \item 
For the helicity and transversity PDFs, knowledge is significantly more limited than for unpolarized PDFs, both
theoretically and experimentally. The standard accuracy for global analyses of these PDFs is currently NLO and LO, respectively. The quality and quantity of data limit their sensitivity to only a limited subset of partons in a fairly restricted kinematic region. This state of affairs will be overcome by the advent of an EIC.

    \item 
For the nuclear PDFs, the uncertainties are significantly larger than those for the proton PDFs. 
The nuclear PDF analyses have more degrees of freedom, but typically have less data points than the 
proton PDFs; hopefully, this situation will be remedied by the proposed experimental facilities. 
Additionally, new Lattice QCD efforts are beginning to explore light nuclei; this could be 
helpful in describing nPDFs as simple $A$-dependent interpolations in this region may be insufficient. %
As the nPDF precision increases, we can being to investigate collective effects and other phenomena
that are present for nuclei but may not be evident in the proton.

\item 
In lattice QCD, calculations of PDF moments have improved for both unpolarized, helicity and transversity moments. Additionally, new ideas on QCD factorizable and lattice-QCD calculable quantities, such as quasi- and pseudo-PDFs and correlations of currents, offer new avenues of inquiry and the possibility to extract PDFs over a broader range of $x$. Theoretical efforts in identifying new QCD factorizable quantities, especially those difficult to measure experimentally, would allow us to leverage the the power of lattice QCD to map out unknown structures.

\item
The study of GPDs offers insight into the 3D structure of the nucleon
via exclusive processes such as DVCS and meson production.
A significant challenge of these measurements is that they involve a
larger number of variables as compared to collinear PDFs.
As such, progress on this topic will require a multifaceted effort
combining new experimental data with a variety of theoretical methods
including Mellin moments, quasi- and pseudo-distributions, and Machine Learning
to provide a 3D tomographic image of the nucleon structure.

\pagebreak[4]
    \item 
The extension of collinear PDFs to TMD PDFs provides a new and fertile area for research,
and there are a number of topics which are particularly well suited to the Lattice QCD approach. 
Recent progress has enabled these calculations to scan the $(b\cdot P)$-dependence of the hadronic matrix elements to access the $x$-dependence of the TMD PDFs. Further extensions of the calculations 
to nonzero transverse momentum transfer offer the possibility to measure the quark orbital angular momentum. 
These advances, combined with recent work that provides a rigorous theoretical formalism for the TMD PDF framework, indicate a promising future for this line of research. 

\end{itemize}

\subsection{Conclusions}

This report is the outcome of a 2019 workshop which brought the
perturbative and non-perturbative PDF communities together to address
how a combined effort from these complementary perspectives can
advance our knowledge of the nucleon structure.
In addition to providing an update of the traditional collinear PDFs,
we have extended our analysis to include both the GPDs and TMD PDFs
which, when taken together, can provide a complete description of both
the transverse motion and position of the partons inside a nucleon.

The broader scope of our investigations highlights areas where
perturbative global PDF analyses and the Lattice QCD calculations can
benefit each other.
For example, the global analyses of the unpolarized
collinear PDFs are very mature and can thus provide constraints which
can be utilized by the Lattice QCD calculations.  Conversely, for some
of the polarized and generalized PDF measurements, 
the Lattice QCD approach can provide important inputs.

With the prospect of new facilities and experiments in the near
future, there is a urgent need for improved theoretical tools to keep
pace with the experimental data.
This work is ongoing, and the techniques and methods described here
will not only form the basis of next-generation PDFs, but also afford
a deeper understanding of the QCD theory which will facilitate
precision measurements and new discoveries for future experiments.